\documentclass[11pt]{article}
\usepackage[utf8]{inputenc}
\usepackage[a4paper, total={6in, 8in}]{geometry}
\usepackage{mathrsfs}
\usepackage{amssymb}
\usepackage{braket}
\usepackage{amsmath}
\usepackage{graphicx}
\usepackage{comment} 
\usepackage{longtable}
\usepackage{dsfont}
\usepackage{pdflscape}
\usepackage[vcentermath]{youngtab}
\usepackage{color} 

\usepackage[
backend=bibtex,
style=phys,
biblabel=brackets,
abbreviate=false,
doi=false,
url=false,
isbn=false,
eprint=true,
]{biblatex}
\AtEveryBibitem{%
\clearfield{note}%
}

\usepackage[T1]{fontenc}
\usepackage{lmodern}
\usepackage{morefloats}
\usepackage[svgnames]{xcolor}
\usepackage{tikz}
\usetikzlibrary{calc,decorations.markings, shapes.misc, decorations.pathmorphing}
\tikzset{cross/.style={cross out, draw, 
	minimum size=2*(#1-\pgflinewidth), 
	inner sep=0pt, outer sep=0pt}} 
\usepackage{tqft}
\definecolor{lstbgcolor}{rgb}{0.9,0.9,0.9} 
\usepackage{listings}
\usepackage{fancyvrb}

\newcommand{\normord}[1]{%
{{#1}}%
}

\newcommand{\bea}{\begin{eqnarray}}
\newcommand{\eea}{\end{eqnarray}}
\newcommand{\bg}{\begin{gathered}}
\newcommand{\eg}{\end{gathered}}

\newcommand\scalemath[2]{\scalebox{#1}{\mbox{\ensuremath{\displaystyle #2}}}}

\numberwithin{equation}{section}
\allowdisplaybreaks
\usepackage{subcaption}

\usepackage{amsthm}
\usepackage{rotating}
\usepackage{physics}
\usepackage{tensor}
\makeatletter
\newcommand{\bigtimes}{%
\DOTSB\mathop{\mathpalette\mattos@bigtimes\relax}\slimits@
}
\newcommand\mattos@bigtimes[2]{%
\vcenter{\hbox{%
		\sbox\z@{$#1\sum$}%
		\resizebox{!}{0.9\dimexpr\ht\z@+\dp\z@}{\raisebox{\depth}{$\m@th#1\times$}}%
}}%
\vphantom{\sum}%
}
\makeatother

\newcommand{\merge}{\vee}

\newcommand{\Hilbertspace}{\mathcal{H}_{\text{inv}}}

\newcommand{\DimSN}[1]{\Dim V^{S_N}_{#1}}
\newcommand{\DimSk}[1]{\Dim V^{S_k}_{#1}}
\newcommand{\DimPk}[1]{\Dim V^{P_k(N)}_{#1}}
\newcommand{\SWdual}[1]{\bar{#1}}
\newcommand{\Adj}[1]{\operatorname{Ad}(#1)}

\DeclareMathOperator{\con}{con}

\DeclareMathOperator{\Dim}{Dim}
\DeclareMathOperator{\Mult}{Mult}

\DeclareMathOperator{\End}{End}

\newcommand{\Span}{\mathrm{Span}}
\newcommand{\Sym}{\mathrm{Sym}}
\newcommand{\rotaterelation}[1]{\rotatebox[origin=c]{90}{$\mathstrut#1$}}
\definecolor{GREEN}{rgb}{0.0,0.70,0.24}
\definecolor{BLUE}{rgb}{0.0,0.24,0.70}
\definecolor{2}{rgb}{0.9, 0.17, 0.31}
\definecolor{5}{rgb}{0.05, 0.5, 0.06}
\definecolor{6}{rgb}{0.0,0.24,0.70}
\definecolor{3}{rgb}{1.0, 0.75, 0.0}
\definecolor{4}{rgb}{1.0, 0.97, 0.0}

\def\mC{ \mathbb{C} } 
\def\cL{ \mathcal{L} } 
\def\cYS{ \mathcal{Y_S}}

\newcommand{\PAdiagram}[3][]{\;\begin{tikzpicture}[baseline={([yshift=-.5ex]current bounding box.center)}]
	\def \n {#2};
	\def \edges {#3};
	\def \arcs {#1}
	\def \sep {0.5};
	\foreach \v in {1,...,\n}
	{
		\pgfmathparse{(\v-1)*\sep};
		\coordinate (v\v) at (\pgfmathresult,0.25);
		\node[circle,fill,inner sep=1pt] at (v\v) {};
	}
	\foreach \v in {1,...,\n}
	{
		\pgfmathparse{(\v-1)*\sep};
		\coordinate (v-\v) at (\pgfmathresult,-0.25);
		\node[circle,fill,inner sep=1pt] at (v-\v) {};
	}
	\foreach \endpointOne/\endpointTwo in \edges
	{
		\draw[] (v\endpointOne) -- (v\endpointTwo);
	}
	\foreach \endpointOne/\endpointTwo in \arcs
	{
		\draw[] (v\endpointOne) to[bend left] (v\endpointTwo);
	}
\end{tikzpicture}\;}

\newcommand{\SPAdiagram}[3][]{\qty[\PAdiagram[#1]{#2}{#3}]}

\newcommand{\PAdiagramLabeled}[3][]{\;\begin{tikzpicture}[baseline={([yshift=-.5ex]current bounding box.center)}]
	\def \n {#2};
	\def \edges {#3};
	\def \arcs {#1}
	\def \sep {0.5};
	\foreach \v in {1,...,\n}
	{
		\pgfmathparse{(\v-1)*\sep};
		\coordinate (v\v) at (\pgfmathresult,0.25);
		\node[circle,fill,inner sep=1pt, label=above:{\tiny $\v'$}] at (v\v) {};
	}
	\foreach \v in {1,...,\n}
	{
		\pgfmathparse{(\v-1)*\sep};
		\coordinate (v-\v) at (\pgfmathresult,-0.25);
		\node[circle,fill,inner sep=1pt, label=below:{\tiny $\v$}] at (v-\v) {};
	}
	\foreach \endpointOne/\endpointTwo in \edges
	{
		\draw[] (v\endpointOne) -- (v\endpointTwo);
	}
	\foreach \endpointOne/\endpointTwo in \arcs
	{
		\draw[] (v\endpointOne) to[bend left] (v\endpointTwo);
	}
\end{tikzpicture}\;}

\newcommand{\PAdiagramLabeledTwo}[3][]{\;\begin{tikzpicture}[baseline={([yshift=-.5ex]current bounding box.center)}]
	\def \n {#2};
	\def \edges {#3};
	\def \arcs {#1}
	\def \sep {0.5};
	\foreach \v in {1,...,\n}
	{
		\pgfmathparse{(\v-1)*\sep};
		\coordinate (v\v) at (\pgfmathresult,0.25);
		\node[circle,fill,inner sep=1pt, label=above:{\tiny $\v$}] at (v\v) {};
	}
	\foreach \v in {1,...,\n}
	{
		\pgfmathparse{(\v-1)*\sep};
		\coordinate (v-\v) at (\pgfmathresult,-0.25);
		\node[circle,fill,inner sep=1pt, label=below:{\tiny $\v$}] at (v-\v) {};
	}
	\foreach \endpointOne/\endpointTwo in \edges
	{
		\draw[] (v\endpointOne) -- (v\endpointTwo);
	}
	\foreach \endpointOne/\endpointTwo in \arcs
	{
		\draw[] (v\endpointOne) to[bend left] (v\endpointTwo);
	}
\end{tikzpicture}\;}

\newcommand{\PAdiagramOrbit}[3][]{\;\begin{tikzpicture}[baseline={([yshift=-.5ex]current bounding box.center)}]
	\def \n {#2};
	\def \edges {#3};
	\def \arcs {#1}
	\def \sep {0.5};
	\foreach \v in {1,...,\n}
	{
		\pgfmathparse{(\v-1)*\sep};
		\coordinate (v\v) at (\pgfmathresult,0.25);
		\node[circle,draw,inner sep=1pt] at (v\v) {};
	}
	\foreach \v in {1,...,\n}
	{
		\pgfmathparse{(\v-1)*\sep};
		\coordinate (v-\v) at (\pgfmathresult,-0.25);
		\node[circle,draw,inner sep=1pt] at (v-\v) {};
	}
	\foreach \endpointOne/\endpointTwo in \edges
	{
		\draw[] (v\endpointOne) -- (v\endpointTwo);
	}
	\foreach \endpointOne/\endpointTwo in \arcs
	{
		\draw[] (v\endpointOne) to[bend left] (v\endpointTwo);
	}
\end{tikzpicture}\;}

\newcommand{\PAdiagramcurve}[3][]{\;\begin{tikzpicture}[baseline={([yshift=-.5ex]current bounding box.center)}]
	\def \n {#2};
	\def \edges {#3};
	\def \arcs {#1}
	\def \sep {0.5};
	\foreach \v in {1,...,\n}
	{
		\pgfmathparse{(\v-1)*\sep};
		\coordinate (v\v) at (\pgfmathresult,0.25);
		\node[circle,fill,inner sep=1pt] at (v\v) {};
	}
	\foreach \v in {1,...,\n}
	{
		\pgfmathparse{(\v-1)*\sep};
		\coordinate (v-\v) at (\pgfmathresult,-0.25);
		\node[circle,fill,inner sep=1pt] at (v-\v) {};
	}
	\foreach \endpointOne/\endpointTwo in \edges
	{
		\draw[] (v\endpointOne) -- (v\endpointTwo);
	}
	\foreach \endpointOne/\endpointTwo in \arcs
	{
		\draw[] (v\endpointOne) to[bend left] (v\endpointTwo);
	}
\end{tikzpicture}\;}

\def\ls[#1]{ {}_{#1}}
\def\us[#1]{ \mbox{\tiny{#1}}}

\usepackage{hyperref}
\hypersetup{
colorlinks,
linkcolor={red!50!black},
citecolor={blue!50!black},
urlcolor={blue!80!black}
}
\bibliography{bibliography}

\title{\Large \bf Permutation symmetry in large N   \\ Matrix Quantum Mechanics and Partition Algebras}
\date{}
\author{{George Barnes$^{a,}$\footnote{g.barnes@qmul.ac.uk}, Adrian Padellaro$^{a,}$\footnote{a.k.s.padellaro@qmul.ac.uk}, Sanjaye Ramgoolam$^{a , b,}$\footnote{s.ramgoolam@qmul.ac.uk}  }}
\begin{document}
\begin{flushright}
	QMUL-PH-22-19
\end{flushright}
{\let\newpage\relax\maketitle}
\vspace{-2.5em}
\begin{center}
	$^{a}${\em  Centre for Theoretical Physics}, {\em School of Physical and Chemical Sciences}, \\
	{\em Queen Mary University of London}, \\
	{\em London E1 4NS, United Kingdom }\\
	
	\medskip
	$^{b}${\em  School of Physics and Mandelstam Institute for Theoretical Physics,} \\   
	{\em University of Witwatersrand}, \\ 
	{\em Wits, 2050, South Africa} \\
	\medskip	
\end{center}
\begin{abstract}
	We describe the implications of permutation symmetry for the state space and dynamics of quantum mechanical systems of matrices of general size $N$. We solve the general 11-parameter  permutation invariant quantum matrix harmonic oscillator Hamiltonian  and calculate the  canonical partition function.  The permutation invariant sector of the Hilbert space, for general Hamiltonians, can be described using partition algebra diagrams forming the bases of a tower of partition algebras $P_k(N)$. The integer $k$ is interpreted as the degree of matrix oscillator polynomials in the quantum mechanics. 
Families of interacting Hamiltonians are described which are diagonalised by a representation theoretic basis for the permutation invariant subspace which we construct for $ N \ge 2k $. These include Hamiltonians for which  the low-energy states are permutation invariant and can  give rise to large  ground state degeneracies related to the dimensions of partition algebras. A symmetry-based mechanism for quantum many body scars discussed in the literature can be realised in these  matrix systems with permutation symmetry.  A mapping of the matrix index values to lattice sites allows a realisation of the mechanism in the context of modified Bose-Hubbard models. Extremal correlators analogous to those studied in AdS/CFT are shown to obey selection rules based on Clebsch-Gordan multiplicites (Kronecker coefficients) of symmetric groups.
\end{abstract}
\newpage
\tableofcontents
\section{Introduction}

Systems with matrix degrees of freedom transforming in the adjoint or bifundamental representation of a group $G$, such as $U(N)$, $SU(N)$, $SO(N)$, $SP(N)$, are ubiquitous in physics. The group $G$ is often a gauge symmetry and physical states or operators of interest are $G$-invariant. The large $N$ limit has been known, since the work of 't Hooft \cite{tHooft}, to exhibit important simplifications related to the combinatorics of string worldsheets.  
Notable examples of gauge-string duality based on such large $N$ properties include: the duality between low-dimensional non-critical strings and matrix models \cite{Douglas:1989ve,Brezin1990,GrossMigdal},  between two-dimensional Yang-Mills theories and Hurwitz spaces \cite{1993Gross_1, Minahan1993,  GrTa, SCHNITZER1993, Gross1993_3, MP1993, Horava1996, CMR, Kimura:2008gs}; the AdS/CFT correspondence \cite{Malda, Witten1998, GKP}; the correspondence  between Gaussian matrix theories and Belyi maps \cite{ITZYK, RobSanj, Gopak2011, dMKLN}. Random matrix theories have also been used to model statistical properties of complex systems \cite{WignerRMT,Dyson,mehta2004RMT,Beenakker_1997,Guhr_1998,Edelman2013}. In zero-dimensional matrix models, invariance is not forced upon us by any gauge symmetry. However, it is still a fruitful perspective to consider the invariant sectors as computationally tractable sectors which encode significant properties of complex systems. This was the perspective taken in \cite{LMT, PIGMM, Ramgoolam2019, Ramgoolam2022}, which used zero-dimensional matrix models with permutation symmetry to model the statistics of words in computational linguistics \cite{coecke2010mathematical, maillard2014type, Baroni2014FregeIS, Grefenstette2015ConcreteMA}.

Large discrete groups, e.g.  the symmetric groups $S_N$ of all permutations of $N$ objects, also play a central role in holography. Two dimensional CFTs for orbifolds $M^{N}/S_N$, for some CFT $M$,  provide the CFTs in ${\rm AdS}_3/{\rm CFT}_2 $ dualities \cite{AGMOO}. 
These orbifold CFTs have recently provided the setting for a derivation of holographic duality \cite{Eberhardt:2019ywk}. It is natural to ask if matrix systems   with discrete symmetries such as $S_N$ have holographic duals. Recent results on large $N$ factorisation in permutation invariant matrix models  \cite{PIMO_Factor} are encouraging for this prospect. By regarding matrix models as zero dimensional quantum field theories, in this paper we take the natural next step of considering one-dimensional QFTs, i.e. matrix quantum mechanical systems  with permutation symmetry. We pay particular attention to methods which are applicable for general $N$ and allow large $N$ expansions. We give a general  description of the permutation invariant subspace in matrix quantum mechanical systems, drawing on relevant results from the mathematical literature on partition algebras. This is followed by a discussion of  interesting Hamiltonians  for many-body quantum physics. This is motivated by the vibrant interplay  between holography and many-body quantum mechanical systems  which manifests itself, for example, in the connection between free fermions and large $N$ two-dimensional Yang  Mills theory \cite{DouglasFerms};  free fermions and the half-BPS sector of  $\mathcal{N} =4$ SYM \cite{CJR,Berenstein0403}; free fermions and supersymmetric indices \cite{Sameer}, bosons in a 3D harmonic oscillator and eighth BPS states in $\mathcal{N}=4$ SYM \cite{BerEmergent,KMMR,CLBSR2020}; quantum mechanical spin matrix theory which is used 
as a simplified set-up to study the emergence mechanisms of AdS/CFT \cite{Harmark:2014mpa, 
Baiguera:2021hky}.  This interplay is also visible in the prominent role of 
coherent states, a technique widely used in many body quantum physics,  in the study of large $N$  systems. This theme appears in early work on large $N$ (e.g. \cite{JevSak,Yaffe}) as well as more recent developments (e.g. \cite{BerWang,HolWang,Lin}). 
% supergravity in AdS gives a new approach to phases in condensed matter systems in the AdS/CMT correspondence (see \cite{nlabAdSCMT,HLS2016} for overviews of the subject). 
%In this paper we develop the implications of permutation invariance in matrix quantum mechanics, drawing inspiration both from holography and from condensed matter physics. 

Many aspects of large $N$ simplifications in matrix systems  are consequences of Schur-Weyl duality.   The standard instance of Schur-Weyl duality \cite{FultonHarris} concerns the tensor product $ V^{ \otimes k } $ of the fundamental representation $V$ of $U(N)$. The symmetric group $S_k$ of all permutations of $k$ objects acts on $ V^{ \otimes k }$ by permuting the factors of the tensor product. Schur-Weyl duality states that  the algebra of operators commuting with the standard $U(N)$ action on the tensor product $V^{ \otimes k }$ is the group algebra $\mathbb{C}[S_k]$. 
This has important implications for the classification of $U(N)$ gauge invariant polynomial functions of matrix variables, where a matrix $X $ transforms as 
$ X \rightarrow U X U^{\dagger} $ for $ U \in U(N)$. Schur-Weyl duality relates this problem to the rich  combinatorics and representation theory of symmetric groups (see e.g. \cite{Hamermesh1962}). For example, the gauge invariant polynomial functions  of degree $k$ for one matrix of size $N$, taking $N > k $ for simplicity, are labelled by conjugacy classes of $S_k$. Finite $N$ effects are captured with the use of Young diagrams. Schur-Weyl duality has been used as a powerful tool in the construction of gauge invariant observables in one-matrix and multi-matrix systems in connection with the  AdS/CFT correspondence. This played an important role in identifying the CFT duals \cite{BBNS,CJR,Berenstein0403} of giant gravitons \cite{mst,HHI2000,GMT2000} in the AdS/CFT correspondence. The Schur-Weyl duality framework has been further applied to the computation of 1-matrix and multi-matrix correlators \cite{CJR,Kimura2007a, Brown2008, Bhattacharyya2008, Bhattacharyya2008b, Kimura2008, Brown2009,QuivCalc,CDD1301,Ber1504,KRS,CLBSR2018,ADHSSS,LY2107}.
A short review is  \cite{Ramgoolam2016}. These  multi-matrix applications involve dual algebras beyond the symmetric group algebras. For example Brauer algebras, which have a basis of diagrams, are used in  \cite{Kimura2007a}. The symmetric group algebra 
$\mathbb{C}[S_k]$ can also be viewed as a diagram algebra with multiplication given by the composition of diagrams. For example the following six diagrams give a basis of $\mathbb{C}[ S_3]$, the corresponding permutations are given in cycle notation 
\begin{align} \nonumber \label{eq: S_3 basis diagrams}
	(1)(2)(3) &= \PAdiagramLabeledTwo[]{3}{1/-1,2/-2,3/-3}, \hspace{10pt} (12)(3) = \PAdiagramLabeledTwo[]{3}{1/-2,2/-1,3/-3}, \hspace{10pt} (13)(2) = \PAdiagramLabeledTwo[]{3}{1/-3,2/-2,3/-1}, \\
	(1)(23) &= \PAdiagramLabeledTwo[]{3}{1/-1,2/-3,3/-2}, \hspace{10pt} (132) = \PAdiagramLabeledTwo[]{3}{1/-3,2/-1,3/-2}, \hspace{10pt} (123) = \PAdiagramLabeledTwo[]{3}{1/-2,2/-3,3/-1}.
\end{align}

The same general philosophy can be applied to the case where we are considering polynomial functions of a matrix $X$ invariant under the transformation
 $ X \rightarrow M_{ \sigma }  X M_{ \sigma}^T$, where $M_{\sigma } $ is a
matrix representing the permutation $\sigma \in S_N$ in the $N$-dimensional natural representation of $S_N$, satisfying $M_{ \sigma}^T = M_{ \sigma}^{-1}$. This problem in the invariant theory of matrices arises in the application of permutation invariant matrix models to language data \cite{LMT,PIGMM} and 
 Schur-Weyl duality was used to study these invariants in \cite{PIMO_Factor}. 
  The algebra dual to $S_N$ acting on $V_N^{\otimes k}$, where $V_N$ is the natural representation of the symmetric group $S_N$, is called the partition algebra $P_k(N)$. Partition algebras were first introduced in \cite{Martin1994, Jones1994, Martin1996} in application to the statistical mechanics of Potts models (see \cite{Halverson2004} for a  survey of partition algebras). Partition algebras $P_k(N)$ are diagram algebras with a basis labeled by diagrams corresponding to set partitions of $2k$ objects. These include the diagrams corresponding to elements of $\mathbb{C}[S_k]$ as well as  more general diagrams. For example, in addition to the diagrams in \eqref{eq: S_3 basis diagrams}, the following are elements of $P_3(N)$
\begin{align}
\PAdiagram[]{3}{}, \hspace{4pt} \PAdiagram[]{3}{1/-1,2/-2}, \hspace{4pt} \PAdiagram[2/1]{3}{2/-2}, \hspace{4pt} \PAdiagram[]{3}{1/-2, 3/-2, 2/-3, 2/-1}, \hspace{4pt} \PAdiagram[2/1, 3/2, -1/-2, -2/-3]{3}{1/-1}.
\end{align}
We will discuss these diagrams in more detail in section \ref{sec: Perm subspace}. Matrix systems with $S_N$ symmetry together with partition algebras allow us to study large $N$ simplifications in the case of discrete (finite) groups. Partition algebras and their relation to the representation theory of symmetric groups is an active area of mathematical research \cite{Halverson2001,Enyang2012,HalversonBenk2017,Halverson2018}.

An algebraic description of permutation invariant matrix polynomials of degree $k$ was given in \cite{PIMO_Factor} using symmetrised partition algebras $SP_k(N)$. $SP_k(N)$ consists of equivalence classes of elements in $P_k(N)$. The equivalence is defined using the $\mathbb{C}[S_k]$ subalgebra of $P_k(N)$ and accounts for the commuting nature of matrix variables. The work in \cite{PIMO_Factor} showed that distinct permutation invariant matrix polynomials in the diagram basis satisfy a factorization property at large $N$. The diagram basis is the analogue of the trace basis for $U(N)$ invariant matrix polynomials. Partition algebras have also been used to study permutation invariant random matrix distributions from the point of view of mathematical statistics \cite{gabrielcombinatorial1, gabrielcombinatorial2, gabrielcombinatorial3}.

Polynomials in matrix variables $M^i_j$  are closely related to quantum mechanical states constructed from matrix oscillators $(a^{\dagger} )^i_j$. This allows us to translate the technology developed for zero-dimensional matrix models \cite{LMT,PIGMM,PIG2MM,PIMO_Factor}  to the setting of matrix quantum mechanics. We will give a detailed description of the space of $S_N$ invariant states constructed from matrix oscillators. Polynomials in matrix oscillators can be organised by the degree of the polynomials. At degree $k$, the state space is isomorphic to an $S_k$ symmetric subspace $\mathcal{H}^{ (k) }$ of ${\rm End } ( V_{ N }^{ \otimes k  } ) $:
\bea 
\mathcal{H}^{ (k) } \rightarrow {\rm End } ( V_{ N }^{ \otimes k  } ). 
\eea
There is a one-to-one correspondence between tensors 
\bea 
\langle e^{ i_1} \cdots e^ {i_k }  | T | e_{ j_1} \cdots e_{ j_k } \rangle
= T^{ i_1 \cdots i_k }_{ j_1 \cdots j_k }  
\eea
and elements in ${\rm End } ( V_{ N }^{ \otimes k  } )$. The bosonic symmetry of the oscillators 
imposes an invariance under simultaneous re-ordering of the upper and lower indices. Commuting with the $S_k$ action is the $S_N$ action on $V_{N}^{ \otimes k } $ which we denote $\cL (\sigma )$. The $S_N$ permutation invariance translates to an invariance of $T$ under an adjoint action  
\bea 
\Adj{\sigma} [ T ] = \cL ( \sigma )  T  \cL ( \sigma^{-1} ) 
\eea 
Many of our results on the $S_N$ invariant state space of matrix oscillators, particularly in sections \ref{sec: Perm subspace} and \ref{sec: charges}
are independent of the Hamiltonian. They can be viewed as a detailed account of the $S_N$ invariant subspace in matrix quantum mechanics using partition algebras and representation theory. The use of the partition algebra $P_k(N)$  to study operators and quantum states in $\mathcal{H}^{(k)}$ allows us to take advantage of simplifications in the limit where $k$ is kept fixed as $N \rightarrow \infty$.

The representation theoretic approach allows the construction of solvable algebraic Hamiltonians where the $S_N$ invariant states are resolved according to representation theoretic characteristics. Sections 
\ref{sec: PIMQM} and \ref{sec: deformations} discuss different classes of solvable $S_N$ invariant Hamiltonians obeying 
\bea\label{invtHamilt} 
\Adj{\sigma}H = H \hspace{-2pt} \Adj{\sigma}
\eea
We build on this discussion in section \ref{sec: scars}, using Hamitonians of the form 
$ ( H + H_s ) $:  $H$ obeys \eqref{invtHamilt} while $H_s $ is subject to a restriction defined in terms of permutation invariant states. 

The paper is organised as follows. For concreteness, section \ref{sec: MQM} contains a review of the simplest quantum mechanical model with matrix degrees of freedom. This is the free matrix quantum harmonic oscillator. It is a model containing $N^2$ decoupled harmonic oscillators $X_{ij}, i,j=1,\dots,N$ with a global $U(N^2)$ symmetry. The Hilbert space of this model is a Fock space $\mathcal{H}$ of states constructed using matrix oscillators $(a^{\dagger})^i_j$. This model also serves as a good place to introduce the diagram notation that we will use in the rest of the paper.

In section \ref{sec: Perm subspace} we consider the $S_N$ invariant subspace $\Hilbertspace$ of the total Hilbert space $\mathcal{H}$ of a general quantum mechanics matrix system. This is the subspace of states invariant under $a^{ \dagger}  \rightarrow M_{ \sigma } a^{ \dagger} M_{ \sigma }^{ T } $, where $M_{ \sigma} $ is a permutation matrix of size $N$. We explain the correspondence between permutation invariant matrix states of degree $k$ and partition algebras $P_k(N)$. The partition algebras have three natural bases, and each one gives rise to a different basis for $\Hilbertspace$. The diagram basis is natural when discussing inner and outer products. The factorization property in \cite{PIMO_Factor} translates to orthogonality of the diagram basis at large $N$. The so-called orbit basis gives rise to an orthogonal basis for all $N$. We call the third basis the representation basis. In the mathematical literature, the representation basis is called a complete set of matrix units. The product in the matrix unit basis is a generalization of the product for elementary matrices for matrix algebras. The representation basis can be constructed using Fourier transformation on $P_k(N)$ and is a direct analogue of the Schur basis for $U(N)$ invariants. Appendix \ref{apx: Semi-Simple Algebra Technology} gives the necessary background for Fourier transforms on semi-simple algebras, closely following \cite{AR90DissertCh1} but with some modifications that are important for our application. Physically, the representation basis can be understood as a basis that diagonalizes a set of algebraic commuting charges.

Section \ref{sec: charges} is devoted to the construction and diagonalization of these charges, which can be used to give the explicit transformation from the diagram basis to the representation basis at large $N$. We illustrate the method for small $k$ and large $N$. These are tabulated in appendix \ref{apx: matrix units}. The representation basis forms an energy eigenbasis for the Hamiltonian of the free matrix quantum harmonic oscillator presented in section \ref{sec: MQM}.

In section \ref{sec: PIMQM} we introduce an $11$ parameter family of exactly solvable quantum matrix systems. The potential in these systems is the most general permutation invariant quadratic function of the matrix variables. These quantum systems can therefore  be viewed as  general matrix harmonic oscillator systems compatible with permutation symmetry.  We find the spectrum for general choices of the parameters by adapting the representation theoretic techniques which have been used to compute correlators in permutation invariant Gaussian matrix models \cite{PIGMM}.  Further, we write the canonical partition function in a simple closed form.  The representation basis states from section \ref{sec: Perm subspace} do not form an eigenbasis for the general Hamiltonians considered here. The action of the Hamiltonians on the representation basis states leads to a mixing which is constrained by Clebsch-Gordan multiplicities for the symmetric groups. We briefly discuss this mixing. 
%We discuss the restricted form of mixing that occurs when diagonalising the general permutation invariant Hamiltonian on $\Hilbertspace$.

In section \ref{sec: deformations} we discuss interacting Hamiltonians, parametrised by a positive integer $K$, constructed using partition algebra elements,  with the property that the  ground states are all permutation invariant states and have degeneracies controlled by a sequence of partition algebras $ P_{ k } ( N ) $ for $ k \in \{ 0 , 1, \dots , K \}$. The energy gap between the ground states and the lowest excited state is also determined by $K$. By deforming these Hamiltonians with other partition algebra elements, we design Hamiltonians where the  degeneracy of the  ground states is broken by small amounts - these two scenarios are illustrated in figure \ref{fig: spectrum scenarios}. We also include a general  description of permutation invariant Hamiltonians, finding an interesting relation to the counting of 2-matrix permutation invariants of the kind considered in \cite{PIG2MM}. 
We conclude this section with an interpretation of the oscillators $(a^{\dagger})^j_i$ as creation operators on a  square lattice with sites labelled $( i , j )$.

% whose spectrum is exactly solvable in the invariant subspace $\Hilbertspace$. The Hamiltonians are diagonal in the representation basis. We describe how to 
% Using motivations from holography, we explain how to construct Hamiltonians for which states in  the invariant subspace  have finite energy while non-invariant states are decoupled, i.e. have infinite energy. 

%  Section \ref{sec: deformations} constructs solvable Hamiltonians which decouple the invariant states from non-invariant states and at the same time allow  non-trivial resolution of states within the invariant sector. In degenerate limits, all the invariant states are ground states.
  
  % We describe a lattice interpretation 
%of the quantum mechanical systems where $(a^{ \dagger})^i_j$ creates a bosonic particle in on a square lattice at sites labelled $(i,j)$. The degenerate ground states have the characteristics of long-range quantum entanglement.  

Subspaces of invariant states play an important role in the group-theoretic proposal  \cite{PPPK2020, PPPK2021} for a mechanism of weak ergodicity breaking, experimentally discovered in \cite{RydbergSim}, now known as quantum many-body scars \cite{RydbergScars}. In section \ref{sec: scars} we discuss how the permutation invariant state space in this paper can be turned into a scar subspace. Adapting the ideas in \cite{PPPK2020, PPPK2021} 
for the realisation of group-theoretic scar states, we describe Hamiltonians which exhibit the revival properties characteristic of scars. The lattice interpretation of the matrix oscillators from section \ref{sec: deformations} allows us to interpret these Hamiltonians as deformations of Bose Hubbard models. 

We compute a set of two- and  three-point correlators of invariant operators in section \ref{sec: extremal correlators}.  The two-point correlators have a large $N$ factorisation property described in the context of matrix models in \cite{PIMO_Factor}.  The three-point functions are similar to extremal correlators which are relevant to quantum mechanical models considered in AdS/CFT. The extremal correlators are shown to obey selection rules based on Clebsch-Gordan multiplicities (Kronecker coefficients) of symmetric groups.

\section{Review: matrix harmonic oscillator} \label{sec: MQM}

This section is a review of the simplest matrix quantum harmonic oscillator. The Lagrangian \eqref{eq: free lagrangian} describes $N^2$ free harmonic oscillators. The corresponding Hamiltonian has a global $U(N^2)$ symmetry. This has a $U(N) \times U(N)$ subgroup 
of unitary matrices acting by left and right multiplication.  There is also a smaller $S_N \times S_N$ subgroup of the $U(N) \times U(N)$ which plays an important role in subsequent sections.
The simplest, non-interacting $U(N^2)$ invariant model will serve as a very good set-up to introduce the notation used in the rest of the paper.
In particular, we describe how to construct states and operators in $\mathcal{H}$, the Hilbert space of the theory, by considering the oscillators $a_{ij}, a_{ij}^{\dagger}$ as endomorphisms on $V_N$ (an $N$ dimensional vector space). We will frequently have this view in mind when manipulating states and operators and it is often practical to employ diagrammatic notation in order to do so. The basics of this diagrammatic notation we introduce at the end of this section.

The simplest matrix harmonic oscillator is described by the Lagrangian
\begin{equation} \label{eq: free lagrangian}
	L_0 = \frac{1}{2}\Bigg( \sum_{i,j=1}^N \partial_t X_{ij} \partial_t X_{ij} - X_{ij} X_{ij} \Bigg).
\end{equation}
It describes $N^2$ decoupled oscillators. The conjugate momenta are 
\begin{equation}
	\Pi_{ij} = \pdv{L_0}{(\partial_t X_{ij})}= \pdv{t} X_{ij}. 
\end{equation}
The Hamiltonian corresponding to $L_0$ is
\begin{equation} \label{eq: free H}
	H_0 = \frac{1}{2} \Bigg( \sum_{i,j=1}^N \Pi_{ij}\Pi_{ij} + X_{ij}X_{ij} \Bigg).
\end{equation}
The canonical commutation relations are
\begin{equation}
	\comm{X_{ij}}{\Pi_{kl}} = i\delta_{ik}\delta_{jl}.
\end{equation}

The Hamiltonian given in \eqref{eq: free H} is diagonalized in the usual way - introducing oscillators $a^{\dagger}_{ij}, a_{ij}$ defined by
\begin{equation}
	\begin{aligned}
		X_{ij} &= \sqrt{\frac{1}{2}}\qty(a^\dagger_{ij} + a_{ij}), \\
		\Pi_{ij} &= i\sqrt{\frac{1}{2}}\qty(a^\dagger_{ij} - a_{ij}),
	\end{aligned} \label{eq: free oscillators}
\end{equation}
with commutation relations
\begin{equation}
	\comm{a_{ij}}{a^\dagger_{kl}} = \delta_{ik}\delta_{jl}. \label{eq: simplest oscillators}
\end{equation}
Normal ordering $H_0$ gives
\begin{equation}
	H_0 = \sum_{i,j=1}^N a_{ij}^{\dagger} a_{ij}, \label{eq: simplest hamiltonian}
\end{equation}
which is just a number operator. We now show that $H_0$ is invariant under a $U(N^2)$ symmetry that acts on oscillators as
\begin{align}
	a_{ij} &\rightarrow \sum_{k,l = 1}^N U_{ij; kl} a_{kl}, \\
	a^{\dagger}_{ij} &\rightarrow \sum_{k,l = 1}^N U^{\dagger}_{kl ; ij} a^{\dagger}_{kl},
\end{align}
with $U_{ij; kl}$ an $N^2 \times N^2$ unitary matrix satisfying
\begin{align}
	\sum_{k,l = 1}^N U_{ij; kl} U^{\dagger}_{kl; mn} = \delta_{im} \delta_{jn}.
\end{align}
Under the $U(N^2)$ transformation $H_0$ is invariant,
\begin{align} \nonumber
	H_0 \rightarrow &\sum_{i,j,k,l,m,n} U^{\dagger}_{kl ; ij} U_{ij ; mn} a^{\dagger}_{kl} a_{mn} \\ \nonumber
	= &\sum_{k,l,m,n} \delta_{km} \delta_{ln} a^{\dagger}_{kl} a_{mn} \\
	= &\sum_{k,l} a^{\dagger}_{kl} a_{kl}.
\end{align}

The oscillator states
\begin{equation}
	\prod_{i,j}\frac{(a^\dagger_{ij})^{k_{ij}}}{\sqrt{{k_{ij}!}}} \ket{0} \label{eq: H0 eigenbasis}
\end{equation}
labelled by non-negative integers $k_{ij}$ with $i,j=1,\dots,N$ are energy eigenstates of $H_0$.
The total Hilbert (Fock) space $\mathcal{H}$ decomposes into subspaces $\mathcal{H}^{(k)}$ with fixed number of oscillators (degree) $k$,
\begin{equation}
	\mathcal{H} \cong \bigoplus_{k=0}^{\infty} \mathcal{H}^{(k)}.
\end{equation}
The subset of states with $k=\sum_{i,j} k_{ij}$ form an eigenbasis for the subspace $\mathcal{H}^{(k)}$ and have energy $k$.
In general the spectrum is highly degenerate. The number of states with energy $k$ is
\begin{equation}
	\Dim \mathcal{H}^{(k)} = \binom{N^2+k-1}{k} = \frac{N^2(N^2 +1 ) \dots (N^2+k-1)}{k!}.
\end{equation}
This is the number of ways to choose $k$ elements from a set of $N^2$ when repetition is allowed.
It is also the dimension of the symmetric part of a $k$-fold tensor product of a vector space with dimension $N^2$.
Equivalently, it is the dimension of the vector space of states composed of $k$ bosonic oscillators $a^\dagger_{ij}$. For fixed $k$ and $N \gg  2k$ the dimension grows as $N^{2k}$.

\subsection{Diagram notation}
Throughout this paper we will use diagrammatic notation to describe states and operators in $\mathcal{H}^{(k)}$.
For this purpose, it is useful to introduce the following matrices of oscillators $(a^{\dagger})^i_j = a^{\dagger}_{ji}$ and $a^i_j = a_{ij}$ which satisfy
\begin{equation}
	\comm{a^i_j}{(a^{\dagger})^l_k} = \delta^i_k \delta_j^l. \label{eq: matrix basis comm relations}
\end{equation}
Let $V_N$ be an $N$-dimensional vector space with basis $\{e_1, \dots, e_N\}$. The matrices of oscillators can be viewed as (operator-valued) elements in $\End(V_N)$, where $\End(V_N)$ is the set of all linear maps $V_N \rightarrow V_N$. In this language, the above oscillators are matrix elements,
\begin{equation}
	a^{\dagger}(e_i) = \sum_{j=1}^N (a^{\dagger})^j_i e_j \qq{and} a(e_i) = \sum_{j=1}^N a^j_i e_j.
\end{equation}
Consequently, a general degree one state in $\mathcal{H}$ can be written as
\begin{equation}
	\Tr_{V_N}(T a^{\dagger})\ket{0} = \sum_{i,j=1}^N T^i_j (a^{\dagger})^j_i  \ket{0} \equiv \ket{T},
\end{equation}
where $T \in \End(V_N)$ (an $N$-by-$N$ matrix) and the last equality is a definition of $\ket{T}$.

The degree $k$ subspace is given by
\begin{equation} \label{eq: H^k span}
	\mathcal{H}^{(k)}  \cong \Span_{\mathbb{C}}\,\qty{ (a^{\dagger})^{i_1}_{j_1} \dots (a^{\dagger})^{i_k}_{j_k}\ket{0} },
\end{equation}
therefore general states are parametrised by tensors $T^{j_1 \dots j_k}_{i_1 \dots i_k}$.
It is convenient to view these tensors as elements of $\End(V_N^{\otimes k})$, where $V_N^{\otimes k}$ is the $k$th tensor product of $V_N$.
That is, in the usual basis for tensor product spaces
\begin{equation}
	T(e_{i_1} \otimes e_{i_2} \otimes {\dots} \otimes e_{i_k}) = \sum_{j_1,j_2,\dots,j_k =1}^N T^{j_1 \dots j_k}_{i_1 \dots i_k} e_{j_1} \otimes e_{j_2} \otimes {\dots} \otimes e_{j_k}.
\end{equation}
Analogous to the degree one case, a general state $\ket{T} \in \mathcal{H}^{(k)}$ can be written as a trace
\begin{equation} \label{eq: state definition}
	\boxed{\ket{T} =  \Tr_{V_N^{\otimes k}}(T(a^\dagger)^{\otimes k}) \ket{0} = \sum_{\substack{i_1 , \dots , i_k \\ j_1 , \dots , j_k}}T^{j_1 \dots j_k}_{i_1 \dots i_k}  (a^{\dagger})^{i_1}_{j_1} \dots (a^{\dagger})^{i_k}_{j_k}\ket{0},}
\end{equation}
for $T \in \End(V_N^{\otimes k})$ and $(a^\dagger)^{\otimes k} = a^{\dagger} \otimes \dots \otimes a^{\dagger}$ with matrix elements
\begin{equation}
	(a^\dagger)^{\otimes k}(e_{j_1} \otimes \dots \otimes e_{j_k}) = \sum_{i_1, \dots, i_k} (a^{\dagger})^{i_1}_{j_1} \dots (a^{\dagger})^{i_k}_{j_k}e_{i_1} \otimes \dots \otimes e_{i_k}.
\end{equation}
It should be emphasized that, due to the bosonic symmetry of the oscillators, $T^{j_1 \dots j_k}_{i_1 \dots i_k}$ is a symmetric tensor (under simultaneous permutations of upper and lower indices), for example  $T^{j_1 j_2 \dots j_k}_{i_1 i_2 \dots i_k} = T^{j_2 j_1 \dots j_k}_{i_2 i_1 \dots i_k}$.

It is useful to formulate this restriction in terms of $S_k$ invariance. An element $\tau \in S_k$, viewed as a bijective map $\tau: \{1,\dots,k\} \rightarrow \{1,\dots,k\}$ defines a linear operator $\mathcal{L}_{\tau^{-1}}$ which acts on $V_N^{\otimes k}$ as
\begin{equation}
	\mathcal{L}_{\tau^{-1}}(e_{i_1}  \otimes  \dots \otimes e_{i_k}) =  e_{i_{\tau(1)}} \otimes  \dots \otimes e_{i_{\tau(k)}}.
\end{equation}
The symmetry of $T$ is equivalent to the statement
\begin{equation}
\boxed{ 	\mathcal{L}_{\tau}T \mathcal{L}_{\tau^{-1}} = T, \quad \forall \tau \in S_k, }  \label{eq: Sk invariance of state tensors}
\end{equation}
or in index notation
\begin{equation}
	T^{j_{\tau(1)} \dots j_{\tau(k)}}_{i_{\tau(1)} \dots i_{\tau(k)}} = T^{j_1 \dots j_k}_{i_1 \dots i_k}, \quad \forall \tau \in S_k.
\end{equation}
Therefore, states in $\mathcal{H}^{(k)}$ are in one-to-one correspondence with elements $T \in \End_{S_k}(V_N^{\otimes k})$, the subspace of linear maps that commute with the action of $S_k$.

We introduce diagrammatic notation to simplify manipulations involving tensor equations. A map $T \in \End(V_N^{\otimes k})$ is represented by a box
\begin{equation}	
	\tikzset{every picture/.style={line width=0.75pt}} %set default line width to 0.75pt        
	T^{j_1 \dots j_k}_{i_1 \dots i_k}=\vcenter{\hbox{
	\begin{tikzpicture}[x=0.75pt,y=0.75pt,yscale=-.8,xscale=.8]
		%uncomment if require: \path (0,437); %set diagram left start at 0, and has height of 437
		
		%Shape: Rectangle [id:dp2658424183937629] 
		\draw   (80,340) -- (140,340) -- (140,380) -- (80,380) -- cycle ;
		%Straight Lines [id:da5689928301130718] 
		\draw    (110,320) -- (110,340) ;
		%Straight Lines [id:da5494449553365999] 
		\draw    (110,380) -- (110,400) ;
		
		% Text Node
		\draw (100,351) node [anchor=north west][inner sep=0.75pt]   [align=left] {$\displaystyle T$};
		% Text Node
		\draw (84,297) node [anchor=north west][inner sep=0.75pt]    {$j_{1} \dotsc j_{k}$};
		% Text Node
		\draw (83,397) node [anchor=north west][inner sep=0.75pt]    {$i_{1} \dotsc i_{k}$};
	\end{tikzpicture}}}
\end{equation}
where the edges correspond to states in $V_N^{\otimes k}$.
Internal lines in a diagram correspond to contracted indices.
For example, the state $\ket{T} \in \mathcal{H}^{(k)}$ can be represented diagrammatically as
\begin{equation} \label{eq: state from endo diagram}
	\ket{T} = 
	\tikzset{every picture/.style={line width=0.75pt}} %set default line width to 0.75pt        
	\vcenter{ \hbox{
			\begin{tikzpicture}[x=0.75pt,y=0.75pt,yscale=-.8,xscale=.8]
				%Straight Lines [id:da12866535061418038] 
				\draw    (200,170) -- (200,200) ;
				%Shape: Rectangle [id:dp25724550757973197] 
				\draw   (170,130) -- (230,130) -- (230,170) -- (170,170) -- cycle ;
				%Straight Lines [id:da836231452695215] 
				\draw    (190,200) -- (210,200) ;
				%Straight Lines [id:da5478045134437806] 
				\draw    (200,100) -- (200,130) ;
				%Shape: Rectangle [id:dp6653027352210579] 
				\draw   (170,61) -- (230,61) -- (230,100) -- (170,100) -- cycle ;
				%Straight Lines [id:da897728786421] 
				\draw    (200,30) -- (200,60) ;
				%Straight Lines [id:da6608062404181438] 
				\draw    (190,30) -- (210,30) ;
				% Text Node
				\draw (190,144) node [anchor=north west][inner sep=0.75pt]   [align=left] {$ T$};
				% Text Node
				\draw (175,69) node [anchor=north west][inner sep=0.75pt]   [align=left] {$ (a^{\dagger })^{\otimes k}$};
	\end{tikzpicture}}}
	\ket{0}.
\end{equation} 
The horizontal lines identify the top edge with the bottom edge to give a trace, and the line between the $(a^{\dagger })^{\otimes k}$ and $T$ boxes signifies that the corresponding indices are identified and summed over. This diagram should be compared to \eqref{eq: state definition}.

\section{Permutation invariant sectors for quantum matrix systems} \label{sec: Perm subspace}

In this section we consider the action of $S_N$ on the subspace  $\mathcal{H}^{(k)}$,  
spanned by degree $k$ polynomials in matrix oscillators $ (a^{\dagger})^i_j$ acting on the vacuum.  The adjoint action of permutations $\sigma \in S_N$  on the quantum mechanical matrix variables
\bea 
\sigma : X^{ i }_j \rightarrow ( M_{ \sigma } X M_{ \sigma^{-1} } )^i_j = X^{ \sigma ( i ) }_{ \sigma (j) } 
\eea 
translates into action on oscillators 
\bea\label{adjoscact} 
 \sigma :  (a^{ \dagger})^{i}_j \rightarrow 
(a^{\dagger} )^{ \sigma (i) }_{ \sigma (j) } \, . 
\eea 
We  turn our attention to the subspace $\Hilbertspace^{(k)} \subset \mathcal{H}^{(k)}$ of $S_N$ invariant states constructed from polynomials in these oscillators.  We will construct bases for $\Hilbertspace^{(k)}$, for general $k$, taking inspiration from \cite{PIMO_Factor}. There, a basis for the space of $S_N$ invariant polynomials in matrix indeterminates $X^i_j$ of degree $k$ was given in terms of elements of the diagrammatic partition algebra $P_{k}(N)$ \cite{Halverson2004}. With the identification
\begin{equation}
	X^i_j \leftrightarrow (a^\dagger)^i _j,
\end{equation}
we can employ these techniques to construct $S_N$ invariant states in matrix quantum mechanics.

The algebra $\End_{S_N}(V_N^{\otimes k})$,  of linear operators on $V_N^{\otimes k}$ that commute with $S_N$,  is of central importance in understanding the implications of permutation invariance in quantum mechanical matrix systems. For $ N \ge 2k $ this algebra is isomorphic to the partition algebra $P_k(N)$  \cite{Halverson2004}: 
\bea 
\End_{S_N}(V_N^{\otimes k}) \cong P_k (N ) \, . 
\eea
The Hilbert space 	$\Hilbertspace^{(k)} $ spanned by degree $k$ polynomials in the oscillators is isomorphic to an $S_k$ invariant subalgebra of $P_k(N)$ : 
\begin{equation}
	\Hilbertspace^{(k)} \cong \End_{S_N \times S_k}(V_N^{\otimes k}) \subseteq \End_{S_N}(V_N^{\otimes k}),
\end{equation}
The partition algebras are finite-dimensional associative algebras with dimension $B(2k)$, the Bell numbers. The Bell numbers $B(k)$ count the number of possible set partitions of a set of $k$ elements. Notably, $B(2k)$ does not depend on $N$. Consequently, $\Dim \Hilbertspace^{(k)}$ does not grow with $N $ for $N \ge 2k$. This is in contrast to $\Dim \mathcal{H}^{(k)}$, which grows like $N^{2k}$ for $N \gg 2k $. 

We have chosen to construct states using the oscillators $(a^{\dagger})^i_{j}$. This produces a basis for $\Hilbertspace$ that is simultaneously an energy eigenbasis of $H_0$. However, it is worth emphasising that the resulting description of $\Hilbertspace$ is applicable to any quantum matrix system, not only the system with Hamiltonian $H_0$. For example, the description of $\Hilbertspace$ in terms of partition algebras holds equally well if the Hamiltonian is a perturbation of $H_0$ by a polynomial in the matrix creation and annihilation operators. 

We begin this section in \ref{sec: PA} by reviewing the connection between partition algebras and states in $\Hilbertspace$. The basic algebraic structure of partition algebras is reviewed in section \ref{sec: diagram basis}. The partition algebras are introduced in the most geometrical basis, the diagram basis, where multiplication is given by diagram concatenation. In section \ref{sec: ON basis} we introduce the representation basis, so called because it is labelled by a set of representation theoretic data. This basis uses Fourier transforms \cite{AR90DissertCh1} on $P_k(N)$ to construct an all-orders orthogonal basis for $N \geq 2k$, which diagonalises a set of algebraic charges. These charges are discussed in detail in section \ref{sec: charges}, and used in section \ref{sec: deformations} to construct algebraic Hamiltonians with interesting spectra.

\subsection{Partition algebras and invariant tensors}\label{sec: PA}
For any $\sigma \in S_N$ we have a linear operator $\mathcal{L}(\sigma) \in \End(V_N^{\otimes k})$ defined by
\begin{equation}
	\mathcal{L}(\sigma^{-1})(e_{i_1} \otimes e_{i_2} \otimes {\dots} \otimes e_{i_k}) = e_{\sigma(i_1)} \otimes e_{\sigma(i_2)} \otimes {\dots} \otimes e_{\sigma(i_k)}. \label{eq: left action SN}
\end{equation}
Here $\sigma \in S_N$ is a bijective map $\{1, \dots, N\} \rightarrow \{1,\dots,N\}$.
Group multiplication is given by composition of maps $\sigma_1 \sigma_2(i) = \sigma_2(\sigma_1(i))$ for $\sigma_1 , \sigma_2 \in S_N$.
This is used to define the adjoint action $\Adj{\sigma}$ of $\sigma \in S_N$ on states $\ket{T} \in \mathcal{H}^{(k)}$,
\begin{equation}
\begin{aligned}
		\Adj{\sigma} \ket{T} &=\Tr_{V_N^{\otimes k}}[\mathcal{L}(\sigma)T \mathcal{L}(\sigma^{-1}) (a^\dagger)^{\otimes k} ] \ket{0} \\
		&= \sum_{\substack{i_1 , \dots , i_k \\ j_1 , \dots , j_k}}T^{j_1 \dots j_k}_{i_1 \dots i_k}  (a^{\dagger})^{\sigma^{-1}(i_1)}_{\sigma^{-1}(j_1)} \dots (a^{\dagger})^{\sigma^{-1}(i_k)}_{\sigma^{-1}(j_k)}\ket{0} \\
		&=\sum_{\substack{i_1, \dots , i_k \\ j_1 , \dots , j_k}}T^{\sigma(j_1) \dots \sigma(j_k)}_{\sigma(i_1) \dots \sigma(i_k)}  (a^{\dagger})^{i_1}_{j_1} \dots (a^{\dagger})^{i_k}_{j_k}\ket{0}. \label{eq: diagonally permutation action on hilbert}
\end{aligned}
\end{equation}
This adjoint action on the tensor coefficients of the oscillators corresponds to the 
adjoint action on the oscillators which follows from  \eqref{adjoscact}. 
States $\ket{T} \in \Hilbertspace^{(k)}$ are called $S_N$ invariant because they satisfy
\begin{equation}
	\Adj{\sigma} \ket{T} = \ket{T}.
\end{equation}
That is, all states in $\Hilbertspace^{(k)}$ can be constructed from tensors satisfying
\begin{equation}
	T^{\sigma(j_1) \dots \sigma(j_k)}_{\sigma(i_1) \dots \sigma(i_k)} = T^{j_1 \dots j_k}_{i_1 \dots i_k}, \quad \forall \sigma \in S_N, 
\end{equation}
or
\begin{equation} \label{eq: SN invariant tensors}
 \boxed{ \mathcal{L}(\sigma) T \mathcal{L}(\sigma^{-1}) = T } 
\end{equation}  
 
The vector space of $S_N$ invariant linear maps on $V_N^{\otimes k}$ is denoted $\End_{S_N}(V_N^{\otimes k})$.
For $N \geq 2k$, $\End_{S_N}(V_N^{\otimes k})$ is isomorphic to the partition algebra $P_k(N)$:
\begin{equation}
	\End_{S_N}(V_N^{\otimes k}) = \Span_{\mathbb{C}}\{T \in \End(V_N^{\otimes k}) \, : \, \mathcal{L}(\sigma) T \mathcal{L}(\sigma^{-1}) = T ,\;  \forall \sigma \in S_N\} \cong P_k(N).
\end{equation}
For tensors labelling states we have further $S_k$ invariance.
The vector space of $S_N \times S_k$ invariant linear maps is denoted
\begin{equation}
	\begin{aligned}
		&\End_{S_N \times S_k}(V_N^{\otimes k}) = \\
	&\Span_{\mathbb{C}}\{T \in \End(V_N^{\otimes k}) \, : \, \mathcal{L}(\sigma) T \mathcal{L}(\sigma^{-1}) = \mathcal{L}_{\tau} T \mathcal{L}_{\tau^{-1}} = T, \; \forall \sigma \in S_N, \tau \in S_k\}.
	\end{aligned}
\end{equation}
and we have the correspondence
\begin{equation}
	\Hilbertspace^{(k)} \cong \End_{S_N \times S_k}(V_N^{\otimes k}). 
\end{equation}
The partition algebra $P_k(N)$ contains a subalgebra $SP_k(N)$, spanned by elements that commute with $\mathbb{C}[S_k] \subset P_k(N)$, called a symmetrized partition algebra. 
For $N \geq 2k$, $SP_k(N)$ is isomorphic to $\End_{S_N \times S_k}(V_N^{\otimes k})$, and by extension $\Hilbertspace^{(k)}$:
\begin{align}
\boxed{ 	\Hilbertspace^{(k)} \cong \End_{S_N \times S_k}(V_N^{\otimes k}) \cong SP_k(N). } 
\end{align}
This motivates the next subsection, where we study $P_k(N)$ and its symmetrized subalgebra $SP_k(N)$.

To summarise the above steps in words, we are investigating the adjoint action of permutations in $S_N$  on $N \times N$  quantum mechanical matrix variables $X^i_j$. The corresponding oscillators inherit the  adjoint $S_N$ action. Oscillator states with $k$ oscillators correspond to tensors $T$ with $k$ upper and lower indices, subject to an $S_k$ symmetry  permuting the $k$ upper-lower index pairs  along the tensor. This  $S_k$ symmetry arises from the bosonic nature of the  oscillators. The $S_N$ invariant $k$-oscillator states correspond to tensors having $k$ upper and $k$ lower indices, subject to an $S_N \times S_k$ invariance. This subspace of tensors can be described as a symmetrized sub-algebra $SP_k(N) $ of the partition algebra $P_k(N)$.

\subsection{Diagram basis} \label{sec: diagram basis}
We introduce the partition algebras in the diagram basis following the treatment in \cite{Halverson2004}. This is a nice starting point because it gives the most straight-forward description of multiplication in $P_k(N)$. As we will see in section \ref{sec: extremal correlators}, the diagram basis also gives a simple description of an outer product in $P_k(N)$, which is relevant to the the discussion of extremal correlators.

The partition algebra $P_k(N)$ is an algebra of dimension $B(2k)$. The Bell number $B(2k)$ is the number of possible partitions of a set with $2k$ distinct elements. Bell numbers can be computed from the generating function
\begin{equation}
	\sum_{k=0}^\infty \frac{B(k)}{k!}x^k = e^{e^x -1},
\end{equation}
from which one finds $B(2k) = 2, 15, 203, 4140$ for $k=1,2,3,4$.

A set partition $\pi$ of a set $S$ is a set of disjoint subsets of $S$ such that their union is all of $S$. The diagram basis for $P_k(N)$ is labelled by set partitions of the set $\{1, \dots, k, 1',\dots,k'\}$. The set of all set partitions of $\{1, \dots, k, 1',\dots,k'\}$ is denoted $\Pi_{2k}$. For example, the set $\Pi_{4}$ contains the following $B(4) = 15$ set partitions (subsets are separated by a vertical bar)
\begin{equation}
	\begin{aligned}
		&1|2|1'|2', \\
		&11'|2|2', \quad 12'|1'|2, \quad 12|1'|2', \quad 1'2'|1|2, \quad 1'2|1|2', \quad 22'|1'|1, \\
		&11'2'|2, \quad 121'|2', \quad 122'|1', \quad 1'2'2|1, \quad 11'|22', \quad 12'|1'2, \quad 12|1'2', \\
		&121'2'. \label{eq: example set partitions}
	\end{aligned}
\end{equation}

Each $\pi \in \Pi_{2k}$ labels an element of the diagram basis of $P_k(N)$. We write $d_{\pi}$ for the diagram basis element corresponding to $\pi \in \Pi_{2k}$. As the name suggests, $d_\pi$ should be thought of as a diagram. It is a diagram with $2k$ vertices divided into two rows. The bottom vertices are labelled $1,\dots,k$ from left to right and the vertices of the top row are labelled $1',\dots,k'$ from left to right. Two vertices are connected by an edge if they belong to the same subset of $\pi$. The diagrams corresponding to the set partitions in \eqref{eq: example set partitions} are
\begin{equation} 
	\begin{aligned}
		&\PAdiagramLabeled[]{2}{}, \\
		&\PAdiagramLabeled[]{2}{-1/1}, \PAdiagramLabeled[]{2}{-1/2}, \PAdiagramLabeled[-1/-2]{2}{}, \PAdiagramLabeled[2/1]{2}{}, \PAdiagramLabeled[]{2}{1/-2}, \PAdiagramLabeled[]{2}{-2/2},\\
		&\PAdiagramLabeled[2/1]{2}{-1/1}, \PAdiagramLabeled[-1/-2]{2}{1/-1}, \PAdiagramLabeled[-1/-2]{2}{2/-2}, \PAdiagramLabeled[2/1]{2}{2/-2}, \PAdiagramLabeled[]{2}{-1/1,-2/2}, \PAdiagramLabeled[]{2}{-1/2,-2/1}, \PAdiagramLabeled[-1/-2,2/1]{2}{}, \\
		&\PAdiagramLabeled[-1/-2,2/1]{2}{-1/1,-2/2}.
	\end{aligned}
\end{equation}
There is a redundancy in the diagram picture.
The redundancy arises from the fact that we are free to choose any set of edges, as long as every vertex in a subset of the set partition can be reached from any other vertex in the same subset, by a path along the edges.
For example, the following pairs of diagrams correspond to the same element in $P_3(N)$
\begin{equation}
	\PAdiagram[3/2]{3}{3/-3} = \PAdiagram[]{3}{2/-3,3/-3} \qq{and } \PAdiagram[]{3}{2/-1,2/-3} = \PAdiagram[-1/-3]{3}{-3/2}.
\end{equation}

The partition algebras are so-called diagram algebras because multiplication can be defined through diagram concatenation (in the diagram basis).
The product in $P_k(N)$ is independent of the choice of representative diagram.
Let $d_{\pi}$ and $d_{\pi'}$ be two diagrams in $P_k(N)$.
The composition $d_{\pi''} = d_{\pi} d_{\pi'}$ is constructed by placing $d_{\pi}$ above $d_{\pi'}$ and identifying the bottom vertices of $d_{\pi}$ with the top vertices of $d_{\pi'}$. The diagram is simplified by following the edges connecting the bottom vertices of $d_{\pi'}$ to the top vertices of $d_{\pi}$. Any connected components within  the middle rows are removed and we multiply by $N^c$, where $c$ is the number of these complete blocks removed.
For example,
\begin{equation}
	\begin{aligned}
		&\PAdiagram[-1/-2]{3}{-3/2} \\
		&\PAdiagram[2/1,-2/-3]{3}{-3/3}
	\end{aligned} = N \PAdiagram[-2/-3]{3}{-3/2}
	\qq{and} 
	\begin{aligned}
		&\PAdiagram[]{3}{-1/1,-2/3,-3/2} \\
		&\PAdiagram[-2/-3]{3}{-1/2,-3/3}
	\end{aligned} = \PAdiagram[-2/-3]{3}{-1/3,-3/2},
\end{equation}
where the factor of $N$ in the first equation comes from removing the middle component at vertex $1$ and $2$.
For linear combinations of diagrams, multiplication is defined by linear extension.

The subset of diagrams with $k$ edges, each connecting a vertex at the top to a vertex at the bottom, where every vertex has exactly one incident edge, span a subalgebra. This subalgebra is isomorphic to the symmetric group algebra $\mathbb{C}[S_k]$. For example, there is a one-to-one correspondence between permutations in $S_3$ and the following set of diagrams in $P_3(N)$
\begin{equation}
	\PAdiagram[]{3}{1/-1,2/-2,3/-3}, \hspace{2pt} \PAdiagram[]{3}{1/-2,2/-1,3/-3}, \hspace{2pt} \PAdiagram[]{3}{1/-3,2/-2,3/-1}, \hspace{2pt} \PAdiagram[]{3}{1/-1,2/-3,3/-2}, \hspace{2pt} \PAdiagram[]{3}{1/-3,2/-1,3/-2}, \hspace{2pt} \PAdiagram[]{3}{1/-2,2/-3,3/-1}. \label{eq: CS3 subalgebra diagrams}
\end{equation}
In the language of set partitions, these diagrams correspond to set partitions with subsets of the form $\{i j'\}$ for $i,j \in {1,\dots,k}$.
We denote the diagrams forming a basis for $\mathbb{C}[S_k]$ by $\tau$. 

The diagram $d_{\pi} \in P_k(N)$ corresponds to an element of $\End(V_N^{\otimes k})$ through the following action
\begin{equation}
	d_{\pi}(e_{i_1} \otimes \dots \otimes e_{i_k}) = \sum_{i_{1'}, \dots , i_{k'}} (d_{\pi})^{i_{1'} \dots i_{k'}}_{i_1 \dots i_k} e_{i_{1'}} \otimes \dots \otimes e_{i_{k'}}. \label{eq: P_k(N) to End(VN otimes k)}
\end{equation}
The matrix elements $(d_{\pi})^{i_{1'} \dots i_{k'}}_{i_1 \dots i_k}$ correspond to the diagram representation by associating a Kronecker delta to every  edge connecting a pair of vertices. For example,
\begin{equation}
	\PAdiagramLabeled[2/1, -1/-2]{2}{-1/1,-2/2} = \delta_{i_1 i_2} \delta_{i_2}^{i_{2'}} \delta^{i_{2'} i_{1'}} \quad \text{and} \quad \PAdiagramLabeled[-1/-2]{2}{-1/1} = \delta_{i_1 i_2} \delta_{i_1}^{i_{1'}}. \label{eq: examples of diagram tensors}
\end{equation}
Every diagram corresponds to an $S_N$ invariant tensor in the sense of equation \eqref{eq: SN invariant tensors}. As mentioned previously, this gives a basis for $\End_{S_N}(V_N^{\otimes k})$ for $N \geq 2k$ \cite{Halverson2004}.

Due to the bosonic symmetry of the oscillators, the invariant states are not in one-to-one correspondence with elements in $P_k(N)$.
Instead, every state in $\Hilbertspace^{(k)}$ corresponds to an element in the $S_k$ invariant sub-algebra of $P_k(N)$, which we call the symmetrised partition algebra and denote $SP_k(N)$. Consider the action of $S_k$ on the diagrams given, for any $ \tau \in S_k , d_{ \pi } \in P_k ( N )   $, by 
\begin{equation}\label{Skaction}  
	\tau : d_{ \pi } \rightarrow \tau d_{ \pi } \tau^{-1} \, . 
\end{equation}
A basis for $SP_k(N)$ is labeled by distinct orbits under this action. 
We denote as  $[d_\pi] \in SP_k(N)$ the invariant element obtained by averaging over the $S_k$ orbit of $d_{ \pi } $ : 
\begin{equation} \label{eq: SP_kN basis}
	[d_{\pi}] = \frac{1}{k!} \sum_{ \tau \in S_k } \tau d_{ \pi } \tau^{-1} = 
 \frac{1}{\abs{[d_{ \pi} ]}} 		\sum_{d_{\pi'} \in [d_\pi]} d_{\pi'} ,  
\end{equation}
where $\abs{[d_{ \pi} ]}$ is the size of the orbit. The equality follows because $\abs{[d_{ \pi} ]} $ 
is equal to $k!$ divided by the number of permutations $\tau$ leaving $d_{ \pi }$ fixed (orbit stabilizer theorem).  It follows that a basis for $\Hilbertspace^{(k)}$ is labeled by $[d_{\pi}] \in SP_k(N)$  through the correspondence
\begin{equation}
	\ket{ [ d_{\pi}] }  = \Tr_{V_N^{\otimes k}}([d_{\pi}](a^\dagger)^{\otimes k})\ket{0} = \sum_{\substack{i_1, \dots, i_k \\ i_{1'}, \dots, i_{k'}}} ([d_{\pi}])^{i_{1'} \dots i_{k'}}_{i_1 \dots i_k} (a^{\dagger})^{i_1}_{i_{1'}} \dots (a^{\dagger})^{i_k}_{i_{k'}}\ket{0}. \label{eq: diagram state def}
\end{equation}
Note that
\begin{align}
\ket{d_{\pi}} = \Tr_{V_N^{\otimes k}}(d_{\pi}(a^\dagger)^{\otimes k})\ket{0} = \ket{[d_{\pi}]},
\end{align}
for the sake of notational efficiency we will often label states with $d_{\pi}$ instead of $[d_{\pi}]$. Examples are 
\begin{equation}
	\ket{ \qty[ \PAdiagram[2/1, -1/-2]{2}{-1/1,-2/2} ]} = \ket{\PAdiagram[2/1, -1/-2]{2}{-1/1,-2/2}} = \sum_{i} (a^\dagger)^{i}_{i} (a^\dagger)^{i}_{i} \ket{0},
\end{equation}
and
\begin{equation}
	\ket{ \qty[ \PAdiagram[-1/-2]{2}{-1/1} ]} = \Big| \frac{1}{2} \qty( \PAdiagram[-1/-2]{2}{-1/1} + \PAdiagram[-1/-2]{2}{-2/2} ) \Big\rangle = \ket{ \PAdiagram[-1/-2]{2}{-1/1}} = \sum_{i,j} (a^\dagger)^{i}_{i} (a^\dagger)_{j}^{i}\ket{0}.
\end{equation}

States obtained by acting with the annihilation operators $a^i_j$ on the dual vacuum $ \langle 0 |$ can also be labelled by partition algebra diagrams as follows : 
\begin{align} \nonumber
	\bra{d_{\pi}} &= \bra{0} \Tr_{V_N^{\otimes k}}(d_{\pi}^T a^{\otimes k}) \\
	&= \bra{0} \Tr_{V_N^{\otimes k}}([d_{\pi}^T] a^{\otimes k}) = \bra{0} \sum_{\substack{i_1, \dots, i_k \\ i_{1'}, \dots, i_{k'}}} ([d_{\pi}])^{i_{1'} \dots i_{k'}}_{i_1 \dots i_k} a^{i_{1'}}_{i_{1}} \dots a^{i_{k'}}_{i_{k}} =  \bra{0} \sum_{\substack{i_1, \dots, i_k \\ i_{1'}, \dots, i_{k'}}} ([d_{\pi}^T])^{i_{1} \dots i_{k}}_{i_{1'} \dots i_{k'}} a^{i_{1'}}_{i_{1}} \dots a^{i_{k'}}_{i_{k}}, \label{eq: dual state}
\end{align}
where $d_{\pi}^T$ is the transpose of $d_{\pi}$. As a diagram, $d_{\pi}^T$ is the reflection of $d_{\pi}$ across a horizontal line, for example
\begin{align} 
\qty( \PAdiagram[2/1]{3}{-1/1, 2/-3} )^T = \PAdiagram[-1/-2]{3}{-1/1, -2/3} \label{eq: transpose of diagram}.
\end{align}
The use of  the transpose in this definition is motivated by the orthonormality property below \eqref{ortholargeN}.  Using the commutation relations in equation \eqref{eq: matrix basis comm relations}, the inner product can be written as a trace of products of elements in $SP_k(N)$,
\begin{equation}
	\bra{d_{\pi}}\ket{d_{\pi'}} = \sum_{\tau \in S_k} (d_{\pi}^T\tau d_{\pi'}\tau^{-1})^{i_1 \dots i_k}_{i_1 \dots i_k} = \sum_{\tau \in S_k}\Tr_{V_N^{\otimes k}}(d_{\pi}^T \tau d_{\pi'} \tau^{-1}). \label{eq: inner product is trace}
\end{equation}

The large $N$ factorization result in \cite{PIMO_Factor} implies that the normalized states
\begin{equation} \label{eq: noramlised state definition}
	\ket*{\hat{d}_{\pi}} = \frac{1}{\sqrt{\bra{d_{\pi}}\ket{d_{\pi}}}}\ket{d_{\pi}},
\end{equation}
are orthonormal at large $N$ (to leading order in $1/\sqrt{N}$)
\begin{equation}\label{ortholargeN} 
	\bra*{\hat{d}_{\pi}}\ket*{\hat{d}_{\pi'}} = 
	\begin{cases}
		1 + O(1/\sqrt{N}) \qq{if $[d_{\pi}] = [d_{\pi'}]$,} \\
		0 + O(1/\sqrt{N}) \qq{otherwise.}
	\end{cases}
\end{equation}

\subsection{Representation basis} \label{sec: ON basis}

The connection between $S_N$ invariant states and partition algebras gives rise to a natural basis, labelled by representation theoretic data. The representation basis diagonalises a set of commuting algebraic charges that we introduce in section \ref{sec: charges}. This observation gives a concrete construction algorithm for the change of basis matrix (from diagram basis to representation basis). We now describe how the representation theoretic basis for $SP_k(N)$ arises using Schur-Weyl duality between $S_N $ and $P_k(N)$, along with the implementation of the invariance under the $S_k$ action of \eqref{Skaction} in the representation theoretic basis.  The transition from a combinatorial basis of diagrams in an algebra defined by physical constraints (in this case a  bosonic symmetry of matrix oscillators) to a representation theoretic basis is an example of Fourier transformation which has been useful in multi-matrix as well as tensor systems of interest in AdS/CFT and holography (a short review of these applications is in \cite{Ramgoolam2016}). The proofs of some statements quoted here are in appendix \ref{apx: Semi-Simple Algebra Technology}. 

From the point of view of representation theory, the correspondence between permutation invariant states and partition algebras should be understood as a consequence of Schur-Weyl duality. In particular, Schur-Weyl duality says that the decomposition (see section 2.5 in \cite{Halverson2018})
\begin{equation}
	V_N^{\otimes k} \cong \bigoplus_{l=0}^{k}\bigoplus_{\substack{\Lambda_1^\# \vdash l}} V_{\qty[N-l,\Lambda_1^\#]}^{S_N} \otimes V_{\qty[N-l,\Lambda_1^\#]}^{P_k(N)}
	 \label{eq: Vn otimes k}
\end{equation}
is multiplicity free in terms of irreducible representations of $S_N$ and $P_k(N)$. The Young diagram $\Lambda_1 = [N-l,\Lambda_1^\#]$, which is an integer partition of $N$, is constructed by placing the diagram $\Lambda_1^\#$ (having $l$ boxes) below a first row of $N-l$ boxes. Requiring $\Lambda_1 $ to be a valid Young diagram imposes a condition on the first row length of   $ r_1(  \Lambda_1^{\#} )  \le  N - l $.  This condition is non-trivial for $ N < 2 k $, while it is trivially satisfied for all  $\Lambda_1^{\#} $ having up to $k$ boxes  for $ N \ge 2k$. The latter is called the stable limit. In this limit we can write the decomposition \eqref{eq: Vn otimes k} in a simplified form
\begin{align} \label{eq: schur-weyl simple}
V_N^{\otimes k} = \bigoplus_{\Lambda_1  \in \cYS ( k ) } V^{S_N}_{\Lambda_1} \otimes V^{P_k(N)}_{\Lambda_1},
\end{align}
in which the sum can be labelled by the set of all Young diagrams  $\Lambda_1^{\#} $ having up to $k$ boxes: these are inserted below a first row to form Young diagrams with $N$ boxes. This stable set of Young diagrams having $N$ boxes is denoted $ \cYS ( k ) $.  With the exception of appendix \ref{sec: orbit basis} this is the limit within which we will work. 

Equation \eqref{eq: diagonally permutation action on hilbert} implies that we can identify
\begin{equation}
	\End(V_N^{\otimes k}) \cong V_N^{\otimes k} \otimes V_N^{\otimes k},
\end{equation}
as a representation of $S_N$. We use Schur-Weyl duality \eqref{eq: schur-weyl simple} to decompose each factor on the RHS as
\begin{equation}
	 V_N^{\otimes k} \otimes V_N^{\otimes k} = \qty(\bigoplus_{\Lambda_1 \in \cYS (k)  } V_{\Lambda_1}^{S_N} \otimes V^{P_k(N)}_{\Lambda_1}) \otimes  \qty(\bigoplus_{\Lambda_1' \in \cYS (k) } V_{\Lambda_1'}^{S_N} \otimes V^{P_k(N)}_{\Lambda_1'}),
\end{equation}
where we are assuming the stable limit.
Projecting to $S_N$ invariants on both sides gives
\begin{equation}
	P_k(N) \cong \End_{S_N}(V_N^{\otimes k}) \cong \bigoplus_{\Lambda_1 \in \cYS ( k ) } V^{P_k(N)}_{{\Lambda_1}} \otimes V^{P_k(N)}_{{\Lambda_1}}. \label{eq: AW EndSN}
\end{equation}
This follows because the decomposition of $V_{\Lambda_1}^{S_N} \otimes V_{\Lambda_1'}^{S_N}$ contains an invariant if and only if $\Lambda_1 = \Lambda_1'$.

The RHS of \eqref{eq: AW EndSN} reflects a  decomposition of $P_k(N)$ into a direct sum of matrix algebras. Such a decomposition always exists for a semi-simple algebra by the Artin-Wedderburn theorem. This implies that there exists a basis of generalized elementary matrices (also called a complete set of matrix units) for $P_k(N)$. A complete set of matrix units is a basis
\begin{equation} 
	Q^{\Lambda_1}_{\alpha \beta}, \quad \Lambda_1 \in \cYS ( k ), \quad \alpha,\beta \in \{1, \dots, \Dim(V_{\Lambda_1}^{P_k(N)})\}, 
\end{equation}
with the property
\begin{equation}
	Q^{\Lambda_1}_{\alpha \beta} Q^{\Lambda'_1}_{\alpha' \beta'} = \delta^{\Lambda_1 \Lambda_1'} \delta_{\beta \alpha'} Q^{\Lambda_1}_{\alpha \beta'}.
\end{equation}
In other words, $P_k(N)$ can be realized as block-diagonal matrices, with each block labelled by an irreducible representation $\Lambda_1$ of $P_k(N)$. The Artin-Wedderburn decomposition implies
\begin{equation}
	\Dim \qty( P_k(N) )  = B(2k) = \sum_{\Lambda_1 \in \cYS (k) } \qty(\Dim V_{\Lambda_1}^{P_k(N)})^2,
\end{equation}
which is analogous to the expression
\begin{equation}
	\abs{G} = \sum_{R \in \text{Rep}(G)} \qty(\Dim V_{R}^G)^2
\end{equation}
for the order of a finite group $G$ in terms of its irreducible representations $R$.

As we prove in \ref{apx: subsec matrix unit PkN}, the following set of linear combinations of elements in $P_k(N)$ form a complete set of matrix units for $P_k(N)$,
\begin{equation}	
	\boxed{ Q_{\alpha \beta}^{\Lambda_1} = \sum_{i=1}^{B(2k)} \Dim(V_{\Lambda_1}^{S_N}) D_{\beta \alpha}^{\Lambda_1}((b_i^*)^T)b_i.  } \label{eq: Fourier Basis P_k(N)}
\end{equation}
The coefficients $D_{\beta \alpha}^{\Lambda_1}(d)$ are matrix elements of the representation of $P_k(N)$, labelled by $\Lambda_1 \vdash N$. The sum is over a basis $b_i, i\in\{1,\dots,B(2k)\}$ for $P_k(N)$ (for example the diagram basis).
The element $b_i^*$ is called the dual of $b_i$. It has an explicit construction in terms of the inverse of the Gram matrix defined by
\begin{equation}
	g_{i j} = \Tr_{V_N^{\otimes k}}(b_i b_j^T).
\end{equation}
The dual of $b_i$ is
\begin{equation}
	b_i^* = \sum_{j=1}^{B(2k)} g^{-1}_{ij}b_j,
\end{equation}
and the inverse of the Gram matrix in the diagram basis can be written as a series expansion in $N$ (see equation \eqref{eq: inverse metric}).

To construct a representation basis for $\Hilbertspace^{(k)}$, we need to construct matrix units for $SP_k(N)$. They can be constructed from matrix units for $P_k(N)$ as follows. The partition algebra $P_k(N)$ contains a subalgebra $\mathbb{C}[S_k]$. Consequently, we can restrict an irreducible representation $V_{\Lambda_1}^{P_k(N)}$ to a representation of $\mathbb{C}[S_k]$, which in general is reducible. Letting $V_{\Lambda_2}^{\mathbb{C}[S_k]} $ be an irreducible representation of $\mathbb{C}[S_k]$ labelled by a Young diagram $\Lambda_2$  with $k$ boxes, we have 
\begin{equation}
	V_{\Lambda_1}^{P_k(N)} \cong \bigoplus_{\Lambda_2 \vdash k} V_{\Lambda_2}^{\mathbb{C}[S_k]} \otimes V_{\Lambda_1 \Lambda_2}^{P_k(N) \rightarrow \mathbb{C}[S_k]}. \label{eq: Pk to Sk}
\end{equation}
The dimension of $V_{\Lambda_1 \Lambda_2}^{P_k(N) \rightarrow \mathbb{C}[S_k]}$ is the branching multiplicity
\begin{equation}
	\text{Dim} \Big( V_{\Lambda_1 \Lambda_2}^{P_k(N) \rightarrow \mathbb{C}[S_k]} \Big) = \text{Mult} \Big( V^{P_k(N)}_{\Lambda_1} \rightarrow V^{\mathbb{C}[S_k]}_{\Lambda_2} \Big),
\end{equation}
 In the rest of the paper we will use $\Lambda_1$ to label irreducible representations of $S_N$ and $P_k(N)$. Irreducible representations of $S_k$ are denoted by $\Lambda_2$.
Inserting the decomposition \eqref{eq: Pk to Sk} into equation \eqref{eq: AW EndSN} and projecting to $S_k$ invariants gives
\begin{equation}
\boxed{ 
	\Hilbertspace^{(k)} \cong \End_{S_N \times S_k}(V_N^{{\otimes k}}) \cong \bigoplus_{\substack{\Lambda_1 \in \cYS ( k )  \\ \Lambda_2 \vdash k}} V_{\Lambda_1 \Lambda_2}^{P_k(N) \rightarrow \mathbb{C}[S_k]} \otimes V_{\Lambda_1 \Lambda_2}^{P_k(N) \rightarrow \mathbb{C}[S_k]}. }  \label{eq: Artin Decomposition SPkN}
\end{equation}
This should be understood as an Artin-Wedderburn decomposition of $SP_k(N)$.

Equation \eqref{eq: Pk to Sk} points us towards a construction of matrix units for $SP_k(N)$ from matrix units of $P_k(N)$. On the LHS we have a basis
\begin{equation}
	E^{\Lambda_1}_{\alpha}, \quad \alpha \in \{1, \dots \Dim(V_{\Lambda_1}^{P_k(N)})\},
\end{equation}
where the representation of $d \in P_k(N)$ is irreducible,
\begin{equation}
	d(E^{\Lambda_1}_{\alpha}) = \sum_\beta D_{\beta \alpha}^{\Lambda_1}(d)E^{\Lambda_1}_{\beta}. \label{eq: PkN irrep}
\end{equation}
The RHS has a basis
\begin{equation}
	\begin{aligned}
		E^{\Lambda_1, \mu}_{\Lambda_2, p}, \quad &p \in \{1, \dots, \Dim(V_{\Lambda_1}^{\mathbb{C}[S_k]}) \}, \\
		&\mu \in \{1,\dots, \Dim(V_{\Lambda_1 \Lambda_2}^{P_k(N) \rightarrow \mathbb{C}[S_k]})\}, 
	\end{aligned}
\end{equation}
where $\mu$ is a multiplicity label for $V_{\Lambda_2}^{\mathbb{C}[S_k]}$ in the decomposition.
We demand that the representation of $\tau \in \mathbb{C}[S_k]$ is irreducible in this basis,
\begin{equation}
	\tau(E^{\Lambda_1, \mu}_{\Lambda_2, p}) = \sum_q D_{qp}^{\Lambda_2}(\tau)E^{\Lambda_1, \mu}_{\Lambda_2, q},\label{eq: Sk branching irrep}
\end{equation}
where $D_{qp}^{\Lambda_2}(\tau)$ is an irreducible representation of $\tau \in \mathbb{C}[S_k]$. The change of basis coefficients are called Branching coefficients
\begin{equation}
	E^{\Lambda_1, \mu}_{\Lambda_2, p} = \sum_{\alpha}B^{P_k(N)\rightarrow \mathbb{C}[S_k]}_{\Lambda_1,\alpha \rightarrow \Lambda_2, p; \mu} E^{\Lambda_1}_{\alpha}, \label{eq: branching coeff}
\end{equation}
or in braket notation
\begin{equation}
	B^{P_k(N)\rightarrow \mathbb{C}[S_k]}_{\Lambda_1,\alpha \rightarrow \Lambda_2, p; \mu} = \bra{E_\alpha^{\Lambda_1}}\ket{E^{\Lambda_1, \mu}_{\Lambda_2, p}}.
\end{equation}
%Requiring equations \eqref{eq: PkN irrep}  and \eqref{eq: Sk branching irrep} to be compatible constrains the branching coefficients, but does not fully determine them. This is particularly true when the multiplicity is greater than 1.

The elements
\begin{equation}
\boxed{ 	Q^{\Lambda_1}_{\Lambda_2, \mu \nu} = \sum_{\alpha, \beta, p} Q^{\Lambda_1}_{\alpha \beta}B^{P_k(N) \rightarrow \mathbb{C}[S_k]}_{\Lambda_1, \alpha \rightarrow \Lambda_2, p; \mu}B^{P_k(N) \rightarrow \mathbb{C}[S_k]}_{\Lambda_1, \beta \rightarrow \Lambda_2, p; \nu},  } \label{eq: Fourier Basis SP_k(N)}
\end{equation}
form a complete set of matrix units for $SP_k(N)$. The sum over $p$ implements the projection to $S_k$ invariants.
The above elements satisfy (see equation \eqref{eq: SPkN matrix unit product})
\begin{equation}
	Q^{\Lambda_1}_{\Lambda_2, \mu \nu}Q^{\Lambda_1'}_{\Lambda_2' ,\mu' \nu'} = \delta^{\Lambda_1 \Lambda_1'}\delta_{\Lambda_2 \Lambda_2'}\delta_{\nu \mu'}Q^{\Lambda_1}_{\Lambda_2 \mu \nu'}, \label{eq: SPkN matrix units}
\end{equation}
and orthogonality of states
\begin{equation}
	\ket*{Q^{\Lambda_1}_{\Lambda_2, \mu \nu}} = \Tr_{V_N^{\otimes k}}(Q^{\Lambda_1}_{\Lambda_2, \mu \nu} (a^{\dagger})^{\otimes k}). \label{eq: rep basis states}
\end{equation}
follows from the form of the inner product \eqref{eq: inner product is trace}.
The proof goes as follows
\begin{equation}
	\begin{aligned}
		\bra{Q^{\Lambda_1}_{\Lambda_2, \mu \nu}}\ket{Q^{\Lambda_1'}_{\Lambda_2', \mu' \nu'}} &= \sum_{\tau \in S_k} \Tr_{V_N^{\otimes k}}(Q^{\Lambda_1}_{\Lambda_2, \mu \nu} \tau \qty(Q^{\Lambda_1'}_{\Lambda_2', \mu' \nu'})^T \tau^{-1} ) \\
		&= \sum_{\tau \in S_k} \Tr_{V_N^{\otimes k}}(Q^{\Lambda_1}_{\Lambda_2, \mu \nu} \tau Q^{\Lambda_1'}_{\Lambda_2', \nu' \mu' } \tau^{-1} ) \\ 
		&= k! \Tr_{V_N^{\otimes k}}(Q^{\Lambda_1}_{\Lambda_2, \mu \nu} Q^{\Lambda_1'}_{\Lambda_2', \nu' \mu' } ) \\
		&= k! \delta^{{\Lambda_1} {\Lambda_1'}}\delta_{{\Lambda_2} {\Lambda_2'}} \delta_{{\nu} {\nu'}} \Tr_{V_N^{\otimes k}}(Q^{\Lambda_1}_{\Lambda_2, \mu \mu'}).
	\end{aligned}
\end{equation}
In the second equality we used $\qty(Q^{\Lambda_1'}_{\Lambda_2', \mu' \nu'})^T = Q^{\Lambda_1'}_{\Lambda_2', \nu' \mu' }$ which follows from equation \eqref{eq: transpose diagram is transpose matrix element}. Note that
\begin{equation}
	\begin{aligned}
		\Tr_{V_N^{\otimes k}}(Q^{\Lambda_1}_{\Lambda_2, \mu \mu'})
		&= \Tr_{V_N^{\otimes k}}(Q^{\Lambda_1}_{\Lambda_2, \mu 1}Q^{\Lambda_1}_{\Lambda_2, 1 \mu'}) \\
		&= \Tr_{V_N^{\otimes k}}(Q^{\Lambda_1}_{\Lambda_2, 1 \mu'} Q^{\Lambda_1}_{\Lambda_2, \mu 1}) \\
		&= \delta_{\mu \mu'}\Tr_{V_N^{\otimes k}}(Q^{\Lambda_1}_{\Lambda_2, 11}) \\
		&= \delta_{\mu \mu'}\mathcal{N}_{\Lambda_1 \Lambda_2},
	\end{aligned}
\end{equation}
such that the normalization (see equation \eqref{eq: normalization constant as dimensions})
\begin{equation}
	\mathcal{N}_{\Lambda_1 \Lambda_2} = \Dim V_{\Lambda_1}^{S_N} \Dim V_{\Lambda_2}^{S_k}, \label{eq: Normalization Constant Two Point Function}
\end{equation}
only depends on irreducible representations $\Lambda_1, \Lambda_2$, which proves orthogonality.

To summarize, we have shown that there exists an orthogonal basis for $\Hilbertspace^{(k)}$ labelled by representation theoretic data, for arbitrary $N \geq 2k$, using Fourier transforms on semi-simple algebras.
In appendix \ref{apx: Semi-Simple Algebra Technology} we provide the detailed proofs of these results.
In the next section we will provide explicit formulas for the change of basis from the diagram basis to the basis of matrix units. 
We leave the elucidation of finite $N$ effects (the case $N < 2k$ which lies beyond the stable limit)  in the representation basis for future work.

\section{Representation basis and algebraic  charges}
\label{sec: charges}

In this section we discuss the  construction of the representation basis elements $Q^{\Lambda_1}_{\Lambda_2, \mu \nu}$ as linear combinations of diagrams in $P_k(N)$. 
These can, in principle, be computed using equation \eqref{eq: Fourier Basis SP_k(N)} by first computing the branching coefficients. The computation of these requires explicit choices of basis in the representations $V_{\Lambda_1}^{P_k(N)} $ and $ V_{\Lambda_2}^{\mathbb{C}[S_k]}$. Such choices can be bypassed. The basic idea is to find
the  $Q^{\Lambda_1}_{\Lambda_2, \mu \nu}$  as eigenvectors of appropriate elements of  $P_k(N)$ which can be viewed as operators on  $P_k(N)$ acting by the algebra  multiplication.  The  subspaces labelled by $ \Lambda_1 , \Lambda_2 $, associated with irreducible representations of $S_N$ and $S_k$ respectively, are identified using central elements (Casimirs)  in the group algebras $\mathbb{C}[S_N] $ and $ \mathbb{C}[S_k]$. These Casimirs  can be expressed as elements of $P_k(N)$ using Schur-Weyl duality. This is particularly useful in the large $N$ limit where 
 $k$ is kept fixed and $ N \gg k$, since the dimension of $P_k(N)$ does not grow with $N$. 
The more refined determination of  subspaces labelled by $ \mu $ and $\nu$ is achieved by picking non-central elements of $P_k(N)$ which nevertheless generate a  maximally commuting subalgebra.

We explicitly construct the change of basis for the special cases of degree $k=1,2,3$. Tables of these basis elements are found in appendix \ref{apx: matrix units}. The expansion coefficients are given as functions of $N$ and are therefore valid for all $N \geq 2k$.

%In the language of algebras, the general idea is to define a set of pairwise commuting elements (abelian subalgebra) $\{T, T', T'', \dots \}$ in $SP_k(N)$ whose eigenvalues distinguish the labels on matrix units $Q^{\Lambda_1}_{\Lambda_2, \mu \nu}$. Physically, the commuting elements will correspond to Hermitian operators acting on $\mathcal{H}^{(k)}$ whose eigenvectors with distinct eigenvalues are orthogonal. The simultaneous eigenvectors of the commuting set will correspond to states labelled by a complete set of matrix units.

Analogous constructions  in multi-matrix systems with continuous gauge symmetry, relevant to AdS/CFT,  are given in  \cite{Kimura2008,BPSChargesSymGroup2019}. They also played a role, using  developments in tensor models with $U(N)$ symmetries,  in  \cite{QMRibbonGraphs2021}  in giving a combinatorial interpretation of Kronecker coefficients. 
%Below, we have restricted the discussion to commuting operators relevant for $k=1,2,3$.
%The operators distinguishing $\Lambda_1$ and multiplicity indices for general $k$ are discussed in \cite{Halverson2004}. We found theorem 3.10 in \cite{Enyang2012} particularly useful.

\subsection{Central elements in the  partition algebra }    \label{subsec: central elements in SPkN}
For a fixed pair $\Lambda_1, \Lambda_2$, the linear span of $Q^{\Lambda_1}_{\Lambda_2, \mu \nu}$ for $\mu, \nu = 1, \dots, \Dim V_{\Lambda_1 \Lambda_2}^{P_k(N)\rightarrow \mathbb{C}[S_k]}$ forms a subspace of $SP_k(N)$. We will now describe how this subspace can be identified with simultaneous eigenspaces of Casimirs associated with $\mathbb{C}[S_N]$ and $\mathbb{C}[S_k]$.

%More generally, a semi-simple algebra is isomorphic to a direct sum of matrix algebras (not necessarily of the same dimensions).
%For $SP_k(N)$, the distinct matrix algebras are labelled by the pair $(\Lambda_1, \Lambda_2)$ of irreducible representations of $S_N$ and $S_k$.
%For a fixed pair, the matrix units $Q^{\Lambda_1}_{\Lambda_2, \mu \nu}$ correspond to elementary matrices for an algebra of matrices of size $\Dim V_{\Lambda_1 \Lambda_2}^{P_k(N)\rightarrow \mathbb{C}[S_k]}$. Below we describe how the labels $\Lambda_1, \Lambda_2$ can be distinguished by central elements of $SP_k(N)$ and $\mathbb{C}[S_k]$, respectively. Central elements of $SP_k(N)$ can be constructed from central elements of $\mathbb{C}[S_N]$ using Schur-Weyl duality.

First, we will define Casimirs of $\mathbb{C}[S_N]$, and explain their relation to $P_k(N)$.
The center $\mathcal{Z}(\mathbb{C}[S_N])$ of $\mathbb{C}[S_N]$ consists of elements
\begin{equation}
	\mathcal{Z}(\mathbb{C}[S_N]) = \{z \in \mathbb{C}[S_N] : z\sigma = \sigma z, \quad \forall \sigma \in \mathbb{C}[S_N]\}.
\end{equation}
Elements in the center are called central elements. For a central element $z$,  the homomorphism property of representations implies 
\bea 
\mathcal{L}(z)\mathcal{L}(\sigma)=\mathcal{L}(\sigma)\mathcal{L}(z), \quad \forall \sigma \in S_N,
\eea
 and it follows that $\mathcal{L}(z)$ is an element of the algebra of operators acting on  $V_N^{\otimes k}$ which commutes with  $S_N$. This algebra is denoted $\End_{S_N}(V_N^{\otimes k}) $.  

As we reviewed in the previous section, $P_k(N) \cong \End_{S_N}(V_N^{\otimes k})$ for $N \geq 2k$.
This establishes a connection between $\mathcal{Z}(\mathbb{C}[S_N])$ and $P_k(N)$ as linear operators acting on $V_N^{ \otimes k }$. In particular, for every $z \in \mathcal{Z}(\mathbb{C}[S_N])$, there exists an element $\SWdual{z} \in P_k(N)$ defined by
\begin{equation}
\boxed{ 
	\SWdual{z}(e_{i_1} \otimes \dots \otimes e_{i_k}) = \mathcal{L}(z)(e_{i_1} \otimes \dots \otimes e_{i_k}).  } \label{eq: ZCSN PkN homomorphism}
\end{equation}
Note that the definition of $\SWdual{z}$ depends on $k$.
Further, observe that
\begin{equation}
	\mathcal{L}(z)d(e_{i_1} \otimes \dots \otimes e_{i_k}) = d\mathcal{L}(z)(e_{i_1} \otimes \dots \otimes e_{i_k}),
\end{equation}
for all $d \in P_k(N)$ because $P_k(N) $ and $\mathbb{C}[S_N] $ are mutual commutants in $\End ( V_N^{ \otimes k } )$. 
%
%
%$z \in \mathbb{C}[S_N]$ and therefore
%\begin{equation}
%	\SWdual{z}d (e_{i_1} \otimes \dots \otimes e_{i_k}) = d \SWdual{z}(e_{i_1} \otimes %\dots \otimes e_{i_k}),
%\end{equation}
%for all $d \in P_k(N)$ by \eqref{eq: ZCSN PkN homomorphism}.
This implies that $\SWdual{z}$ is automatically in the center of $P_k(N)$, which we denote $\mathcal{Z}(P_k(N))$.
In other words, equation \eqref{eq: ZCSN PkN homomorphism} defines a homomorphism from $\mathcal{Z}(\mathbb{C}[S_N])$ to $\mathcal{Z}(P_k(N))$.
As a particular case of being central in $P_k(N)$, $\SWdual{z}$ commutes with $\mathbb{C}[S_k] \subset P_k(N)$.
% That is, $\SWdual{z}$ is a central element of $SP_k(N)$.

Central elements play a special role in representation theory. Schur's lemma implies that an irreducible matrix representation of a central element is proportional to the identity matrix. The proportionality constant is a normalized character. In particular we have
\begin{equation}
	D^{\Lambda_1}_{ab}(z) = \hat{\chi}^{\Lambda_1}(z)\delta_{ab}, \label{eq: schur lemma T2}
\end{equation}
where we have introduced the short-hand
\begin{equation}
	\hat{\chi}^{\Lambda_1}(z) = \frac{\chi^{\Lambda_1}(z)}{\DimSN{\Lambda_1}} 
\end{equation}
for normalized characters. In this sense central elements are Casimirs, they act by constants on irreducible subspaces, and the constants can be used to determine the particular representation.

The element of $\mathbb{C}[S_N]$ formed by summing over all elements in a distinct conjugacy class of $S_N$ is central.
For example, we define the element $T_2 \in \mathcal{Z}(\mathbb{C}[S_N])$ as
\begin{equation}
	T_2 = \sum_{1 \leq i < j \leq N}  (ij),
\end{equation} 
where the sum is over all transpositions. 
By the argument in the previous paragraph, there exists an element $\SWdual{T}_2^{(k)} \in \mathcal{Z}(P_k(N))$ such that
\begin{equation}
	\begin{aligned}
		\SWdual{T}_2^{(k)}(e_{i_1} \otimes \dots \otimes e_{i_k})  &= \mathcal{L}(T_2)(e_{i_1} \otimes \dots \otimes e_{i_k}) \\ 
		\rotaterelation{=}\quad \quad \quad \quad & \quad \quad \quad \quad \quad \quad \rotaterelation{=} \\
		\sum_{i_{1'} {\dots } i_{k'}}(\SWdual{T}_2^{(k)})^{i_{1'} {\dots } i_{k'}}_{i_1 \dots i_k}e_{i_{1'}} \otimes \dots \otimes e_{i_{k'}} &=
		\sum_{\substack{\sigma = (ij) \\ 1 \leq i<j \leq N}}e_{\sigma^{-1}(i_1)} \otimes \dots \otimes e_{\sigma^{-1}(i_k)}.\label{eq: Center of S_N inside PkN}
	\end{aligned}
\end{equation}
As we will explain, the eigenvalues of the central element $\SWdual{T}_2^{(k)}$ can be used to distinguish the label $\Lambda_1$ on matrix units.
Since $\SWdual{T}_2^{(k)}$ is an element of $SP_k(N)$, it has an expansion in terms of diagrams (see \cite[Equation 3.32, Theorem 3.35]{Halverson2004})
\begin{equation} \boxed{
	\SWdual{T}_2^{(k)} = \sum_{\pi \in \Pi_{2k}} (\SWdual{T}_2^{(k)})^{\pi} d_{\pi}. } \label{eq: T2 expansion in terms of diagrams}
\end{equation}
The equality in \eqref{eq: Center of S_N inside PkN} implies a radical simplification for large $N$. The element $T_2$ contains order $N^2$ transpositions, while $\SWdual{T}_2^{(k)}$ contains at most $B(2k)$ diagrams. The dependence on $N$ is incorporated in the coefficients $(\SWdual{T}_2^{(k)})^{\pi}$, which are polynomial functions of $N$. Explicit examples are in \eqref{eq: T2 P1N}, \eqref{eq:T2bar2} and \eqref{eq:T2bar3}. 

 There exist similar elements $t_2^{(k)} \in \mathcal{Z}(\mathbb{C}[S_k]) \subset \mathcal{Z}(P_k(N))$ defined by summing over transposition diagrams. For example,
\begin{equation}
	t_2^{(2)} = \PAdiagram{2}{-1/2,-2/1}, \quad t_2^{(3)} = \PAdiagram{3}{-1/2,-2/1, 3/-3} + \PAdiagram{3}{-1/1, -2/3,2/-3} + \PAdiagram{3}{-1/3,-2/2,-3/1}.
\end{equation}
The eigenvalues of $t_2^{(k)}$ will be used to distinguish the label $\Lambda_2$.

Equation \eqref{eq: Center of S_N inside PkN} together with equation \eqref{eq: schur lemma T2} gives
\begin{equation}
	D^{\Lambda_1}_{\alpha \beta}(\SWdual{T}_2^{(k)}) =  \frac{\chi^{\Lambda_1}(\SWdual{T}_2^{(k)})}{\DimPk{\Lambda_1}}\delta_{\alpha \beta} = \hat{\chi}^{\Lambda_1}(T_2)\delta_{\alpha \beta},
\end{equation}
where the distinction between the two characters is
\begin{equation}
	\chi^{\Lambda_1}(\SWdual{T}_2^{(k)}) = \sum_{\alpha=1}^{\DimPk{\Lambda_1}} D^{\Lambda_1}_{\alpha \alpha}(\SWdual{T}_2^{(k)}), \qq{and} \chi^{\Lambda_1}(T_2) = \sum_{a=1}^{\DimSN{\Lambda_1}} D^{\Lambda_1}_{aa}(T_2).
\end{equation}
That is, the first character is a character of $P_k(N)$, the second is a character of $\mathbb{C}[S_N]$.
Similarly,
\begin{equation}
	D^{\Lambda_2}_{pq}(t_2^{(k)}) = \hat{\chi}^{\Lambda_2}(t_2^{(k)})\delta_{pq},
\end{equation}
where
\begin{equation}
	\hat{\chi}^{\Lambda_2}(t_2^{(k)}) = \frac{\chi^{\Lambda_2}(t_2^{(k)})}{\DimSk{\Lambda_2}}.
\end{equation}

Normalized characters of $T_2$ and $t_2^{(k)}$ can be expressed in terms of combinatorial quantities (known as the contents) of boxes of  Young diagrams (see example 7 in section I.7 of \cite{Macdonald1998}).
Let $Y_{\Lambda_1}, Y_{\Lambda_2}$ be the Young diagrams corresponding to integer partitions $\Lambda_1 \in \cYS ( k ), \Lambda_2 \vdash k$. Then
\begin{equation}
	\hat{\chi}^{\Lambda_1}(T_2) = \sum_{(i,j) \in Y_{\Lambda_1}} (j-i),  \quad \hat{\chi}^{\Lambda_2}(t_2^{(k)}) = \sum_{(i,j) \in Y_{\Lambda_2}} (j-i),  \label{eq: norm characters of T2s}
\end{equation}
where $(i,j)$ corresponds to the cell in the $i$th row and $j$th column of the Young diagram (the top left box has coordinate $(1,1)$).

With the above facts at hand, we can understand how the $\Lambda_1, \Lambda_2$ labels correspond to eigenvalues of $\SWdual{T}_2^{(k)}, t_2^{(k)}$.
As we prove in appendix \ref{apx: subsec matrix unit PkN} the $P_k(N)$ matrix units have the property
\begin{equation}
	d Q^{\Lambda_1}_{\alpha \beta} =\sum_{\gamma} D^{\Lambda_1}_{\gamma \alpha}(d)Q^{\Lambda_1}_{\gamma \beta}, \qq{for $d \in P_k(N)$,} \label{eq: Pk action of matrix units}
\end{equation}
and therefore
\begin{equation} \label{eq: T_2 eigenvec of Q}
	Q^{\Lambda_1}_{\alpha \beta} \SWdual{T}_2^{(k)} = \SWdual{T}_2^{(k)} Q^{\Lambda_1}_{\alpha \beta} = \sum_{\gamma} D^{\Lambda_1}_{\gamma \alpha}(\SWdual{T}_2^{(k)})Q^{\Lambda_1}_{\gamma \beta} = \hat{\chi}^{\Lambda_1}(T_2)Q^{\Lambda_1}_{\alpha \beta}.
\end{equation}
We derive a similar equation for $t_2^{(k)}$ acting on $Q^{\Lambda_1}_{\Lambda_2, \mu\nu}$ using the definition in \eqref{eq: Fourier Basis SP_k(N)}. From the definition we have
\begin{equation}
	\begin{aligned}
		t_2^{(k)}Q^{\Lambda_1}_{\Lambda_2, \mu \nu} &= \sum_{\alpha, \beta, p} t_2^{(k)}Q^{\Lambda_1}_{\alpha \beta}B^{P_k(N) \rightarrow \mathbb{C}[S_k]}_{\Lambda_1, \alpha \rightarrow \Lambda_2, p; \mu}B^{P_k(N) \rightarrow \mathbb{C}[S_k]}_{\Lambda_1, \beta \rightarrow \Lambda_2, p; \nu} \\
	&= \sum_{\alpha, \beta, \gamma, \gamma', p} D_{\gamma \alpha}^{\Lambda_1}(t_2^{(k)})\delta_{\gamma \gamma'}Q^{\Lambda_1}_{\gamma' \beta}B^{P_k(N) \rightarrow \mathbb{C}[S_k]}_{\Lambda_1, \alpha \rightarrow \Lambda_2, p; \mu}B^{P_k(N) \rightarrow \mathbb{C}[S_k]}_{\Lambda_1, \beta \rightarrow \Lambda_2, p; \nu}.
	\end{aligned} \label{eq: t2 on Q step 1}
\end{equation}
We re-write the Kronecker delta using the completeness relation
\begin{equation}
	\sum_{\Lambda_2', p', \mu'} B^{P_k(N) \rightarrow \mathbb{C}[S_k]}_{\Lambda_1, \gamma \rightarrow \Lambda_2', p'; \mu'} B^{P_k(N) \rightarrow \mathbb{C}[S_k]}_{\Lambda_1, \gamma' \rightarrow \Lambda_2', p'; \mu'} = \delta_{\gamma \gamma'}.
\end{equation}
Inserting this into \eqref{eq: t2 on Q step 1} gives
\begin{equation}
	\begin{aligned}
		&\sum_{\alpha, \beta, \gamma, \gamma', p} D_{\gamma \alpha}^{\Lambda_1}(t_2^{(k)})\delta_{\gamma \gamma'}Q^{\Lambda_1}_{\gamma' \beta}B^{P_k(N) \rightarrow \mathbb{C}[S_k]}_{\Lambda_1, \alpha \rightarrow \Lambda_2, p; \mu}B^{P_k(N) \rightarrow \mathbb{C}[S_k]}_{\Lambda_1, \beta \rightarrow \Lambda_2, p; \nu}= \\
		&\sum_{\alpha, \beta, \gamma,\gamma', p} \sum_{\Lambda_2', p', \mu'} D_{\gamma \alpha}^{\Lambda_1}(t_2^{(k)})  B^{P_k(N) \rightarrow \mathbb{C}[S_k]}_{\Lambda_1, \gamma \rightarrow \Lambda_2', p'; \mu'} B^{P_k(N) \rightarrow \mathbb{C}[S_k]}_{\Lambda_1, \gamma' \rightarrow \Lambda_2', p'; \mu'}  Q^{\Lambda_1}_{\gamma' \beta}B^{P_k(N) \rightarrow \mathbb{C}[S_k]}_{\Lambda_1, \alpha \rightarrow \Lambda_2, p; \mu}B^{P_k(N) \rightarrow \mathbb{C}[S_k]}_{\Lambda_1, \beta \rightarrow \Lambda_2, p; \nu}.
	\end{aligned}  \label{eq: t2 on Q step 2}
\end{equation}
Now note that
\begin{equation}
	\sum_{\gamma, \alpha}D_{\gamma \alpha}^{\Lambda_1}(t_2^{(k)}) B^{P_k(N) \rightarrow \mathbb{C}[S_k]}_{\Lambda_1, \gamma \rightarrow \Lambda_2', p'; \mu'} B^{P_k(N) \rightarrow \mathbb{C}[S_k]}_{\Lambda_1, \alpha \rightarrow \Lambda_2, p; \mu} = \delta_{\Lambda_2 \Lambda_2'} \delta_{\mu' \mu}D_{p' p}^{\Lambda_2}(t_2^{(k)}) =\delta_{\Lambda_2 \Lambda_2'} \delta_{\mu' \mu} \delta_{p'p} \hat{\chi}^{\Lambda_2}(t_2^{(k)}).
\end{equation}
We substitute this into \eqref{eq: t2 on Q step 2} and find
\begin{equation}
	\begin{aligned}
		&\sum_{\alpha, \beta, \gamma,\gamma', p} \sum_{\Lambda_2', p', \mu'} D_{\gamma \alpha}^{\Lambda_1}(t_2^{(k)})  B^{P_k(N) \rightarrow \mathbb{C}[S_k]}_{\Lambda_1, \gamma \rightarrow \Lambda_2', p'; \mu'} B^{P_k(N) \rightarrow \mathbb{C}[S_k]}_{\Lambda_1, \gamma' \rightarrow \Lambda_2', p'; \mu'}  Q^{\Lambda_1}_{\gamma' \beta}B^{P_k(N) \rightarrow \mathbb{C}[S_k]}_{\Lambda_1, \alpha \rightarrow \Lambda_2, p; \mu}B^{P_k(N) \rightarrow \mathbb{C}[S_k]}_{\Lambda_1, \beta \rightarrow \Lambda_2, p; \nu} = \\
		&\sum_{\beta,\gamma', p} \sum_{\Lambda_2', p', \mu'} \delta_{\Lambda_2 \Lambda_2'} \delta_{\mu' \mu} \delta_{p'p} \hat{\chi}^{\Lambda_2}(t_2^{(k)})  B^{P_k(N) \rightarrow \mathbb{C}[S_k]}_{\Lambda_1, \gamma' \rightarrow \Lambda_2', p'; \mu'}  Q^{\Lambda_1}_{\gamma' \beta} B^{P_k(N) \rightarrow \mathbb{C}[S_k]}_{\Lambda_1, \beta \rightarrow \Lambda_2, p; \nu} = \\
		&\sum_{\beta,\gamma', p} \hat{\chi}^{\Lambda_2}(t_2^{(k)})    Q^{\Lambda_1}_{\gamma' \beta} B^{P_k(N) \rightarrow \mathbb{C}[S_k]}_{\Lambda_1, \gamma' \rightarrow \Lambda_2, p; \mu} B^{P_k(N) \rightarrow \mathbb{C}[S_k]}_{\Lambda_1, \beta \rightarrow \Lambda_2, p; \nu} = \hat{\chi}^{\Lambda_2}(t_2^{(k)})  Q^{\Lambda_1}_{\Lambda_2,\mu \nu},
	\end{aligned}  \label{eq: t2 on Q step 3}
\end{equation}
which proves the analogue of \eqref{eq: T_2 eigenvec of Q} in the case of $t_2^{(k)}$.

We define linear operators  on $SP_k(N)$ using multiplication by  $\SWdual{T}_2^{(k)}, t_2^{(k)}$
\begin{equation}
\boxed{ 	\SWdual{T}_2^{(k)}(Q^{\Lambda_1}_{\Lambda_2, \mu \nu}) = \SWdual{T}_2^{(k)}Q^{\Lambda_1}_{\Lambda_2, \mu \nu} = \hat{\chi}^{\Lambda_1}(T_2)Q^{\Lambda_1}_{\Lambda_2, \mu \nu}, } 
\end{equation}
and
\begin{equation}
\boxed{ 	 t_2^{(k)}(Q^{\Lambda_1}_{\Lambda_2, \mu \nu})  = t_2^{(k)}Q^{\Lambda_1}_{\Lambda_2, \mu \nu} = 	\hat{\chi}^{\Lambda_2}(t_2^{(k)})  Q^{\Lambda_1}_{\Lambda_2, \mu \nu}. } 
\end{equation}
That is, the matrix units for  $SP_k(N)$ are eigenvectors of the linear operators associated with $\SWdual{T}_2^{(k)}, t_2^{(k)}$.
%\begin{equation}
%	\SWdual{T}_2Q^{\Lambda_1}_{\Lambda_2, \mu \nu} = \hat{\chi}^{\Lambda_1}(T_2)Q^{\Lambda_1}_{\Lambda_2, \mu \nu}, \qq{and} t_2 Q^{\Lambda_1}_{\Lambda_2, \mu \nu} = 	\hat{\chi}^{\Lambda_2}(t_2)  Q^{\Lambda_1}_{\Lambda_2, \mu \nu}.
%\end{equation}
The eigenvalues are sufficient to determine the subspaces labelled by irreducible representations $\Lambda_1, \Lambda_2$ for $k=1,2,3$ and general $N$.
As discussed in detail in \cite{BPSChargesSymGroup2019}, a larger set of central elements is needed to distinguish different pairs $\Lambda_1, \Lambda_2$ for general $k$ and $N$.

\subsection{Multiplicity labels and maximal commuting subalgebras} 
\label{sec:multiplicity-non-central}   \label{subsec: matrix algebras}
In the previous subsection we described how the subspace spanned by $Q^{ \Lambda_1}_{ \Lambda_2 , \mu  \nu }$ for fixed $\Lambda_1, \Lambda_2$ is a simultaneous eigenspace of central elements $\SWdual{T}_2^{(k)}, t_2^{(k)}$. The subspaces labeled by fixed $\mu, \nu$ are not eigenspaces of any central elements of $SP_k(N)$. Nevertheless, they are eigenspaces of elements that (multiplicatively) generate a maximal commutative subalgebra of $SP_k(N)$.
%
%The multiplicity labels $\mu, \nu$ in the representation basis elements  $Q^{ \Lambda_1}_{ \Lambda_2 , \mu , \nu } $  do not correspond to the eigenvalues of any central element of $SP_k(N)$. For this, it is necessary to pick a set of commuting elements in $SP_k(N)$ that generate a maximal commuting subalgebra. By acting with these elements on the left and right of $SP_k(N)$, in analogy with the linear operators $T^L, T^{R}$ in section \ref{subsec: matrix algebras}, we can distinguish the multiplicity labels.
%For special cases of $\Lambda_1, \Lambda_2$ there is no multiplicity: $\Dim V_{{\Lambda_1} \Lambda_2}^{P_k(N) \rightarrow \mathbb{C}[S_k]} = 1$, and the matrix units $Q^{\Lambda_1}_{\Lambda_2}$ can be constructed by diagonalizing $\SWdual{T_2}$ and $t_2$.
%For $k=1,2$ and general $N \geq 2k$, we will give a set of elements in $SP_k(N)$ whose eigenvalues distinguish all labels.
%The multiplicity free matrix units are constructed for $k=3$.   The role of generating sets for maximal commuting sub-algebras in permutation algebras has been discussed in some generality in \cite{Kimura2008,PCA,BPSChargesSymGroup2019}. The choice of elements that generate a maximal commuting subalgebra for $SP_k( N ) $ for general $k$ is left for the future.
%The matrix units for $SP_k(N)$ can be constructed by diagonalizing $T_p^{P_k(N)}$ and $T_p^{S_k}$

We illustrate this in the simple case of a single matrix algebra.
This is directly relevant, because the matrix units $Q^{ \Lambda_1}_{ \Lambda_2 , \mu  \nu }$ form (are isomorphic to) a matrix algebra $M_n$ with $n=\Dim V_{\Lambda_1 \Lambda_2}^{P_k(N)\rightarrow \mathbb{C}[S_k]}$, for fixed $\Lambda_1, \Lambda_2$.
The matrix algebra $M_n$ has a basis of matrix units $E_{rs}$ for $r,s=1,\dots,n$. These are just the elementary matrices with zeroes everywhere except in row $r$, column $s$ where there is a one. In this explicitly realized algebra, it is straight-forward to verify that
\begin{equation}
	E_{rs}E_{r's'} = \delta_{sr'} E_{rs'}. \label{eq: elem matrix product}
\end{equation} 
It follows from equation \eqref{eq: elem matrix product} that
\begin{equation}
	E_{tt}E_{rs} = \delta_{tr}E_{ts} = \begin{cases}
		E_{rs} \qq{if $r=t$,}\\
		0 \qq{otherwise.}
	\end{cases}
\end{equation}
This fact will be useful in what follows.

We will now define a pair of linear operators acting on $M_n$ whose eigenvalues uniquely determine the indices $r,s$ on $E_{rs}$.
Let
\begin{equation}
	T = 1E_{11} + 2E_{22} + \dots + n E_{nn},
\end{equation}
and $T^L, T^R$ be the linear operators on $M_n$ defined by left and right action of $T$ respectively,
\begin{equation}
	T^L(E_{rs}) = T E_{rs}, \quad T^R (E_{rs}) = E_{rs}T.
\end{equation}
The $n^2 \times n^2$ matrix $(T^L)_{rs}^{tu}$ associated with the linear operator $T^L$ has eigenvalues $\{1, 2, \dots, n\}$ (each one is $n$-fold degenerate) with eigenvectors $E_{rs}$,
\begin{equation}
	\sum_{t,u} (T^L)_{rs}^{tu}E_{tu} = T^{L} (E_{rs}) = r E_{rs}.
\end{equation}
Similarly for the matrix $(T^R)_{rs}^{tu}$ associated with the linear operator $T^R$,
\begin{equation}
	\sum_{t,u} (T^R)_{rs}^{tu} E_{tu}  = T^R (E_{rs}) = sE_{rs}.
\end{equation}
The operators $T^L$ and $T^R$ commute, and their simultaneous eigenvectors $E_{rs}$ have eigenvalues $r$ and $s$, respectively.

The algebra spanned by $\{E_{11}, E_{22}, \dots, E_{nn}\}$ is a maximal commuting subalgebra of $M_n$.
It is multiplicatively generated by $T$. In particular (see \cite[Lemma~3.3.1]{BPSChargesSymGroup2019} or \cite[Lemma~2.1]{CanonIdemPots})
\begin{equation}
	E_{rr} = \prod_{s \neq r} \frac{(T-s)}{(r-s)}.
\end{equation}

These ideas generalize to $Q^{\Lambda_1}_{\Lambda_2, \mu \nu}$, and in the next section we will give the appropriate operators corresponding to $T^L, T^R$ for $SP_{2}(N)$.

\subsection{Construction of low degree representation bases}
We now use the tools presented in this section to explicitly construct the representation basis elements as sums of diagrams, for $k=1,2,3$ and large $N$.
Tables of the representation basis elements expanded in terms of diagrams are found in appendix \ref{apx: matrix units}.
The associated Sage code can be found together with the arXiv version of this paper.

\subsubsection{Degree one basis}
For $k=1$ it is enough to use $\SWdual{T_2}^{(1)}$ to distinguish the irreducible representations.
We expect to find matrix units
\begin{equation}
	Q^{[N]}_{[1]}, Q^{[N-1,1]}_{[1]},
\end{equation}
since $S_1$ only has the trivial representation and the decomposition in \eqref{eq: schur-weyl simple} only contains irreducible representations $[N]$ and $[N-1,1]$ of $P_1(N)$.

The map
\begin{equation}
	T_2 \mapsto \SWdual{T}_2^{(1)}
\end{equation}
is given by (see the section called Murphy elements for $\mathbb{C}A_k(N)$ in \cite{Halverson2004})
\begin{equation}
	\SWdual{T}_2^{(1)} = \frac{N(N-3)}{2}\PAdiagram[]{1}{1/-1} + \PAdiagram[]{1}{}. \label{eq: T2 P1N}
\end{equation}

It is straight-forward to diagonalize $\SWdual{T}_2^{(1)}$ acting on $P_1(N)$ from the left.
Define
\begin{equation}
	Q^{[N]}_{[1]} = \frac{1}{N}\PAdiagram[]{1}{}, \quad Q^{[N-1,1]}_{[1]} = \PAdiagram[]{1}{1/-1}-\frac{1}{N}\PAdiagram[]{1}{},
\end{equation}
they satisfy
\begin{equation}
	Q^{[N]}_{[1]} Q^{[N-1,1]}_{[1]} = 0, \quad Q^{[N]}_{[1]} Q^{[N]}_{[1]} = Q^{[N]}_{[1]}, \quad Q^{[N-1,1]}_{[1]} Q^{[N-1,1]}_{[1]} = Q^{[N-1,1]}_{[1]}.
\end{equation}
and have eigenvalues
\begin{equation}
	\begin{aligned}
		\SWdual{T}_2^{(1)} Q^{[N]}_{[1]} &= \frac{N(N-1)}{2} Q^{[N]}_{[1]}, \\
		\SWdual{T}_2^{(1)} Q^{[N-1,1]}_{[1]} &= \frac{N(N-3)}{2} Q^{[N-1,1]}_{[1]},
	\end{aligned}
\end{equation}
which are exactly equal to the normalized characters. Note that $S_1$ has no non-trivial representations, and $t_2^{(1)} = 0$, which is consistent with the normalized character $\tfrac{k(k-1)}{2}=0$ of the trivial representation.

The orthogonal basis elements for $\Hilbertspace^{(1)}$, corresponding to these matrix units, are
\begin{equation}
	\ket{Q^{[N]}_{[1]}} = \frac{1}{N}\sum_{i_1, i_{1'}} (a^{\dagger})^{i_{1'}}_{i_1}\ket{0}, \qq{and} 	\ket{Q^{[N-1,1]}_{[1]}} =\sum_{i_1} (a^{\dagger})^{i_{1}}_{i_1} - \frac{1}{N} \sum_{i_1, i_{1'}} (a^{\dagger})^{i_{1'}}_{i_1} \ket{0}.
\end{equation}

\subsubsection{Degree two basis}\label{sec: degree two basis}
The procedure was particularly easy at degree one because $S_1$ is trivial, and there are no multiplicities appearing. For $k=2$ we have the sign representation $[1,1]$ and the trivial representation $[2]$ of $S_2$, and pairs of irreducible representations $\Lambda_1,\Lambda_2$ appear with multiplicity larger than one.
To distinguish multiplicities we will have to introduce non-central elements, as discussed in subsection \ref{sec:multiplicity-non-central}.
%The extra operators are closely related to the operator $\SWdual{T_2}^{(1)} \in \mathcal{Z}(P_1)$ defined in equation \eqref{eq: T2 P1N}, but are now embedded into $SP_2(N)$ by adding edges and symmetrising.

At degree two, the partition algebra element we use to distinguish $\Lambda_1$ is \cite{Halverson2004}
\begin{equation}\label{eq:T2bar2}
	\SWdual{T}_2^{(2)} = \begin{aligned}
		\frac{(N-2)(N-3)-4}{2}&\PAdiagram[]{2}{1/-1,2/-2} + \PAdiagram[]{2}{1/-1} + \PAdiagram[]{2}{2/-2} + \PAdiagram[2/1, -1/-2]{2}{} + \PAdiagram[]{2}{1/-2,2/-1} + N\PAdiagram[2/1]{2}{1/-1,2/-2} \\
		& - \PAdiagram[2/1]{2}{2/-2} - \PAdiagram[2/1]{2}{2/-2} -\PAdiagram[-1/-2]{2}{1/-1} - \PAdiagram[2/1]{2}{1/-1}.
	\end{aligned}
\end{equation}
As a linear map (acting on the left or right) on $P_2(N)$, it has eigenvalues
\begin{align} \nonumber
	\SWdual{T}_2^{(2)}(Q^{[N]}_{\Lambda_2, \mu \nu}) &= \frac{N(N-1)}{2} Q^{[N]}_{\Lambda_2, \mu \nu}, \\ \nonumber
	\SWdual{T}_2^{(2)}(Q^{[N-1,1]}_{\Lambda_2, \mu \nu}) &= \frac{N(N-3)}{2} Q^{[N-1,1]}_{\Lambda_2, \mu \nu}, \\ \nonumber
	\SWdual{T}_2^{(2)}(Q^{[N-2,2]}_{\Lambda_2, \mu \nu}) &= \frac{(N-1)(N-4)}{2} Q^{[N-2,2]}_{\Lambda_2, \mu \nu}, \\
	\SWdual{T}_2^{(2)}(Q^{[N-2,1,1]}_{\Lambda_2, \mu \nu}) &= \frac{N(N-5)}{2} Q^{[N-2,1,1]}_{\Lambda_2, \mu \nu}.
\end{align}
The element we use to distinguish $\Lambda_2$ is
\begin{equation}
	t_2^{(2)} = \PAdiagram[]{2}{1/-2,2/-1}.
\end{equation}
The eigenvalues of the corresponding linear map are $1$ for $[2]$ and $-1$ for $[1,1]$.

The non-central element we will use to distinguish multiplicities is
\begin{equation}
	\SWdual{T}_{2,1}^{(2)} = \PAdiagram[]{2}{-1/1} + \PAdiagram[]{2}{-2/2}. \label{eq: P1N tensor 1a}
\end{equation}
It is closely related to $\SWdual{T}_2^{(1)} \in \mathcal{Z}(P_1)$ in equation \eqref{eq: T2 P1N} because
\begin{equation}
	\SWdual{T}_2^{(1)} \otimes 1 + 1 \otimes \SWdual{T}_2^{(1)} = \PAdiagram[]{2}{-1/1} + \PAdiagram[]{2}{-2/2} + N(N-3)\PAdiagram[]{2}{-2/2,-1/1}. \label{eq: P1N tensor 1b}
\end{equation}
Roughly speaking, $\SWdual{T}_{2,1}^{(2)}$ comes from the embedding of $\SWdual{T}_2^{(1)}$ into $SP_2(N)$ by adding strands. Symmetrization has been used to ensure that we have an element in $SP_2(N)$. 

To determine the multiplicity labels we need to act from the left as well as the right using $\SWdual{T}_{2,1}^{(2)}$.
We define $\SWdual{T}_{2,1}^{(2),L}$ and $\SWdual{T}_{2,1}^{(2), R}$ acting on $d \in P_2(N)$ by
\begin{equation}
	\SWdual{T}_{2,1}^{(2), L} d = \SWdual{T}_{2,1}^{(2)} d, \quad \SWdual{T}_{2,1}^{(2), R}d = d \SWdual{T}_{2,1}^{(2)}. \label{eq: T2 left right action}
\end{equation}

Appendix \ref{apx: matrix units} gives a representation theoretic argument for why these operators fully distinguish all labels on matrix units, together with a complete table of all $k=2$ matrix units.
As an example, we find a matrix unit (see \eqref{eq: k=2 matrix unit example})
\begin{equation}
	(Q^{[N-2,1,1]}_{[1,1]})_{22} = \frac{1}{N} \PAdiagram[]{2}{1/-1} - \frac{1}{N} \PAdiagram[]{2}{2/-1} - \frac{1}{N} \PAdiagram[]{2}{1/-2} + \PAdiagram[]{2}{1/-2, 2/-1} + \frac{1}{N} \PAdiagram[]{2}{2/-2} - \PAdiagram[]{2}{1/-1, 2/-2},
\end{equation}
which corresponds to the (unnormalized) $S_N$ invariant state
\begin{equation}
	\ket{(Q^{[N-2,1,1]}_{[1,1]})_{22} } = \frac{2}{N} \Big( \sum_{i,j,k=1}^N \qty[(a^\dagger)^i_i (a^\dagger)^j_k - (a^\dagger)^j_i (a^\dagger)^i_k] + \sum_{i,j=1}^N \qty[(a^\dagger)^i_j (a^\dagger)^j_i - (a^\dagger)^i_i (a^\dagger)^j_j] \Big) \ket{0}.
\end{equation}

\subsubsection{Degree three basis} \label{sec: degree three basis}
The multiplicity free matrix units for $k=3$ have $\Lambda_1 = [N-3,3], [N-3,2,1], [N-3,1,1,1]$.
To find the corresponding linear combinations of diagrams it is sufficient to find eigenvectors of $\SWdual{T}_2^{(3)}$ defined by
\begin{align} \nonumber
	\frac{1}{3!}\SWdual{T}_2^{(3)} = &\SPAdiagram[2/1, 3/2, -1/-2, -2/-3]{3}{} + \SPAdiagram[2/1, -1/-2, -2/-3]{3}{1/-1} - N \SPAdiagram[2/1, 3/2, -1/-2, -2/-3]{3}{1/-1} - \SPAdiagram[2/1, -1/-2, -2/-3]{3}{3/-3} - \SPAdiagram[2/1, -2/-3]{3}{3/-1, 2/-3} + \SPAdiagram[2/1, 3/2, -2/-3]{3}{3/-3} \\ \nonumber
	- &\SPAdiagram[-1/-3]{3}{1/-1,2/-2} + \SPAdiagram[2/1, -1/-3]{3}{2/-2, 3/-3} + \SPAdiagram[]{3}{1/-3, 2/-2, 3/-1} + \SPAdiagram[2/1, -1/-2]{3}{2/-3, 3/-2} - \SPAdiagram[2/1, 3/2, -1/-2]{3}{3/-3} + \SPAdiagram[2/1, -1/-2]{3}{3/-3} \\
	+ &(N-1) \SPAdiagram[2/1, -1/-2]{3}{1/-1, 3/-3} - \SPAdiagram[2/1]{3}{2/-2, 3/-3} + \SPAdiagram[]{3}{2/-2, 3/-3} + \frac{(N-1)(N-6)}{2} \SPAdiagram[]{3}{1/-1, 2/-2, 3/-3} \label{eq:T2bar3}
\end{align}
with eigenvalues
\begin{align} \nonumber
	 \SWdual{T}_2^{(3)}(Q^{[N-3,3]}_{\Lambda_2, \mu \nu}) &= \frac{(N-3)(N-4)}{2} Q^{[N-3,3]}_{\Lambda_2, \mu \nu}, \\ \nonumber
	 \SWdual{T}_2^{(3)}(Q^{[N-3,2,1]}_{\Lambda_2, \mu \nu}) &= \frac{(N-1)(N-6)}{2} Q^{[N-3,2,1]}_{\Lambda_2, \mu \nu}, \\
	\SWdual{T}_2^{(3)}(Q^{[N-3,1,1,1]}_{\Lambda_2, \mu \nu}) &= \frac{N(N-7)}{2} Q^{[N-3,1,1,1]}_{\Lambda_2, \mu \nu}.
\end{align}
The square brackets in \eqref{eq:T2bar3} denote $S_3$ symmetrization as in equation \eqref{eq: SP_kN basis}. Explicit expansions of these matrix units in terms of diagrams are given in appendix \ref{apx: matrix units}.

\section{Exactly solvable permutation invariant matrix harmonic oscillator}
\label{sec: PIMQM}
The simplest quantum mechanical matrix Hamiltonian we considered in section \ref{sec: MQM} is invariant under the symmetric group action 
\bea 
\sigma : X_{ i j } \rightarrow X_{ \sigma (i) \sigma(j) },  \quad \forall \sigma \in S_N.
\eea
It is also invariant under the much larger symmetry of continuous transformations by $U(N^2)$.  In this section we generalise the quadratic potential to the most general quadratic function 
$V(X)$ invariant under the above permutation symmetry. We will thus  present a quantum mechanical model of $N^2$ matrix variables $X_{ij}$ in a permutation invariant  quadratic potential $V(X)$.  The most general permutation invariant quadratic action in a zero-dimensional matrix model was constructed in \cite{PIGMM} using representation theory. Borrowing these techniques, we explicitly construct an $11$ parameter family of permutation invariant quadratic potentials. The corresponding Hamiltonian is exactly diagonalizable. In general, the diagonalization only involves diagonalizing a $3\times 3$ symmetric matrix and a $2 \times 2$ symmetric matrix. We describe the spectrum of the full Hamiltonian and discuss the degeneracy when the quanta of energy are generic, and when they satisfy integrality properties. In the former case we are able to give a lower bound on the order of the degeneracy, this is given in equation \eqref{eq: lower bound energy degeneracy}.  In the latter case, the degeneracy can be phrased in terms of an integer partition problem. The integer partition problem has a solution in terms of a canonical partition function (generating function) given by equation \eqref{eq: partition function}.  We end this section in \ref{sec: energy eigenbasis} with a brief discussion of the role that the representation basis could play in simplifying the diagonalisation of $H$, given in equation \eqref{eq: hamiltonian}, on $\Hilbertspace$.

\subsection{Construction}
A matrix harmonic oscillator in a potential is described by the Lagrangian
\begin{equation}
	L = \frac{1}{2}\sum_{i,j=1}^N \partial_t X_{ij} \partial_t X_{ij} - \frac{1}{2}V(X). \label{eq: lagrangian}
\end{equation}
We take the potential to be a general quadratic $S_N$ (permutation) invariant potential
\begin{equation}
	V(X_{ij}) = V(X_{\sigma(i)\sigma(j)}). \label{eq: potential invariance}
\end{equation}
The action of $S_N$ on $X_{ij}$ defined in \eqref{eq: potential invariance} corresponds to the diagonal action on the tensor product $V_N \otimes V_N$. This is given in \eqref{eq: left action SN} for general $k$, for the $k=2$ case at hand we have
\begin{equation}
	\mathcal{L}(\sigma^{-1})(e_i \otimes e_j) = e_{\sigma(i)} \otimes e_{\sigma(j)}.
\end{equation}
%Unless there is a possibility of confusion, we will use the notation $\mathcal{L}(\sigma)$ for the action of $\sigma$ on $V_N$ as well as the diagonal action on $V_N \otimes V_N$.

The vector space $V_N \otimes V_N$ is reducible with respect to the diagonal action. There exists an isomorphism
\begin{equation}
	V_N \otimes V_N \cong 2V_{[N]}^{S_N} \oplus 3V_{[N-1,1]}^{S_N} \oplus V_{[N-2,2]}^{S_N} \oplus V_{[N-2,1,1]}^{S_N}, \label{eq: VN tensor VN decomp}
\end{equation}
into irreducible subspaces. The representation $V_{[N]}^{S_N}$ is the one-dimensional trivial representation of $S_N$. The representations $V_{[N-1,1]}^{S_N}, V_{[N-2,2]}^{S_N}, V_{[N-2,1,1]}^{S_N}$ are non-trivial irreducible representations of $S_N$, labelled by integer partitions of $N$. Detailed descriptions, including explicit constructions of irreducible representations of $S_N$ can be found in \cite{Hamermesh1962, Sagan2013}. The dimensions of the non-trivial irreducible representations in \eqref{eq: VN tensor VN decomp} are respectively,
\begin{equation}
	N-1, \hspace{4pt} (N-1)(N-2)/2, \hspace{4pt} N(N-3)/2.
\end{equation}
We take the RHS of the isomorphism \eqref{eq: VN tensor VN decomp} to be a vector space with orthonormal basis $X_{a}^{\Lambda, \alpha}$ labelled by
\begin{equation}\label{Laaranges} 
	\begin{aligned}
		\Lambda &\in \Big\{ [N], [N-1,1], [N-2,2], [N-2,1,1] \Big\}, \\
		a &\in  \qty{1, \dots, \Dim V_{\Lambda}^{S_N}}, \\
		\alpha &\in \qty{1, \dots, \Mult(V_N \otimes V_N \rightarrow V_{\Lambda}^{S_N})}.
	\end{aligned}
\end{equation}
By definition the Clebsch-Gordan coefficients $C_{a,ij}^{\Lambda, \alpha}$ are the matrix elements of the equivariant map between the two sides of equation \eqref{eq: VN tensor VN decomp},
\begin{equation}
	X_{a}^{\Lambda, \alpha} = \sum_{i,j} C_{a,ij}^{\Lambda, \alpha} X_{ij}. \label{eq: matrix to rep position change of basis}
\end{equation}
As a consequence, they have the following property
\begin{equation}
	\sum_{i,j} C_{a,ij}^{\Lambda, \alpha} X_{\sigma^{-1}(i)\sigma^{-1}(j)} = \sum_{b} D^{\Lambda}_{ba}(\sigma)X_{b}^{\Lambda, \beta} \label{eq: CGC equivariance}
\end{equation}
where $D^{\Lambda}(\sigma)$ is an irreducible (unitary and real) matrix representation of $\sigma \in S_N$.

%The construction of the quadratic potential in equation \eqref{eq: lagrangian}, does not require detailed knowledge of the Clebsch-Gordan coefficients.
In the representation basis the potential has a simple form,
\begin{equation}
	V(X) = \sum_{\Lambda, \alpha, \beta, a}  X_{a}^{\Lambda, \alpha} g^{\Lambda}_{\alpha \beta} X_{a}^{\Lambda, \beta},
\end{equation}
where $g^{\Lambda}_{\alpha \beta} $ are symmetric matrices.
To define a system with energy bounded from below they are required to have non-negative eigenvalues.
Translating back to the original basis gives
\begin{equation}
	V(X) =  \sum_{\Lambda, \alpha, \beta, a} \sum_{i,j,k,l} C_{a,ij}^{\Lambda, \alpha}  g^{\Lambda}_{\alpha \beta} C_{a,kl}^{\Lambda, \beta} X_{ij} X_{kl}.
\end{equation}
We define the tensors 
\begin{equation}
	Q_{ijkl}^{\Lambda,\alpha \beta} = \sum_{a} C_{a,ij}^{\Lambda, \alpha} C_{a,kl}^{\Lambda, \beta}
\end{equation}
and write the potential $V(X)$ as
\begin{equation}
	V(X) = \sum_{\Lambda, \alpha, \beta} \sum_{i,j,k,l} Q_{ijkl}^{\Lambda,\alpha \beta}g^{\Lambda}_{\alpha \beta}X_{ij} X_{kl}. \label{eq: invariant potential}
\end{equation}
The tensors $Q_{ijkl}^{\Lambda,\alpha \beta}$ are known explicitly \cite{PIGMM}.
For example,
\begin{align}
	&Q_{ijkl}^{[N],11} = \frac{1}{N^2}, \label{eq: Q example1 }\\
	&Q_{ijkl}^{[N],22} = \frac{1}{N-1}\qty(\delta_{ij}\delta_{kl} - \frac{1}{N}\delta_{ij} - \frac{1}{N}\delta_{kl} - \frac{1}{N^2}).  \label{eq: Q example2}
\end{align}

Their construction using Clebsch-Gordan coefficients means that they satisfy 
\begin{equation}
	Q_{\sigma(i)\sigma(j)\sigma(k)\sigma(l)}^{\Lambda,\alpha \beta} = Q_{ijkl}^{\Lambda,\alpha \beta}.
\end{equation}
This follows from the equivariance property \eqref{eq: CGC equivariance}
\begin{equation}
	\begin{aligned}
		Q_{\sigma(i)\sigma(j)\sigma(k)\sigma(l)}^{\Lambda,\alpha \beta} &= \sum_{a} C_{a,\sigma(i)\sigma(j)}^{\Lambda, \alpha} C_{a,\sigma(k)\sigma(l)}^{\Lambda, \beta} = \sum_{a,b,c} C_{b,ij}^{\Lambda, \alpha} C_{c,kl}^{\Lambda, \beta} D^{\Lambda}_{ab}(\sigma)D^{\Lambda}_{ac}(\sigma) \\
		&=\sum_{b,c} C_{b,ij}^{\Lambda, \alpha} C_{c,kl}^{\Lambda, \beta} \delta_{bc} = Q_{ijkl}^{\Lambda,\alpha \beta}.
	\end{aligned}
\end{equation}
Going to the second line uses $D^{\Lambda}_{ab}(\sigma)=D^{\Lambda}_{ba}(\sigma^{-1})$ which follows from the fact that representation matrices for $S_N$ can be chosen to real and unitary, i.e. orthogonal matrices. 

\subsection{Spectrum}
The full Hamiltonian with quadratic potential given in \eqref{eq: invariant potential} can be diagonalized using oscillators. We will see that diagonalizing the Hamiltonian only requires the diagonalization of a set of small parameter matrices (one $3 \times 3$ and another $2\times 2$), despite having a potentially large number of harmonic oscillators ($N^2$).

The full Lagrangian in the representation basis is
\begin{equation}
	L = \sum_{\Lambda, \alpha,\beta, a} \delta_{\alpha\beta}\partial_t X_{a}^{\Lambda, \alpha} \partial_t X_{a}^{\Lambda, \beta}-X_{a}^{\Lambda, \alpha} g^{\Lambda}_{\alpha \beta} X_{a}^{\Lambda, \beta}.
\end{equation}
It describes a set of coupled harmonic oscillators.
We write the Lagrangian in decoupled form in the usual way. Let $\Omega_{\alpha \beta}^{\Lambda} = (\omega^{\Lambda}_{\alpha})^2 \delta_{\alpha \beta}$ be the diagonal matrix\footnote{We assume the eigenvalues are positive such that the spectrum of the Hamiltonian is bounded from below. Therefore, we may write the eigenvalues as squares without loss of generality.} such that
\begin{equation} \label{eq: full Hamiltonian metric}
	g_{\alpha \beta}^{\Lambda} = \sum_{\gamma, \delta} U_{\alpha \gamma}^{\Lambda} \Omega_{\gamma \delta}^{\Lambda}U_{\beta \delta}^{\Lambda},
\end{equation}
where $U^{\Lambda}$ are orthogonal change of basis matrices. In the decoupled basis
\begin{equation}
	S_a^{\Lambda, \alpha} = \sum_\beta X_a^{\Lambda, \beta}U^{\Lambda}_{\beta \alpha}, \label{eq: decoupled basis}
\end{equation}
we have
\begin{equation}
	L =\sum_{\Lambda, \alpha, a}  \frac{1}{2} \partial_t S_{a}^{\Lambda, \alpha} \partial_t S_{a}^{\Lambda, \alpha} -\frac{1}{2} (\omega^{\Lambda}_{\alpha})^2 S_{a}^{\Lambda, \alpha} S_{a}^{\Lambda, \alpha}.
\end{equation}
The canonical momenta are given by
\begin{equation}
	\Sigma^{\Lambda, \alpha}_a = \partial_t S_{a}^{\Lambda, \alpha}.
\end{equation}
The new canonical coordinates satisfy
\begin{equation}
	\comm{\Sigma^{\Lambda, \alpha}_a}{S^{\Lambda', \beta}_b} = i\delta^{{\Lambda \Lambda'}}\delta^{\alpha \beta}\delta_{ab},
\end{equation}
since $U^{\Lambda}$ are orthogonal matrices.

The corresponding Hamiltonian,
\begin{equation} \label{eq: perm inv Hamiltonian}
	H = \frac{1}{2}\sum_{\Lambda, \alpha, a}  \Sigma_{a}^{\Lambda, \alpha} \Sigma_{a}^{\Lambda, \alpha} + (\omega^{\Lambda}_{\alpha})^2 S_{a}^{\Lambda, \alpha} S_{a}^{\Lambda, \alpha},
\end{equation}
is diagonalized by introducing oscillators
\begin{equation}
	\begin{aligned}
		&S_{a}^{\Lambda, \alpha} = \sqrt{\frac{1}{2\omega^{\Lambda}_{\alpha}}}((A^\dagger)^{\Lambda, \alpha}_a + A^{\Lambda, \alpha}_a), \\
		&\Sigma_{a}^{\Lambda, \alpha} = i\sqrt{\frac{\omega^{\Lambda}_{\alpha}}{2}}((A^\dagger)^{\Lambda, \alpha}_a - A^{\Lambda, \alpha}_a),
	\end{aligned}
\end{equation}
which satisfy
\begin{equation}
	\comm{A^{\Lambda, \alpha}_a}{(A^\dagger)^{\Lambda', \alpha'}_{a'}} = \delta^{\Lambda \Lambda'} \delta^{\alpha \alpha'} \delta_{aa'}.
\end{equation}
In the oscillator basis, the normal ordered Hamiltonian has the form
\begin{equation}\label{eq: hamiltonian}
\boxed{ 	H = \sum_{\Lambda, \alpha, a} \omega^{{\Lambda}}_{\alpha} (A^\dagger)^{\Lambda, \alpha}_a A^{\Lambda, \alpha}_a  \, . } 
\end{equation} 
Defining number operators $\widehat{N}^{\Lambda, \alpha}_a$  and
$ \widehat{N}^{\Lambda, \alpha} $ : 
\begin{equation}
\widehat{N}^{\Lambda, \alpha}_a = (A^\dagger)^{\Lambda, \alpha}_a A^{\Lambda, \alpha}_a \, , 
\end{equation} 
\begin{equation} 
 \widehat{N}^{\Lambda, \alpha} = \sum_{a} \widehat{N}^{\Lambda, \alpha}_a \,  , 
\end{equation} 
we may write 
\bea 
	H = \sum_{\Lambda, \alpha, a} \widehat{N}^{\Lambda, \alpha}_a =
	 \sum_{\Lambda, \alpha} \widehat{N}^{\Lambda, \alpha} \, . 
\eea
The energy quanta $\omega^{\Lambda}_\alpha$ do not depend on the oscillator state index $a$. This is a manifestation of the $S_N$ invariance of the Hamiltonian $H$.

The Hilbert space $\mathcal{H}^{(k)}$ has a basis of energy eigenstates
\begin{equation}
	\prod_{\substack{\Lambda \in \{[N],[N-1,1],[N-2,2],[N-2,1,1]\} \\ \alpha \in \{1,\dots \Mult(V_N \otimes V_N \rightarrow V_{\Lambda}^{S_N})\} \\ a \in \{1, \dots, \Dim V_{\Lambda}^{S_N}\}}} \frac{\qty[(A^\dagger)^{\Lambda, \alpha}_a]^{N^{\Lambda, \alpha}_a}}{\sqrt{N^{\Lambda,\alpha}_a!}}\ket{0} \label{eq: rep basis fock space}
\end{equation}
where $k = \sum_{\Lambda, \alpha, a} N^{\Lambda,\alpha}_a$ is the eigenvalue of the (total) number operator
\begin{equation}
	\widehat{N} = \sum_{\Lambda, \alpha,a} \widehat{N}^{\Lambda, \alpha}_a,
\end{equation}
and $N_a^{\Lambda, \alpha}$ is the eigenvalue of $\widehat{N}^{\Lambda, \alpha}_a$.

Since the Hamiltonian \eqref{eq: hamiltonian} is a linear combination of number operators $\widehat{N}^{\Lambda, \alpha}$, it is natural to organize $\mathcal{H}^{(k)}$ into eigenspaces of $\widehat{N}^{\Lambda, \alpha}$ with eigenvalues $N^{\Lambda,\alpha} = \sum_a N^{\Lambda, \alpha}_a$ satisfying $k = \sum_{\Lambda, \alpha} N^{\Lambda, \alpha}$. Diagonalizing the number operators $\widehat{N}^{\Lambda, \alpha}$ organizes $\mathcal{H}^{(k)}$ into subspaces
\begin{equation}
	\mathcal{H}^{(k)} \cong \bigoplus_{\Sigma N^{\Lambda, \alpha} = k} \hspace{4pt} \bigotimes_{\Lambda, \alpha} \mathcal{H}^{[N^{\Lambda, \alpha}]}, \label{eq: Hk decomposition by number operators}
\end{equation}
where
\begin{equation} \label{eq: number operator decomp}
	\mathcal{H}^{[N^{\Lambda, \alpha}]} \cong \Sym^{N^{\Lambda,\alpha}}(V^{S_N}_{\Lambda}).
\end{equation}
Each summand in \eqref{eq: Hk decomposition by number operators} is a vector space of dimension	
\begin{align}
	\Dim &\qty(\bigotimes_{\Lambda, \alpha} \mathcal{H}^{[N^{\Lambda, \alpha}]})= \prod_{\Lambda, \alpha} \binom{\Dim V_{\Lambda}^{S_N} + N^{\Lambda ,\alpha} - 1}{N^{\Lambda ,\alpha}} = \nonumber \\
	&\binom{1 + N^{[N],1} - 1}{N^{[N],1}}\binom{1 + N^{[N],2} - 1}{N^{[N],2}} \times \nonumber \\
	&\binom{N-1 + N^{[N-1,1],1} - 1}{N^{[N-1,1],1}}\binom{N-1 + N^{[N-1,1],2} - 1}{N^{[N-1,1],2}} \binom{N-1 + N^{[N-1,1],3} - 1}{N^{[N-1,1],3}} \times \nonumber \\
	&\binom{(N-1)(N-2)/2 + N^{[N-2,2]} - 1}{N^{[N-2,2]}}\binom{N(N-3)/2 + N^{[N-2,1,1]} - 1}{N^{[N-2,2]}} = \nonumber \\
	%&\phantom{\qty(\bigotimes_{\Lambda, \alpha} \Sym^{N^{\Lambda,\alpha}}(V^{S_N}_{\Lambda})) }  \nonumber \\
	&\binom{N-2 + N^{[N-1,1],1}}{N^{[N-1,1],1}}\binom{N-2 + N^{[N-1,1],2}}{N^{[N-1,1],2}} \binom{N-2 + N^{[N-1,1],3}}{N^{[N-1,1],3}} \times  \nonumber \\
	&\binom{N(N-3)/2 + N^{[N-2,2]}}{N^{[N-2,2]}}\binom{N(N-3)/2 -1 + N^{[N-2,1,1]}}{N^{[N-2,2]}}. \label{eq: lower bound energy degeneracy}
\end{align} 
The vectors in $\bigotimes_{\Lambda, \alpha} \mathcal{H}^{[N^{\Lambda, \alpha}]}$ have energy
\begin{equation}
	E  ( \{ N^{ \Lambda , \alpha }  \} ) = \sum_{\Lambda, \alpha} N^{\Lambda,\alpha} \omega^{\Lambda}_{\alpha}.
\end{equation}
Equation \eqref{eq: lower bound energy degeneracy} thus gives the degeneracy of energy eigenstates for the specified integers $ \{ N^{ \Lambda , \alpha } \}  $, associated with $\Lambda, \alpha $ as given in \eqref{Laaranges}.   This puts a lower bound on the degeneracy of energy eigenstates. Further degeneracy may occur for particular choices of the constants $\omega^{\Lambda}_{\alpha}$, which can lead to the same  numerical value of 
$E  ( \{ N^{ \Lambda , \alpha } \}  ) $ for different choices of $\{ N^{ \Lambda , \alpha }\} $. 

\subsection{Canonical partition function}
  The canonical partition function  is  defined as 
\begin{equation}
	Z ( \beta ) = { \rm Tr }_{\mathcal{H}} ~ e^{ - \beta H }  = \sum_{ \mathcal{E} } N ( \mathcal{E} ) e^{ - \beta \mathcal{E} }  
\end{equation}
where $N (\mathcal{E} )$ is the degeneracy of eigenstates at energy $\mathcal{E}$ and $ \beta $ is the inverse temperature. 

The binomial factors in \eqref{eq: lower bound energy degeneracy} arise in the expansion of simple rational functions. Defining $ x= e^{- \beta} $ for convenience,   we can therefore write  
\begin{equation}
\begin{aligned}
	Z ( \beta ) = 
		&\frac{1}{(1-x^{\omega^{[N]}_1})(1-x^{\omega^{[N]}_2})}\frac{1}{(1-x^{\omega^{[N-1,1]}_1})^{N-1}(1-x^{\omega^{[N-1,1]}_2})^{N-1}(1-x^{\omega^{[N-1,1]}_3})^{N-1}} \times \\ 
		&\frac{1}{(1-x^{\omega^{[N-2,2]}})^{(N-1)(N-2)/2}(1-x^{\omega^{[N-2,1,1]}})^{N(N-3)/2}}. \label{eq: partition function}
\end{aligned}
\end{equation}
When the quanta of energy ($\omega^{\Lambda}_\alpha$) in \eqref{eq: hamiltonian} are integers, the possible state energies $\mathcal{E}$ are integers and $N(\mathcal{E})$ is related to what we refer to as an integer partition problem. The integer partition problem is the following: Pick any integer $\mathcal{E}$, enumerate the set of solutions (choices of $N^{\Lambda,\alpha}_a$) to
\begin{equation}
	\mathcal{E} = \sum_{\Lambda, \alpha, a} N^{\Lambda,\alpha}_a \omega^{\Lambda}_{\alpha}. \label{eq: degeneracy partition problem}
\end{equation}
The number of solutions is equal to $N(\mathcal{E})$ and a single solution is denoted $N^{\Lambda,\alpha}_a(\mathcal{E})$. This problem depends on $N$ because the state label $a$ ranges over $\{1, \dots, \DimSN{\Lambda}\}$.
Fortunately the $N$-dependence can be factorized, due to the $S_N$ symmetry of the problem, thus greatly simplifying the problem.

To see this, consider the $N$-independent integer partition problem
\begin{equation}
	\mathcal{E} = \sum_{\Lambda, \alpha} N^{\Lambda, \alpha}  \omega^{\Lambda}_{\alpha}, \label{eq: deg part problem 2}
\end{equation}
where a solution is given by a list of seven integers $N^{\Lambda, \alpha}(\mathcal{E})$.
For every solution $N^{\Lambda, \alpha}(\mathcal{E})$ to \eqref{eq: deg part problem 2} the number of solutions to the integer partition problem in \eqref{eq: degeneracy partition problem} is given by 
\begin{equation}
	\Dim \qty(\bigotimes_{\Lambda, \alpha} \mathcal{H}^{[N^{\Lambda, \alpha}(\mathcal{E})]}). 
\end{equation}
In this sense, the $N$-dependence in the problem has factorized: we only need to find solutions to the $N$-independent equation \eqref{eq: deg part problem 2} and multiply each solution by a known $N$-dependent factor. The total number of solutions to \eqref{eq: degeneracy partition problem} is given by
\begin{equation}
	\sum_{N^{\Lambda, \alpha}(\mathcal{E})} \Dim \qty(\bigotimes_{\Lambda, \alpha} \mathcal{H}^{[N^{\Lambda, \alpha}(\mathcal{E})]}),
\end{equation}
where the sum is over the set of solutions to \eqref{eq: deg part problem 2}.

%We refer to the problem of finding the possible $N^{\Lambda,\alpha}_a$ for given $\mathcal{E}$ and $\omega^{\Lambda}_{\alpha}$ as the integer partition problem. The number of solutions to equation \eqref{eq: degeneracy partition problem} is the coefficient in front of $e^{ - \beta \mathcal{E}} $ in the expansion of the generating function $ { \rm Tr } ~ e^{ - \beta H }  $.

\subsection{Energy eigenbases} \label{sec: energy eigenbasis}

We have observed that the oscillator states constructed using partition algebra diagram operators in tensor space contracted with oscillators $(a^{\dagger})^j_{i}$ obeying \eqref{eq: simplest oscillators} are eigenstates of the simplest matrix hamiltonian $H_0$ in \eqref{eq: simplest hamiltonian}. By contracting the representation basis elements in the partition algebra with the oscillators we produce quantum states
\begin{equation}
	\ket{Q^{\Lambda_1}_{\Lambda_2, \mu \nu}} = \Tr_{V_N^{\otimes k}}(Q^{\Lambda_1}_{\Lambda_2, \mu \nu} (a^{\dagger})^{\otimes k}) \ket{0}
\end{equation}
which are eigenstates of $H_0$ and also diagonalize algebraic conserved charges.

The representation basis states are not eigenstates of the general permutation invariant harmonic oscillator Hamiltonians $H$ in \eqref{eq: perm inv Hamiltonian}. There is mixing of the representation basis labels $ ( \Lambda_1 , \Lambda_2 , \mu , \nu ) $ caused by  the different weights for the representations $\Lambda$ appearing in the expansion of the $S_N$ invariant harmonic oscillator Hamiltonian defined in equation \eqref{eq: hamiltonian}. We expect this mixing of the labels in the $ ( \Lambda_1 , \Lambda_2 , \mu , \nu )$ basis to be constrained, for example by the $S_N$ Clebsch-Gordan decompositions of $\Lambda \otimes \Lambda_1 $. Such constrained mixing of representation theory bases for matrix systems arises in Hamiltonians of interest in {AdS/CFT}. A number of representation theory bases for $U(N)$ invariant multi-matrix systems have been described which capture information about finite $N$ effects and are eigenstates of the Hamiltonian (in radial quantization) in the free Yang-Mills limit \cite{Kimura2007a, Brown2008, Bhattacharyya2008, Bhattacharyya2008b, Brown2009, Kimura2008}.
However, the one-loop dilatation operator defines a non-trivial Hamiltonian which is, in general,  not diagonalized by these representation theoretic bases (although there are some interesting exceptions to this statement, see \cite{KimuraQuarter}).
 Representation theoretic constraints on the mixing caused by the one-loop dilatation operator are described in \cite{GGO,SSS,Tom1,KimuraQuarter,DoubCos}, following earlier work on one-loop mixings related to strings attached to giant gravitons, e.g. \cite{KochGSA,BerOpen}.

\section{Algebraic Hamiltonians and permutation invariant ground states }
\label{sec: deformations}
\begin{figure}[h!]
	\centering
	\begin{subfigure}[t]{0.66\textwidth}
		\centering
		\tikzset{every picture/.style={line width=0.75pt}} %set default line width to 0.75pt        
		\begin{tikzpicture}[x=0.75pt,y=0.75pt,yscale=-1,xscale=1]
			%uncomment if require: \path (0,461); %set diagram left start at 0, and has height of 461
			
			%Straight Lines [id:da579588073643597] 
			\draw [color={rgb, 255:red, 74; green, 105; blue, 226 }  ,draw opacity=1 ][fill={rgb, 255:red, 255; green, 255; blue, 255 }  ,fill opacity=1 ][line width=0.75]    (30,272) -- (55.4,272) -- (160,272) ;
			%Straight Lines [id:da20076859522224844] 
			\draw [color={rgb, 255:red, 74; green, 105; blue, 226 }  ,draw opacity=1 ][line width=0.75]    (168,267) -- (300,267) ;
			%Straight Lines [id:da10364227484291555] 
			\draw [color={rgb, 255:red, 74; green, 105; blue, 226 }  ,draw opacity=1 ][line width=0.75]    (168,277) -- (300,277) ;
			%Straight Lines [id:da5287570583376946] 
			\draw [color={rgb, 255:red, 74; green, 105; blue, 226 }  ,draw opacity=1 ][line width=0.75]    (30,203) -- (160,203) ;
			%Straight Lines [id:da7715056050681564] 
			\draw [color={rgb, 255:red, 74; green, 105; blue, 226 }  ,draw opacity=1 ][fill={rgb, 255:red, 255; green, 255; blue, 255 }  ,fill opacity=1 ][line width=0.75]    (29.6,223) -- (55,223) -- (159.6,223) ;
			%Straight Lines [id:da2479209468857584] 
			\draw [color={rgb, 255:red, 74; green, 105; blue, 226 }  ,draw opacity=1 ][fill={rgb, 255:red, 255; green, 255; blue, 255 }  ,fill opacity=1 ][line width=0.75]    (29.6,218) -- (55,218) -- (159.6,218) ;
			%Straight Lines [id:da5664516600768048] 
			\draw [color={rgb, 255:red, 74; green, 105; blue, 226 }  ,draw opacity=1 ][fill={rgb, 255:red, 255; green, 255; blue, 255 }  ,fill opacity=1 ][line width=0.75]    (29.6,213) -- (55,213) -- (159.6,213) ;
			%Straight Lines [id:da12459864106247331] 
			\draw [color={rgb, 255:red, 74; green, 105; blue, 226 }  ,draw opacity=1 ][fill={rgb, 255:red, 255; green, 255; blue, 255 }  ,fill opacity=1 ][line width=0.75]    (29.6,208) -- (55,208) -- (159.6,208) ;
			%Straight Lines [id:da4343432550763191] 
			\draw    (30,180) -- (160,180) ;
			%Straight Lines [id:da994786334534568] 
			\draw    (30,286) -- (30,158) ;
			\draw [shift={(30,156)}, rotate = 90] [color={rgb, 255:red, 0; green, 0; blue, 0 }  ][line width=0.75]    (10.93,-3.29) .. controls (6.95,-1.4) and (3.31,-0.3) .. (0,0) .. controls (3.31,0.3) and (6.95,1.4) .. (10.93,3.29)   ;
			%Straight Lines [id:da15742086992100668] 
			\draw [color={rgb, 255:red, 74; green, 105; blue, 226 }  ,draw opacity=1 ][line width=0.75]    (30.4,178) -- (160.4,178) ;
			%Straight Lines [id:da5071890721993906] 
			\draw [color={rgb, 255:red, 74; green, 105; blue, 226 }  ,draw opacity=1 ][fill={rgb, 255:red, 255; green, 255; blue, 255 }  ,fill opacity=1 ][line width=0.75]    (30,198) -- (55.4,198) -- (160,198) ;
			%Straight Lines [id:da15507871571928855] 
			\draw [color={rgb, 255:red, 74; green, 105; blue, 226 }  ,draw opacity=1 ][fill={rgb, 255:red, 255; green, 255; blue, 255 }  ,fill opacity=1 ][line width=0.75]    (30,193) -- (55.4,193) -- (160,193) ;
			%Straight Lines [id:da5294288554813205] 
			\draw [color={rgb, 255:red, 74; green, 105; blue, 226 }  ,draw opacity=1 ][fill={rgb, 255:red, 255; green, 255; blue, 255 }  ,fill opacity=1 ][line width=0.75]    (30,188) -- (55.4,188) -- (160,188) ;
			%Straight Lines [id:da21469671595984807] 
			\draw [color={rgb, 255:red, 74; green, 105; blue, 226 }  ,draw opacity=1 ][fill={rgb, 255:red, 255; green, 255; blue, 255 }  ,fill opacity=1 ][line width=0.75]    (30,183) -- (55.4,183) -- (152.17,183) -- (160,183) ;
			%Straight Lines [id:da9152148197650085] 
			\draw    (30,176) -- (160,176) ;
			%Straight Lines [id:da5887597963823057] 
			\draw    (30,172) -- (160,172) ;
			%Straight Lines [id:da8879508579804227] 
			\draw    (30,170) -- (160,170) ;
			%Straight Lines [id:da6197587559492612] 
			\draw [color={rgb, 255:red, 74; green, 105; blue, 226 }  ,draw opacity=1 ][line width=0.75]    (168,269) -- (300,269) ;
			%Straight Lines [id:da11834520903730006] 
			\draw [color={rgb, 255:red, 74; green, 105; blue, 226 }  ,draw opacity=1 ][line width=0.75]    (168,271) -- (300,271) ;
			%Straight Lines [id:da7827148312107977] 
			\draw [color={rgb, 255:red, 74; green, 105; blue, 226 }  ,draw opacity=1 ][line width=0.75]    (168,273) -- (300,273) ;
			%Straight Lines [id:da9704953442856352] 
			\draw [color={rgb, 255:red, 74; green, 105; blue, 226 }  ,draw opacity=1 ]   (168,267) -- (160,272) ;
			%Straight Lines [id:da6070011302253904] 
			\draw [color={rgb, 255:red, 74; green, 105; blue, 226 }  ,draw opacity=1 ]   (160,272) -- (168,277) ;
			%Straight Lines [id:da6416985849259336] 
			\draw [color={rgb, 255:red, 74; green, 105; blue, 226 }  ,draw opacity=1 ]   (168,269) -- (160,272) ;
			%Straight Lines [id:da4893447878051198] 
			\draw [color={rgb, 255:red, 74; green, 105; blue, 226 }  ,draw opacity=1 ]   (168,271) -- (160,272) ;
			%Straight Lines [id:da1387258663299229] 
			\draw [color={rgb, 255:red, 74; green, 105; blue, 226 }  ,draw opacity=1 ]   (168,273) -- (160,272) ;
			%Straight Lines [id:da3261660195583127] 
			\draw    (168,287) -- (168,159) ;
			\draw [shift={(168,157)}, rotate = 90] [color={rgb, 255:red, 0; green, 0; blue, 0 }  ][line width=0.75]    (10.93,-3.29) .. controls (6.95,-1.4) and (3.31,-0.3) .. (0,0) .. controls (3.31,0.3) and (6.95,1.4) .. (10.93,3.29)   ;
			%Straight Lines [id:da5895209761609239] 
			\draw    (30,221) -- (160,221) ;
			%Straight Lines [id:da4920834051261098] 
			\draw    (30,216) -- (160,216) ;
			%Straight Lines [id:da08515559247995408] 
			\draw    (30,210) -- (160,210) ;
			%Straight Lines [id:da4852789333556524] 
			\draw    (30,206) -- (160,206) ;
			%Straight Lines [id:da8024693069698978] 
			\draw    (30,200) -- (160,200) ;
			%Straight Lines [id:da983407106212143] 
			\draw    (30,196) -- (160,196) ;
			%Straight Lines [id:da255274913243112] 
			\draw    (30,190) -- (160,190) ;
			%Straight Lines [id:da17847838604615118] 
			\draw    (30,186) -- (160,186) ;
			%Straight Lines [id:da15120545490510073] 
			\draw [color={rgb, 255:red, 74; green, 105; blue, 226 }  ,draw opacity=1 ][line width=0.75]    (167.6,203) -- (297.6,203) ;
			%Straight Lines [id:da3661109484381957] 
			\draw [color={rgb, 255:red, 74; green, 105; blue, 226 }  ,draw opacity=1 ][fill={rgb, 255:red, 255; green, 255; blue, 255 }  ,fill opacity=1 ][line width=0.75]    (167.2,223) -- (192.6,223) -- (297.2,223) ;
			%Straight Lines [id:da4758796705271595] 
			\draw [color={rgb, 255:red, 74; green, 105; blue, 226 }  ,draw opacity=1 ][fill={rgb, 255:red, 255; green, 255; blue, 255 }  ,fill opacity=1 ][line width=0.75]    (167.2,218) -- (192.6,218) -- (297.2,218) ;
			%Straight Lines [id:da9744723872811809] 
			\draw [color={rgb, 255:red, 74; green, 105; blue, 226 }  ,draw opacity=1 ][fill={rgb, 255:red, 255; green, 255; blue, 255 }  ,fill opacity=1 ][line width=0.75]    (167.2,213) -- (192.6,213) -- (297.2,213) ;
			%Straight Lines [id:da0016458634342542133] 
			\draw [color={rgb, 255:red, 74; green, 105; blue, 226 }  ,draw opacity=1 ][fill={rgb, 255:red, 255; green, 255; blue, 255 }  ,fill opacity=1 ][line width=0.75]    (167.2,208) -- (192.6,208) -- (297.2,208) ;
			%Straight Lines [id:da08637904329279356] 
			\draw    (167.6,180) -- (297.6,180) ;
			%Straight Lines [id:da7152719057252641] 
			\draw [color={rgb, 255:red, 74; green, 105; blue, 226 }  ,draw opacity=1 ][line width=0.75]    (168,178) -- (298,178) ;
			%Straight Lines [id:da7535782316302924] 
			\draw [color={rgb, 255:red, 74; green, 105; blue, 226 }  ,draw opacity=1 ][fill={rgb, 255:red, 255; green, 255; blue, 255 }  ,fill opacity=1 ][line width=0.75]    (167.6,198) -- (193,198) -- (297.6,198) ;
			%Straight Lines [id:da22772506518689406] 
			\draw [color={rgb, 255:red, 74; green, 105; blue, 226 }  ,draw opacity=1 ][fill={rgb, 255:red, 255; green, 255; blue, 255 }  ,fill opacity=1 ][line width=0.75]    (167.6,193) -- (193,193) -- (297.6,193) ;
			%Straight Lines [id:da7925147042932634] 
			\draw [color={rgb, 255:red, 74; green, 105; blue, 226 }  ,draw opacity=1 ][fill={rgb, 255:red, 255; green, 255; blue, 255 }  ,fill opacity=1 ][line width=0.75]    (167.6,188) -- (193,188) -- (297.6,188) ;
			%Straight Lines [id:da2972699090452684] 
			\draw [color={rgb, 255:red, 74; green, 105; blue, 226 }  ,draw opacity=1 ][fill={rgb, 255:red, 255; green, 255; blue, 255 }  ,fill opacity=1 ][line width=0.75]    (167.6,183) -- (193,183) -- (289.77,183) -- (297.6,183) ;
			%Straight Lines [id:da05587536502256807] 
			\draw    (167.6,176) -- (297.6,176) ;
			%Straight Lines [id:da002618820871658656] 
			\draw    (167.6,172) -- (297.6,172) ;
			%Straight Lines [id:da8306627097775938] 
			\draw    (167.6,170) -- (297.6,170) ;
			%Straight Lines [id:da7157786354378128] 
			\draw    (167.6,221) -- (297.6,221) ;
			%Straight Lines [id:da48098791796880835] 
			\draw    (167.6,216) -- (297.6,216) ;
			%Straight Lines [id:da5256840711767568] 
			\draw    (167.6,210) -- (297.6,210) ;
			%Straight Lines [id:da4341190307348979] 
			\draw    (167.6,206) -- (297.6,206) ;
			%Straight Lines [id:da7899998591970887] 
			\draw    (167.6,200) -- (297.6,200) ;
			%Straight Lines [id:da1743211195555383] 
			\draw    (167.6,196) -- (297.6,196) ;
			%Straight Lines [id:da062302459697773926] 
			\draw    (167.6,190) -- (297.6,190) ;
			%Straight Lines [id:da49827818426362436] 
			\draw    (167.6,186) -- (297.6,186) ;
			
			% Text Node
			\draw (5,225) node [anchor=north west][inner sep=0.75pt]   [align=left] {E};
			% Text Node
			\draw (195,221) node [anchor=north west][inner sep=0.75pt]    {$H_{K} +H_{\text{res}}$};
			% Text Node
			\draw (84,221) node [anchor=north west][inner sep=0.75pt]    {$H_{K}$};
			% Text Node
			\draw (20,257) node [anchor=north west][inner sep=0.75pt]   [align=left] {{\tiny 0}};
			% Text Node
			\draw (18,211) node [anchor=north west][inner sep=0.75pt]   [align=left] {{\tiny K}};
		
		\end{tikzpicture}
		\caption{}
		\label{fig: spectrum scenarios d}
	\end{subfigure}%
	~
	\begin{subfigure}[t]{0.33\textwidth}
		\centering
		\tikzset{every picture/.style={line width=0.75pt}} %set default line width to 0.75pt        
		\begin{tikzpicture}[x=0.75pt,y=0.75pt,yscale=-1,xscale=1]
			%uncomment if require: \path (0,461); %set diagram left start at 0, and has height of 461
			%Straight Lines [id:da9567567611888961] 
			\draw    (30,92) -- (160,92) ;
			%Straight Lines [id:da7568445871181468] 
			\draw    (30,108) -- (160,108) ;
			%Straight Lines [id:da500866457834461] 
			\draw    (30,96) -- (160,96) ;
			%Straight Lines [id:da3518862281752282] 
			\draw    (30,130) -- (160,130) ;
			%Straight Lines [id:da5773520429872101] 
			\draw [color={rgb, 255:red, 74; green, 105; blue, 226 }  ,draw opacity=1 ][line width=0.75]    (30,100) -- (160,100) ;
			%Straight Lines [id:da33975155999771234] 
			\draw [color={rgb, 255:red, 74; green, 105; blue, 226 }  ,draw opacity=1 ][line width=0.75]    (30,102) -- (160,102) ;
			%Straight Lines [id:da6754995971119875] 
			\draw [color={rgb, 255:red, 74; green, 105; blue, 226 }  ,draw opacity=1 ][line width=0.75]    (30,126) -- (160,126) ;
			%Straight Lines [id:da2335897286621358] 
			\draw    (30,144) -- (160,144) ;
			%Straight Lines [id:da4433298717580514] 
			\draw    (30,142) -- (160,142) ;
			%Straight Lines [id:da6480716390293622] 
			\draw    (30,138) -- (160,138) ;
			%Straight Lines [id:da6867539857044027] 
			\draw [color={rgb, 255:red, 74; green, 105; blue, 226 }  ,draw opacity=1 ][line width=0.75]    (30,48) -- (160,48) ;
			%Straight Lines [id:da0034175548290271696] 
			\draw    (30,58) -- (160,58) ;
			%Straight Lines [id:da21662586017771535] 
			\draw    (30,72) -- (160,72) ;
			%Straight Lines [id:da7701225158253342] 
			\draw    (30,34) -- (160,34) ;
			%Straight Lines [id:da302167329458892] 
			\draw    (30,150) -- (30,22) ;
			\draw [shift={(30,20)}, rotate = 90] [color={rgb, 255:red, 0; green, 0; blue, 0 }  ][line width=0.75]    (10.93,-3.29) .. controls (6.95,-1.4) and (3.31,-0.3) .. (0,0) .. controls (3.31,0.3) and (6.95,1.4) .. (10.93,3.29)   ;
			% Text Node
			\draw (5,89) node [anchor=north west][inner sep=0.75pt]   [align=left] {E};
		\end{tikzpicture}
		\caption{}
		\label{fig: spectrum scenarios a}
	\end{subfigure}%
	\caption{The figure illustrates the type of spectra that can be engineered using the algebraic Hamiltonians discussed in this section. Blue lines correspond to states that are invariant under the adjoint action of $S_N$. Black lines are non-invariant states.}		
	%		Hamiltonians with non-degenerate spectra can be grouped into three physically distinct categories based on where in the spectrum invariant states appear. In figure \ref{fig: spectrum scenarios a} the invariant eigenstates (blue) are mixed with the non-invariant eigenstates (black). Figure \ref{fig: spectrum scenarios b} shows a spectrum where the lowest energy states are invariant states. There is a small gap between the the highest energy invariant state and the lowest energy non-invariant state. In figure \ref{fig: spectrum scenarios c} the invariant and non-invariant states are separated in energy, but now the gap is very large and one may expect the low-energy physics to be well described by the physics of the invariant states. The invariant states are effectively decoupled from the rest of the Hilbert space. The Hamiltonian $H(N)$ \eqref{eq: degenerate invariant ground state hamiltonian} has an infinitely degenerate ground state made out of invariant states as depicted in the left spectrum of figure \ref{fig: spectrum scenarios d}. The first excited states have order $N$ energy, which decouples from the ground states in the large $N$ limit. Deforming this Hamiltonian by $\delta H$ made out of the algebraic charges discussed in section \ref{sec: charges} breaks the degeneracy of the invariant ground states as the right spectrum of figure \ref{fig: spectrum scenarios d}.}
	\label{fig: spectrum scenarios}
\end{figure}
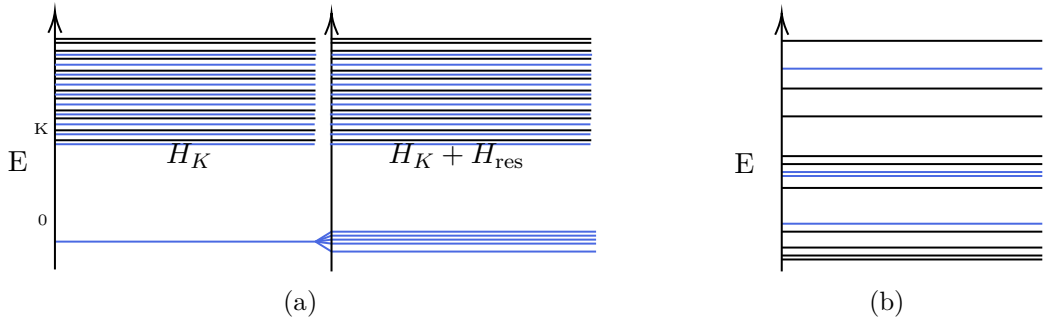

So far our discussion of $S_N$ invariant subspaces in quantum mechanical matrix systems has largely (with the exception of the previous section) been independent of any choice of Hamiltonian  acting on the Hilbert space. It can be viewed as a general description of the kinematics of $S_N$ invariance, independent of the dynamics determined by the Hamiltonian. In this section we present Hamiltonians $H$ which realise the eigenspectrum scenarios depicted in figure \ref{fig: spectrum scenarios}, this includes Hamiltonians for which the low energy eigenstates are permutation invariant states.  

 The Hamiltonians we consider here preserve the $S_N$ invariant subspace $	\Hilbertspace $ defined as 
\begin{equation}
	\Hilbertspace = \{\ket{T} \in \mathcal{H} \qq{s.t.} \Adj{\sigma} \ket{T} = \ket{T} ,\, \forall \sigma \in S_N\}.
\end{equation}
The adjoint action of permutations $ \sigma \in S_N $ on the tensors $T$ labelling the states simultaneously transforms the upper and lower indices of $T$ according to \eqref{eq: diagonally permutation action on hilbert}. 
For any state $\ket{T} \in \Hilbertspace$ the Hamiltonians $H$ obey the condition 
\begin{equation} \label{eq: H_inv}
	H \ket{T} \in \Hilbertspace.
\end{equation}
A sufficient condition for $H$ to satisfy \eqref{eq: H_inv} is for $H$ itself to be $S_N$ invariant or $\comm{\Adj{\sigma}}{H} = 0$ for all $\sigma \in S_N$.

We will show how to construct Hamiltonians $H_K$ of this type, depending on an integer  parameter $K$, with a  finite-dimensional space of $S_N$ invariant ground states. Both the energy gap between the ground states and the lowest non-zero energy level, and the ground state degeneracy depend on $K$ in a way that is determined by the algebraic construction. As sketched in the left-hand figure of \ref{fig: spectrum scenarios d}, $H_K$ has an energy gap of order $K$. The construction of $H_K$ can be viewed as including, in the Hamiltonian, central elements in $\mathbb{C}[S_N]$ acting on $\mathcal{H}^{(k)}$ using $\Adj{\sigma}$ for $ k \le K$.  This can be related to the action of elements of $P_{2k}(N)$ acting on $\mathcal{H}^{(k)}$ for $k \leq K$. We will briefly  mention some analogies between the present construction and the phenomenon of topological degeneracy which is widely studied in condensed matter physics.

The ground state degeneracy of $H_K$ can be resolved by adding a term $H_{\text{res}}$, made from the central algebraic charges discussed in section \ref{sec: charges}. This breaks the degeneracy of the invariant ground states as illustrated in the spectrum on the right of figure \ref{fig: spectrum scenarios d}. The representation basis $\ket*{Q^{\Lambda_1}_{\Lambda_2, \mu \nu}}$ presented in section \ref{sec: ON basis} diagonalizes these Hamiltonians in the invariant subspace, and the state energies depend on labels $\Lambda_1, \Lambda_2$. 

Multiplicity labels $\mu,\nu$ are not distinguished by the central algebraic charges.
Distinguishing multiplicity labels requires more general elements of $P_{k}(N)$, as discussed in \ref{sec:multiplicity-non-central}.
Generalizing the construction of $H_{\text{res}}$ naturally leads to a large class of $S_N$ invariant Hamiltonians related to the left action of elements of $P_{k}(N)$, which can be used to break the degeneracy associated with multiplicity labels.
Hamiltonians of this type can have non-trivial spectra, in which invariant states are distributed across the energy spectrum, with no discernible pattern of difference compared to non-invariant states, as illustrated in figure \ref{fig: spectrum scenarios a}.

The 11-parameter Hamiltonians in section \ref{sec: PIMQM} typically have such non-trivial spectra. Given the non-trivial index contractions in \eqref{eq: invariant potential},
\begin{align}
\sum_{i,j,k,l} Q_{ijkl}^{\Lambda,\alpha \beta} X_{ij} X_{kl} \longrightarrow (a^{\dagger})^i_j a^k_l Q_{ijkl}^{\Lambda,\alpha \beta},
\end{align}
these Hamiltonians are not of the kind involving only the left action of $P_k(N)$. Similarly, $H_K$ is not of this kind. This implies that a more general construction of $S_N$ invariant Hamiltonians exists. We give a description of this more general construction, which involves elements of $P_{2k}(N)$.
We end the section with a lattice interpretation of the matrix oscillators. This sets us up for section \ref{sec: scars} which concerns the non-trivial interplay between the invariant sector and the Hamiltonian and includes realisations based on the lattice interpretation.

\subsection{Partition algebra elements as quantum mechanical operators} \label{sec: conserved charges}
We now translate much of the discussion in section \ref{sec: charges} into the language of quantum mechanical operators on $\mathcal{H}$. Finding representation bases corresponds to the diagonalization of commuting operators on $\mathcal{H}$. Notably, elements of $SP_k(N)$ naturally correspond to operators for fixed $k$, or maps $\mathcal{H}^{(k)} \rightarrow \mathcal{H}^{(k)}$.
However, it will be useful to have expressions for these fixed $k$ operators in terms of oscillators, which act on the entire Hilbert space $\mathcal{H}$.
These two kinds of operators are related by projectors $\mathcal{P}_k: \mathcal{H} \rightarrow \mathcal{H}^{(k)}$ to fixed $k$ subspaces. 
We use this in the construction of Hamiltonians in the remainder of section \ref{sec: deformations}.

For a general state $\ket{T} \in \mathcal{H}^{(k)}$ (see \eqref{eq: state definition}) and element $[d] \in SP_k(N)$ there is a corresponding operator defined as
\begin{equation} \label{eq: partition algebra as operator}
	[d]^L\ket{T} = \ket{[d]T} = \ket{dT},
\end{equation}
where the superscript $L$ stands for left action, and
\begin{equation}
	(dT)^{i_1 \dots i_k}_{i_{1'} {\dots} i_{k'}} = \sum_{j_1, \dots, j_k} d^{i_1 \dots i_k}_{j_{1} \dots j_k} T^{j_1 \dots j_k}_{i_{1'} {\dots} i_{k'}}.
\end{equation}
The second equality in \eqref{eq: partition algebra as operator} follows since
\begin{equation}
\begin{aligned}
		\ket{[d]T} = \Tr_{V_N^{\otimes k}}( [d]T (a^{\dagger})^{\otimes k})\ket{0} &= \frac{1}{k!}\sum_{\gamma \in S_k} \Tr_{V_N^{\otimes k}}( \mathcal{L}_{\gamma}d \mathcal{L}_{\gamma^{-1}}T (a^{\dagger})^{\otimes k})\ket{0} \\
		&= \Tr_{V_N^{\otimes k}}(d T (a^{\dagger})^{\otimes k})\ket{0} = \ket{dT},
\end{aligned}
\end{equation}
where $\mathcal{L}(\sigma)$ is defined in equation \eqref{eq: left action SN}. We have used $\mathcal{L}_{\gamma}T = T\mathcal{L}_{\gamma}$ together with $\mathcal{L}_{\gamma}(a^{\dagger})^{\otimes k} =(a^{\dagger})^{\otimes k}\mathcal{L}_{\gamma}$ to go to the second line.
We may also define operators corresponding to right action,
\begin{equation}
	[d]^R \ket{T} = \ket{T d}.
\end{equation}

We extend $[d]^L$ to an operator on $\mathcal{H}$, expressible in terms of oscillators and projectors $\mathcal{P}_k: \mathcal{H} \rightarrow \mathcal{H}^{(k)}$ as
\begin{equation}
\boxed{ 	[d]^L = \frac{1}{k!}\mathcal{P}_k \Tr_{V_N^{\otimes k}}((a^{\dagger})^{\otimes k}d a^{\otimes k}) \mathcal{P}_k . } 
\end{equation}
Similarly, we can extend $[d]^R$ to an operator on $\mathcal{H}$,
\begin{equation}
	\boxed{ [d]^R = \frac{1}{k!}\mathcal{P}_k \Tr_{V_N^{\otimes k}}(d (a^{\dagger})^{\otimes k} a^{\otimes k}) \mathcal{P}_k. } 
\end{equation}
In what follows we will prove results explicitly for the left action. For the sake of brevity we omit the analogous proofs for the right action. 

The definition of $\mathcal{P}_k$ in the oscillator basis is
\begin{equation} \label{eq: projector oscillator def}
	\mathcal{P}_{k'} (a^{\dagger})^{i_1}_{j_1} \dots (a^{\dagger})^{i_k}_{j_k} \ket{0} = \delta_{kk'} (a^{\dagger})^{i_1}_{j_1} \dots (a^{\dagger})^{i_k}_{j_k} \ket{0}.
\end{equation}
We now prove
\begin{equation} \label{eq: oscillator PkN equivalence}
	\frac{1}{k!}\mathcal{P}_k \Tr_{V_N^{\otimes k}}((a^{\dagger})^{\otimes k}d a^{\otimes k}) \mathcal{P}_k \ket{T} = \ket*{dT^{(k)}},
\end{equation}
where $\ket{T} = \sum_{k=0}^\infty \ket*{T^{(k)}}$ and $\ket*{T^{(k)}} \in \mathcal{H}^{(k)}$.
The projector immediately gives $\mathcal{P}_k \ket{T} = | T^{(k)} \rangle$. It remains to prove
\begin{equation} \label{eq: trace action of PkN}
	\frac{1}{k!}\mathcal{P}_k \Tr_{V_N^{\otimes k}}((a^{\dagger})^{\otimes k}d a^{\otimes k}) | T^{(k)} \rangle = | d T^{(k)} \rangle .
\end{equation}
We prove this diagrammatically, using the state definition in terms of diagrams \eqref{eq: state from endo diagram}
\begin{equation}
	\begin{aligned}
		\frac{1}{k!}\Tr_{V_N^{\otimes k}}((a^{\dagger})^{\otimes k}da^{\otimes k}) | T^{(k)} \rangle &= 
		\tikzset{every picture/.style={line width=0.75pt}} %set default line width to 0.75pt        
		\vcenter{\hbox{ \scalebox{.5}{ 
					\begin{tikzpicture}[x=0.75pt,y=0.75pt,yscale=-1,xscale=1]
						
						%Straight Lines [id:da4899274331518464] 
						\draw    (130,160) -- (130,190) ;
						%Shape: Rectangle [id:dp8786856708315745] 
						\draw   (100,120) -- (160,120) -- (160,160) -- (100,160) -- cycle ;
						%Straight Lines [id:da16802414813996713] 
						\draw    (130,90) -- (130,120) ;
						%Shape: Rectangle [id:dp96997764130762] 
						\draw   (100,51) -- (160,51) -- (160,90) -- (100,90) -- cycle ;
						%Straight Lines [id:da9530701789478033] 
						\draw    (130,20) -- (130,50) ;
						%Straight Lines [id:da2642712928560609] 
						\draw    (120,20) -- (140,20) ;
						%Shape: Rectangle [id:dp05004402835919253] 
						\draw   (100,190) -- (160,190) -- (160,230) -- (100,230) -- cycle ;
						%Straight Lines [id:da24488668974689687] 
						\draw    (130,230) -- (130,260) ;
						%Straight Lines [id:da9333433939914717] 
						\draw    (120,260) -- (140,260) ;
						%Straight Lines [id:da06002972973693188] 
						\draw    (220,230) -- (220,260) ;
						%Shape: Rectangle [id:dp26789553703055136] 
						\draw   (190,190) -- (250,190) -- (250,230) -- (190,230) -- cycle ;
						%Straight Lines [id:da6818368310453649] 
						\draw    (210,260) -- (230,260) ;
						%Straight Lines [id:da7575116362207548] 
						\draw    (220,160) -- (220,190) ;
						%Shape: Rectangle [id:dp8126512465119999] 
						\draw   (190,120) -- (250,120) -- (250,159) -- (190,159) -- cycle ;
						%Straight Lines [id:da7730566058355652] 
						\draw    (220,20) -- (220,120) ;
						%Straight Lines [id:da056066645172053065] 
						\draw    (210,20) -- (230,20) ;
						
						% Text Node
						\draw (125,135) node [anchor=north west][inner sep=0.75pt]  {$ d$};
						% Text Node
						\draw (113,59) node [anchor=north west][inner sep=0.75pt]   {$ (a^{\dagger })^{\otimes k}$};
						% Text Node
						\draw (123,199) node [anchor=north west][inner sep=0.75pt]   {$ a^{\otimes k}$};
						% Text Node
						\draw (203,199) node [anchor=north west][inner sep=0.75pt]   {$ (a^{\dagger })^{\otimes k}$};
						% Text Node
						\draw (210,128) node [anchor=north west][inner sep=0.75pt]   {$ T^{(k)}$};
		\end{tikzpicture}} }}\ket{0} 
		= \frac{1}{k!}\sum_{\gamma \in S_k}
		\tikzset{every picture/.style={line width=0.75pt}} %set default line width to 0.75pt        
		\vcenter{\hbox{ \scalebox{.5}{					
					\begin{tikzpicture}[x=0.75pt,y=0.75pt,yscale=-1,xscale=1]
						%uncomment if require: \path (0,461); %set diagram left start at 0, and has height of 461
						
						%Straight Lines [id:da7124935974588078] 
						\draw    (350,160) -- (350,170) ;
						%Straight Lines [id:da9941619315386976] 
						\draw    (350,90) -- (350,120) ;
						%Shape: Rectangle [id:dp674649909898108] 
						\draw   (320,51) -- (380,51) -- (380,90) -- (320,90) -- cycle ;
						%Straight Lines [id:da09486631047502359] 
						\draw    (350,20) -- (350,50) ;
						%Straight Lines [id:da8367133600667203] 
						\draw    (340,20) -- (360,20) ;
						%Straight Lines [id:da7169964881441306] 
						\draw    (350,240) -- (350,260) ;
						%Straight Lines [id:da6307505074785329] 
						\draw    (340,260) -- (360,260) ;
						%Straight Lines [id:da6976463368286077] 
						\draw    (440,240) -- (440,260) ;
						%Straight Lines [id:da7773305714261438] 
						\draw    (430,260) -- (450,260) ;
						%Straight Lines [id:da9382611653220043] 
						\draw    (440,20) -- (440,120) ;
						%Straight Lines [id:da15040111467996797] 
						\draw    (430,20) -- (450,20) ;
						%Shape: Square [id:dp49183281294591374] 
						\draw   (420,200) -- (460,200) -- (460,240) -- (420,240) -- cycle ;
						%Shape: Rectangle [id:dp8453002955386408] 
						\draw   (320,120) -- (380,120) -- (380,160) -- (320,160) -- cycle ;
						%Straight Lines [id:da7494552059131181] 
						\draw    (350,170) -- (440,190) ;
						%Straight Lines [id:da07821854155703512] 
						\draw    (350,190) -- (440,170) ;
						%Straight Lines [id:da07082319983218888] 
						\draw    (440,160) -- (440,170) ;
						%Shape: Rectangle [id:dp002339023183775435] 
						\draw   (410,120) -- (470,120) -- (470,160) -- (410,160) -- cycle ;
						%Shape: Square [id:dp5738241749488762] 
						\draw   (330,200) -- (370,200) -- (370,240) -- (330,240) -- cycle ;
						%Straight Lines [id:da13011265030874442] 
						\draw    (440,190) -- (440,200) ;
						%Straight Lines [id:da6744057398653815] 
						\draw    (350,190) -- (350,200) ;
						
						% Text Node
						\draw (335,60) node [anchor=north west][inner sep=0.75pt]  {$(a^{\dagger })^{\otimes k}$};
						% Text Node
						\draw (340,133) node [anchor=north west][inner sep=0.75pt] {$d$};
						% Text Node
						\draw (433,129) node [anchor=north west][inner sep=0.75pt]  {$T^{(k)}$};
						% Text Node
						\draw (340,211) node [anchor=north west][inner sep=0.75pt]    {$\mathcal{L}_{\gamma }$};
						% Text Node
						\draw (424,211) node [anchor=north west][inner sep=0.75pt]    {$\mathcal{L}_{\gamma ^{-1}}$};
				\end{tikzpicture}}} }\ket{0} \\
		&= \frac{1}{k!}\sum_{\gamma \in S_k}
		\tikzset{every picture/.style={line width=0.75pt}} %set default line width to 0.75pt        
		\vcenter{\hbox{ \scalebox{.5}{
					\begin{tikzpicture}[x=0.75pt,y=0.75pt,yscale=-1,xscale=1]
						%uncomment if require: \path (0,461); %set diagram left start at 0, and has height of 461
						%Straight Lines [id:da46751625704315103] 
						\draw    (590,10) -- (590,20) ;
						%Straight Lines [id:da23990864404059753] 
						\draw    (580,10) -- (600,10) ;
						%Shape: Rectangle [id:dp4315986804302214] 
						\draw   (560,21) -- (620,21) -- (620,60) -- (560,60) -- cycle ;
						%Shape: Rectangle [id:dp013831176150676638] 
						\draw   (560,80) -- (620,80) -- (620,120) -- (560,120) -- cycle ;
						%Straight Lines [id:da7201589148814074] 
						\draw    (590,60) -- (590,80) ;
						%Straight Lines [id:da6099498457985284] 
						\draw    (590,120) -- (590,140) ;
						%Shape: Rectangle [id:dp1770065853088174] 
						\draw   (560,200) -- (620,200) -- (620,240) -- (560,240) -- cycle ;
						%Straight Lines [id:da7129982640817976] 
						\draw    (590,180) -- (590,200) ;
						%Straight Lines [id:da3881713117156975] 
						\draw    (590,240) -- (590,260) ;
						%Straight Lines [id:da3389826716799813] 
						\draw    (590,300) -- (590,310) ;
						%Straight Lines [id:da012099102206072354] 
						\draw    (600,310) -- (580,310) ;
						%Shape: Square [id:dp36933409530798067] 
						\draw   (570,140) -- (610,140) -- (610,180) -- (570,180) -- cycle ;
						%Shape: Square [id:dp11354438393765065] 
						\draw   (570,260) -- (610,260) -- (610,300) -- (570,300) -- cycle ;
						
						% Text Node
						\draw (574,150) node [anchor=north west][inner sep=0.75pt]    {$\mathcal{L}_{\gamma ^{-1}}$};
						% Text Node
						\draw (577,270) node [anchor=north west][inner sep=0.75pt]    {$\mathcal{L}_{\gamma }$};
						% Text Node
						\draw (570,30) node [anchor=north west][inner sep=0.75pt] {$(a^{\dagger })^{\otimes k}$};
						% Text Node
						\draw (583,91) node [anchor=north west][inner sep=0.75pt]    {$d$};
						% Text Node
						\draw (575,210) node [anchor=north west][inner sep=0.75pt]    {$T^{( k)}$};
						
					\end{tikzpicture}
		}} }\ket{0} \\
		&= | dT^{(k)} \rangle.
	\end{aligned}
\end{equation}
In the second equality we have moved all annihilation operators past the creation operators, giving a sum over contractions.
The sum over $\gamma \in S_k$ encodes the contractions and in the second line we have straightened the diagram.
The last identification follows since $\mathcal{L}_{\gamma^{-1}}T^{(k)}\mathcal{L}_{\gamma} = T^{(k)}$. Because $\ket*{dT^{(k)}} \in \mathcal{H}^{(k)}$ we have $\mathcal{P}_k \ket*{dT^{(k)}} = \ket*{dT^{(k)}}$, which establishes the equality in \eqref{eq: oscillator PkN equivalence}.

As we now show, the Hermitian conjugate of the operator $[d_{\pi}]^L$ is $[d_{\pi}^T]^L$, where $d_\pi^T$ is the element obtained by flipping the diagram $d_\pi$ horizontally. This follows from the inner product
\begin{equation}
	\bra{T'}\ket{T} = \sum_{\gamma \in S_k}\Tr_{V_N^{\otimes k}}((T')^T {\gamma} T {{\gamma^{-1}}}),
\end{equation}
defined in \eqref{eq: inner product is trace}) and
\begin{align} \nonumber
	\bra{T'}\ket{d_{\pi}T} &= \sum_{\gamma \in S_k}\Tr_{V_N^{\otimes k}}((T')^T {\gamma} d_{\pi} T {{\gamma^{-1}}}) = k! \Tr_{V_N^{\otimes k}}((T')^T d_{\pi} T) \\
	&= \sum_{\gamma \in S_k}\Tr_{V_N^{\otimes k}}((d_\pi^T T')^T \gamma^{-1} T \gamma) = \langle d_{\pi}^T T' | T \rangle.
\end{align}

As operators on $\mathcal{H}$, $T_2 \in \mathcal{Z}(\mathbb{C}[S_N]), \SWdual{T_2} \in \mathcal{Z}(P_k(N))$ and $t_2 \in \mathcal{Z}(\mathbb{C}[S_k])$ can be written as oscillators.
From the definition of the action of $T_2$ in \eqref{eq: Center of S_N inside PkN} we have
\begin{equation}
	\begin{aligned}
		{T}_2^{(k), L} &\equiv \frac{1}{k!}\mathcal{P}_k \Tr_{V_N^{\otimes k}}\qty[(a^\dagger)^{\otimes k} \mathcal{L}(T_2)  a^{\otimes k}] \mathcal{P}_k \\
		& = \frac{1}{k!}\mathcal{P}_k\sum_{\substack{\sigma = (ij) \\ 1 \leq i<j \leq N}} \Tr_{V_N^{\otimes k}}\qty[(a^\dagger)^{\otimes k} \mathcal{L}(\sigma)  a^{\otimes k}] \mathcal{P}_k \\
		&=\frac{1}{k!}\mathcal{P}_k\sum_{\substack{\sigma = (ij) \\ 1 \leq i<j \leq N}} \sum_{\substack{i_1 \dots i_k \\ i_{1'} \dots i_{k'}}} (a^\dagger)^{i_{1'}}_{\sigma^{-1}(i_{1})} \dots(a^\dagger)^{i_{k'}}_{\sigma^{-1}(i_{k})} a^{i_{1}}_{i_{1'}} \dots a^{i_{k}}_{i_{k'}} \mathcal{P}_k. \label{eq: T2SN Hermitian op}
	\end{aligned}
\end{equation}
Similarly, the fixed $k$ operators corresponding to $\SWdual{T}_2$ are
\begin{equation}
	\begin{aligned}
		\SWdual{T_2}^{(k), L} & = \frac{1}{k!} \mathcal{P}_k\Tr_{V_N^{\otimes k}}\qty[(a^\dagger)^{\otimes k} \SWdual{T_2} a^{\otimes k}] \mathcal{P}_k \\
		&= \frac{1}{k!}\mathcal{P}_k \sum_{\substack{i_1 \dots i_k  \\ j_1 \dots j_k \\ j_{1'} \dots j_{k'}}}  (a^\dagger)^{i_{1}}_{i_{1'}} \dots (a^\dagger)^{i_{k}}_{i_{k'}} (\SWdual{T_2})^{i_{1'} \dots i_{k'}}_{j_1 \dots j_k} a^{j_{1}}_{i_{1}} \dots a^{j_{k}}_{i_{k}} \mathcal{P}_k,
	\end{aligned} \label{eq: T2PkN Hermitian op}
\end{equation}
where $ \SWdual{T_2}$ can be expanded in in the diagram basis as in \eqref{eq: T2 expansion in terms of diagrams}.
Finally, the fixed $k$ operators corresponding to $t_2$ are
\begin{equation}
	\begin{aligned}\label{eq: T2Sk Hermitian op}
		t_2^{(k), L} &= \frac{1}{k!}\mathcal{P}_k\sum_{\substack{\tau = (ij) \\ 1 \leq i<j \leq k}} \Tr_{V_N^{\otimes k}}\qty[(a^\dagger)^{\otimes k} \mathcal{L}_{\tau^{-1}} a^{\otimes k}] \mathcal{P}_k \\
		&= \frac{1}{k!}\mathcal{P}_k\sum_{\substack{\tau = (ij) \\ 1 \leq i<j \leq k}}\sum_{\substack{i_1 \dots i_k \\ i_{1'} \dots i_{k'}}}  (a^\dagger)^{i_{1'}}_{i_{\tau(1)}} \dots(a^\dagger)^{i_{k'}}_{i_{\tau(k)}} a^{i_{1}}_{i_{1'}} \dots a^{i_{k}}_{i_{k'}} \mathcal{P}_k.		
	\end{aligned}
\end{equation}

These operators are Hermitian, because $(T_2)^T = T_2$ and $(t_2)^T = t_2$, and consequently their eigenvectors with distinct eigenvalues are orthogonal. They are difficult to diagonalize over the entirety of $\mathcal{H}^{(k)}$, since the dimension grows as $N^{2k}$ for $N \gg k$. But the diagonalization over $\Hilbertspace^k$ is feasible since the dimension is bounded by $B(2k)$, which does not scale with $N$. Further simplification arises when acting on states $\ket{d} \in \Hilbertspace^{(k)}$, since the action can be formulated as multiplication in $SP_k(N)$, thus bypassing the computation of large index contractions. That is, for $\ket{d} \in \Hilbertspace^k$
\begin{equation}
	\SWdual{T_2}^{(k),L} \ket{d} = | \SWdual{T_2}d \rangle, \label{eq: hilbert space action as algebra action}
\end{equation}
where the product $\SWdual{T_2}^{(k)}d$ can be taken in $P_k(N)$.
It follows that,
\begin{equation}
	\SWdual{T_2}^{(k),L}\ket{Q^{\Lambda_1}_{\Lambda_2, \mu \nu}} =  \ket{\SWdual{T_2} Q^{\Lambda_1}_{\Lambda_2, \mu \nu}} = \hat{\chi}^{\Lambda_1}(T_2) \ket{Q^{\Lambda_1}_{\Lambda_2, \mu \nu}}, \label{eq: rep basis is eigenbasis of T2}
\end{equation}
and similarly for $t_2^{(k), L}$.

The free Hamiltonian $H_0$ in equation \eqref{eq: free H} is just the number operator. The above operators conserve the number of particles. Consequently,
\begin{equation}
	\comm{H_0}{T_2^{(k),L}} = \comm{H_0}{\SWdual{T_2}^{(k),L}} = \comm{H_0}{t_2^{(k),L}} = 0,
\end{equation}
and the corresponding charges are conserved.

\subsection{Decoupling invariant sectors and invariant ground states} \label{sec: deformations and ground state}
We now present a Hermitian operator with algebraic origin that can be used to control the energies of states invariant under the adjoint action of $S_N$ on $\mathcal{H}^{(k)}$. We use the operator to construct a Hamiltonian with a large number of invariant ground states.

The adjoint action of $\sigma \in S_N$ on $\mathcal{H}^{(k)}$ is defined in equation \eqref{eq: diagonally permutation action on hilbert} as
\begin{equation} \label{eq: sigma on ket T}
	\Adj{\sigma} \ket{T} = \Tr_{V_N^{\otimes k}}(\mathcal{L}(\sigma)T \mathcal{L}(\sigma^{-1}) (a^{\dagger})^{\otimes k} ) \ket{0}  = 
\sum_{\substack{i_1 , \dots , i_k \\ j_1 , \dots , j_k}}T^{\sigma(j_1) \dots \sigma(j_k)}_{\sigma(i_1) \dots \sigma(i_k)}  (a^{\dagger})^{i_1}_{j_1} \dots (a^{\dagger})^{i_k}_{j_k}\ket{0}.
\end{equation}
We may write $\Adj{\sigma}$ in terms of oscillators and projectors $\mathcal{P}_k: \mathcal{H} \rightarrow \mathcal{H}^{(k)}$ defined in equation \eqref{eq: projector oscillator def}. For $\ket{T} \in \mathcal{H}^{(k)}$,
\begin{equation} \label{eq: adjoint on H_k}
	\Adj{\sigma} \ket{T} = \frac{1}{k!}\mathcal{P}_k\Tr_{V_N^{\otimes k}}(\mathcal{L}(\sigma^{-1}) (a^{\dagger})^{\otimes k} \mathcal{L}(\sigma) a^{\otimes k} )\mathcal{P}_k\ket{T}.
\end{equation}

We note that the the ordering of $a^{\dagger}$ relative to $a$ is understood to be as shown in the above equation.
To understand the equality in \eqref{eq: adjoint on H_k}, we evaluate
\begin{equation}
	\Tr_{V_N^{\otimes k}}(\mathcal{L}(\sigma^{-1}) (a^{\dagger})^{\otimes k} \mathcal{L}(\sigma) a^{\otimes k} ) \ket{T},
\end{equation}
where we take $\ket{T} \in \mathcal{H}^{(k)}$ (there is no loss of generality since $\mathcal{P}_k$ projects to $\mathcal{H}^{(k)}$).
Diagrammatically we have
\begin{equation}
	\begin{aligned}
		\vcenter{\hbox{\scalebox{.5}{
					\tikzset{every picture/.style={line width=0.75pt}} %set default line width to 0.75pt        
					\begin{tikzpicture}[x=0.75pt,y=0.75pt,yscale=-1,xscale=1]
						%uncomment if require: \path (0,437); %set diagram left start at 0, and has height of 437
						%Straight Lines [id:da9999884042539149] 
						\draw    (90,179.46) -- (90,190) ;
						%Shape: Rectangle [id:dp6963795717485375] 
						\draw   (58,140) -- (118,140) -- (118,180) -- (58,180) -- cycle ;
						%Straight Lines [id:da6016942691843352] 
						\draw    (90,80) -- (90,90) ;
						%Straight Lines [id:da9224427609680363] 
						\draw    (80,20) -- (100,20) ;
						%Shape: Rectangle [id:dp5829781159184302] 
						\draw   (58,190) -- (118,190) -- (118,230) -- (58,230) -- cycle ;
						%Straight Lines [id:da4283285237578709] 
						\draw    (90,230) -- (90,260) ;
						%Straight Lines [id:da6350577296900968] 
						\draw    (80,260) -- (100,260) ;
						%Straight Lines [id:da4272354405456895] 
						\draw    (178,230) -- (178,260) ;
						%Shape: Rectangle [id:dp11266324813813111] 
						\draw   (148,190) -- (208,190) -- (208,230) -- (148,230) -- cycle ;
						%Straight Lines [id:da25501837616485723] 
						\draw    (168,260) -- (188,260) ;
						%Straight Lines [id:da023230593851193193] 
						\draw    (178,160) -- (178,190) ;
						%Shape: Rectangle [id:dp32231844506491303] 
						\draw   (148,120) -- (208,120) -- (208,159) -- (148,159) -- cycle ;
						%Straight Lines [id:da22958057266970333] 
						\draw    (178,20) -- (178,120) ;
						%Straight Lines [id:da3127655053248095] 
						\draw    (168,20) -- (188,20) ;
						%Shape: Rectangle [id:dp8184056261648862] 
						\draw   (58,91) -- (118,91) -- (118,130) -- (58,130) -- cycle ;
						%Straight Lines [id:da005400274751374834] 
						\draw    (90,130) -- (90,140.54) ;
						%Shape: Rectangle [id:dp15245524076070183] 
						\draw   (58,40) -- (118,40) -- (118,80) -- (58,80) -- cycle ;
						%Straight Lines [id:da4489634513741527] 
						\draw    (90,20) -- (90,40) ;
						% Text Node
						\draw (73,200) node [anchor=north west][inner sep=0.75pt]  {$a^{\otimes k}$};
						% Text Node
						\draw (160,200) node [anchor=north west][inner sep=0.75pt]  {$ (a^{\dagger })^{\otimes k}$};
						% Text Node
						\draw (171,130) node [anchor=north west][inner sep=0.75pt]   {$T$};
						% Text Node
						\draw (75,150) node [anchor=north west][inner sep=0.75pt]    {$\mathcal{L}(\sigma)$};
						% Text Node
						\draw (73,100) node [anchor=north west][inner sep=0.75pt]  {$ (a^{\dagger })^{\otimes k}$};
						% Text Node
						\draw (65,50) node [anchor=north west][inner sep=0.75pt]    {$\mathcal{L}(\sigma^{-1})$};
		\end{tikzpicture} }}}\ket{0} &= \sum_{\gamma \in S_k}
		\vcenter{\hbox{\scalebox{.5}{
					\begin{tikzpicture}[x=0.75pt,y=0.75pt,yscale=-1,xscale=1]
						%uncomment if require: \path (0,437); %set diagram left start at 0, and has height of 437
						
						%Straight Lines [id:da9612813725872293] 
						\draw    (310,180) -- (310,200) ;
						%Straight Lines [id:da8975884864886576] 
						\draw    (310,260) -- (310,280) ;
						%Straight Lines [id:da7684964934228584] 
						\draw    (300,280) -- (320,280) ;
						%Straight Lines [id:da7560121868707983] 
						\draw    (400,260) -- (400,280) ;
						%Straight Lines [id:da2498965537217166] 
						\draw    (390,280) -- (410,280) ;
						%Straight Lines [id:da6261779508758296] 
						\draw    (400,40) -- (400,140) ;
						%Straight Lines [id:da08778840759811257] 
						\draw    (390,40) -- (410,40) ;
						%Straight Lines [id:da5155944712234881] 
						\draw    (310,200) -- (400,210) ;
						%Straight Lines [id:da24790756947660353] 
						\draw    (310,210) -- (400,200) ;
						%Straight Lines [id:da6293862003732003] 
						\draw    (400,180) -- (400,200) ;
						%Shape: Rectangle [id:dp11889880900300609] 
						\draw   (370,140) -- (430,140) -- (430,180) -- (370,180) -- cycle ;
						%Shape: Rectangle [id:dp8566730827705724] 
						\draw   (280,140) -- (340,140) -- (340,180) -- (280,180) -- cycle ;
						%Straight Lines [id:da53852414483604] 
						\draw    (312,80) -- (312,90) ;
						%Straight Lines [id:da9612354806194261] 
						\draw    (302,20) -- (322,20) ;
						%Shape: Rectangle [id:dp8003041386540706] 
						\draw   (280,91) -- (340,91) -- (340,130) -- (280,130) -- cycle ;
						%Straight Lines [id:da7794256198600251] 
						\draw    (312,130) -- (312,140.54) ;
						%Shape: Rectangle [id:dp10159874773433741] 
						\draw   (280,40) -- (340,40) -- (340,80) -- (280,80) -- cycle ;
						%Straight Lines [id:da9469273313365338] 
						\draw    (312,20) -- (312,40) ;
						%Straight Lines [id:da8064024661797882] 
						\draw    (310,210) -- (310,220.54) ;
						%Straight Lines [id:da5300440009514533] 
						\draw    (400,210) -- (400,220.54) ;
						%Shape: Square [id:dp39146985193933737] 
						\draw   (290,220) -- (330,220) -- (330,260) -- (290,260) -- cycle ;
						%Shape: Square [id:dp6103376725705019] 
						\draw   (380,220) -- (420,220) -- (420,260) -- (380,260) -- cycle ;
						
						% Text Node
						\draw (292,101) node [anchor=north west][inner sep=0.75pt]   [align=left] {$( a^{\dagger })^{\otimes k}$};
						% Text Node
						\draw (300,235) node [anchor=north west][inner sep=0.75pt]    {$\mathcal{L}_{\gamma }$};
						% Text Node
						\draw (385,235) node [anchor=north west][inner sep=0.75pt]    {$\mathcal{L}_{\gamma ^{-1}}$};
						% Text Node
						\draw (284,46) node [anchor=north west][inner sep=0.75pt]    {$\mathcal{L}( \sigma ^{-1})$};
						% Text Node
						\draw (291,150) node [anchor=north west][inner sep=0.75pt]    {$\mathcal{L}( \sigma )$};
						% Text Node
						\draw (393,149) node [anchor=north west][inner sep=0.75pt]   [align=left] {$T$};
				\end{tikzpicture}}}}\ket{0} 
		= \sum_{\gamma \in S_k} \vcenter{\hbox{\scalebox{.5}{
					\tikzset{every picture/.style={line width=0.75pt}} %set default line width to 0.75pt        
					\begin{tikzpicture}[x=0.75pt,y=0.75pt,yscale=-1,xscale=1]
						%uncomment if require: \path (0,437); %set diagram left start at 0, and has height of 437
						
						%Shape: Rectangle [id:dp8867173159064803] 
						\draw   (520,270) -- (580,270) -- (580,310) -- (520,310) -- cycle ;
						%Straight Lines [id:da8547422371513185] 
						\draw    (550,250) -- (550,270) ;
						%Straight Lines [id:da5313723907595258] 
						\draw    (550,310) -- (550,330) ;
						%Straight Lines [id:da34781091524290453] 
						\draw    (550,370) -- (550,380) ;
						%Straight Lines [id:da6999475165281495] 
						\draw    (560,380) -- (540,380) ;
						%Straight Lines [id:da5802647179283027] 
						\draw    (550,190) -- (550,210) ;
						%Shape: Rectangle [id:dp8966106985853026] 
						\draw   (520,150) -- (580,150) -- (580,190) -- (520,190) -- cycle ;
						%Straight Lines [id:da14462355710982266] 
						\draw    (552,90) -- (552,100) ;
						%Straight Lines [id:da28911966576848624] 
						\draw    (540,40) -- (560,40) ;
						%Shape: Rectangle [id:dp5706504569746031] 
						\draw   (520,101) -- (580,101) -- (580,140) -- (520,140) -- cycle ;
						%Straight Lines [id:da06660529545209615] 
						\draw    (552,140) -- (552,150.54) ;
						%Shape: Rectangle [id:dp6044318918552372] 
						\draw   (520,50) -- (580,50) -- (580,90) -- (520,90) -- cycle ;
						%Straight Lines [id:da63088252124867] 
						\draw    (550,40) -- (550,50) ;
						%Shape: Square [id:dp17908436634242042] 
						\draw   (530,330) -- (570,330) -- (570,370) -- (530,370) -- cycle ;
						%Shape: Square [id:dp7193671459649069] 
						\draw   (530,210) -- (570,210) -- (570,250) -- (530,250) -- cycle ;
						
						% Text Node
						\draw (537,340) node [anchor=north west][inner sep=0.75pt]    {$\mathcal{L}_{\gamma }$};
						% Text Node
						\draw (532,220) node [anchor=north west][inner sep=0.75pt]    {$\mathcal{L}_{\gamma ^{-1}}$};
						% Text Node
						\draw (531,159) node [anchor=north west][inner sep=0.75pt]    {$\mathcal{L}( \sigma )$};
						% Text Node
						\draw (524,56) node [anchor=north west][inner sep=0.75pt]    {$\mathcal{L}(\sigma^{-1})$};
						% Text Node
						\draw (533,111) node [anchor=north west][inner sep=0.75pt]   [align=left] {$( a^{\dagger })^{\otimes k}$};
						% Text Node
						\draw (543,280) node [anchor=north west][inner sep=0.75pt]    {$T$};
				\end{tikzpicture}}}} \ket{0} \\
		&= k! \Tr_{V_N^{\otimes k}}(\mathcal{L}(\sigma)T \mathcal{L}(\sigma^{-1}) (a^{\dagger})^{\otimes k}) \ket{0}.
	\end{aligned}
\end{equation}
The first equality follows by encoding the contraction of annihilation/creation operators in a sum over $\gamma \in S_k$, and the last equality follows by $\mathcal{L}_{\gamma}T=T\mathcal{L}_{\gamma}$. This establishes the equality \eqref{eq: adjoint on H_k}.

We are now in a position to define the Hermitian operator of interest.
Let $C_3^{(k)}$ be the operator defined to act on $\ket{T} \in \mathcal{H}^{(k)}$ as
\begin{equation}
	C_3^{{(k)}}\ket{T} = \frac{1}{3} \sum_{\substack{\sigma = (ijk) \\ 1 \leq i \neq j \neq k \leq N}} \Adj{\sigma}\ket{T} = \frac{1}{3} \sum_{\substack{\sigma = (ijk) \\ 1 \leq i \neq j \neq k \leq N}} \vcenter{\hbox{\scalebox{.5}{				
				\begin{tikzpicture}[x=0.75pt,y=0.75pt,yscale=-1,xscale=1]
					%uncomment if require: \path (0,498); %set diagram left start at 0, and has height of 498
					
					%Straight Lines [id:da626375140590903] 
					\draw    (130,90) -- (130,110) ;
					%Straight Lines [id:da25443912327523255] 
					\draw    (130,210) -- (130,230) ;
					%Straight Lines [id:da05825802010376102] 
					\draw    (130,330) -- (130,350) ;
					%Straight Lines [id:da9573371321555422] 
					\draw    (140,350) -- (120,350) ;
					%Shape: Rectangle [id:dp9131718681325762] 
					\draw   (100,170) -- (160,170) -- (160,210) -- (100,210) -- cycle ;
					%Straight Lines [id:da40892125721536865] 
					\draw    (120,90) -- (140,90) ;
					%Shape: Rectangle [id:dp4112020291829763] 
					\draw   (100,111) -- (160,111) -- (160,150) -- (100,150) -- cycle ;
					%Straight Lines [id:da5224816222521644] 
					\draw    (130,150) -- (130,170) ;
					%Straight Lines [id:da6692383747657287] 
					\draw    (130,270) -- (130,290) ;
					%Shape: Rectangle [id:dp5615429718704468] 
					\draw   (100,230) -- (160,230) -- (160,270) -- (100,270) -- cycle ;
					%Shape: Rectangle [id:dp24010794177670425] 
					\draw   (100,290) -- (160,290) -- (160,330) -- (100,330) -- cycle ;
					
					% Text Node
					\draw (123,239) node [anchor=north west][inner sep=0.75pt]   [align=left] {$T$};
					% Text Node
					\draw (107,120) node [anchor=north west][inner sep=0.75pt]   [align=left] {$( a^{\dagger })^{\otimes k}$};
					% Text Node
					\draw (103,300) node [anchor=north west][inner sep=0.75pt]    {$\mathcal{L}(\sigma^{-1})$};
					% Text Node
					\draw (111,180) node [anchor=north west][inner sep=0.75pt]    {$\mathcal{L}(\sigma)$};				
			\end{tikzpicture}}}} \ket{0}, \label{eq: C3k action}
\end{equation}
where the sum is over all $3$-cycles. It commutes with the adjoint action of $S_N$,
\begin{equation}
	\Adj{\gamma}C_3^{(k)} = C_3^{(k)}\Adj{\gamma}, \quad \forall \gamma \in S_N,
\end{equation}
because $C_3^{(k)}$ is a sum over an entire conjugacy class. We now use a sequence of diagrammatic manipulations to show that the action of $C_3^{(k)}$ can equivalently be expressed using an element $\SWdual{T}_3^{(2k)} \in P_{2k}(N)$. A useful way to rewrite the diagram in \eqref{eq: C3k action} is
\begin{equation}
	\frac{1}{3} \sum_{\substack{\sigma = (ijk) \\ 1 \leq i \neq j \neq k \leq N}} \vcenter{\hbox{\scalebox{.5}{
				\begin{tikzpicture}[x=0.75pt,y=0.75pt,yscale=-1,xscale=1]
					%uncomment if require: \path (0,498); %set diagram left start at 0, and has height of 498
					
					%Straight Lines [id:da626375140590903] 
					\draw    (130,90) -- (130,110) ;
					%Straight Lines [id:da25443912327523255] 
					\draw    (130,210) -- (130,230) ;
					%Straight Lines [id:da05825802010376102] 
					\draw    (130,330) -- (130,350) ;
					%Straight Lines [id:da9573371321555422] 
					\draw    (140,350) -- (120,350) ;
					%Shape: Rectangle [id:dp9131718681325762] 
					\draw   (100,170) -- (160,170) -- (160,210) -- (100,210) -- cycle ;
					%Straight Lines [id:da40892125721536865] 
					\draw    (120,90) -- (140,90) ;
					%Shape: Rectangle [id:dp4112020291829763] 
					\draw   (100,111) -- (160,111) -- (160,150) -- (100,150) -- cycle ;
					%Straight Lines [id:da5224816222521644] 
					\draw    (130,150) -- (130,170) ;
					%Straight Lines [id:da6692383747657287] 
					\draw    (130,270) -- (130,290) ;
					%Shape: Rectangle [id:dp5615429718704468] 
					\draw   (100,230) -- (160,230) -- (160,270) -- (100,270) -- cycle ;
					%Shape: Rectangle [id:dp24010794177670425] 
					\draw   (100,290) -- (160,290) -- (160,330) -- (100,330) -- cycle ;
					
					% Text Node
					\draw (123,239) node [anchor=north west][inner sep=0.75pt]   [align=left] {$T$};
					% Text Node
					\draw (107,120) node [anchor=north west][inner sep=0.75pt]   [align=left] {$( a^{\dagger })^{\otimes k}$};
					% Text Node
					\draw (103,300) node [anchor=north west][inner sep=0.75pt]    {$\mathcal{L}(\sigma^{-1})$};
					% Text Node
					\draw (111,180) node [anchor=north west][inner sep=0.75pt]    {$\mathcal{L}(\sigma)$};				
			\end{tikzpicture}}}} \ket{0} = \frac{1}{3} \sum_{\substack{\sigma = (ijk) \\ 1 \leq i \neq j \neq k \leq N}}  \vcenter{\hbox{\scalebox{.5}{ 
				\begin{tikzpicture}[x=0.75pt,y=0.75pt,yscale=-1,xscale=1]
					%uncomment if require: \path (0,498); %set diagram left start at 0, and has height of 498
					
					%Shape: Rectangle [id:dp4315773755404646] 
					\draw   (190,230) -- (250,230) -- (250,270) -- (190,270) -- cycle ;
					%Straight Lines [id:da31820613266726294] 
					\draw    (220,210) -- (220,230) ;
					%Straight Lines [id:da33345203582425564] 
					\draw    (220,331.47) -- (220,350) ;
					%Straight Lines [id:da032862907394472396] 
					\draw    (330,350) -- (210,350) ;
					%Shape: Rectangle [id:dp06477490702230271] 
					\draw   (190,170) -- (250,170) -- (250,210) -- (190,210) -- cycle ;
					%Shape: Rectangle [id:dp8120222571679907] 
					\draw   (190,110) -- (250,110) -- (250,150) -- (190,150) -- cycle ;
					%Shape: Rectangle [id:dp6931884017060568] 
					\draw   (290,170) -- (350,170) -- (350,210) -- (290,210) -- cycle ;
					%Straight Lines [id:da12155176203516183] 
					\draw    (220,270) -- (220,290) ;
					%Curve Lines [id:da1612773701777097] 
					\draw    (220,290) .. controls (219.95,314.82) and (320.95,314.15) .. (320,290) ;
					%Curve Lines [id:da1389481559097987] 
					\draw    (220,331.47) .. controls (220.28,310.15) and (320.95,309.49) .. (320,330) ;
					%Straight Lines [id:da679182819806502] 
					\draw    (320,210) -- (320,290) ;
					%Straight Lines [id:da7420900511569184] 
					\draw    (320,330) -- (320,350) ;
					%Straight Lines [id:da6557178866590176] 
					\draw    (210,90) -- (330,90) ;
					%Straight Lines [id:da6911798812038479] 
					\draw    (320,90) -- (320,170) ;
					%Straight Lines [id:da7832161310481847] 
					\draw    (220,150) -- (220,170) ;
					%Straight Lines [id:da627609193650722] 
					\draw    (220,90) -- (220,110) ;
					
					% Text Node
					\draw (213,241) node [anchor=north west][inner sep=0.75pt]   [align=left] {$T$};
					% Text Node
					\draw (299,179) node [anchor=north west][inner sep=0.75pt]    {$\mathcal{L}( \sigma )$};
					% Text Node
					\draw (199,179) node [anchor=north west][inner sep=0.75pt]    {$\mathcal{L}( \sigma )$};
					% Text Node
					\draw (197,119) node [anchor=north west][inner sep=0.75pt]    {$(a^{\dagger } )^{\otimes k}$};
			\end{tikzpicture} }}}\ket{0},
\end{equation}
where we have gone from a trace in $V_N^{\otimes k}$ to a trace in $V_N^{\otimes 2k}$. By arguments analogous to those in section \ref{subsec: central elements in SPkN}, the action of
\begin{equation}
	\frac{1}{3} \sum_{\substack{\sigma = (ijk) \\ 1 \leq i \neq j \neq k \leq N}} \mathcal{L}(\sigma),
\end{equation}
on $V_N^{\otimes 2k}$ is related to an element in $P_{2k}(N)$, that we call $\SWdual{T}_3^{(2k)}$. Diagrammatically, this is understood from the following sequence of identifications,
\begin{equation}
	\frac{1}{3} \sum_{\substack{\sigma = (ijk) \\ 1 \leq i \neq j \neq k \leq N}}  \vcenter{\hbox{\scalebox{.5}{ 
				\begin{tikzpicture}[x=0.75pt,y=0.75pt,yscale=-1,xscale=1]
					%uncomment if require: \path (0,498); %set diagram left start at 0, and has height of 498
					
					%Shape: Rectangle [id:dp4315773755404646] 
					\draw   (190,230) -- (250,230) -- (250,270) -- (190,270) -- cycle ;
					%Straight Lines [id:da31820613266726294] 
					\draw    (220,210) -- (220,230) ;
					%Straight Lines [id:da33345203582425564] 
					\draw    (220,331.47) -- (220,350) ;
					%Straight Lines [id:da032862907394472396] 
					\draw    (330,350) -- (210,350) ;
					%Shape: Rectangle [id:dp06477490702230271] 
					\draw   (190,170) -- (250,170) -- (250,210) -- (190,210) -- cycle ;
					%Shape: Rectangle [id:dp8120222571679907] 
					\draw   (190,110) -- (250,110) -- (250,150) -- (190,150) -- cycle ;
					%Shape: Rectangle [id:dp6931884017060568] 
					\draw   (290,170) -- (350,170) -- (350,210) -- (290,210) -- cycle ;
					%Straight Lines [id:da12155176203516183] 
					\draw    (220,270) -- (220,290) ;
					%Curve Lines [id:da1612773701777097] 
					\draw    (220,290) .. controls (219.95,314.82) and (320.95,314.15) .. (320,290) ;
					%Curve Lines [id:da1389481559097987] 
					\draw    (220,331.47) .. controls (220.28,310.15) and (320.95,309.49) .. (320,330) ;
					%Straight Lines [id:da679182819806502] 
					\draw    (320,210) -- (320,290) ;
					%Straight Lines [id:da7420900511569184] 
					\draw    (320,330) -- (320,350) ;
					%Straight Lines [id:da6557178866590176] 
					\draw    (210,90) -- (330,90) ;
					%Straight Lines [id:da6911798812038479] 
					\draw    (320,90) -- (320,170) ;
					%Straight Lines [id:da7832161310481847] 
					\draw    (220,150) -- (220,170) ;
					%Straight Lines [id:da627609193650722] 
					\draw    (220,90) -- (220,110) ;
					
					% Text Node
					\draw (213,241) node [anchor=north west][inner sep=0.75pt]   [align=left] {$T$};
					% Text Node
					\draw (299,179) node [anchor=north west][inner sep=0.75pt]    {$\mathcal{L}( \sigma )$};
					% Text Node
					\draw (199,179) node [anchor=north west][inner sep=0.75pt]    {$\mathcal{L}( \sigma )$};
					% Text Node
					\draw (197,119) node [anchor=north west][inner sep=0.75pt]    {$(a^{\dagger } )^{\otimes k}$};
			\end{tikzpicture}}}} \ket{0} = \frac{1}{3} \sum_{\substack{\sigma = (ijk) \\ 1 \leq i \neq j \neq k \leq N}}  \vcenter{\hbox{\scalebox{.5}{ 
				\begin{tikzpicture}[x=0.75pt,y=0.75pt,yscale=-1,xscale=1]
					%uncomment if require: \path (0,498); %set diagram left start at 0, and has height of 498
					
					%Shape: Rectangle [id:dp5165061403912805] 
					\draw   (390,230) -- (510,230) -- (510,270) -- (390,270) -- cycle ;
					%Straight Lines [id:da8103046944928609] 
					\draw    (400,210) -- (400,230) ;
					%Straight Lines [id:da5921728447157892] 
					\draw    (400,331.47) -- (400,350) ;
					%Straight Lines [id:da06127266179691748] 
					\draw    (510,350) -- (390,350) ;
					%Shape: Rectangle [id:dp7692552045238563] 
					\draw   (390,169) -- (510,169) -- (510,209) -- (390,209) -- cycle ;
					%Shape: Rectangle [id:dp4217744622064057] 
					\draw   (390,110) -- (510,110) -- (510,149) -- (390,149) -- cycle ;
					%Straight Lines [id:da03327284699993416] 
					\draw    (400,270) -- (400,290) ;
					%Curve Lines [id:da5308854816459345] 
					\draw    (400,290) .. controls (399.95,314.82) and (500.95,314.15) .. (500,290) ;
					%Curve Lines [id:da7616061333471067] 
					\draw    (400,331.47) .. controls (400.28,310.15) and (500.95,309.49) .. (500,330) ;
					%Straight Lines [id:da5962011654337684] 
					\draw    (500,210) -- (500,230) ;
					%Straight Lines [id:da6761833388632805] 
					\draw    (500,330) -- (500,350) ;
					%Straight Lines [id:da6098485782966487] 
					\draw    (390,90) -- (510,90) ;
					%Straight Lines [id:da7487267920110341] 
					\draw    (500,90) -- (500,110) ;
					%Straight Lines [id:da45211450013142307] 
					\draw    (400,150) -- (400,170) ;
					%Straight Lines [id:da35534162395655144] 
					\draw    (400,90) -- (400,110) ;
					%Straight Lines [id:da4029525198052937] 
					\draw    (500,270) -- (500,290) ;
					%Shape: Rectangle [id:dp18113372861967592] 
					\draw   (390,290) -- (510,290) -- (510,330) -- (390,330) -- cycle ;
					%Straight Lines [id:da4412201045083659] 
					\draw    (500,150) -- (500,170) ;
					
					% Text Node
					\draw (428,241) node [anchor=north west][inner sep=0.75pt]   [align=left] {$T\otimes 1$};
					% Text Node
					\draw (413,118) node [anchor=north west][inner sep=0.75pt]   [align=left] {$ (a^{\dagger } )^{\otimes k} \otimes 1$};
					% Text Node
					\draw (429,178) node [anchor=north west][inner sep=0.75pt]    {$\mathcal{L}( \sigma )$};
			\end{tikzpicture}}}}\ket{0} = \vcenter{\hbox{\scalebox{.5}{  
				\begin{tikzpicture}[x=0.75pt,y=0.75pt,yscale=-1,xscale=1]
					%uncomment if require: \path (0,498); %set diagram left start at 0, and has height of 498
					
					%Shape: Rectangle [id:dp6708722458609166] 
					\draw   (530,230) -- (650,230) -- (650,270) -- (530,270) -- cycle ;
					%Straight Lines [id:da7029185876202912] 
					\draw    (540,210) -- (540,230) ;
					%Straight Lines [id:da3677364915779926] 
					\draw    (540,331.47) -- (540,350) ;
					%Straight Lines [id:da12205302106878979] 
					\draw    (650,350) -- (530,350) ;
					%Shape: Rectangle [id:dp6525885937972771] 
					\draw   (530,169) -- (650,169) -- (650,209) -- (530,209) -- cycle ;
					%Shape: Rectangle [id:dp274378537235193] 
					\draw   (530,110) -- (650,110) -- (650,149) -- (530,149) -- cycle ;
					%Straight Lines [id:da053545346761818324] 
					\draw    (540,270) -- (540,290) ;
					%Curve Lines [id:da8766309823300324] 
					\draw    (540,290) .. controls (539.95,314.82) and (640.95,314.15) .. (640,290) ;
					%Curve Lines [id:da24617431445761873] 
					\draw    (540,331.47) .. controls (540.28,310.15) and (640.95,309.49) .. (640,330) ;
					%Straight Lines [id:da9301117888899537] 
					\draw    (640,210) -- (640,230) ;
					%Straight Lines [id:da5685991349479873] 
					\draw    (640,330) -- (640,350) ;
					%Straight Lines [id:da3007908102970831] 
					\draw    (530,90) -- (650,90) ;
					%Straight Lines [id:da8176488940917512] 
					\draw    (640,90) -- (640,110) ;
					%Straight Lines [id:da05003780900524113] 
					\draw    (540,150) -- (540,170) ;
					%Straight Lines [id:da4499491074557558] 
					\draw    (540,90) -- (540,110) ;
					%Straight Lines [id:da5576577117353909] 
					\draw    (640,270) -- (640,290) ;
					%Shape: Rectangle [id:dp596901702293569] 
					\draw   (530,290) -- (650,290) -- (650,330) -- (530,330) -- cycle ;
					%Straight Lines [id:da7022548493893712] 
					\draw    (640,150) -- (640,170) ;
					
					% Text Node
					\draw (568,241) node [anchor=north west][inner sep=0.75pt]   [align=left] {$T\otimes 1$};
					% Text Node
					\draw (569,176) node [anchor=north west][inner sep=0.75pt]    {$T_{2}^{( 2k)}$};
					% Text Node
					\draw (553,118) node [anchor=north west][inner sep=0.75pt]   [align=left] {$ (a^{\dagger } )^{\otimes k} \otimes 1$};
			\end{tikzpicture}}}} \ket{0}. \label{eq: C3 to P2k}
\end{equation}
That is, we have
\begin{equation}
	C_3^{(k)}\ket{T} = \Tr_{V_N^{\otimes 2k}} \big( c (T \otimes 1) \SWdual{T}_3^{(2k)} ((a^\dagger)^{\otimes k} \otimes 1) ) \big) \ket{0},
\end{equation}
where $c \in P_{2k}(N)$ is the bottom box in the diagram on the RHS of \eqref{eq: C3 to P2k} and
\begin{equation}
	(c)^{i_1 \dots i_{2k}}_{j_1 \dots {j_{2k}}} = \delta^{i_1 i_{k+1}} \dots \delta^{i_k i_{2k}} \delta_{j_1 j_{k+1}} \dots \delta_{j_k j_{2k}}.
\end{equation}
The explicit formula for $\SWdual{T}_3^{(2k)}$ could be derived using steps similar to the derivation of the relation between $\SWdual{T}_2^{(k)}$ and $T_2^{(k)}$ in section \ref{subsec: central elements in SPkN}. Relating $C_3^{(k)}$ to an element $\SWdual{T}_3^{(2k)}$ using $P_{2k}(N)$ allows for two kinds of large $N$ simplification. Firstly, in place of $N! / (N-3)!3!$ terms in $C_3^{(k)}$ we have no more than $B({2k})$ terms in $\SWdual{T}_3^{(2k)}$, where $B({2k})$ are the Bell numbers. Additionally, index contractions ranging over $N$ can be replaced by multiplication in the partition algebra $P_{2k}(N)$ when $\ket{T} \in \Hilbertspace$, the complexity of this multiplication scales with $k$.

We now move on to discuss the spectrum of $C_3^{(k)}$.
Since $\mathcal{H}^{(k)}$ is reducible with respect to the adjoint action of $S_N$, it decomposes into irreducible representations of $S_N$, labeled by Young diagrams $Y$ with $N$ boxes. By Schur's lemma the action of $C_3^{(k)}$ on each irreducible subspace of this decomposition is proportional to the identity. The constant of proportionality is the normalized character of $C_3^{(k)}$ in the irreducible representation $Y$,
\begin{equation}
	\hat{\chi}_Y(C_3^{(k)}) = \frac{\chi_Y(C_3^{(k)})}{\DimSN{Y}}.
\end{equation}
Normalized characters of $C_3^{(k)}$ are known \cite[Theorem 4]{Lassalle2007} to equal
\begin{equation} \label{eq: norm char C_3}
	\hat{\chi}_Y(C_3^{(k)}) = \sum_{(p,q) \in Y} (q-p)^2 - \frac{N(N-1)}{2},
\end{equation}
where the sum is over all cells in the Young diagram $Y$, using coordinates $(p,q)$ for rows and columns respectively. For example, the largest eigenvalue of $C_3^{(k)}$ corresponds to the trivial representation (Young diagram with all $N$ boxes in the first row) where
\begin{equation}
	\sum_{(p,q) \in Y} (q-p)^2 = 0^2 + 1^2 + 2^2 + \dots + (N-1)^2 = \frac{N(N-1)(2N-1)}{6},
\end{equation}
which gives the eigenvalue $\tfrac{N(N-1)(N-2)}{3}$ in \eqref{eq: norm char C_3}. In what follows it will be useful to shift the eigenvalue of the trivial representation to zero by considering the operator
\begin{equation}
	\hat{C}_3^{(k)} = \frac{N(N-1)(N-2)}{3}-C_3^{(k)}.
\end{equation}
In terms of oscillators and projectors, $\hat{C}_3^{(k)}$ is written as
\begin{equation} \label{eq: C3k}
	\hat{C}_3^{{(k)}} = \frac{1}{k!}\mathcal{P}_k\qty[\frac{N(N-1)(N-2)}{3} - \sum_{\substack{\sigma = (ijk) \\ 1 \leq i \neq j \neq k \leq N}} \Tr_{V_N^{\otimes k}}(\mathcal{L}(\sigma^{-1}) (a^{\dagger})^{\otimes k} \mathcal{L}(\sigma) a^{\otimes k} )]\mathcal{P}_k.
\end{equation}

We can use $\hat{C}_3^{(k)}$ to construct Hamiltonians with interesting spectra.
Consider the family of Hamiltonians (depending on $K$)
\begin{equation}
	\boxed{ H_{K} = \sum_{k=0}^K \hat{C}_3^{(k)}H_0 + \sum_{k=K+1}^\infty \mathcal{P}_kH_0 
	\, , } 
	\label{eq: degenerate invariant ground state hamiltonian}
\end{equation}
where $H_0$ is the free Hamiltonian (number operator) defined in \eqref{eq: free H}. In this model, all invariant states of degree $k\leq K$ have zero energy, while non-invariant states have energies that scale with $N$. For example, degree $k\leq K$ states in the representation $[N-1,1]$ (a Young diagram with $N-1$ boxes in the first row and a single box in the second row) of $S_N$ have energies $kN(N-2)$. More generally, degree $k \leq K$ states in the representation $[N-a,a]$ for $1 \leq a < \lfloor N/2 \rfloor$ have energy $k(N-a+1)(N-2)a$.
States of degree $k > K$ have energy $k$.
The spectrum of $H_K$ is illustrated on the left hand side of figure \ref{fig: spectrum scenarios d}. Taking $ N \gg K$, there is a $K$-dependent degeneracy of invariant ground states and a gap of order $K$.
%The excited states with energies of order $K$ are $S_N$ invariant. For finite $N$ the energy of non-invariant states is finite but scales with $N$. In the limit $N \rightarrow \infty$ the non-invariant states decouple.
In this scenario, the subspace of ground states has dimension
\begin{equation}
	\sum_{k=0}^K \Dim \Hilbertspace^{(k)} = 1 + \sum_{k=1}^K \Dim SP_k(N),
\end{equation}
where $\Hilbertspace^{(k)}$ is the degree $k$ subspace of $\Hilbertspace$ (see equation B.11 in \cite{LMT} for explicit formulas computing $\Dim \Hilbertspace^{(k)}$). By taking 
$N \gg K \gg 1$, we can have a large degeneracy of ground states alongside the interesting correlations between the degeneracy of ground states and the energy gap. A large ground state degeneracy associated with elements of a diagrammatic algebra, in this case the partition algebras $SP_{ k } ( N  ) $ for $k \le K$, is reminiscent of topological degeneracy and its links to anyons \cite{WenNiu1990, KitaevAnyons2006}. We leave a more detailed investigation of the analogies between the present algebraic constructions and topological degeneracy for the future.

\subsection{Resolving the invariant spectrum} \label{sec: charges hamiltonian}
In the previous section we discussed a Hamiltonian \eqref{eq: degenerate invariant ground state hamiltonian} with degenerate ground state. We will now use the commuting algebraic charges $\SWdual{T}_2^{(k)}, t_2^{(k)} \in P_{k}(N)$, constructed in section \ref{sec: charges}, to resolve this degeneracy. Note that the charges commute with $\Adj{\sigma}$ and in particular they commute with $\hat{C}_3^{(k)}$. We prove this in the next subsection, where we consider more general operators coming from elements of $P_k(N)$. Note that because $\SWdual{T}_2^{(k)}$ and $t_2^{(k)}$ are central elements of $P_k(N)$, and the representation basis states $\ket*{Q^{\Lambda_1}_{\Lambda_2, \mu \nu}}$ correspond to elements in $P_k(N)$, the charge's left and right actions are equivalent on these basis states.

%Recall (see equation \eqref{eq: T2SN Hermitian op} and \eqref{eq: T2Sk Hermitian op}) that $\SWdual{T}_2^{(k)}, t_2^{(k)}$ are related to the lower(left) action of $\sigma \in S_N$
%\begin{equation}
%	\mathcal{L}(\sigma^{-1})\ket{T} = \sum_{\substack{i_1 \dots i_k \\ j_1 \dots j_k}}T^{j_1 \dots j_k}_{i_1 \dots i_k}  (a^{\dagger})^{i_1}_{\sigma(j_1)} \dots (a^{\dagger})^{i_k}_{\sigma(j_{k})}\ket{0},
%\end{equation}
%and $\tau \in S_k$ on $\mathcal{H}^{(k)}$
%\begin{equation}
%	\mathcal{L}_{\tau}\ket{T} = \sum_{\substack{i_1 \dots i_k \\ j_1 \dots j_k}}T^{j_1 \dots j_k}_{i_1 \dots i_k}  (a^{\dagger})^{i_1}_{j_{\tau(1)}} \dots (a^{\dagger})^{i_k}_{j_{\tau(k)}}\ket{0}.
%\end{equation}
%respectively.
The algebraic charges can be written in terms of oscillators and projectors as in \eqref{eq: T2PkN Hermitian op} and \eqref{eq: T2Sk Hermitian op}.
Importantly, the representation basis states $\ket*{Q^{\Lambda_1}_{\Lambda_2, \mu \nu}}$ are eigenstates of $\SWdual{T}_2^{(k),L}, t_2^{(k),L}$. The eigenvalues are normalized characters of the representations $\Lambda_1$ of $S_N$ and $\Lambda_2$ of $S_k$ respectively (see \eqref{eq: rep basis is eigenbasis of T2}). That is
\begin{align}
	&\SWdual{T}_2^{(k),L}\ket{Q^{\Lambda_1}_{\Lambda_2, \mu \nu}} = \ket{\SWdual{T}_2^{(k)} Q^{\Lambda_1}_{\Lambda_2, \mu \nu}} = \SWdual{T}_2^{(k),R}\ket{Q^{\Lambda_1}_{\Lambda_2, \mu \nu}} =  \hat{\chi}^{\Lambda_1}(T_2) \ket{Q^{\Lambda_1}_{\Lambda_2, \mu \nu}},\\
	&t_2^{(k),L}\ket{Q^{\Lambda_1}_{\Lambda_2, \mu \nu}} = \ket{t_2^{(k)} Q^{\Lambda_1}_{\Lambda_2, \mu \nu}} = t_2^{(k),R}\ket{Q^{\Lambda_1}_{\Lambda_2, \mu \nu}} = \hat{\chi}^{\Lambda_2}(t_2) \ket{Q^{\Lambda_1}_{\Lambda_2, \mu \nu}},
\end{align}
where the normalized characters $\hat{\chi}$ are defined in \eqref{eq: norm characters of T2s}. Note that the eigenvalues of the operator $t_2^{(k), L}$ range between $\pm \tfrac{k(k-1)}{2}$, and those of $\SWdual{T}_2^{(k), L}$ between $\pm \tfrac{N(N-1)}{2}$, including an infinite number of such operators in a Hamiltonian may result in a spectrum that is not bounded from below.
By adding these algebraic charges to the Hamiltonian \eqref{eq: degenerate invariant ground state hamiltonian} the energy of the states $\ket*{Q^{\Lambda_1}_{\Lambda_2, \mu \nu}}$ labeled by distinct pairs $\Lambda_1, \Lambda_2$ will split. As discussed in section \ref{sec:multiplicity-non-central}, the multiplicity labels $\mu, \nu$ are not distinguished by these central algebraic charges. Hamiltonians that resolve more detailed information such as multiplicity labels are discussed in the next subsection.

For concreteness we consider the spectrum of the Hamiltonian
\begin{align} \label{eq: partially broken ground state degen H} \nonumber
	H_K' &= H_K + H_{\text{res}} \\ \nonumber
	&= H_K - \frac{2}{N(N-1)}\sum_{k=1}^K \SWdual{T}_2^{(k),L} \\
	&=\sum_{k=0}^K \hat{C}_3^{(k)}H_0 + \sum_{k=K+1}^\infty \mathcal{P}_k H_0 - \frac{2}{N(N-1)}\sum_{k=1}^K \SWdual{T}_2^{(k),L}.
\end{align}
The ground state degeneracy is reduced compared to $H_K$. The lowest energy states are degree $k \leq K$ states $\ket*{Q^{[N]}_{\Lambda_2, \mu \nu}}$ with energy $-1$. The highest energy state with degree $k \leq K$ is $\ket*{Q^{[N-K,1^K]}_{[1^K]}}$, it has degree $K$ and energy $-\tfrac{(N-2K-1)}{(N-1)}$. The gap of order $K$ remains, as illustrated on the right of figure \ref{fig: spectrum scenarios d}. The label $\Lambda_2$ can be resolved by including $t_2^{(k),L}$ in the Hamiltonian.

To fully resolve the labels $\Lambda_1, \Lambda_2$ for general $k$ and $N$, new charges are necessary. Detailed discussions of the problem of using such charges in the centre of the symmetric group algebra $ \mC [S_n ] $, with motivations coming from a model for information loss in AdS/CFT \cite{IILoss}, are given in \cite{BPSChargesSymGroup2019, RamgoSharpe}.  It can be proved that $ \{ T_2 , T_3 , \cdots, T_n \} $ provide an adequate set of charges and these also provide a multiplicative generating set for the centre of the group algebra. Typically, a smaller set 
$ \{ T_2 , T_3 , \cdots , T_{ k_* ( n )} \}$ suffices. For example $ k_* ( 5 ) = 2 , k_* ( 14 ) = 3, k_*(80 ) = 6 $. In the present discussion these results can be applied by choosing $n = k $ and $ n = N $ respectively.

\subsection{Precision resolution of the invariant spectrum}
%In section \ref{sec: PA} we found that there is a close correspondence between $S_N$ invariant states and partition algebras.
In the previous section we presented Hamiltonians involving commuting algebraic charges, constructed from central elements in $P_k(N)$, that resolve the representation labels $\Lambda_1, \Lambda_2$ of representation basis elements $\ket{Q^{\Lambda_1}_{\Lambda_2, \mu \nu}}$.
As discussed in section \ref{sec:multiplicity-non-central}, and illustrated in an explicit example in section \ref{sec: degree two basis}, more general elements of $SP_k(N)$ are necessary to resolve the multiplicity labels $\mu, \nu$. We will use this observation to construct $S_N$ invariant Hamiltonians, involving operators $[d]^L$ and $[d]^R$ constructed from non-central elements $[d] \in SP_k(N)$, with non-degenerate eigenvalues.

%As we mentioned in section \ref{sec: conserved charges}, partition algebra elements can be used to construct quantum mechanical operators which are bilinear in oscillators. We use this correspondence to build $S_N$ invariant Hamiltonians.

Since we want to construct Hamiltonians $H$ satisfying $\comm{\Adj{\sigma}}{H} = 0$, built from operators $[d]^L, [d]^R$, we will now prove that $\comm*{\Adj{\sigma}}{[d]^L} = \comm*{\Adj{\sigma}}{[d]^R}= 0$.
To show that $[d]^L \Adj{\sigma} = \Adj{\sigma}[d]^L$ we combine equation \eqref{eq: partition algebra as operator} with equation \eqref{eq: sigma on ket T}
\begin{equation}
	\begin{aligned}
	\Adj{\sigma}[d]^L\ket{T} &= \Tr_{V_N^{\otimes k}}(\mathcal{L}(\sigma) dT \mathcal{L}(\sigma^{-1}) (a^{\dagger})^{\otimes k}  )\ket{0} \\
		&= \Tr_{V_N^{\otimes k}}(d \mathcal{L}(\sigma)T \mathcal{L}(\sigma^{-1}) (a^{\dagger})^{\otimes k} ) \ket{0} \\
		&=[d]^L\Adj{\sigma} \ket{T},
	\end{aligned} \label{eq: hamiltonians from partition algebras}
\end{equation}
where the second line follows since $\mathcal{L}(\sigma)d = d \mathcal{L}(\sigma)$ as elements of $\End(V_N^{\otimes k})$ (linear maps $V_N^{\otimes k} \rightarrow V_N^{\otimes k}$). The argument is identical for $[d]^R\Adj{\sigma} = \Adj{\sigma}[d]^R$. 

To construct Hamiltonians $H$, using the above operators, we need to ensure that any operator we include in $H$ is Hermitian.
The operators $[d]^L$, $[d]^R$ are not Hermitian in general, unless $[d^T] = [d]$.
Taking this into account, we can parametrise a large family of $S_N$ invariant Hamiltonians using the diagram basis for $P_k(N)$. We write
\begin{equation}
	H = \frac{1}{2}\sum_{k=1}^{\infty} \sum_{[d_{\pi}]} \qty(L_{k,\pi}[d_{\pi}]^L+L_{k,\pi}^*[d^T_{\pi}]^L + R_{k,\pi}[d_{\pi}]^R+R_{k,\pi}^*[d^T_{\pi}]^R ) , \label{eq: inv hamiltonians}
\end{equation}
where the sum over $[d_\pi]$ runs over a basis for $SP_k(N)$ and $L_{k,\pi}, R_{k,\pi}$ are complex parameters with the constraint $L_{k,\pi}^* = L_{k, \pi'}$ and $R_{k,\pi}^* = R_{k, \pi'}$ if $d_\pi^T = d_{\pi'}$.
The equivalent expression for $H$ in terms of oscillators and projectors is
\begin{equation}
	\begin{aligned}
		H = &\frac{1}{2}\sum_{k=1}^{\infty} \sum_{[d_{\pi}]} \mathcal{P}_k \Tr_{V_N^{\otimes k}}\qty((a^{\dagger})^{\otimes k}\frac{L_{k,\pi} d_{\pi}+ L_{k,\pi}^*d^T_{\pi}}{k!}a^{\otimes k}) \mathcal{P}_k \\
		&+\frac{1}{2}\sum_{k=1}^{\infty} \sum_{[d_{\pi}]} \mathcal{P}_k \Tr_{V_N^{\otimes k}}\qty(\frac{R_{k,\pi} d_{\pi}+ R_{k,\pi}^*d^T_{\pi}}{k!}(a^{\dagger})^{\otimes k}a^{\otimes k}) \mathcal{P}_k.
	\end{aligned}
\end{equation}
Progressively turning on parameters in equation \eqref{eq: inv hamiltonians} will tend to break degeneracy in the spectrum.
Eventually, the spectrum may take the form in figure \ref{fig: spectrum scenarios a} where invariant and non-invariant states are mixed, and most of the degeneracy is broken.
%Given a Hamiltonian $H$ with such a spectrum, we can use $\hat{C}_3^{(k)}$ to control the energy of invariant states. Depending on the details of how $\hat{C}_3^{(k)}$ is included in the Hamiltonian, the gap between invariant states and non-invariant states can be either finite, as in \ref{fig: spectrum scenarios b}, or infinite as in \ref{fig: spectrum scenarios c}.

\subsection{General invariant Hamiltonians from partition algebras}
The Hamiltonian $H$ in \eqref{eq: inv hamiltonians} is not the most general Hamiltonian satisfying $\comm{H}{\Adj{\sigma}}=0$.
For example, it does not include the Hamiltonian \eqref{eq: hamiltonian} constructed in section \ref{sec: PIMQM} nor $H_{K}$ in \eqref{eq: degenerate invariant ground state hamiltonian}.
As we noticed in \eqref{eq: C3 to P2k}, $C_3^{(k)}$ is related to an element in $P_{2k}(N)$.
We now generalize this observation to give a construction of general $S_N$ invariant operators from elements in $P_{2k}(N)$.

General degree preserving operators that commute with $\Adj{\sigma}$ can be constructed from elements $d \in P_{2k}(N)$ as
\begin{equation}
	\frac{1}{k!}\mathcal{P}_k \Tr_{V_N^{\otimes 2k}}(d (a^{\dagger})^{\otimes k} \otimes a^{\otimes k}) \mathcal{P}_k \leftrightarrow \frac{1}{k!}\mathcal{P}_k \vcenter{\hbox{
			\begin{tikzpicture}[x=0.75pt,y=0.75pt,yscale=-1,xscale=1]
				%uncomment if require: \path (0,364); %set diagram left start at 0, and has height of 364
				
				%Straight Lines [id:da21805089538341282] 
				\draw    (35,20) -- (35,25) ;
				%Straight Lines [id:da301906696027314] 
				\draw    (35,50) -- (35,60) ;
				%Shape: Rectangle [id:dp7273966817703077] 
				\draw   (0,25) -- (50,25) -- (50,50) -- (0,50) -- cycle ;
				%Straight Lines [id:da7343601056611742] 
				\draw    (70,20) -- (70,25) ;
				%Straight Lines [id:da9135600405686415] 
				\draw    (30,20) -- (75,20) ;
				%Straight Lines [id:da3400939727150323] 
				\draw    (70,50) -- (70,60) ;
				%Shape: Rectangle [id:dp7887032586866076] 
				\draw   (55,25) -- (105,25) -- (105,50) -- (55,50) -- cycle ;
				%Shape: Rectangle [id:dp40109170752284906] 
				\draw   (20,60) -- (85,60) -- (85,85) -- (20,85) -- cycle ;
				%Straight Lines [id:da5961734923038404] 
				\draw    (35,85) -- (35,95) ;
				%Straight Lines [id:da4842339912791658] 
				\draw    (70,85) -- (70,95) ;
				%Straight Lines [id:da8475203450398408] 
				\draw    (30,95) -- (75,95) ;
				
				% Text Node
				\draw (44,62) node [anchor=north west][inner sep=0.75pt]    {$d$};
				% Text Node
				\draw (3,30) node [anchor=north west][inner sep=0.75pt]    {$(a^{\dagger } )^{\otimes k}$};
				% Text Node
				\draw (67,30) node [anchor=north west][inner sep=0.75pt]    {$a^{\otimes k}$};
		\end{tikzpicture}}} \mathcal{P}_k 
\end{equation}
The action of these operators on $\ket{T} \in \mathcal{H}^{(k)}$ is
\begin{equation}
	\frac{1}{k!}\sum_{\gamma \in S_k}\vcenter{\hbox{
				\begin{tikzpicture}[x=0.75pt,y=0.75pt,yscale=-1,xscale=1]
					%uncomment if require: \path (0,482); %set diagram left start at 0, and has height of 482
					
					%Straight Lines [id:da687266933404951] 
					\draw    (185,185) -- (185,215) ;
					%Straight Lines [id:da10853156239798323] 
					\draw    (185,240) -- (185,250) ;
					%Shape: Rectangle [id:dp8765010034440197] 
					\draw   (160,215) -- (210,215) -- (210,240) -- (160,240) -- cycle ;
					%Straight Lines [id:da5516430691693646] 
					\draw    (220,185) -- (220,190) ;
					%Straight Lines [id:da7057883562312772] 
					\draw    (180,185) -- (225,185) ;
					%Straight Lines [id:da6216799796975958] 
					\draw    (220,235) -- (220,250) ;
					%Shape: Rectangle [id:dp4385014007736292] 
					\draw   (170,250) -- (235,250) -- (235,275) -- (170,275) -- cycle ;
					%Straight Lines [id:da7756394343304152] 
					\draw    (185,275) -- (185,285) ;
					%Straight Lines [id:da9810059433557479] 
					\draw    (220,275) -- (220,285) ;
					%Straight Lines [id:da1158596426750933] 
					\draw    (180,285) -- (225,285) ;
					%Straight Lines [id:da024213537911054628] 
					\draw    (275,235) -- (275,250) ;
					%Straight Lines [id:da11769750397928824] 
					\draw    (285,285) -- (265,285) ;
					%Straight Lines [id:da8295128011082207] 
					\draw    (265,185) -- (285,185) ;
					%Straight Lines [id:da7482779144810288] 
					\draw    (275,185) -- (275,190) ;
					%Straight Lines [id:da033486054874460436] 
					\draw    (275,275) -- (275,285) ;
					%Shape: Rectangle [id:dp8441656152841235] 
					\draw   (255,250) -- (295,250) -- (295,275) -- (255,275) -- cycle ;
					%Curve Lines [id:da3745161375251429] 
					\draw    (275,235) .. controls (275.05,224.99) and (219.76,224.61) .. (220,210) ;
					%Curve Lines [id:da009760498261240258] 
					\draw    (220,235) .. controls (219.96,225.35) and (275.05,225.35) .. (275,210) ;
					%Shape: Rectangle [id:dp9161946993745367] 
					\draw   (205,190) -- (230,190) -- (230,210) -- (205,210) -- cycle ;
					%Shape: Rectangle [id:dp7944090532437658] 
					\draw   (254,190) -- (289,190) -- (289,210) -- (254,210) -- cycle ;
					
					% Text Node
					\draw (160,215) node [anchor=north west][inner sep=0.75pt]   [align=left] {$ (a^{\dagger } )^{\otimes k}$};
					% Text Node
					\draw (194,252) node [anchor=north west][inner sep=0.75pt]    {$d$};
					% Text Node
					\draw (267,255) node [anchor=north west][inner sep=0.75pt]   [align=left] {$T$};
					% Text Node
					\draw (205,190) node [anchor=north west][inner sep=0.75pt]    {$\mathcal{L}_{\gamma }$};
					% Text Node
					\draw (254,190) node [anchor=north west][inner sep=0.75pt]    {$\mathcal{L}_{\gamma^{-1}}$};
			\end{tikzpicture}}} \ket{0} = \vcenter{\hbox{
			\begin{tikzpicture}[x=0.75pt,y=0.75pt,yscale=-1,xscale=1]
				%uncomment if require: \path (0,482); %set diagram left start at 0, and has height of 482
				
				%Straight Lines [id:da9580215477884384] 
				\draw    (340,245) -- (340,265) ;
				%Straight Lines [id:da7604251044313071] 
				\draw    (340,290) -- (340,300) ;
				%Straight Lines [id:da3452759972783912] 
				\draw    (335,245) -- (380,245) ;
				%Straight Lines [id:da7454509251255632] 
				\draw    (375,290) -- (375,300) ;
				%Shape: Rectangle [id:dp9841543572259148] 
				\draw   (325,300) -- (390,300) -- (390,325) -- (325,325) -- cycle ;
				%Straight Lines [id:da5046653468558031] 
				\draw    (340,325) -- (340,335) ;
				%Straight Lines [id:da44040561211003304] 
				\draw    (375,325) -- (375,335) ;
				%Straight Lines [id:da07355664826753028] 
				\draw    (335,335) -- (380,335) ;
				%Straight Lines [id:da9358249371744021] 
				\draw    (375,245) -- (375,265) ;
				%Shape: Rectangle [id:dp7828963122962675] 
				\draw   (360,265) -- (390,265) -- (390,290) -- (360,290) -- cycle ;
				%Shape: Rectangle [id:dp07055631816373165] 
				\draw   (305,265) -- (355,265) -- (355,290) -- (305,290) -- cycle ;
				
				% Text Node
				\draw (349,305) node [anchor=north west][inner sep=0.75pt]    {$d$};
				% Text Node
				\draw (307,265) node [anchor=north west][inner sep=0.75pt]    {$(a^{\dagger } )^{\otimes k}$};
				% Text Node
				\draw (370,267) node [anchor=north west][inner sep=0.75pt]    {$T$};
		\end{tikzpicture}}}\ket{0}. \label{eq: P2k on T}
\end{equation}
Commutativity with $\Adj{\sigma}$ follows from the following diagrammatic manipulations
\begin{align}
		\frac{1}{k!}\vcenter{\hbox{
				\begin{tikzpicture}[x=0.75pt,y=0.75pt,yscale=-1,xscale=1]
					%uncomment if require: \path (0,482); %set diagram left start at 0, and has height of 482
					%Straight Lines [id:da21805089538341282] 
					\draw    (60,35) -- (60,40) ;
					%Straight Lines [id:da301906696027314] 
					\draw    (60,65) -- (60,75) ;
					%Shape: Rectangle [id:dp7273966817703077] 
					\draw   (25,40) -- (75,40) -- (75,65) -- (25,65) -- cycle ;
					%Straight Lines [id:da7343601056611742] 
					\draw    (95,35) -- (95,40) ;
					%Straight Lines [id:da9135600405686415] 
					\draw    (55,35) -- (100,35) ;
					%Straight Lines [id:da3400939727150323] 
					\draw    (95,65) -- (95,75) ;
					%Shape: Rectangle [id:dp7887032586866076] 
					\draw   (80,40) -- (130,40) -- (130,65) -- (80,65) -- cycle ;
					%Shape: Rectangle [id:dp40109170752284906] 
					\draw   (45,75) -- (110,75) -- (110,100) -- (45,100) -- cycle ;
					%Straight Lines [id:da5961734923038404] 
					\draw    (60,100) -- (60,110) ;
					%Straight Lines [id:da4842339912791658] 
					\draw    (95,100) -- (95,110) ;
					%Straight Lines [id:da8475203450398408] 
					\draw    (55,110) -- (100,110) ;
					%Straight Lines [id:da7744541137189511] 
					\draw    (160,10) -- (160,15) ;
					%Straight Lines [id:da4631868138197308] 
					\draw    (160,75) -- (160,85) ;
					%Straight Lines [id:da13214416104459015] 
					\draw    (160,155) -- (160,165) ;
					%Straight Lines [id:da549314601275114] 
					\draw    (170,165) -- (150,165) ;
					%Shape: Rectangle [id:dp10040196262790935] 
					\draw   (140,50) -- (180,50) -- (180,75) -- (140,75) -- cycle ;
					%Straight Lines [id:da8534022939071009] 
					\draw    (150,10) -- (170,10) ;
					%Straight Lines [id:da5387308552645484] 
					\draw    (160,40) -- (160,50) ;
					%Straight Lines [id:da6977473044491955] 
					\draw    (160,110) -- (160,120) ;
					%Shape: Rectangle [id:dp29083292978209574] 
					\draw   (140,85) -- (180,85) -- (180,110) -- (140,110) -- cycle ;
					%Shape: Rectangle [id:dp5520275187341732] 
					\draw   (130,120) -- (190,120) -- (190,155) -- (130,155) -- cycle ;
					%Shape: Rectangle [id:dp1124220352296963] 
					\draw   (140,15) -- (185,15) -- (185,40) -- (140,40) -- cycle ;
					% Text Node
					\draw (69,77) node [anchor=north west][inner sep=0.75pt]    {$d$};
					% Text Node
					\draw (150,87) node [anchor=north west][inner sep=0.75pt]   [align=left] {$T$};
					% Text Node
					\draw (135,125) node [anchor=north west][inner sep=0.75pt]    {$\mathcal{L} (\sigma^{-1} )$};
					% Text Node
					\draw (145,51) node [anchor=north west][inner sep=0.75pt]    {$\mathcal{L}( \sigma )$};
					% Text Node
					\draw (27,40) node [anchor=north west][inner sep=0.75pt]    {$(a^{\dagger } )^{\otimes k}$};
					% Text Node
					\draw (91,40) node [anchor=north west][inner sep=0.75pt]    {$a^{\otimes k}$};
					% Text Node
					\draw (139,15) node [anchor=north west][inner sep=0.75pt]    {$(a^{\dagger } )^{\otimes k}$};
			\end{tikzpicture}}}\ket{0} = 
		\vcenter{\hbox{
		\begin{tikzpicture}[x=0.75pt,y=0.75pt,yscale=-1,xscale=1]
		%uncomment if require: \path (0,482); %set diagram left start at 0, and has height of 482
		%Straight Lines [id:da44447540157369825] 
		\draw    (50,270) -- (50,320) ;
		%Straight Lines [id:da9982781651935868] 
		\draw    (50,345) -- (50,385) ;
		%Shape: Rectangle [id:dp3368152196049514] 
		\draw   (30,320) -- (75,320) -- (75,345) -- (30,345) -- cycle ;
		%Straight Lines [id:da6091246839375903] 
		\draw    (85,270) -- (85,280) ;
		%Straight Lines [id:da7721580306431195] 
		\draw    (45,270) -- (90,270) ;
		%Straight Lines [id:da17014985801757754] 
		\draw    (85,345) -- (85,385) ;
		%Shape: Rectangle [id:dp11516910728461927] 
		\draw   (35,385) -- (100,385) -- (100,410) -- (35,410) -- cycle ;
		%Straight Lines [id:da11874476601752582] 
		\draw    (50,410) -- (50,465) ;
		%Straight Lines [id:da7508685305462874] 
		\draw    (85,410) -- (85,465) ;
		%Straight Lines [id:da4770192672793119] 
		\draw    (45,465) -- (90,465) ;
		%Straight Lines [id:da8639064756993591] 
		\draw    (140,375) -- (140,385) ;
		%Straight Lines [id:da39757023480639675] 
		\draw    (140,455) -- (140,465) ;
		%Straight Lines [id:da5612893362397544] 
		\draw    (150,465) -- (130,465) ;
		%Shape: Rectangle [id:dp4156660438145803] 
		\draw   (120,350) -- (160,350) -- (160,375) -- (120,375) -- cycle ;
		%Straight Lines [id:da7225999073807765] 
		\draw    (130,270) -- (150,270) ;
		%Straight Lines [id:da9856518724820837] 
		\draw    (140,270) -- (140,280) ;
		%Straight Lines [id:da8140912748777653] 
		\draw    (140,410) -- (140,420) ;
		%Shape: Rectangle [id:dp24489940076145067] 
		\draw   (120,385) -- (160,385) -- (160,410) -- (120,410) -- cycle ;
		%Shape: Rectangle [id:dp9176850338497813] 
		\draw   (110,420) -- (170,420) -- (170,455) -- (110,455) -- cycle ;
		%Curve Lines [id:da045699893948025316] 
		\draw    (140,345) .. controls (140.05,334.99) and (84.76,324.61) .. (85,310) ;
		%Curve Lines [id:da6730886237882334] 
		\draw    (85,345) .. controls (84.96,335.35) and (140.05,325.35) .. (140,310) ;
		%Shape: Rectangle [id:dp859877909103721] 
		\draw   (120,280) -- (160,280) -- (160,305) -- (120,305) -- cycle ;
		%Shape: Rectangle [id:dp02124154033522907] 
		\draw   (65,280) -- (105,280) -- (105,305) -- (65,305) -- cycle ;
		%Straight Lines [id:da795851907126546] 
		\draw    (140,305) -- (140,310) ;
		%Straight Lines [id:da7397767854436907] 
		\draw    (85,305) -- (85,310) ;
		%Straight Lines [id:da42627486463454956] 
		\draw    (140,345) -- (140,350) ;
		% Text Node
		\draw (28,320) node [anchor=north west][inner sep=0.75pt]   [align=left] {$(a^{\dagger } )^{\otimes k}$};
		% Text Node
		\draw (59,387) node [anchor=north west][inner sep=0.75pt]    {$d$};
		% Text Node
		\draw (130,387) node [anchor=north west][inner sep=0.75pt]   [align=left] {$T$};
		% Text Node
		\draw (111,425) node [anchor=north west][inner sep=0.75pt]    {$\mathcal{L}( \sigma^{-1})$};
		% Text Node
		\draw (121,351) node [anchor=north west][inner sep=0.75pt]    {$\mathcal{L}( \sigma )$};
		% Text Node
		\draw (73,285) node [anchor=north west][inner sep=0.75pt]    {$\mathcal{L}_{\gamma }$};
		% Text Node
		\draw (122,285) node [anchor=north west][inner sep=0.75pt]    {$\mathcal{L}_{\gamma ^{-1}}$};
		\end{tikzpicture}}} 
		= \vcenter{\hbox{ 
				\begin{tikzpicture}[x=0.75pt,y=0.75pt,yscale=-1,xscale=1]
					%uncomment if require: \path (0,482); %set diagram left start at 0, and has height of 482
					%Straight Lines [id:da4539473518154926] 
					\draw    (315,20) -- (315,25) ;
					%Straight Lines [id:da29108825365075797] 
					\draw    (315,50) -- (315,175) ;
					%Straight Lines [id:da6969673694967105] 
					\draw    (310,20) -- (355,20) ;
					%Straight Lines [id:da6190594820577555] 
					\draw    (350,165) -- (350,175) ;
					%Shape: Rectangle [id:dp899495375413405] 
					\draw   (300,175) -- (365,175) -- (365,200) -- (300,200) -- cycle ;
					%Straight Lines [id:da7862226129974734] 
					\draw    (315,200) -- (315,210) ;
					%Straight Lines [id:da03037895670839874] 
					\draw    (350,200) -- (350,210) ;
					%Straight Lines [id:da7884182964165198] 
					\draw    (310,210) -- (355,210) ;
					%Straight Lines [id:da2390725864391734] 
					\draw    (350,85) -- (350,95) ;
					%Shape: Rectangle [id:dp26382399169664006] 
					\draw   (330,60) -- (370,60) -- (370,85) -- (330,85) -- cycle ;
					%Straight Lines [id:da9414039320650014] 
					\draw    (350,20) -- (350,60) ;
					%Straight Lines [id:da012512169904822734] 
					\draw    (350,120) -- (350,130) ;
					%Shape: Rectangle [id:dp46828478769359494] 
					\draw   (330,95) -- (370,95) -- (370,120) -- (330,120) -- cycle ;
					%Shape: Rectangle [id:dp45280325749021744] 
					\draw   (320,130) -- (380,130) -- (380,165) -- (320,165) -- cycle ;
					%Shape: Rectangle [id:dp9468682080431237] 
					\draw   (290,25) -- (335,25) -- (335,50) -- (290,50) -- cycle ;
					% Text Node
					\draw (324,177) node [anchor=north west][inner sep=0.75pt]    {$d$};
					% Text Node
					\draw (325,137) node [anchor=north west][inner sep=0.75pt]    {$\mathcal{L} (\sigma ^{-1} )$};
					% Text Node
					\draw (331,61) node [anchor=north west][inner sep=0.75pt]    {$\mathcal{L}( \sigma )$};
					% Text Node
					\draw (290,25) node [anchor=north west][inner sep=0.75pt]    {$(a^{\dagger } )^{\otimes k}$};
					% Text Node
					\draw (343,100) node [anchor=north west][inner sep=0.75pt]    {$T$};
		\end{tikzpicture}}}\ket{0}\nonumber& \\
		=
		\vcenter{\hbox{
				\begin{tikzpicture}[x=0.75pt,y=0.75pt,yscale=-1,xscale=1]
					%uncomment if require: \path (0,482); %set diagram left start at 0, and has height of 482
					%Straight Lines [id:da8820557982672885] 
					\draw    (420,25) -- (420,30) ;
					%Straight Lines [id:da12331113950155714] 
					\draw    (420,165) -- (420,175) ;
					%Shape: Rectangle [id:dp5319115475069578] 
					\draw   (395,30) -- (440,30) -- (440,55) -- (395,55) -- cycle ;
					%Straight Lines [id:da768805900316168] 
					\draw    (415,25) -- (460,25) ;
					%Straight Lines [id:da6240606099479864] 
					\draw    (455,165) -- (455,175) ;
					%Shape: Rectangle [id:dp9788933552535319] 
					\draw   (405,175) -- (470,175) -- (470,200) -- (405,200) -- cycle ;
					%Straight Lines [id:da7386785560531006] 
					\draw    (420,200) -- (420,210) ;
					%Straight Lines [id:da9493912107021063] 
					\draw    (455,200) -- (455,210) ;
					%Straight Lines [id:da043056639414554176] 
					\draw    (415,210) -- (460,210) ;
					%Straight Lines [id:da47660762166929427] 
					\draw    (455,90) -- (455,100) ;
					%Shape: Rectangle [id:dp049739061732437806] 
					\draw   (405,65) -- (470,65) -- (470,90) -- (405,90) -- cycle ;
					%Straight Lines [id:da5708862415574842] 
					\draw    (455,25) -- (455,65) ;
					%Straight Lines [id:da317578889347244] 
					\draw    (455,125) -- (455,135) ;
					%Shape: Rectangle [id:dp24060249279362744] 
					\draw   (435,100) -- (475,100) -- (475,125) -- (435,125) -- cycle ;
					%Shape: Rectangle [id:dp6151114635340624] 
					\draw   (405,135) -- (470,135) -- (470,165) -- (405,165) -- cycle ;
					%Straight Lines [id:da3002588701751483] 
					\draw    (420,55) -- (420,65) ;
					%Straight Lines [id:da6312613782995171] 
					\draw    (420,90) -- (420,135) ;
					% Text Node
					\draw (429,177) node [anchor=north west][inner sep=0.75pt]    {$d$};
					% Text Node
					\draw (411,137) node [anchor=north west][inner sep=0.75pt]    {$\mathcal{L} (\sigma ^{-1} )$};
					% Text Node
					\draw (416,67) node [anchor=north west][inner sep=0.75pt]    {$\mathcal{L}( \sigma )$};
					% Text Node
					\draw (445,102) node [anchor=north west][inner sep=0.75pt]    {$T$};
					% Text Node
					\draw (394,31) node [anchor=north west][inner sep=0.75pt]    {$(a^{\dagger } )^{\otimes k}$};
			\end{tikzpicture}}} \ket{0} = \vcenter{\hbox{
		\begin{tikzpicture}[x=0.75pt,y=0.75pt,yscale=-1,xscale=1]
			%uncomment if require: \path (0,482); %set diagram left start at 0, and has height of 482
			%Straight Lines [id:da3712768524940553] 
			\draw    (525,130) -- (525,175) ;
			%Straight Lines [id:da43081594610430063] 
			\draw    (520,25) -- (565,25) ;
			%Straight Lines [id:da42006780349044437] 
			\draw    (560,165) -- (560,175) ;
			%Shape: Rectangle [id:dp6951832418415191] 
			\draw   (510,175) -- (575,175) -- (575,200) -- (510,200) -- cycle ;
			%Straight Lines [id:da7235439285947718] 
			\draw    (525,200) -- (525,210) ;
			%Straight Lines [id:da47012548921255615] 
			\draw    (560,200) -- (560,210) ;
			%Straight Lines [id:da021142216417466386] 
			\draw    (520,210) -- (565,210) ;
			%Straight Lines [id:da5402827223233408] 
			\draw    (560,130) -- (560,140) ;
			%Shape: Rectangle [id:dp8968538777678365] 
			\draw   (510,105) -- (575,105) -- (575,130) -- (510,130) -- cycle ;
			%Straight Lines [id:da8175608881223522] 
			\draw    (560,60) -- (560,105) ;
			%Shape: Rectangle [id:dp002508643287704837] 
			\draw   (540,140) -- (580,140) -- (580,165) -- (540,165) -- cycle ;
			%Shape: Rectangle [id:dp5876120633149555] 
			\draw   (510,30) -- (575,30) -- (575,60) -- (510,60) -- cycle ;
			%Straight Lines [id:da5796911728961744] 
			\draw    (525,95) -- (525,105) ;
			%Straight Lines [id:da2881348648710562] 
			\draw    (525,60) -- (525,70) ;
			%Straight Lines [id:da37234248141307247] 
			\draw    (525,25) -- (525,30) ;
			%Straight Lines [id:da20319489183785122] 
			\draw    (560,25) -- (560,30) ;
			%Shape: Rectangle [id:dp24216669558845294] 
			\draw   (505,70) -- (550,70) -- (550,95) -- (505,95) -- cycle ;
			% Text Node
			\draw (534,177) node [anchor=north west][inner sep=0.75pt]    {$d$};
			% Text Node
			\draw (516,33) node [anchor=north west][inner sep=0.75pt]    {$\mathcal{L} (\sigma^{-1} )$};
			% Text Node
			\draw (521,106) node [anchor=north west][inner sep=0.75pt]    {$\mathcal{L}( \sigma )$};
			% Text Node
			\draw (503,71) node [anchor=north west][inner sep=0.75pt]    {$(a^{\dagger } )^{\otimes k}$};
			% Text Node
			\draw (552,142) node [anchor=north west][inner sep=0.75pt]    {$T$};
			\end{tikzpicture}}} \ket{0}= \vcenter{\hbox{%
				\begin{tikzpicture}[x=0.75pt,y=0.75pt,yscale=-1,xscale=1]
					%uncomment if require: \path (0,482); %set diagram left start at 0, and has height of 482
					%Straight Lines [id:da23822104143131928] 
					\draw    (630,130) -- (630,175) ;
					%Straight Lines [id:da4708495650825337] 
					\draw    (625,25) -- (670,25) ;
					%Straight Lines [id:da6217792995703362] 
					\draw    (665,165) -- (665,175) ;
					%Shape: Rectangle [id:dp04709183223021807] 
					\draw   (615,175) -- (680,175) -- (680,200) -- (615,200) -- cycle ;
					%Straight Lines [id:da7129603799658324] 
					\draw    (630,200) -- (630,210) ;
					%Straight Lines [id:da027643530442898134] 
					\draw    (665,200) -- (665,210) ;
					%Straight Lines [id:da3290401574569599] 
					\draw    (625,210) -- (670,210) ;
					%Straight Lines [id:da6391076970412173] 
					\draw    (665,130) -- (665,140) ;
					%Shape: Rectangle [id:dp7325747367813824] 
					\draw   (615,105) -- (655,105) -- (655,130) -- (615,130) -- cycle ;
					%Straight Lines [id:da6293623117005904] 
					\draw    (665,30) -- (665,130) ;
					%Shape: Rectangle [id:dp8957916497119203] 
					\draw   (645,140) -- (685,140) -- (685,165) -- (645,165) -- cycle ;
					%Straight Lines [id:da3936439545116437] 
					\draw    (630,95) -- (630,105) ;
					%Straight Lines [id:da21877873601915065] 
					\draw    (630,60) -- (630,70) ;
					%Straight Lines [id:da8170494502086483] 
					\draw    (630,25) -- (630,30) ;
					%Straight Lines [id:da6404832950157071] 
					\draw    (665,25) -- (665,30) ;
					%Shape: Rectangle [id:dp47288185584365094] 
					\draw   (600,30) -- (660,30) -- (660,60) -- (600,60) -- cycle ;
					%Shape: Rectangle [id:dp6028678737157505] 
					\draw   (600,70) -- (645,70) -- (645,95) -- (600,95) -- cycle ;%
					% Text Node
					\draw (639,177) node [anchor=north west][inner sep=0.75pt]    {$d$};
					% Text Node
					\draw (655,142) node [anchor=north west][inner sep=0.75pt]   [align=left] {$T$};
					% Text Node
					\draw (615,106) node [anchor=north west][inner sep=0.75pt]    {$\mathcal{L}( \sigma )$};
					% Text Node
					\draw (599,71) node [anchor=north west][inner sep=0.75pt]    {$(a^{\dagger } )^{\otimes k}$};
					% Text Node
					\draw (603,33) node [anchor=north west][inner sep=0.75pt]    {$\mathcal{L} (\sigma^{-1} )$};
			\end{tikzpicture}}}\ket{0}&, 
\end{align}
where the first equality uses equation \eqref{eq: P2k on T}. The second line introduces an identity operator of the form $\mathcal{L}(\sigma^{-1})\mathcal{L}(\sigma)$ acting on the left-hand vector space $V_N^{\otimes k}$. The third equality follows from $\mathcal{L}(\sigma^{-1})d = d \mathcal{L}(\sigma^{-1})$ and the cyclicity of the trace. The last line removes the identity operator $\mathcal{L}(\sigma)\mathcal{L}(\sigma^{-1})$ acting on the right-hand vector space $V_N^{\otimes k}$.
The last diagram is equal to
\begin{equation}
	\Adj{\sigma} \frac{1}{k!} \Tr_{V_N^{\otimes 2k}}(d (a^{\dagger})^{\otimes k} \otimes a^{\otimes k}) \ket{T},
\end{equation}
which proves that they commute.

The construction readily generalizes to operators that do not preserve the degree of states. Consider
\begin{equation}
	\frac{1}{{k_1}!}\mathcal{P}_{k_2} \Tr_{V_N^{\otimes 2k}}(d (a^{\dagger})^{\otimes k_2} \otimes a^{\otimes k_1}) \mathcal{P}_{k_1},
\end{equation}
this gives a map $d: \mathcal{H}^{(k_1)} \rightarrow \mathcal{H}^{(k_2)}$ labeled by elements $d \in P_{k_1+k_2}(N)$.
Note that these operators have a $S_{k_2} \times S_{k_1}$ symmetry, which permutes the creation operators and annihilation operators separately.
Therefore, the dimension of the space of these operators is related to the counting of 2-matrix permutation invariants, which was studied in section 2 of \cite{PIG2MM}.

\subsection{Bosons on a lattice} \label{sec: lattice}
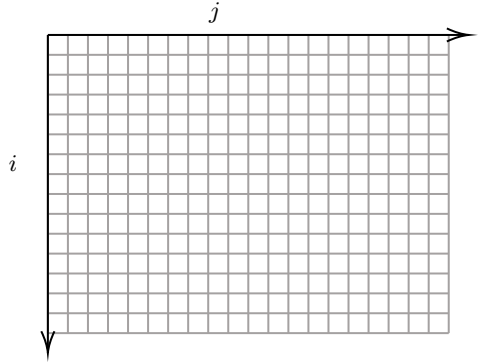
\begin{figure}[h!]
	\centering
	\scalebox{1.0}{
		\tikzset{every picture/.style={line width=0.75pt}} %set default line width to 0.75pt        
		\begin{tikzpicture}[x=0.75pt,y=0.75pt,yscale=-1,xscale=1]
			
			%Straight Lines [id:da5262234960688332] 
			\draw [color={rgb, 255:red, 164; green, 161; blue, 161 }  ,draw opacity=1 ]   (100,240) -- (300,240) ;
			%Straight Lines [id:da23693446073649138] 
			\draw [color={rgb, 255:red, 164; green, 161; blue, 161 }  ,draw opacity=1 ]   (300,90) -- (300,160) -- (300,240) ;
			%Shape: Grid [id:dp30331998876397703] 
			\draw  [draw opacity=0] (100,90) -- (300,90) -- (300,240) -- (100,240) -- cycle ; \draw  [color={rgb, 255:red, 164; green, 161; blue, 161 }  ,draw opacity=1 ] (100,90) -- (100,240)(110,90) -- (110,240)(120,90) -- (120,240)(130,90) -- (130,240)(140,90) -- (140,240)(150,90) -- (150,240)(160,90) -- (160,240)(170,90) -- (170,240)(180,90) -- (180,240)(190,90) -- (190,240)(200,90) -- (200,240)(210,90) -- (210,240)(220,90) -- (220,240)(230,90) -- (230,240)(240,90) -- (240,240)(250,90) -- (250,240)(260,90) -- (260,240)(270,90) -- (270,240)(280,90) -- (280,240)(290,90) -- (290,240) ; \draw  [color={rgb, 255:red, 164; green, 161; blue, 161 }  ,draw opacity=1 ] (100,90) -- (300,90)(100,100) -- (300,100)(100,110) -- (300,110)(100,120) -- (300,120)(100,130) -- (300,130)(100,140) -- (300,140)(100,150) -- (300,150)(100,160) -- (300,160)(100,170) -- (300,170)(100,180) -- (300,180)(100,190) -- (300,190)(100,200) -- (300,200)(100,210) -- (300,210)(100,220) -- (300,220)(100,230) -- (300,230) ; \draw  [color={rgb, 255:red, 164; green, 161; blue, 161 }  ,draw opacity=1 ]  ;
			%Straight Lines [id:da7722903141549615] 
			\draw    (100,90) -- (100,248) ;
			\draw [shift={(100,250)}, rotate = 270] [color={rgb, 255:red, 0; green, 0; blue, 0 }  ][line width=0.75]    (10.93,-3.29) .. controls (6.95,-1.4) and (3.31,-0.3) .. (0,0) .. controls (3.31,0.3) and (6.95,1.4) .. (10.93,3.29)   ;
			%Straight Lines [id:da5559351677806554] 
			\draw    (100,90) -- (308,90) ;
			\draw [shift={(310,90)}, rotate = 180] [color={rgb, 255:red, 0; green, 0; blue, 0 }  ][line width=0.75]    (10.93,-3.29) .. controls (6.95,-1.4) and (3.31,-0.3) .. (0,0) .. controls (3.31,0.3) and (6.95,1.4) .. (10.93,3.29)   ;
			
			% Text Node
			\draw (79,149) node [anchor=north west][inner sep=0.75pt]  [font=\footnotesize] [align=left] {$i$};
			% Text Node
			\draw (179,72) node [anchor=north west][inner sep=0.75pt]  [font=\footnotesize] [align=left] {$j$};
	\end{tikzpicture}}
	\caption{Matrix oscillators are naturally associated with a $N$-by-$N$ square lattice. The creation operator $(a^\dagger)_i^j$ creates a quanta of excitation at row $i$ column $j$ in the lattice.}
	\label{fig: lattice}
\end{figure}

The Fock space of matrix oscillators can be interpreted as the Fock space of bosons on a two-dimensional lattice of size $N^2$.
The lattice is parameterised by ordered pairs $(i,j)$ for $i,j=1,\dots,N$ which label the site in the $i$th row, $j$th column as in figure \ref{fig: lattice}.
The creation operator $(a^{\dagger})^j_i$ creates a quantum of excitation at the site $(i,j)$. In our conventions, $a_j^i$ annihilates a quantum at site $(i,j)$. Permutation invariant states naturally contain excitations spread throughout the entire lattice. For example, the state
\begin{equation}
	\ket{\PAdiagram[]{1}{1/-1}} = \sum_{i=1}^N (a^{\dagger})^{j}_j\ket{0},
\end{equation}
contains an excitation of every site on the diagonal, and the state
\begin{equation}
	\ket{\PAdiagram[]{1}{}} - \ket{\PAdiagram[]{1}{1/-1}} = \sum_{i \neq j} (a^\dagger)^{i}_j \ket{0},
\end{equation}
contains an excitation on every off-diagonal site.

Most choices of $S_N$ invariant Hamiltonians constructed in equation \eqref{eq: inv hamiltonians} contain non-local interactions, connecting sites at opposite sides of the lattice. Note that the left acting terms in the Hamiltonian \eqref{eq: inv hamiltonians} leave the columns fixed while the right acting terms fix the rows. An example of the non-locality is seen by considering
\begin{equation}
	H = P_{1} \Tr_{V_N}(a^{{\dagger}} \PAdiagram[]{1}{} a) P_1 = P_1 \sum_{i,j,k=1}^N (a^{\dagger})^i_j (a)^k_i P_1.
\end{equation}
This interaction moves a single excitation at site $(i,j)$ to every row in column $j$. In particular,
\begin{equation}
	H (a^{\dagger})^{1}_1 \ket{0} = \sum_{i=1}^N (a^{\dagger})^{1}_i \ket{0},
\end{equation}
contains the state $(a^{\dagger})^1_{N}$.

We can enumerate a set of diagrams that give local $S_N$ invariant terms, through left and right action, as follows.
First note that the identity element in $P_k(N)$ gives a local term. For example, in $k=2$
\begin{equation}
	\Tr_{V_N^{\otimes 2}}((a^{\dagger})^{\otimes 2} \PAdiagram[]{2}{-1/1,-2/2} a^{\otimes 2}) = \sum_{i_1,i_2,j_1,j_2=1}^N (a^{\dagger})^{i_1}_{j_1} (a^{\dagger})^{i_2}_{j_2}(a)^{j_1}_{i_1}(a)^{j_2}_{i_2}.
\end{equation}
It follows that any diagram that can be constructed from the identity element by adding additional edges is local.
For example
\begin{equation}
	\Tr_{V_N^{\otimes 2}}((a^{\dagger})^{\otimes 2} \PAdiagram[]{2}{-1/1,1/2,2/-2,-2/-1} a^{\otimes 2}) = \sum_{i_1,i_2,j=1}^N (a^{\dagger})^{i_1}_j (a^{\dagger})^{i_2}_j (a)^j_{i_1}(a)^j_{i_2},
\end{equation}
which is still local.

\section{Permutation invariant quantum scars} \label{sec: scars}
Energy eigenstates in quantum ergodic many-body systems are expected to thermalize, in accordance with the eigenstate thermalization hypothesis \cite{Deutsch1991, Srednicki1994}, which says that such systems are well described by statistical mechanical ensembles.
Integrable systems, and systems which exhibit many-body localization \cite{MBL} are known exceptions to the eigenstate thermalization hypothesis.
This is a consequence of the existence of a large number of conserved quantities, which leads to non-ergodicity.
A weak form of non-ergodicity was recently observed in experiments involving Rydberg-atom quantum simulators \cite{RydbergSim}.
For some initial states, the behaviour was as expected from an ergodic system, while other states would exhibit periodic revival.
This is unexpected, since the experiment is described by a system without any conserved charges or disorder \cite{ScarsReview}.
The term quantum many-body scars was coined in \cite{RydbergScars} to describe these non-ergodic states embedded in a large space of ergodic states.
Many mechanisms and construction schemes for systems and states that exhibit quantum many-body scars have been discussed in the theoretical literature \cite{ScarsProjEmbedding,ScarsSpectrumGenAlg,ScarsXYMagnet,ScarsLieAlgebras, ScarsQuasiSym}.

In this section, we will follow the group theoretic scheme invented in \cite{PPPK2020, PPPK2021} for constructing Hamiltonians which have many-body scars. Two basic ingredients are required in this scheme: a group $G$ acting on a Hilbert space $\mathcal{H}$, and a subspace $\Hilbertspace \subset \mathcal{H}$ of states that are invariant under the action of $G$. To promote $\Hilbertspace$ to a space of many-body scars, the prescription is as follows. First, find a Hamiltonian $H$ such that for all states $\ket{d} \in \Hilbertspace$
\begin{equation}
		H \ket{d} \in \Hilbertspace, \label{eq: H closed on inv}
\end{equation}
and the time-evolution of $\ket{d}$ using $H$ is periodic. This condition is discussed in section \ref{subsec: revival}. Note that $H$ commuting with the action of $g \in G$ is sufficient to satisfy \eqref{eq: H closed on inv}.
Now we break the symmetry of $H$, while retaining the many-body scars, by constructing a total Hamiltonian
\begin{equation} \label{eq: H_tot}
	H_{\text{tot}} = H + H_{\text{s}}.
\end{equation}
The new term will completely break the symmetry of $H$, but is required to satisfy
\begin{equation}
	H_{\text{s}} \ket{d} = 0 \qq{for all $\ket{d} \in \Hilbertspace$.} \label{eq: Hs kills inv}
\end{equation}
This ensures that time-evolution of $\ket{d}$ using $H_{\text{tot}}$ is equivalent to time-evolution using $H$, which was periodic by construction.
Since $H_{\text{tot}}$ has no remaining symmetry the non-invariant states in $\mathcal{H}$, which are not annihilated by $H_{\text{s}}$, will be ergodic and therefore thermalize.
The group theoretic construction of $H_{\text{s}}$ is reviewed in section \ref{subsec: scar hamiltonians}.

By combining the technology presented in this paper with the above scheme, we can construct models with many-body scars for $G=S_N$ acting on the Fock space $\mathcal{H}$ of matrix oscillators.
In particular, section \ref{sec: PA} contains a detailed description of the $S_N$ invariant subspace $\Hilbertspace \subset \mathcal{H}$ and the Hamiltonians in section \ref{sec: deformations} can be used for $H$ in \eqref{eq: H_tot}.
We gave a lattice interpretation of the matrix oscillators in \ref{sec: lattice}, which we will use to construct a lattice model with many-body scars.
The model will be a modified version of the Bose-Hubbard model \cite{BoseHubbard}, which is relevant for physics of cold atoms in an optical lattice \cite{ColdAtomsOpticalLattice}.

%This section is divided into three subsections. In the first subsection we review the integrality conditions on $H$ that give rise to periodic time-evolution of scar states. In the second subsection we construct large families of deformations $H_{\text{s}}$, related to partition algebra diagrams, that break the $S_N$ symmetry of $H$ to smaller subgroups.

% In the most generic case, the states in $\Hilbertspace$ were embedded into the spectrum in a manner that was indistinguishable from the rest of the eigenstates in $\mathcal{H}$. However, since $H$ is $S_N$ invariant, states in $\Hilbertspace$ only mix among themselves under time-evolution. Hamiltonians with this property, together with group invariant sectors, played an important role in a recent scheme  \cite{PPPK2020, PPPK2021} for constructing quantum many-body scar systems described by Hamiltonians without any symmetry. In this section we will follow the setup explained in these papers to turn states in $\Hilbertspace$ into quantum-many body scar states. We explore a modified Bose-Hubbard model that admits the form
%\begin{equation} \label{eq: H_tot}
%	H_{\text{tot}} = H + H_{\text{s}},
%\end{equation}
%where $H$ is a $S_N$ invariant Hamiltonian and $H_{\text{s}}$ breaks the $S_N$ symmetry. The second term satisfies
%\begin{equation}
%	H_{\text{s}} \ket{d} = 0
%\end{equation}
%for any $\ket{d} \in \Hilbertspace$. This setup gives a Hamiltonian $H_{\text{tot}}$ with no particular symmetry in general, where the dynamics of states in $\Hilbertspace$ is decoupled from the dynamics of the rest of the Hilbert space.

\subsection{Periodic time-evolution and revival}\label{subsec: revival}
The Hamiltonian in \eqref{eq: H_tot} contains two pieces, but the dynamics (time evolution) of invariant states is governed by $H$ alone.
In this subsection we will focus on $H$, and give a sufficient condition for it to give rise to periodic time evolution in the invariant subspace, turning the subspace into a many-body scar space.

Let $\ket{d}$ be a (normalized) state in $\Hilbertspace$. Since $H$ is $S_N$ invariant we have $H\ket{d} \in \Hilbertspace$ and we can construct an orthonormal energy eigenbasis $\ket{e_i}$ for $\Hilbertspace$ with eigenvalues $E_i$,
\begin{equation}
	H \ket{e_i} = E_i \ket{e_i}.
\end{equation}
The state $\ket{d}$ exhibits revival with periodicity $T$ if the quantum fidelity (return probability) \cite{RydbergScars2}
\begin{equation}
	f(t) = \abs{\mel{d}{e^{-iHt}}{d}}^2,
\end{equation}
satisfies $f(mT) = 1$ for $m=0,1,\dots$.
Expanding $\ket{d}$ in the eigenbasis
\begin{equation}
	\ket{d} = \sum_i d_i \ket{e_i},
\end{equation}
and computing $f(t)$ gives
\begin{equation}
	f(t) = \abs{\mel{d}{e^{-iHt}}{d}}^2 = \sum_{i,j} \abs{d_i}^2 \abs{d_j}^2 e^{-i(E_i-E_j)t}. \label{eq: revival}
\end{equation}
If all energy differences $\Delta E_{ij} = E_i - E_j$ have a greatest common divisor $E$, that is
\begin{equation}
	\Delta E_{ij} = E_i - E_j = E (\varepsilon_i - \varepsilon_j)	
\end{equation}
and $\varepsilon_i - \varepsilon_j$ is an integer for all $i,j$, then $f(mT) = 1$ for $T = 2\pi / E$. Note that trading $H$ for $H_{\text{tot}}$ in \eqref{eq: revival} does not change the argument above since $H_{\text{s}} \ket{d} = 0$ by construction. That is, the time-evolution of states in $\Hilbertspace$ is determined by $H$.
As a special case, $f(t)$ is periodic if the energies $E_i$ of the states $\ket{e_i}$ relevant to the expansion of $\ket{d}$ are integers.

%This is the case for the Hamiltonian in equation \eqref{eq: resolved hamiltonian} and the deformations in \eqref{eq: general charge deformation} for integer coefficients. Their eigenvalues are related to the normalized characters of $T_2, t_2$ that are known to be integers. This integrality follows from general properties of symmetric group characters and is explained in the context of a quantum mechanics of bi-partite ribbon graphs in section 3.1 in  \cite{JBSR2020}).

\subsection{Scar Hamiltonians}  \label{subsec: scar hamiltonians}
We now turn to the construction of the second part of the Hamiltonian \eqref{eq: H_tot}, using the group theoretic scheme introduced in \cite{PPPK2020, PPPK2021}.
In order to implement this scheme in the present setup we observe
\begin{equation}
	\qty(1-\Adj{\sigma})\ket{d} = 0, \quad \forall \sigma \in S_N, \, \ket{d} \in \Hilbertspace.
\end{equation}
This follows from $\Adj{\sigma}\ket{d}=\ket{d}$, and we will use it to construct $H_{\text{s}}$.

As we show below, the Hermitian conjugate of $\Adj{\sigma}$ is $\Adj{\sigma^{-1}}$.
This will be important because, at the end of the day, we want $H_{\text{s}}$ to be Hermitian.
Starting from the definition of the inner product we have
\begin{equation}
\begin{aligned}
		\bra{T'}\ket{\Adj{\sigma} T} &=\sum_{\gamma \in S_k}\Tr_{V_N^{\otimes k}}((T')^T \mathcal{L}_{\gamma} \mathcal{L}(\sigma)T\mathcal{L}(\sigma^{-1})\mathcal{L}_{\gamma^{-1}} ) \\
		 &= \sum_{\gamma \in S_k}\Tr_{V_N^{\otimes k}}(\mathcal{L}(\sigma^{-1})(T')^T \mathcal{L}(\sigma) \mathcal{L}_{\gamma} T\mathcal{L}_{\gamma^{-1}} ) \\
		 &=  \sum_{\gamma \in S_k}\Tr_{V_N^{\otimes k}}((\mathcal{L}(\sigma^{-1})T'\mathcal{L}(\sigma) )^T \mathcal{L}_{\gamma} T\mathcal{L}_{\gamma^{-1}} ) \\
		&=\bra{\Adj{\sigma^{-1}}T'}\ket{ T},
\end{aligned}
\end{equation}
where the second equality uses $\mathcal{L}_{\gamma}  \mathcal{L}(\sigma) =  \mathcal{L}(\sigma)  \mathcal{L}_{\gamma}$ and the third equality follows from 
\begin{equation}
	(\mathcal{L}(\sigma^{-1})T'\mathcal{L}(\sigma) )^T = \mathcal{L}(\sigma^{-1})(T')^T \mathcal{L}(\sigma).
\end{equation}

Consequently, an operator of the form
\begin{equation}
	H_\sigma = (1-\Adj{\sigma^{-1}})h_\sigma (1-\Adj{\sigma}), \label{eq: scar ham}
\end{equation}
where $h_\sigma$ is any Hermitian operator, is itself Hermitian and satisfies $H_\sigma \ket{d} = 0$, in accordance with the setup in \eqref{eq: H_tot}.
In general, we can write $H_{\text{s}}$ in the form
\begin{equation} \label{eq: scar deformation}
	H_{\text{s}} = \sum_{\sigma \in S_N} c_\sigma H_\sigma,
\end{equation}
where $c_\sigma \in \mathbb{R}$ is a parameter for every $\sigma \in S_N$.

The real dimension of the space of independent Hermitian operators (candidate choices for $h_\sigma$) can be counted as follows.
We organize general (normal ordered) $k$-oscillator operators in terms of the number of creation and annihilation operators ($k_1, k_2$ respectively). They have the form
\begin{equation}
	O = O^{i_1 \dots i_k}_{j_1 \dots j_k} (a^{\dagger})^{j_1}_{i_1} \dots (a^\dagger)^{j_{k_1}}_{i_{k_1}} a^{j_{k_1+1}}_{i_{k_1+1}} \dots a^{j_k}_{i_k}. \label{eq: general k-oscillator operator}
\end{equation}
Their adjoints contain $k_2$ creation and $k_1$ annihilation operators
\begin{equation}
	\begin{aligned} \label{eq: general k-oscillator dagger operator}
		O^{\dagger} &= (O^*)^{i_1 \dots i_{k_1}i_{k_1+1} \dots i_k }_{j_1 \dots j_{k_1}j_{k_1+1} \dots  j_k}  (a^{\dagger})_{j_{k_1+1}}^{i_{k_1+1}} \dots (a^{\dagger})_{j_k}^{i_k}a_{j_1}^{i_1} \dots a_{j_{k_1}}^{i_{k_1}}.  \\
	\end{aligned}
\end{equation}
For every $k = k_1 + k_2$ oscillator operator $O$ with $k_1 > k_2$ there is a Hermitian operator $O +O^{\dagger}$. The real dimension of the independent  Hermitian operators of this form can be counted in terms of the dimensions of symmetric tensor product spaces
\begin{equation} \label{eq: k_1 neq k_2 Hermitian operator counting}
	2\Dim \qty(\Sym^{k_1}(V_N \otimes V_N) \otimes \Sym^{k_2}(V_N \otimes V_N))=2\binom{N^2+k_1-1}{k_1} \binom{N^2+k_2-1}{k_2}
\end{equation}
which follows by identifying operators $(a^{\dagger})^{j_1}_{i_1}$ and $a^{j_1}_{i_1}$ with the vector space $V_N \otimes V_N$. The factor of two comes from the fact that $O^{i_1 \dots i_k}_{j_1 \dots j_k}$ are complex numbers. The Hermitian operators associated with $k_2 < k_1$ oscillator operators are accounted for in \eqref{eq: k_1 neq k_2 Hermitian operator counting} as their conjugates are the $k_1 < k_2$ oscillator operators.

The remaining Hermitian operators to count are those with equal numbers of creation and annihilation oscillators, i.e. those with $k_1 = k_2$. Some of these will be self-adjoint, while the remaining operators can be paired with their adjoints to construct Hermitian operators as before. Inspecting equations \eqref{eq: general k-oscillator operator} and \eqref{eq: general k-oscillator dagger operator} we see that for an operator to be equal to its own adjoint it must be real, with $k_1 = k_2$ and
\begin{align} \label{eq: self-adjoint operator condition}
O^{i_1 \dots i_{k_1}i_{k_1+1} \dots i_k }_{j_1 \dots j_{k_1}j_{k_1+1} \dots  j_k} = (O^*)^{j_{k_1+1} \dots  j_k j_1 \dots j_{k_1}}_{ i_{k_1+1} \dots i_k  i_1 \dots i_{k_1}}.
\end{align}
As they are real, the number of these operators is equal to their real dimension
\begin{align}
	\Dim \qty(\Sym^{k_1}(V_N \otimes V_N))= \binom{N^2+k_1-1}{k_1}.
\end{align}
This counting can be understood as there being exactly one choice of $\big\{ \{i_{k_1 +1}, \dots, i_{2k_1}\}, \{j_{k_1 +1}, \dots, j_{2k_1} \} \big\}$ for which each choice of $\big\{ \{i_1, \dots i_{k_1} \}, \{j_1, \dots j_{k_1} \} \big\}$ satisfies \eqref{eq: self-adjoint operator condition}. The remaining number of operators is 
\begin{align} \nonumber \label{eq: k_1 = k_2 non self dual counting}
\Dim \qty(\Sym^{k_1}(V_N \otimes V_N)) \times \Big[ \Dim \Big( &\Sym^{k_1}(V_N \otimes V_N) \Big) -1 \Big] \\
&= \binom{N^2+k_1-1}{k_1} \times \qty[ \binom{N^2+k_1-1}{k_1} - 1 ].
\end{align}
The factor of two due to $O^{i_1 \dots i_k}_{j_1 \dots j_k}$ being complex is cancelled by the factor of a half introduced when forming Hermitian operators. The real dimension of Hermitian operators of type $k_1 = k_2$ is then
\begin{align}
	\binom{N^2+k_1-1}{k_1} \times \binom{N^2+k_1-1}{k_1}.
\end{align}

\subsection{Modified Bose-Hubbard on a square lattice}\label{subsec: BH model}
Having discussed the general setup, we will now implement the scheme in a specific example.
In section \ref{sec: lattice} we gave a lattice interpretation of the matrix oscillators $(a^{\dagger})^j_i$, where they played the role of bosonic creation operators on the lattice site labeled $(i,j)$.
For simplicity, we will consider a square lattice of dimension $N$.
A simple model of interacting bosons on a lattice is the Bose-Hubbard model \cite{BoseHubbard}, relevant for the physics of cold atoms in an optical lattice \cite{ColdAtomsOpticalLattice}.
The Bose-Hubbard Hamiltonian $H_{\text{BH}}$ contains hopping (kinetic) terms, on-site interactions and chemical potentials.
In the simplest case, the model contains three parameters $t, U, \mu$, and in terms matrix of oscillators it takes the form
\begin{equation}
	H_{\text{BH}} = -t \sum_{<(i,j),(k,l)>} (a^{\dagger})_{i}^j a^{k}_l + \frac{U}{2}\sum_{i,j=1}^N (a^{\dagger})_i^j a_j^i ((a^{\dagger})_i^j a_j^i -1) - \mu \sum_{i,j=1}^N (a^{\dagger})_i^j a_j^i, \label{eq: BH ham}
\end{equation}
where the sum in the first term is over neighbouring sites. For a square lattice this implies the restriction $(i,j)=(k \pm 1,l)$ or $(i,j) = (k, l \pm 1)$.
On-site interactions are implemented using the operators $N_{ij}= (a^{\dagger})_i^j a_j^i $, which count the number of excitations on site $(i,j)$.

%\begin{equation}
%	\begin{aligned}
%		\sum_{<(i,j),(k,l)>} (a^{\dagger})^{i}_j a^{k}_l = \sum_{i,j=2}^{N-1} &\qty[(a^{\dagger})^{i+1}_j  + (a^{\dagger})^{i-1}_j  + (a^{\dagger})^{i}_{j+1} + (a^{\dagger})^{i}_{j-1} ]a^{j}_i + \dots, \\
%		%		&+\sum_{i=2}^{N-1}   (a^{\dagger})^{i}_j a^i_1
%		%		&+(a^{\dagger})^{1}_2 a^{1}_{1}  + (a^{\dagger})^{2}_1  a^{1}_{1} + (a^{\dagger})^{N-1}_N a^{N}_{N}  + (a^{\dagger})^{N}_{N-1}  a^{N}_{N}
%	\end{aligned}
%\end{equation} 

Our aim is to construct a modified Bose-Hubbard Hamiltonian $H_{\text{BH}}'$ such that
\begin{equation}
	H_{\text{BH}}' = H + H_{\text{s}},
\end{equation}
where $\comm{\Adj{\sigma}}{H} = 0$ for all $\sigma \in S_N$, and $H_{\text{s}} \ket{d} = 0$ for all $\ket{d} \in \Hilbertspace$, as per the construction in \eqref{eq: H_tot}.
To this end, we observe that the second and third term in \eqref{eq: BH ham} are $S_N$ invariant. That is,
\begin{equation}
	H = \frac{U}{2}\sum_{i,j=1}^N (a^{\dagger})^i_j a^j_i ((a^{\dagger})^i_j a^j_i -1) - \mu \sum_{i,j=1}^N (a^{\dagger})^i_j a^j_i, \label{eq: BH ham part 2}
\end{equation}
satisfies $\comm{\Adj{\sigma}}{H} = 0$.

The hopping term
\begin{equation}
	h_t = -t\sum_{<(i,j),(k,l)>} (a^{\dagger})_{i}^j a^{k}_l
\end{equation}
is not $S_N$ invariant, but the combination 
\begin{equation}
	H_{\text{s}} = (1- \Adj{\sigma^{-1}}) h_t (1- \Adj{\sigma})
\end{equation}
satisfies $H_{\text{s}} \ket{d} = 0$ for any choice of $\sigma \in S_N$ by the construction in \eqref{eq: scar ham}.

To keep $H_{\text{s}}$ as local as possible, $\sigma$ should not permute distant sites.
With this restriction in mind, a simple choice is $\sigma = (23)$ (for the choice $\sigma = (12)$ we have to consider additional complications from being near the boundary of the square lattice).
This defines our modified Bose-Hubbard Hamiltonian $H_{\text{BH}}'$,
\begin{equation}
	\boxed{ \begin{aligned}
		H_{\text{BH}}' = &\frac{U}{2}\sum_{i,j=1}^N (a^{\dagger})^i_j a^j_i ((a^{\dagger})^i_j a^j_i -1) - \mu \sum_{i,j=1}^N (a^{\dagger})^i_j a^j_i \\
		&-t \sum_{<(i,j),(k,l)>}	\qty(1-\Adj{(23)}) (a^{\dagger})_{i}^j a^{k}_l \qty(1-\Adj{(23)}).  \label{eq: modified BH}
	\end{aligned}} 
\end{equation}
It can be written as
\begin{equation}
	\begin{aligned}
		H_{\text{BH}}' = H_{BH} &- t \sum_{<(i,j),(k,l)>}	\Adj{(23)}) (a^{\dagger})_{i}^j a^{k}_l \Adj{(23)} \\
		&+ t\sum_{<(i,j),(k,l)>}	\Adj{(23)} (a^{\dagger})_{i}^j a^{k}_l \\
		&+ t \sum_{<(i,j),(k,l)>} (a^{\dagger})_{i}^j a^{k}_l \Adj{(23)}.
	\end{aligned}
\end{equation}
We now investigate the conditions on $U$ and $\mu$ for which $H'_{BH}$ exhibits revival.
It is useful to rewrite equation \eqref{eq: BH ham part 2} as
\begin{equation}
	H = \frac{U}{2}\sum_{i,j=1}^N (a^{\dagger})^i_j a^j_i\qty[ (a^{\dagger})^i_j a^j_i - 1 - \frac{2\mu}{U}].
\end{equation}
For integer values of $\frac{2\mu}{U}$ the eigenvalues of $H$ are integer multiples of $\frac{U}{2}$, and similarly for differences of eigenvalues. By the argument given in section \ref{subsec: revival} we therefore expect $H_{\text{BH}}'$ to have many-body scars that revive with period $T= \frac{4\pi}{U}$.
Similarly, we may write \eqref{eq: BH ham part 2} as
\begin{equation}
	H = \frac{\mu}{2}\sum_{i,j=1}^N (a^{\dagger})^i_j a^j_i\qty[\frac{U}{\mu} (a^{\dagger})^i_j a^j_i - \frac{U}{\mu} - 2],
\end{equation}
from which we conclude that revival is possible when $\tfrac{U}{\mu}$ is an integer as well, with an expected revival time of $ \frac{4\pi}{ \mu }$. In the special case $ U = 2 \mu$ where both integrality conditions are satisfied, the revival time is $ T = \min ( \frac{4\pi}{ \mu } , \frac{4\pi}{ U } ) = \frac{4\pi}{ U } $. 

In the subspace where there is a single excitation, the operator $\Adj{(23)}$ takes the form
\begin{equation}
	\begin{aligned}
			\Adj{(23)} =& (a^{\dagger})^{3}_{3} a^{2}_{2} + (a^{\dagger})^{2}_{2} a^{3}_{3} + \sum_{i \neq 2,3}\qty( (a^{\dagger})^{i}_{3} a^{2}_{i} + (a^{\dagger})^{i}_{2} a^{3}_{i} + (a^{\dagger})^{3}_{i}  a^{i}_{2} + (a^{\dagger})^{2}_{i} a^{i}_{3})+\sum_{i,j\neq 2,3} (a^\dagger)_i^j a^i_j.
	\end{aligned}
\end{equation}
The first term is a diagonal hopping term, between sites $(2,2)$ and $(3,3)$.
Consequently, the modified Bose-Hubbard Hamiltonian \eqref{eq: modified BH} will contain some hopping terms beyond nearest-neighbours.

\section{AdS/CFT inspired extremal correlators in matrix quantum mechanics} \label{sec: extremal correlators}

Extremal correlators in $\mathcal{N}=4$ SYM form interesting sectors having non-renormalization properties \cite{NonRenorm}. They are closely connected to representation theoretic quantities such as Littlewood-Richardsson coefficients, and form a crucial set of examples for checking the AdS/CFT correspondence. In the quantum mechanical model presented in this paper, vacuum expectation values similar to extremal correlators can be computed exactly. In this section we make use of a recent factorisation result concerning the two-point function of permutation invariant matrix observables \cite{PIMO_Factor} - this is used to demonstrate that a similar factorisation property holds for quantum mechanical permutation invariant states. We then compute an expression for extremal three-point correlators associated with $S_N$ invariant states, which are simple in the diagram basis, and obey representation theoretic selection rules.

\subsection{Two-point correlators}
The equation \eqref{eq: diagram state def} can be interpreted as a quantum mechanical operator-state correspondence for $S_N$ invariant states labelled by $[d_\pi] \in SP_k(N)$,
\begin{equation}
	\ket{d_\pi} \longleftrightarrow \mathcal{O}_{\pi} = \Tr_{V_N^{\otimes k}}( [d_\pi] (a^\dagger)^{\otimes k}).
\end{equation}
From equation \eqref{eq: dual state} we have
\begin{equation}
	\mathcal{O}_{\pi}^\dagger = \Tr_{V_N^{\otimes k}}( [d_\pi^T] a^{\otimes k}),
\end{equation}
where the transpose $d_\pi^T$ is the diagram obtained by reflecting $d_\pi$ across a horizontal line, as illustrated in \eqref{eq: transpose of diagram}.
The time-dependent operators are given by
\begin{equation}
	\mathcal{O}_{\pi}(t) = \mathrm{e}^{-iH_0t}\mathcal{O}_{\pi}\mathrm{e}^{iH_0t} = \mathrm{e}^{-ikt} \mathcal{O}_{\pi},
\end{equation}
where $H_0$ is the free Hamiltonian, defined in equation \eqref{eq: simplest hamiltonian}.

In \cite{PIMO_Factor} the two-point function of permutation invariant matrix observables was shown to factorise in the large $N$ limit. Here we use this result to show an equivalent factorisation property for the two-point function of permutation invariant quantum mechanical states.
Let $[d_{\pi_1}] \in SP_{k_1}(N)$, $[d_{\pi_2}] \in SP_{k_2}(N)$, and define the two-point correlator to be the vacuum expectation value
\begin{equation}
	\begin{aligned}		
	\bra{0} \mathcal{O}_{\pi_1}^\dagger(t_1) \mathcal{O}_{\pi_2}(t_2) &\ket{0} = \\
	&\mathrm{e}^{ik_1t_1 - i k_2t_2}\bra{0} \Tr_{V_N^{\otimes k_1}}( [d_{\pi_1}^T] a^{\otimes k_1}) \Tr_{V_N^{\otimes k_2}}( [d_{\pi_2}] (a^{\dagger})^{\otimes k_2}) \ket{0}.
	\end{aligned}\label{eq: two-point correlator}
\end{equation}
Ignoring the trivial time dependence and taking normalised operators $[\hat{d}_{\pi_1}],[\hat{d}_{\pi_2}]$, as defined in \eqref{eq: noramlised state definition}, in the large $N$ limit we have
\begin{equation}
\begin{aligned}
		\bra{0} \Tr_{V_N^{\otimes k_1}}( [\hat{d}_{\pi_1}^T] a^{\otimes k_1}) \Tr_{V_N^{\otimes k_2}}( [\hat{d}_{\pi_2}] (a^{\dagger})^{\otimes k_2}) \ket{0}
		&= \delta_{k_1 k_2} \sum_{\gamma \in S_{k_1}} \Tr_{V_N^{\otimes k_2}}(\gamma^{-1} \hat{d}_{\pi_1}^T \gamma \hat{d}_{\pi_2}) \\
		&= 
	\begin{cases}
		1 + O(1/\sqrt{N}) \qq{if $[d_{\pi_1}] = [d_{\pi_2}]$,} \\
		0 + O(1/\sqrt{N}) \qq{otherwise.}
	\end{cases}
\end{aligned}
\end{equation}
In the first line we have absorbed the $S_{k_1}$ averaging into the sum over $\gamma \in S_{k_1}$ arising from the Wick contractions of $a$ and $a^{\dagger}$. In the second line we have used the factorisation result of \cite{PIMO_Factor}.

\subsection{Three-point correlators}

Let $[d_{\pi_1}] \in SP_{k_1}(N), [d_{\pi_2}] \in SP_{k_2}(N), [d_\pi] \in SP_k(N)$, and define the extremal three-point correlator to be the vacuum expectation value
\begin{equation}
	\begin{aligned}		
	&\bra{0} \mathcal{O}_{\pi_1}^\dagger(t_1) \mathcal{O}^{\dagger}_{\pi_2}(t_2) \mathcal{O}_{\pi}(t) \ket{0} = \\
	&\mathrm{e}^{i k_1t_1 + i k_2t_2 - i  k t }\bra{0} \Tr_{V_N^{\otimes k_1}}( [d_{\pi_1}^T] a^{\otimes k_1}) \Tr_{V_N^{\otimes k_2}}( [d_{\pi_2}^T] a^{\otimes k_2}) \Tr_{V_N^{\otimes k}}( [d_{\pi}] (a^\dagger)^{\otimes k})\ket{0},  
	\end{aligned}\label{eq: extremal correlator}
\end{equation}
with the constraint that $k = k_2 + k_1$.
As we now show, extremal correlators are simple in the diagram basis. We compute \eqref{eq: extremal correlator} by Wick contractions, which are encoded in a sum over $\gamma \in S_{k}$. Ignoring the trivial time-dependence we have
\begin{equation}
\begin{aligned}
		&\bra{0} \Tr_{V_N^{\otimes k_1}}( [d_{\pi_1}^T] a^{\otimes k_1}) \Tr_{V_N^{\otimes k_2}}( [d_{\pi_2}^T] a^{\otimes k_2}) \Tr_{V_N^{\otimes k}}( [d_{\pi}] (a^\dagger)^{\otimes k})\ket{0} \\
		&= \sum_{\gamma \in S_k} \Tr_{V_N^{\otimes k}}(\gamma^{-1}(d_{\pi_1}^T \otimes d_{\pi_2}^T) \gamma d_{\pi}) \\
		&=\sum_{\gamma \in S_k} N^{c(\gamma^{-1} (d_{\pi_1} \otimes d_{\pi_2})  \gamma  \merge d_{\pi} )}.
\end{aligned}
\end{equation}
The tensor product $d_{\pi_1} \otimes d_{\pi_2}$ is the diagram obtained by horizontal concatenation of $d_{\pi_1}$ and $d_{\pi_2}$, for example
\begin{align}
	\PAdiagram[2/1, -1/-2]{2}{} \otimes \PAdiagram[3/1]{3}{1/-2} = \PAdiagram[2/1, -1/-2, 5/3]{5}{3/-4}.
\end{align}
This can be viewed as an outer product on partition algebra diagrams which maps 
$P_{ k_1} ( N ) \times P_{ k_2} ( N ) $ to $P_{ k_1 + k_2 } ( N ) $. 
It is a diagram with $2k_1+2k_2$ vertices. The join $d_{\pi_1} \merge d_{\pi_2}$ of two diagrams, each with $2k$ vertices, is obtained by adding all the edges of $d_{\pi_1}$ to the edges of $d_{\pi_2}$ (or vice versa), for example 
\begin{align}
	\PAdiagram[3/2]{3}{1/-1} \merge \PAdiagram[-1/-2]{3}{2/-3} = \PAdiagram[3/2, -1/-2]{3}{1/-1, 3/-3}.
\end{align}
The resulting diagram also has $2k$ vertices. For general elements (linear combinations of diagram basis elements) the two operations are defined by linear extension.

We will now derive a set of representation theoretic selection rules for the extremal correlators. To state the result we are going to prove, we define the operators
\begin{equation}
	\mathcal{O}^{\Lambda_1}_{\Lambda_2, \mu \nu} = \Tr_{V_N^{\otimes k}}( Q^{\Lambda_1}_{\Lambda_2, \mu \nu} (a^\dagger)^{\otimes k}),
\end{equation}
associated with representation basis elements $Q^{\Lambda_1}_{\Lambda_2, \mu \nu} \in SP_{k}(N)$.
Consider the extremal correlator (time-independent part)
\begin{equation}
	\bra{0} (\mathcal{O}^{\Lambda_1}_{\Lambda_2, \mu \nu})^\dagger (\mathcal{O}^{\Lambda_1'}_{\Lambda_2', \mu' \nu'})^\dagger \mathcal{O}^{\Lambda_1''}_{\Lambda_2'', \mu'' \nu''} \ket{0} = k! \Tr_{V_N^{\otimes k}}\Big(\big(Q^{\Lambda_1}_{\Lambda_2,\nu \mu}\otimes Q^{\Lambda_1'}_{\Lambda_2',\nu' \mu'} \big) Q^{\Lambda_1''}_{\Lambda_2'',\mu'' \nu''} \Big), \label{eq: rep basis 3pt SPk}
\end{equation}
for $Q^{\Lambda_1}_{\Lambda_2, \mu \nu} \in SP_{k_1}(N), Q^{\Lambda_1'}_{\Lambda_2', \mu' \nu'} \in SP_{k_2}(N), Q^{\Lambda_1''}_{\Lambda_2'', \mu'' \nu''} \in SP_{k}(N)$. The factor of $k!$ follows since the matrix units for $SP_k(N)$ are invariant under conjugation by $S_k$. Note that the multiplicity labels are exchanged under diagram transposition, which follows from \eqref{eq: transpose diagram is transpose matrix element}.
The selection rule that we will find says that the trace in \eqref{eq: rep basis 3pt SPk} vanishes if $C(\Lambda_1, \Lambda_1', \Lambda_1'') = 0$, where $C(\Lambda_1, \Lambda_1', \Lambda_1'')$ is the Kronecker coefficient for tensor products of irreducible representations of $S_N$.

We start with the simpler, but analogous expression for matrix units of $P_k(N)$,
\begin{equation}
	\Tr_{V_N^{\otimes k}}((Q^{\Lambda_1}_{\beta \alpha}\otimes Q^{\Lambda_1'}_{\beta' \alpha'}) Q^{\Lambda_1''}_{\alpha'' \beta''} ) = 
	\tikzset{every picture/.style={line width=0.75pt}} %set default line width to 0.75pt       
	\vcenter{ \hbox{ 
			\begin{tikzpicture}[x=0.75pt,y=0.75pt,yscale=-0.8,xscale=0.8]
				%Straight Lines [id:da18814113513857578] 
				\draw    (60,160) -- (60,190) ;
				%Straight Lines [id:da9251133232468041] 
				\draw    (60,90) -- (60,120) ;
				%Shape: Rectangle [id:dp8725265158857538] 
				\draw   (30,50) -- (160,50) -- (160,90) -- (30,90) -- cycle ;
				%Straight Lines [id:da15630711407442344] 
				\draw    (90,20) -- (90,50) ;
				%Straight Lines [id:da20001396742416788] 
				\draw    (50,20) -- (140,20) ;
				%Shape: Rectangle [id:dp1475354726195257] 
				\draw   (100,120) -- (160,120) -- (160,160) -- (100,160) -- cycle ;
				%Straight Lines [id:da22938491975955944] 
				\draw    (130,160) -- (130,190) ;
				%Shape: Rectangle [id:dp4054835639828456] 
				\draw   (30,120) -- (90,120) -- (90,160) -- (30,160) -- cycle ;
				%Straight Lines [id:da3084100754602861] 
				\draw    (130,90) -- (130,120) ;
				%Straight Lines [id:da3525452869573087] 
				\draw    (50,190) -- (140,190) ;
				% Text Node
				\draw (73,55) node [anchor=north west][inner sep=0.75pt]   [align=left] {$ Q_{\alpha ''\beta ''}^{\Lambda ''_{1}}$};
				% Text Node
				\draw (43,125.67) node [anchor=north west][inner sep=0.75pt]   [align=left] {$ Q_{\alpha \beta }^{\Lambda _{1}}$};
				% Text Node
				\draw (113,125) node [anchor=north west][inner sep=0.75pt]   [align=left] {$ Q_{\alpha '\beta '}^{\Lambda '_{1}}$};
			\end{tikzpicture}
	}}
\end{equation}
Using (see e.g. equation \eqref{eq: Pk action of matrix units})
\begin{equation}
	(Q^{\Lambda_1}_{\beta \alpha}\otimes Q^{\Lambda_1'}_{\beta' \alpha'} )Q^{\Lambda_1''}_{\alpha'' \beta''}  = \sum_{\gamma''} D^{\Lambda_1''}_{\gamma'' \alpha''}(Q^{\Lambda_1}_{\beta \alpha}\otimes Q^{\Lambda_1'}_{\beta' \alpha'} )Q^{\Lambda_1''}_{\gamma'' \beta''},
\end{equation}
we have
\begin{equation}
	\begin{aligned}
		\Tr_{V_N^{\otimes k}}((Q^{\Lambda_1}_{\beta \alpha}\otimes Q^{\Lambda_1'}_{\beta' \alpha'}) Q^{\Lambda_1''}_{\alpha'' \beta''} )  &= \sum_{\gamma''} D^{\Lambda_1''}_{\gamma'' \alpha''}(Q^{\Lambda_1}_{\beta \alpha}\otimes Q^{\Lambda_1'}_{\beta' \alpha'} )\Tr_{V_N^{\otimes k}}(Q^{\Lambda_1''}_{\gamma'' \beta''}) \\
		&= D^{\Lambda_1''}_{\beta'' \alpha''}(Q^{\Lambda_1}_{\beta \alpha}\otimes Q^{\Lambda_1'}_{\beta' \alpha'} ) \DimSN{\Lambda_1''}\\
		&= \DimSN{\Lambda_1''} 
		\tikzset{every picture/.style={line width=0.75pt}} %set default line width to 0.75pt        
		\vcenter{ \hbox{
				\begin{tikzpicture}[x=0.75pt,y=0.75pt,yscale=-0.8,xscale=0.8]
					
					%Straight Lines [id:da9251133232468041] 
					\draw    (90,70) -- (90,120) ;
					%Shape: Rectangle [id:dp7973284926553135] 
					\draw   (30,120) -- (160,120) -- (160,160) -- (30,160) -- cycle ;
					%Straight Lines [id:da6961071314388567] 
					\draw    (90,160) -- (90,210) ;
					
					% Text Node
					\draw (51,123.67) node [anchor=north west][inner sep=0.75pt]   [align=left] {$ Q_{\alpha \beta }^{\Lambda _{1}} \otimes Q_{\alpha '\beta '}^{\Lambda '_{1}}$};
					% Text Node
					\draw (100,79) node [anchor=north west][inner sep=0.75pt]   [align=left] {$ \Lambda''_{1}$};
					% Text Node
					\draw (100,169) node [anchor=north west][inner sep=0.75pt]   [align=left] {$ \Lambda''_{1}$};
					% Text Node
					\draw (79,209) node [anchor=north west][inner sep=0.75pt]   [align=left] {$ \beta''$};
					% Text Node
					\draw (79,47) node [anchor=north west][inner sep=0.75pt]   [align=left] {$ \alpha''$};
				\end{tikzpicture}
		}}
	\end{aligned} \label{eq: rep basis 3pt step 2}
\end{equation}
The second equality uses \eqref{eq: trace of matrix units}.

To further simplify, we want to turn the RHS into a product of matrix elements. This is achieved by inserting a resolution of the identity using representations of $P_{k_1}(N) \otimes P_{k_2}(N)$.
This resolves to a set of branching coefficients for $P_{k}(N) \rightarrow P_{k_1}(N) \otimes P_{k_2}(N)$. We denote these by
\begin{equation}
	B^{\Lambda_1'' \rightarrow \widetilde{\Lambda}_1 \otimes \widetilde{\Lambda}_1' , \xi}_{ \gamma'' \rightarrow \gamma \gamma'}
\end{equation}
where it is implicit that $k=k_1 + k_2$. The ranges of the labels are
\begin{align} \nonumber
	&\gamma \in  [1, \dots, \Dim \big( V_{\widetilde{\Lambda}_1}^{P_{k_1}(N)} \big) ], \\ \nonumber
	&\gamma' \in  [1, \dots, \Dim \big( V_{\widetilde{\Lambda}_1'}^{P_{k_2}(N)} \big) ], \\ \nonumber
	&\gamma'' \in  [ 1, \dots, \Dim \big( V_{\Lambda_1''}^{P_{k}(N)} \big) ], \\
	&\xi \in [1, \dots, \Mult \big( V_{\Lambda_1''}^{P_{k}(N)} \rightarrow V_{\widetilde{\Lambda}_1}^{P_{k_1}(N)} \otimes V_{\widetilde{\Lambda}_1'}^{P_{k_2}(N)} \big)],
\end{align}
the final label, $\xi$, gives the multiplicity of $\Lambda_1''$ in the decomposition.
Branching coefficients are represented by the following diagrams
\begin{equation}
	B^{\Lambda_1'' \rightarrow \widetilde{\Lambda}_1 \otimes \widetilde{\Lambda}_1' , \xi}_{ \gamma'' \rightarrow \gamma \gamma'} = 
	\tikzset{every picture/.style={line width=0.75pt}} %set default line width to 0.75pt        
	\vcenter{\hbox{\begin{tikzpicture}[x=0.75pt,y=0.75pt,yscale=-0.8,xscale=0.8]
		%uncomment if require: \path (0,672); %set diagram left start at 0, and has height of 672
		
		%Straight Lines [id:da20689736432290773] 
		\draw    (534,537.65) -- (534,490) ;
		\draw [shift={(534,540)}, rotate = 270] [color={rgb, 255:red, 0; green, 0; blue, 0 }  ][line width=0.75]      (0, 0) circle [x radius= 3.35, y radius= 3.35]   ;
		%Straight Lines [id:da8604740222295812] 
		\draw    (564,560) -- (535.96,541.3) ;
		\draw [shift={(534,540)}, rotate = 213.69] [color={rgb, 255:red, 0; green, 0; blue, 0 }  ][line width=0.75]      (0, 0) circle [x radius= 3.35, y radius= 3.35]   ;
		%Straight Lines [id:da8956003254158647] 
		\draw    (504,560) -- (532.04,541.3) ;
		\draw [shift={(534,540)}, rotate = 326.31] [color={rgb, 255:red, 0; green, 0; blue, 0 }  ][line width=0.75]      (0, 0) circle [x radius= 3.35, y radius= 3.35]   ;
		%Straight Lines [id:da9369290875413263] 
		\draw    (564,580) -- (564,560) ;
		%Straight Lines [id:da07444227932528902] 
		\draw    (504,580) -- (504,560) ;
		
		% Text Node
		\draw (543,507) node [anchor=north west][inner sep=0.75pt]   [align=left] {$\displaystyle \Lambda ''_{1}$};
		% Text Node
		\draw (523,469) node [anchor=north west][inner sep=0.75pt]   [align=left] {$\displaystyle \gamma ''$};
		% Text Node
		\draw (479,539) node [anchor=north west][inner sep=0.75pt]   [align=left] {$\displaystyle \widetilde{\Lambda _{1}}$};
		% Text Node
		\draw (563,539) node [anchor=north west][inner sep=0.75pt]   [align=left] {$\displaystyle \widetilde{\Lambda '_{1}}$};
		% Text Node
		\draw (513,519) node [anchor=north west][inner sep=0.75pt]   [align=left] {$\displaystyle \xi $};
		% Text Node
		\draw (494,579) node [anchor=north west][inner sep=0.75pt]   [align=left] {$\displaystyle \gamma $};
		% Text Node
		\draw (555,579) node [anchor=north west][inner sep=0.75pt]   [align=left] {$\displaystyle \gamma '$};
	\end{tikzpicture}}}
\end{equation}
It is worth noting that by Schur-Weyl duality the branching multiplicities for partition algebras are related   to the multiplicities $ C(\widetilde{\Lambda}_1, \widetilde{\Lambda}_1', \Lambda_1'') $, known as Kronecker coefficients, of irreducible 
representations $ \Lambda_1'' $ in tensor products of $S_N$ representations $ \widetilde{\Lambda}_1 \otimes \widetilde{\Lambda}_1' $ (see eq. (3.1.3) of \cite{Bowman2012ThePA})  
\begin{equation}
	\Mult \Big( V_{\Lambda_1''}^{P_{k}(N)} \rightarrow V_{\widetilde{\Lambda}_1}^{P_{k_1}(N)} \otimes V_{\widetilde{\Lambda}_1'}^{P_{k_2}(N)} \Big) = C(\widetilde{\Lambda}_1, \widetilde{\Lambda}_1', \Lambda_1'').
\end{equation}
For simiplicity we are assuming $ N \ge (2 k_1 + 2 k_2)$. For comparison, in Schur-Weyl duality between $U(N)$ and $\mathbb{C}[S_k]$, Littlewood-Richardson coefficients are branching multiplicities for $S_{k_1 + k_2} \rightarrow S_{k_1} \times S_{k_2}$ but correspond to decomposition of tensor products of $U(N)$ representations.

Branching coefficients are equivariant:
\begin{equation} \label{eq: branching coefficients equivariance}
	D^{\Lambda_1''}_{\gamma'' \delta''}(d_{\pi_1} \otimes d_{\pi_2}) = \sum_{\widetilde{\Lambda}_1, \widetilde{\Lambda}_1', \gamma, \delta, \gamma', \delta', \xi} B^{ \Lambda_1'' \rightarrow \widetilde{\Lambda}_1 \otimes \widetilde{\Lambda}_1' , \xi}_{\gamma'' \rightarrow \gamma \gamma'} D^{\widetilde{\Lambda}_1}_{\gamma \delta}(d_{\pi_1}) D^{\widetilde{\Lambda}_1'}_{\gamma' \delta'}(d_{\pi_2}) B^{\Lambda_1'' \rightarrow \widetilde{\Lambda}_1 \otimes \widetilde{\Lambda}_1' , \xi}_{\delta'' \rightarrow \delta \delta' },
\end{equation}
for $d_{\pi_1} \in P_{k_1}(N), d_{\pi_2} \in P_{k_2}(N)$.
Setting $d_{\pi_1} = Q^{\Lambda_1}_{\alpha \beta}, d_{\pi_2} = Q^{{\Lambda_1'}}_{\alpha' \beta'}$, equation \eqref{eq: branching coefficients equivariance} corresponds to the diagram identity
\begin{equation}
	\tikzset{every picture/.style={line width=0.75pt}} %set default line width to 0.75pt        
	\vcenter{ \hbox{
			\begin{tikzpicture}[x=0.75pt,y=0.75pt,yscale=-0.8,xscale=0.8]
				
				%Straight Lines [id:da9251133232468041] 
				\draw    (90,70) -- (90,120) ;
				%Shape: Rectangle [id:dp7973284926553135] 
				\draw   (30,120) -- (160,120) -- (160,160) -- (30,160) -- cycle ;
				%Straight Lines [id:da6961071314388567] 
				\draw    (90,160) -- (90,210) ;
				
				% Text Node
				\draw (51,123.67) node [anchor=north west][inner sep=0.75pt]   [align=left] {$ Q_{\alpha \beta }^{\Lambda _{1}} \otimes Q_{\alpha '\beta '}^{\Lambda '_{1}}$};
				% Text Node
				\draw (100,79) node [anchor=north west][inner sep=0.75pt]   [align=left] {$ \Lambda ''_{1}$};
				% Text Node
				\draw (100,169) node [anchor=north west][inner sep=0.75pt]   [align=left] {$ \Lambda ''_{1}$};
				% Text Node
				\draw (79,209) node [anchor=north west][inner sep=0.75pt]   [align=left] {$ \beta ''$};
				% Text Node
				\draw (79,47) node [anchor=north west][inner sep=0.75pt]   [align=left] {$ \alpha ''$};
			\end{tikzpicture}
	}}
	=	\sum_{\widetilde{\Lambda _{1}}, \widetilde{\Lambda '_{1}}, \xi}
	\tikzset{every picture/.style={line width=0.75pt}} %set default line width to 0.75pt  
	\vcenter{ \hbox{      
			\begin{tikzpicture}[x=0.75pt,y=0.75pt,yscale=-0.8,xscale=0.8]
				
				%Straight Lines [id:da07394160232212732] 
				\draw    (290,222.35) -- (290,270) ;
				\draw [shift={(290,220)}, rotate = 90] [color={rgb, 255:red, 0; green, 0; blue, 0 }  ][line width=0.75]      (0, 0) circle [x radius= 3.35, y radius= 3.35]   ;
				%Straight Lines [id:da9449305088810864] 
				\draw    (260,200) -- (288.04,218.7) ;
				\draw [shift={(290,220)}, rotate = 33.69] [color={rgb, 255:red, 0; green, 0; blue, 0 }  ][line width=0.75]      (0, 0) circle [x radius= 3.35, y radius= 3.35]   ;
				%Shape: Rectangle [id:dp8972054122474298] 
				\draw   (230,150) -- (280,150) -- (280,180) -- (230,180) -- cycle ;
				%Shape: Rectangle [id:dp6257782231358877] 
				\draw   (300,150) -- (350,150) -- (350,180) -- (300,180) -- cycle ;
				%Straight Lines [id:da9798111705275792] 
				\draw    (320,200) -- (291.96,218.7) ;
				\draw [shift={(290,220)}, rotate = 146.31] [color={rgb, 255:red, 0; green, 0; blue, 0 }  ][line width=0.75]      (0, 0) circle [x radius= 3.35, y radius= 3.35]   ;
				%Straight Lines [id:da2380521112234799] 
				\draw    (260,180) -- (260,200) ;
				%Straight Lines [id:da28828765133752876] 
				\draw    (320,180) -- (320,200) ;
				%Straight Lines [id:da6553738692581448] 
				\draw    (290,107.65) -- (290,60) ;
				\draw [shift={(290,110)}, rotate = 270] [color={rgb, 255:red, 0; green, 0; blue, 0 }  ][line width=0.75]      (0, 0) circle [x radius= 3.35, y radius= 3.35]   ;
				%Straight Lines [id:da06376373711273398] 
				\draw    (320,130) -- (291.96,111.3) ;
				\draw [shift={(290,110)}, rotate = 213.69] [color={rgb, 255:red, 0; green, 0; blue, 0 }  ][line width=0.75]      (0, 0) circle [x radius= 3.35, y radius= 3.35]   ;
				%Straight Lines [id:da629210840226569] 
				\draw    (260,130) -- (288.04,111.3) ;
				\draw [shift={(290,110)}, rotate = 326.31] [color={rgb, 255:red, 0; green, 0; blue, 0 }  ][line width=0.75]      (0, 0) circle [x radius= 3.35, y radius= 3.35]   ;
				%Straight Lines [id:da7953689670470372] 
				\draw    (320,150) -- (320,130) ;
				%Straight Lines [id:da45157374119453375] 
				\draw    (260,150) -- (260,130) ;
				
				% Text Node
				\draw (300,227) node [anchor=north west][inner sep=0.75pt]   [align=left] {$ \Lambda ''_{1}$};
				% Text Node
				\draw (279,267) node [anchor=north west][inner sep=0.75pt]   [align=left] {$ \beta ''$};
				% Text Node
				\draw (239,149) node [anchor=north west][inner sep=0.75pt]   [align=left] {$ Q_{\alpha \beta }^{\Lambda _{1}}$};
				% Text Node
				\draw (305,149) node [anchor=north west][inner sep=0.75pt]   [align=left] {$ Q_{\alpha '\beta '}^{\Lambda'_{1}}$};
				% Text Node
				\draw (299,77) node [anchor=north west][inner sep=0.75pt]   [align=left] {$ \Lambda ''_{1}$};
				% Text Node
				\draw (279,39) node [anchor=north west][inner sep=0.75pt]   [align=left] {$ \alpha ''$};
				% Text Node
				\draw (235,109) node [anchor=north west][inner sep=0.75pt]   [align=left] {$ \widetilde{\Lambda _{1}}$};
				% Text Node
				\draw (235,189) node [anchor=north west][inner sep=0.75pt]   [align=left] {$ \widetilde{\Lambda _{1}}$};
				% Text Node
				\draw (319,109) node [anchor=north west][inner sep=0.75pt]   [align=left] {$ \widetilde{\Lambda '_{1}}$};
				% Text Node
				\draw (319,189) node [anchor=north west][inner sep=0.75pt]   [align=left] {$ \widetilde{\Lambda '_{1}}$};
				% Text Node
				\draw (269,219) node [anchor=north west][inner sep=0.75pt]   [align=left] {$ \xi $};
				% Text Node
				\draw (269,89) node [anchor=north west][inner sep=0.75pt]   [align=left] {$ \xi $};		
			\end{tikzpicture}
	}}
\end{equation}

Inserting this into equation \eqref{eq: rep basis 3pt step 2} gives
\begin{align} \nonumber
	\Tr_{V_N^{\otimes k}}(& ( Q^{\Lambda_1}_{\beta \alpha} \otimes Q^{\Lambda_1'}_{\beta' \alpha'} ) ~  Q^{\Lambda_1''}_{\alpha'' \beta''} ) = \\
	&\DimSN{\Lambda_1''} \sum_{\widetilde{\Lambda}_1, \widetilde{\Lambda}_1', \gamma, \eta, \gamma', \eta', \xi} B^{\Lambda_1'' \rightarrow \widetilde{\Lambda}_1 \otimes \widetilde{\Lambda}_1' , \xi}_{\gamma'' \rightarrow \gamma \gamma' } D^{\widetilde{\Lambda}_1}_{\gamma \eta}(Q^{\Lambda_1}_{\alpha \beta}) D^{\widetilde{\Lambda}_1'}_{\gamma' \eta'}(Q^{\Lambda_1'}_{\alpha' \beta'}) B^{ \Lambda_1'' \rightarrow \widetilde{\Lambda}_1 \otimes \widetilde{\Lambda}_1' , \xi}_{ \eta'' \rightarrow \eta \eta' }. \label{eq: rep basis 3pt step 3}
\end{align}
Matrix elements of irreducible representations are orthogonal (see equation \eqref{eq: orthogonality Pk matrix elements}). This implies
\begin{equation}
	D^{\widetilde{\Lambda}_1''}_{\eta'' \gamma''}(Q^{\Lambda_1''}_{\alpha'' \beta''}) = \delta^{\widetilde{\Lambda}_1'' \Lambda_1''} \delta_{\eta'' \beta''} \delta_{\gamma'' \alpha''}
\end{equation}
or the equivalent diagrammatic expression
\begin{equation}
	\tikzset{every picture/.style={line width=0.75pt}} %set default line width to 0.75pt    
	\vcenter{ \hbox{    
			\begin{tikzpicture}[x=0.75pt,y=0.75pt,yscale=-0.8,xscale=0.8]
				%Straight Lines [id:da31635586008883254] 
				\draw    (420,40) -- (420,90) ;
				%Shape: Rectangle [id:dp04565121773081482] 
				\draw   (390,90) -- (450,90) -- (450,130) -- (390,130) -- cycle ;
				%Straight Lines [id:da8742404601606097] 
				\draw    (420,130) -- (420,180) ;
				% Text Node
				\draw (397.6,95.2) node [anchor=north west][inner sep=0.75pt]   [align=left] {$ Q_{\alpha ''\beta ''}^{\Lambda ''_{1}}$};
				% Text Node
				\draw (409,177) node [anchor=north west][inner sep=0.75pt]   [align=left] {$ \eta ''$};
				% Text Node
				\draw (409,15) node [anchor=north west][inner sep=0.75pt]   [align=left] {$ \gamma ''$};
				% Text Node
				\draw (425,49) node [anchor=north west][inner sep=0.75pt]   [align=left] {$ \widetilde{\Lambda _{1}}$};
				% Text Node
				\draw (425,139) node [anchor=north west][inner sep=0.75pt]   [align=left] {$ \widetilde{\Lambda _{1}}$};
			\end{tikzpicture} 
	}} =  \delta^{\Lambda_1'' \widetilde{\Lambda''}_1}
	\tikzset{every picture/.style={line width=0.75pt}} %set default line width to 0.75pt       
	\vcenter{ \hbox{ 
			\begin{tikzpicture}[x=0.75pt,y=0.75pt,yscale=-0.8,xscale=0.8]
				%Straight Lines [id:da757783758262732] 
				\draw    (370,60) -- (370,110) ;
				%Straight Lines [id:da23505941153761722] 
				\draw    (370,170) -- (370,220) ;
				% Text Node
				\draw (359,217) node [anchor=north west][inner sep=0.75pt]   [align=left] {$ \eta''$};
				% Text Node
				\draw (359,35) node [anchor=north west][inner sep=0.75pt]   [align=left] {$ \gamma ''$};
				% Text Node
				\draw (359,149) node [anchor=north west][inner sep=0.75pt]   [align=left] {$ \beta ''$};
				% Text Node
				\draw (359,109) node [anchor=north west][inner sep=0.75pt]   [align=left] {$ \alpha ''$};
			\end{tikzpicture}
	}}.
\end{equation}
Substituting this identity into \eqref{eq: rep basis 3pt step 3} reduces it to
\begin{align}
	\sum_{\xi} \Dim V_{ \Lambda_1''}^{S_N} B^{{\Lambda}_1'' \rightarrow {\Lambda}_1 \otimes {\Lambda}_1' , \xi}_{ \alpha'' \rightarrow \alpha \alpha'} B^{ {\Lambda}_1'' \rightarrow {\Lambda}_1 \otimes {\Lambda}_1' , \xi}_{ \beta'' \rightarrow \beta \beta' } = \sum_{\xi} \Dim V_{ \Lambda_1''}^{S_N} 
	\tikzset{every picture/.style={line width=0.75pt}} %set default line width to 0.75pt      
	\vcenter{ \hbox{  
			\begin{tikzpicture}[x=0.75pt,y=0.75pt,yscale=-0.8,xscale=0.8]
				%uncomment if require: \path (0,643); %set diagram left start at 0, and has height of 643
				%Straight Lines [id:da3075939944148587] 
				\draw    (300,542.35) -- (300,590) ;
				\draw [shift={(300,540)}, rotate = 90] [color={rgb, 255:red, 0; green, 0; blue, 0 }  ][line width=0.75]      (0, 0) circle [x radius= 3.35, y radius= 3.35]   ;
				%Straight Lines [id:da7431275815945515] 
				\draw    (270,520) -- (298.04,538.7) ;
				\draw [shift={(300,540)}, rotate = 33.69] [color={rgb, 255:red, 0; green, 0; blue, 0 }  ][line width=0.75]      (0, 0) circle [x radius= 3.35, y radius= 3.35]   ;
				%Straight Lines [id:da10387798395248948] 
				\draw    (330,520) -- (301.96,538.7) ;
				\draw [shift={(300,540)}, rotate = 146.31] [color={rgb, 255:red, 0; green, 0; blue, 0 }  ][line width=0.75]      (0, 0) circle [x radius= 3.35, y radius= 3.35]   ;
				%Straight Lines [id:da13861700085350726] 
				\draw    (270,500) -- (270,520) ;
				%Straight Lines [id:da5120392716467388] 
				\draw    (330,500) -- (330,520) ;
				%Straight Lines [id:da5560752368542436] 
				\draw    (300,407.65) -- (300,360) ;
				\draw [shift={(300,410)}, rotate = 270] [color={rgb, 255:red, 0; green, 0; blue, 0 }  ][line width=0.75]      (0, 0) circle [x radius= 3.35, y radius= 3.35]   ;
				%Straight Lines [id:da7102204228848794] 
				\draw    (330,430) -- (301.96,411.3) ;
				\draw [shift={(300,410)}, rotate = 213.69] [color={rgb, 255:red, 0; green, 0; blue, 0 }  ][line width=0.75]      (0, 0) circle [x radius= 3.35, y radius= 3.35]   ;
				%Straight Lines [id:da6697053392310997] 
				\draw    (270,430) -- (298.04,411.3) ;
				\draw [shift={(300,410)}, rotate = 326.31] [color={rgb, 255:red, 0; green, 0; blue, 0 }  ][line width=0.75]      (0, 0) circle [x radius= 3.35, y radius= 3.35]   ;
				%Straight Lines [id:da6694321287780429] 
				\draw    (330,450) -- (330,430) ;
				%Straight Lines [id:da40087215914109264] 
				\draw    (270,450) -- (270,430) ;
				% Text Node
				\draw (310,547) node [anchor=north west][inner sep=0.75pt]   [align=left] {$ \Lambda ''_{1}$};
				% Text Node
				\draw (289,587) node [anchor=north west][inner sep=0.75pt]   [align=left] {$ \beta ''$};
				% Text Node
				\draw (309,377) node [anchor=north west][inner sep=0.75pt]   [align=left] {$ \Lambda ''_{1}$};
				% Text Node
				\draw (289,339) node [anchor=north west][inner sep=0.75pt]   [align=left] {$ \alpha ''$};
				% Text Node
				\draw (245,409) node [anchor=north west][inner sep=0.75pt]   [align=left] {$ \Lambda _{1}$};
				% Text Node
				\draw (279,539) node [anchor=north west][inner sep=0.75pt]   [align=left] {$ \xi $};
				% Text Node
				\draw (279,389) node [anchor=north west][inner sep=0.75pt]   [align=left] {$ \xi $};
				% Text Node
				\draw (335,409) node [anchor=north west][inner sep=0.75pt]   [align=left] {$ \Lambda '_{1}$};
				% Text Node
				\draw (329,517) node [anchor=north west][inner sep=0.75pt]   [align=left] {$ \Lambda '_{1}$};
				% Text Node
				\draw (245,517) node [anchor=north west][inner sep=0.75pt]   [align=left] {$ \Lambda _{1}$};
				% Text Node
				\draw (259,448) node [anchor=north west][inner sep=0.75pt]   [align=left] {$ \alpha $};
				% Text Node
				\draw (319,448) node [anchor=north west][inner sep=0.75pt]   [align=left] {$ \alpha '$};
				% Text Node
				\draw (259,478) node [anchor=north west][inner sep=0.75pt]   [align=left] {$ \beta $};
				% Text Node
				\draw (319,479) node [anchor=north west][inner sep=0.75pt]   [align=left] {$ \beta '$};
			\end{tikzpicture}
	}}. \label{eq: rep basis 3pt step 4}
\end{align}
This gives the final result for matrix units of $P_k(N)$.

The full expression for \eqref{eq: rep basis 3pt SPk} - extremal three-point correlators in the representation basis - is given by \eqref{eq: rep basis 3pt step 4} together with branching coefficients from the partition algebras to symmetric group algebras (see \eqref{eq: branching coeff}),
\begin{align} \label{eq: three-point correlator expression} \nonumber
	\Tr_{V_N^{\otimes k}}&\Big( \big(Q^{\Lambda_1}_{\Lambda_2,\nu \mu}\otimes Q^{\Lambda_1'}_{\Lambda_2',\nu' \mu'} \big) Q^{\Lambda_1''}_{\Lambda_2'',\mu'' \nu''} \Big) = \Dim V_{ \Lambda_1''}^{S_N} \sum_{\substack{\alpha, \beta, \alpha', \beta', \alpha'', \beta'',\\ p,p',p'', \xi}} 
	B^{ {\Lambda}_1'' \rightarrow {\Lambda}_1 \otimes {\Lambda}_1' , \xi}_{\alpha'' \rightarrow \alpha \alpha' } B^{{\Lambda}_1'' \rightarrow {\Lambda}_1 \otimes {\Lambda}_1' , \xi}_{\beta'' \rightarrow \beta \beta' } \\
	&B^{P_{k_1}(N) \rightarrow \mathbb{C}[S_{k_1}]}_{\Lambda_1, \alpha \rightarrow \Lambda_2, p; \mu}B^{P_{k_1}(N) \rightarrow \mathbb{C}[S_{k_1}]}_{\Lambda_1, \beta \rightarrow \Lambda_2, p; \nu} 
	B^{P_{k_2}(N) \rightarrow \mathbb{C}[S_{k_2}]}_{\Lambda_1', \alpha' \rightarrow \Lambda_2', p'; \mu'}B^{P_{k_2}(N) \rightarrow \mathbb{C}[S_{k_2}]}_{\Lambda_1', \beta' \rightarrow \Lambda_2', p'; \nu'} 
	B^{P_{k}(N) \rightarrow \mathbb{C}[S_{k}]}_{\Lambda_1'', \alpha'' \rightarrow \Lambda_2'', p''; \mu''}B^{P_{k}(N) \rightarrow \mathbb{C}[S_{k}]}_{\Lambda_1'', \beta'' \rightarrow \Lambda_2'', p''; \nu''}.
\end{align}
Introducing the following diagram representation of these branching coefficients,
\begin{equation}
	B^{P_{k}(N) \rightarrow \mathbb{C}[S_{k}]}_{\Lambda_1, \alpha \rightarrow \Lambda_2, p; \mu} = 
	\tikzset{every picture/.style={line width=0.75pt}} %set default line width to 0.75pt        
	\vcenter{ \hbox{
			\begin{tikzpicture}[x=0.75pt,y=0.75pt,yscale=-0.8,xscale=0.8]
				%uncomment if require: \path (0,643); %set diagram left start at 0, and has height of 643
				
				%Straight Lines [id:da8724499506044037] 
				\draw    (390,430) -- (390,480) ;
				%Straight Lines [id:da998542994191449] 
				\draw    (390,430) -- (390,380) ;
				\draw [shift={(390,430)}, rotate = 270] [color={rgb, 255:red, 0; green, 0; blue, 0 }  ][fill={rgb, 255:red, 0; green, 0; blue, 0 }  ][line width=0.75]      (0, 0) circle [x radius= 3.35, y radius= 3.35]   ;
				
				% Text Node
				\draw (365,447) node [anchor=north west][inner sep=0.75pt]   [align=left] {$ \Lambda _{1}$};
				% Text Node
				\draw (369,419) node [anchor=north west][inner sep=0.75pt]   [align=left] {$ \mu $};
				% Text Node
				\draw (365,389) node [anchor=north west][inner sep=0.75pt]   [align=left] {$ \Lambda _{2}$};
				% Text Node
				\draw (382.2,365) node [anchor=north west][inner sep=0.75pt]   [align=left] {$ p$};
				% Text Node
				\draw (381.4,483.2) node [anchor=north west][inner sep=0.75pt]   [align=left] {$ \alpha$};
			\end{tikzpicture}
	}},
\end{equation}
we can write \eqref{eq: three-point correlator expression} as the following diagram
\begin{equation}
\Tr_{V_N^{\otimes k}} \Big( \big(Q^{\Lambda_1}_{\Lambda_2,\nu \mu}\otimes Q^{\Lambda_1'}_{\Lambda_2',\nu' \mu'} \big) Q^{\Lambda_1''}_{\Lambda_2'',\mu'' \nu''} \Big) =
\vcenter{\hbox{
	\tikzset{every picture/.style={line width=0.75pt}} %set default line width to 0.75pt        
	\begin{tikzpicture}[x=0.75pt,y=0.75pt,yscale=-0.8,xscale=0.8]
		%uncomment if require: \path (0,643); %set diagram left start at 0, and has height of 643
		%Straight Lines [id:da3075939944148587] 
		\draw    (90,572.35) -- (90,620) ;
		\draw [shift={(90,570)}, rotate = 90] [color={rgb, 255:red, 0; green, 0; blue, 0 }  ][line width=0.75]      (0, 0) circle [x radius= 3.35, y radius= 3.35]   ;
		%Straight Lines [id:da7431275815945515] 
		\draw    (60,550) -- (88.04,568.7) ;
		\draw [shift={(90,570)}, rotate = 33.69] [color={rgb, 255:red, 0; green, 0; blue, 0 }  ][line width=0.75]      (0, 0) circle [x radius= 3.35, y radius= 3.35]   ;
		%Straight Lines [id:da10387798395248948] 
		\draw    (120,550) -- (91.96,568.7) ;
		\draw [shift={(90,570)}, rotate = 146.31] [color={rgb, 255:red, 0; green, 0; blue, 0 }  ][line width=0.75]      (0, 0) circle [x radius= 3.35, y radius= 3.35]   ;
		%Straight Lines [id:da13861700085350726] 
		\draw    (60,510) -- (60,550) ;
		%Straight Lines [id:da5560752368542436] 
		\draw    (90,397.65) -- (90,350) ;
		\draw [shift={(90,400)}, rotate = 270] [color={rgb, 255:red, 0; green, 0; blue, 0 }  ][line width=0.75]      (0, 0) circle [x radius= 3.35, y radius= 3.35]   ;
		%Straight Lines [id:da7102204228848794] 
		\draw    (120,420) -- (91.96,401.3) ;
		\draw [shift={(90,400)}, rotate = 213.69] [color={rgb, 255:red, 0; green, 0; blue, 0 }  ][line width=0.75]      (0, 0) circle [x radius= 3.35, y radius= 3.35]   ;
		%Straight Lines [id:da6697053392310997] 
		\draw    (60,420) -- (88.04,401.3) ;
		\draw [shift={(90,400)}, rotate = 326.31] [color={rgb, 255:red, 0; green, 0; blue, 0 }  ][line width=0.75]      (0, 0) circle [x radius= 3.35, y radius= 3.35]   ;
		%Straight Lines [id:da08807351490338933] 
		\draw    (60,510) -- (60,460) ;
		\draw [shift={(60,510)}, rotate = 270] [color={rgb, 255:red, 0; green, 0; blue, 0 }  ][fill={rgb, 255:red, 0; green, 0; blue, 0 }  ][line width=0.75]      (0, 0) circle [x radius= 3.35, y radius= 3.35]   ;
		%Straight Lines [id:da4915587049391228] 
		\draw    (60,470) -- (60,490) ;
		%Straight Lines [id:da41712390813615685] 
		\draw    (60,460) -- (60,420) ;
		\draw [shift={(60,460)}, rotate = 270] [color={rgb, 255:red, 0; green, 0; blue, 0 }  ][fill={rgb, 255:red, 0; green, 0; blue, 0 }  ][line width=0.75]      (0, 0) circle [x radius= 3.35, y radius= 3.35]   ;
		%Straight Lines [id:da10010616859529087] 
		\draw    (120,510) -- (120,550) ;
		%Straight Lines [id:da9493492042409282] 
		\draw    (120,510) -- (120,460) ;
		\draw [shift={(120,510)}, rotate = 270] [color={rgb, 255:red, 0; green, 0; blue, 0 }  ][fill={rgb, 255:red, 0; green, 0; blue, 0 }  ][line width=0.75]      (0, 0) circle [x radius= 3.35, y radius= 3.35]   ;
		%Straight Lines [id:da251996477212062] 
		\draw    (120,470) -- (120,490) ;
		%Straight Lines [id:da04913851577718242] 
		\draw    (120,460) -- (120,420) ;
		\draw [shift={(120,460)}, rotate = 270] [color={rgb, 255:red, 0; green, 0; blue, 0 }  ][fill={rgb, 255:red, 0; green, 0; blue, 0 }  ][line width=0.75]      (0, 0) circle [x radius= 3.35, y radius= 3.35]   ;
		%Straight Lines [id:da16209060867104585] 
		\draw    (220,510) -- (220,550) ;
		%Straight Lines [id:da43645886058808014] 
		\draw    (220,510) -- (220,460) ;
		\draw [shift={(220,510)}, rotate = 270] [color={rgb, 255:red, 0; green, 0; blue, 0 }  ][fill={rgb, 255:red, 0; green, 0; blue, 0 }  ][line width=0.75]      (0, 0) circle [x radius= 3.35, y radius= 3.35]   ;
		%Straight Lines [id:da818498562694443] 
		\draw    (220,470) -- (220,490) ;
		%Straight Lines [id:da371021613845651] 
		\draw    (220,460) -- (220,420) ;
		\draw [shift={(220,460)}, rotate = 270] [color={rgb, 255:red, 0; green, 0; blue, 0 }  ][fill={rgb, 255:red, 0; green, 0; blue, 0 }  ][line width=0.75]      (0, 0) circle [x radius= 3.35, y radius= 3.35]   ;
		%Curve Lines [id:da3134870020273035] 
		\draw    (90,620) .. controls (90.05,650.26) and (220.05,592.56) .. (220,550) ;
		%Curve Lines [id:da9572045552157762] 
		\draw    (90,350) .. controls (89.65,311.37) and (220.45,378.22) .. (220,420) ;
		
		% Text Node
		\draw (100,577) node [anchor=north west][inner sep=0.75pt]   [align=left] {$ \Lambda ''_{1}$};
		% Text Node
		\draw (99,367) node [anchor=north west][inner sep=0.75pt]   [align=left] {$ \Lambda ''_{1}$};
		% Text Node
		\draw (35,399) node [anchor=north west][inner sep=0.75pt]   [align=left] {$ \Lambda _{1}$};
		% Text Node
		\draw (69,569) node [anchor=north west][inner sep=0.75pt]   [align=left] {$ \xi $};
		% Text Node
		\draw (69,379) node [anchor=north west][inner sep=0.75pt]   [align=left] {$ \xi $};
		% Text Node
		\draw (125,399) node [anchor=north west][inner sep=0.75pt]   [align=left] {$ \Lambda '_{1}$};
		% Text Node
		\draw (119,547) node [anchor=north west][inner sep=0.75pt]   [align=left] {$ \Lambda '_{1}$};
		% Text Node
		\draw (35,547) node [anchor=north west][inner sep=0.75pt]   [align=left] {$ \Lambda _{1}$};
		% Text Node
		\draw (39,499) node [anchor=north west][inner sep=0.75pt]   [align=left] {$ \nu $};
		% Text Node
		\draw (39,449) node [anchor=north west][inner sep=0.75pt]   [align=left] {$ \mu $};
		% Text Node
		\draw (35,477) node [anchor=north west][inner sep=0.75pt]   [align=left] {$ \Lambda _{2}$};
		% Text Node
		\draw (127,499) node [anchor=north west][inner sep=0.75pt]   [align=left] {$ \nu '$};
		% Text Node
		\draw (125,449) node [anchor=north west][inner sep=0.75pt]   [align=left] {$ \mu '$};
		% Text Node
		\draw (125,477) node [anchor=north west][inner sep=0.75pt]   [align=left] {$ \Lambda '_{2}$};
		% Text Node
		\draw (227,499) node [anchor=north west][inner sep=0.75pt]   [align=left] {$ \nu ''$};
		% Text Node
		\draw (225,448) node [anchor=north west][inner sep=0.75pt]   [align=left] {$ \mu ''$};
		% Text Node
		\draw (229,477) node [anchor=north west][inner sep=0.75pt]   [align=left] {$ \Lambda ''_{2}$};	
	\end{tikzpicture}}}
\end{equation}
From the above formula we see that the extremal correlator vanishes if the Kronecker coefficient $C(\Lambda_1, \Lambda_1', \Lambda_1'') = 0$. Analogous results for extremal correlators in general quiver gauge theories are described in \cite{QuivCalc}.

\section{Summary and Outlook }

In this paper we investigated the effects of permutation symmetry on the state space and dynamics of quantum mechanical systems of $N \times N$ matrix variables. 
After  a brief review of  the matrix harmonic oscillator and introduction of notation in section \ref{sec: MQM}, we began in section \ref{sec: Perm subspace} by investigating the $S_N$ invariant Hilbert space $\Hilbertspace$ of generic matrix quantum mechanics systems at large $N$. We found that there is a one-to-one correspondence between $S_N$ invariant states of degree $k$ and elements in the symmetrised partition algebra $SP_k(N)$. Two bases of $SP_k(N)$ were discussed: the diagram basis and the representation basis. A construction of the latter was explained in section \ref{sec: charges} in terms of diagonalizing commuting algebraic charges. Having discussed the $S_N$ invariant state space, we moved on to interesting invariant Hamiltonians. The general permutation invariant harmonic matrix oscillator was described and solved (diagonalized) in section \ref{sec: PIMQM}. This was achieved with the introduction of oscillators labelled by representation theoretic quantities, as in \eqref{eq: hamiltonian}. In section \ref{sec: deformations} we described a set of algebraic Hamiltonians for matrix quantum mechanics that preserve the $S_N$ invariant subspace of the Hilbert space. These Hamiltonians, given by equations \eqref{eq: degenerate invariant ground state hamiltonian}, \eqref{eq: partially broken ground state degen H} and \eqref{eq: inv hamiltonians} realise the three dynamical scenarios illustrated on the left hand side of figure \ref{fig: spectrum scenarios d}, the right hand side of figure \ref{fig: spectrum scenarios d}, and figure \ref{fig: spectrum scenarios a} respectively. The representation basis introduced in section \ref{sec: ON basis} diagonalizes all of these algebraic Hamiltonians. We provided a lattice interpretation of the matrix oscillators in section \ref{sec: lattice}. In section \ref{sec: scars} we constructed Hamiltonians which turn the $S_N$ invariant state space into quantum many-body scars. Following the ideas in \cite{PPPK2020,PPPK2021}, we gave Hamiltonians \eqref{eq: H_tot} of the form $H+ H_s$ where $H$ is $S_N$ invariant and $H_s$ annihilates states in the $S_N$ invariant subspace. We noted that the Hamiltonians in section \ref{sec: deformations} are suitable candidates for $H$ if their energies satisfy an integrality condition. As an example, we used the lattice interpretation to give a modified Bose-Hubbard Hamiltonian which exhibits $S_N$ invariant quantum many-body scars.
The diagram basis is the most efficient basis for describing inner and outer products. As a consequence extremal correlators, defined in \eqref{eq: extremal correlator}, which are analogues of three-point extremal correlators in $\mathcal{N} = 4$ SYM are simple in the diagram basis. The extremal correlators satisfy representation theoretic selection rules, based on Kronecker coefficients, which were derived in the representation basis. The selection rules are based on exact expressions for extremal correlators, involving Kronecker coefficients and Littlewood-Richardson coefficients, given in equation \eqref{eq: three-point correlator expression}. 

The representation theoretic basis $Q^{\Lambda_1}_{ \Lambda_2 , \mu \nu} $ for the 
 $S_N$ invariant Hilbert space $\Hilbertspace$ constructed as linear  
 combinations of symmetrised partition algebra elements in $SP_k(N)$ 
  in section   \ref{sec: ON basis} is an eigenstate basis for the free Hamiltonian $H_0$ of section   \ref{sec: MQM} as well as the algebraic Hamiltonians constructed in section 
 \ref{sec: deformations}. The action of the general permutation invariant harmonic oscillator Hamiltonian given in  \eqref{eq: hamiltonian} of section \ref{sec: PIMQM} however causes a non-trivial mixing of the representation labels. This mixing was discussed briefly in section \ref{sec: energy eigenbasis}. Diagonalising the general harmonic oscillator Hamiltonians in $\Hilbertspace$  is an interesting, unsolved problem of finding appropriate linear combinations of the $Q^{\Lambda_1}_{ \Lambda_2 , \mu \nu} $ which are invariant, up to scaling, under the mixing. 
 
%  that diagonalise the simplest matrix quantum mechanics Hamiltonian $H_0$. Building on this we showed that it was possible to perform the equivalent diagonalization for a family of Hamiltonians, using the representation basis. What we leave for future work is the diagonalisation of the full $S_N$ invariant Hamiltonian, given by equation \eqref{eq: hamiltonian}, on the $S_N$ invariant Hilbert space $\Hilbertspace$. In the representation basis, this will involve a limited form of mixing, constrained by the tensor product decomposition of irreducible $S_N$ representations $\Lambda, \Lambda_1$.

With the exception of the $P_k(N)$ orbit basis discussion in appendix \ref{sec: orbit basis} we have assumed $N \geq 2k$, known as the stable limit. This simplified the construction of a basis for the $S_N$ invariant subspace $\Hilbertspace$, a simplification related to the existence of a kernel free map from $P_k(N)$ to $\text{End}(V_N^{\otimes k})$. However, it would be interesting to uncover any finite $N$ effects appearing in these permutation invariant quantum mechanical matrix systems. 
At finite $N$ the diagrams in $P_k(N)$ provide an over complete basis of operators. That is, there are some linear relations between operators. The precise form of these relations can be found using the orbit basis. The question remains of how to use this knowledge in order to construct a representation theoretic basis for $2k < N$. We leave this for future work, but note here that it would involve a detailed study of the Artin-Wedderburn decomposition in \eqref{eq: Artin Decomposition SPkN} below the stable limit. The detailed study includes putting constraints on the irreducible representations appearing in the decomposition below the stable limit, as well as computing the dimension of the multiplicity spaces.

In section \ref{sec: lattice} we gave one interpretation of our model in terms of bosonic excitations $(a^{\dagger})_i^j$ on an $N$-by-$N$ lattice with sites labelled by $(i,j)$. It is natural to ask if the $S_N$ invariant Hamiltonians described by \eqref{eq: inv hamiltonians} interpreted in this way can be realised in experiments. In the real world interactions tend to be local. The demand for these Hamiltonians to be local places restrictions on the sets of permissible terms. In section \ref{subsec: BH model} we used this interpretation to construct a modified Bose-Hubbard Hamiltonian exhibiting $S_N$ invariant quantum many-body scars. More generally, combining the lattice interpretation of matrix oscillators with the group theoretic scheme given in \cite{PPPK2020,PPPK2021}, as was done in \ref{subsec: revival} and \ref{subsec: scar hamiltonians}, provides a useful framework for describing systems with many-body scars in $2+1$ dimensions.

A very interesting avenue towards applications of the Hilbert spaces and Hamiltonians considered here is to find systems where the permutation invariant sectors described using partition algebras are naturally selected by the physics. For example, in 
a Bose-Einstein condensate composed of $N$ identical bosons, excited by  vibrational modes between pairs of particles, oscillators $ (a^{\dagger})^j_i$  exciting the pair $(i,j)$ of particles with $i , j \in \{ 1, \cdots , N \}$ would naturally be subject to the kind of $S_N$ invariance we have considered here. This would provide links between the theoretical application of  partition algebras as considered here with the phenomenological modelling of Bose-Einstein physics, e.g. along the lines of \cite{dipolar}. 

As a closing remark, we note that much of the initial study of the representation theory of partition algebras $P_k(N)$ was done with physical motivations coming from classical statistical models (Potts models) where $k$ is the number of lattice sites and $N$ is the number of classical states for each lattice site. The transfer matrix of the classical statistical model plays a crucial role in these studies \cite{Martin1994,Martin1996,Jones1994}. The present application of partition algebras looks substantially different: we have quantum mechanical matrix oscillators, with matrix size $N$  possible values and $k$ specifies the sector of quantum states with $k$ oscillators acting on the vacuum.  Exploring potential dualities relating systems of the kind studied earlier  and the matrix quantum systems defined here is a fascinating future direction.

\vskip.5cm 

\begin{center} 
	{ \bf Acknowledgements} 
\end{center} 
SR is supported by the STFC consolidated grant ST/T000686/1 ``Amplitudes, Strings and Duality” and a Visiting Professorship at the University of the Witwatersrand. We are pleased to acknowledge useful conversations on the subject of this paper with Matthew Buican, Robert de Mello Koch, Adam Denchfield, Masanori Hanada,  Costis Papageorgakis, Rajath Radhakrishnan.

\appendix
\section{Matrix units and Fourier inversion from inner product}
\label{apx: Semi-Simple Algebra Technology}
In this appendix we prove the results in section \ref{sec: ON basis} on representation bases. We review the construction of matrix units for semi-simple algebras closely following \cite{AR90DissertCh1}. We focus in particular on the partition algebras  $P_k(N)$ and the symmetrised partition algebras  $SP_k(N)$.  The appendix is divided into four subsections. We start by discussing non-degenerate bilinear forms on algebras and how they define dual elements through \eqref{eq: algebra dual}. The existence of dual elements allows us to prove orthogonality of matrix elements of irreducible representations of $P_k(N)$, as stated in \eqref{eq: orthogonality Pk matrix elements}. Orthogonality is essential for the construction of matrix units of $P_k(N)$ using the Fourier inversion formula \eqref{eq: fourier inversion}. Matrix units for $SP_k(N)$ are constructed using branching coefficients, as in \eqref{eq: spkn units}. Minor modifications to the construction in \cite{AR90DissertCh1}, which defines a non-degenerate bilinear using the trace in the regular representation of $P_k(N)$, are necessary. In this paper, the physical trace relevant to the inner product \eqref{eq: inner product is trace} and two point function, is a trace in $V_N^{\otimes k}$. This induces minor changes to the basic formulas. The two traces are related in \eqref{eq: traces relation omega factor}, through a so-called $\Omega$-factor.

%For completeness, this appendix reviews the abstract construction of matrix units for a semi-simple algebra (but in particular the partition algebras $P_k(N)$ and their symmetrised version $SP_k(N)$). It closely follows \cite{AR90DissertCh1}, which we found to be useful background. The appendix is divided into four subsections, each one containing a result that is central in the construction. The first subsection discusses the notion of non-degenerate bilinear forms on algebras and their relation to semi-simplicity. The main takeaway will be the definition of dual elements of an algebra which uses the trace in $V_N^{\otimes k}$. The second subsection reviews orthogonality of matrix elements of irreducible representations. A minor modification to the standard formula for orthogonality is necessary, given our definition of dual elements. The third subsection gives the so-called Fourier inversion formula for matrix units $Q^{\Lambda_1}_{\alpha \beta}$ of $P_k(N)$, and proves their multiplication properties. The last subsection uses the matrix units for $P_k(N)$ together with branching coefficients to construct matrix units for the symmetrised partition algebras $SP_k(N)$. We also compute the normalization constant for the inner product of two states corresponding to matrix units of $SP_k(N)$.

\subsection{Schur-Weyl duality and non-degenerate bilinear forms}
The construction of matrix units for $P_k(N)$ relies on the existence of a non-degenerate bilinear form on $P_k(N)$. The bilinear form used in \cite{AR90DissertCh1} is defined using the trace in the regular representation of $P_k(N)$. In this paper the physical trace, associated with inner products, is a trace in $V_N^{\otimes k}$ including a transposition as in equation \eqref{eq: inner product is trace}. The aim of this subsection is to prove that this trace defines a non-degenerate bilinear form as well. The outline of the proof is as follows. The trace in the regular representation is related to the trace on $V_N^{\otimes k}$ by the insertion of a central element. Given this relation, non-degeneracy of the bilinear form defined by the trace on $V_N^{\otimes k}$ follows by the non-degeneracy of the bilinear form defined by the trace in the regular representation.

Let $\mathcal{B} = \{b_1,\dots,b_{B(2k)}\}$ be a basis for $P_k(N)$. The regular representation of $P_k(N)$ is defined by the left action of $P_k(N)$ on itself. The matrix representation of $b_i$ is defined by the structure constants $C_{ij}^k$
\begin{equation}
	b_i b_j = \sum_{k=1}^{B(2k)}C_{ij}^k b_k.
\end{equation}
Consequently, the trace in the regular representation can be written as
\begin{equation}
	\tr(b_i) = \sum_{j=1}^{B(2k)} C_{ij}^j = \sum_{j=1}^{B(2k)} \mathrm{Coeff}(b_j, b_i b_j),
\end{equation}
where $\mathrm{Coeff}(b_j, d)$ is the coefficient of $b_j$ in the expansion of $d \in P_k(N)$ in the basis $\mathcal{B}$.

For $N \geq 2k$, $P_k(N)$ is semi-simple (see \cite[Theorem~3.27]{Halverson2004}) and therefore,
\begin{equation}
	G_{ij} \equiv \tr(b_i b_j) \label{eq: gram for reg rep}
\end{equation}
is an invertible matrix. We say that the trace in the regular representation defines a non-degenerate bilinear form on $P_k(N)$ (see \cite[Equation~5.9]{Halverson2004}). It will be useful to use the following equivalent definition of non-degeneracy in what follows. A bilinear form on $P_k(N)$ is non-degenerate if there exists no non-zero element $d \in P_k(N)$ such that
\begin{equation}
	\tr(b_i d) = 0 \quad \forall i=1,\dots,B(2k).
\end{equation}

The regular representation of $P_k(N)$ has a decomposition (see statements in proof of \cite[Proposition~5.7]{Halverson2004})
\begin{equation}
	V^{\text{reg}} = \bigoplus_{\Lambda_1} V^{P_k(N)}_{{\Lambda_1}} \otimes V^{P_k(N)}_{{\Lambda_1}}. \label{eq: pkn reg rep decomp}
\end{equation}
It follows that the trace of $d\in P_k(N)$ in the regular representation can be decomposed as
\begin{equation}
	\tr(d) = \sum_{\Lambda_1} \tr_{{\Lambda_1}}(d) \tr_{{\Lambda_1}}(1) = \sum_{\Lambda_1} \DimPk{\Lambda_1} \chi^{{\Lambda_1}}(d), \label{eq: reg trace decomposition}
\end{equation}
where the sum is over all irreducible representations of $P_k(N)$, $\DimPk{\Lambda_1}$ is the dimension of the representation $\Lambda_1$ and $\chi^{{\Lambda_1}}$ is the corresponding character.

The characters can be extracted from the trace by means of projection operators $p_{\Lambda_1} \in P_k(N)$,
\begin{equation}
	\tr(p_{\Lambda_1}d) = \DimPk{\Lambda_1} \chi^{{\Lambda_1}}(d). \label{eq: reg trace decomp}
\end{equation}
This can be seen as a consequence of character orthogonality (see \cite[Theorem~3.8, Theorem~3.9]{AR90DissertCh1} )
\begin{equation}
	\sum_{i,j=1}^{B(2k)} \DimPk{\Lambda_1} \chi^{\Lambda_1}(b_i)(G^{-1})_{ij}\chi^{\Lambda_1'}(b_j)  = \delta^{\Lambda_1 \Lambda_1'},
\end{equation}
and the fact that projectors can be written as
\begin{equation}
	p_{\Lambda_1} = \sum_{i,j=1}^{B(2k)} \DimPk{\Lambda_1} \chi^{\Lambda_1}(b_i)(G^{-1})_{ij}b_j,
\end{equation}
where $(G^{-1})_{ij}$ is the inverse of the matrix $G_{ij}$ in \eqref{eq: gram for reg rep}. Alternatively, it follows from the decomposition \eqref{eq: pkn reg rep decomp}.

We now move on to the trace in $V_N^{\otimes k}$.
As we have reviewed in section \ref{sec: PA}, $P_k(N)~\cong~\End_{S_N}(V_N^{\otimes k})$ when $N \geq 2k$, where $\End(V_N^{\otimes k})$ is the vector space of linear maps $V_N^{\otimes k} \rightarrow V_N^{\otimes k}$ and $\End_{S_N}(V_N^{\otimes k})$ is the subspace of maps that commute with the action of $S_N$.
Note that we use the same symbol for elements $d \in P_k(N)$ and the corresponding element in $d \in \End_{S_N}(V_N^{\otimes k})$ in what follows. It will be clear from context if $d$ is acting on $V_N^{\otimes k}$.

By Schur-Weyl duality \eqref{eq: Vn otimes k}, the trace in $V_N^{\otimes k}$ decomposes as
\begin{equation}
	\Tr_{V_N^{\otimes k}}(d) = \sum_{\Lambda_1} \DimSN{\Lambda_1} \chi^{{\Lambda_1}}(d), \label{eq: traces V_N otimes k decomposition}
\end{equation}
where the sum is over the irreducible representations that appear in equation \eqref{eq: Vn otimes k}.
Consequently, we can relate the two traces by substituting \eqref{eq: reg trace decomp} into each summand of \eqref{eq: traces V_N otimes k decomposition}
\begin{equation}
	\Tr_{V_N^{\otimes k}}(d) = \sum_{\Lambda_1} \DimSN{\Lambda_1} \chi^{{\Lambda_1}}(d) =  \sum_{\Lambda_1} \frac{\DimSN{\Lambda_1}}{\DimPk{\Lambda_1}} \tr(p_{\Lambda_1}d). \label{eq: Vn trace as reg trace}
\end{equation}
It is convenient to define
\begin{equation}
	\Omega = \sum_{\Lambda_1} \frac{\DimSN{\Lambda_1}}{\DimPk{\Lambda_1}} p_{\Lambda_1}. \label{eq: omega}
\end{equation}
such that equation \eqref{eq: Vn trace as reg trace} becomes
\begin{equation}
	\boxed{\Tr_{V_N^{\otimes k}}(d) = \tr(\Omega d).} \label{eq: traces relation omega factor}
\end{equation}

We can now prove that the bilinear form $( - , - ): P_k(N) \times P_k(N) \rightarrow \mathbb{C}$ given by
\begin{equation}
	( b_i, b_j ) = \Tr_{V_N^{\otimes k}}(b_i b_j)
\end{equation}
is non-degenerate. We give a proof by contradiction.
Suppose there exists a non-zero $d \in P_k(N)$ such that
\begin{equation}
	(b_i, d) = 0, \quad \forall i=1,\dots,B(2k).
\end{equation}
From above, it follows that $d$ is such that
\begin{equation}
	(b_i, d) = \Tr_{V_N^{\otimes k}}(b_i d) = \tr(\Omega b_i d) = 0, \quad \forall i=1,\dots,B(2k).
\end{equation}
However, this implies that the element $d' = d \Omega \in P_k(N)$ is such that
\begin{equation}
	\tr(b_i d') = 0, \quad \forall i=1,\dots,B(2k),
\end{equation}
which contradicts the fact that the trace in the regular representation of $P_k(N)$ defines a non-degenerate bilinear form.

It immediately follows (use proof by contradiction again) that the bilinear form given by
\begin{equation}
	\langle b_i, b_j \rangle = \Tr_{V_N^{\otimes k}}(b_i b_j^{T}) \equiv g_{ij}, \label{eq: bilinear form V_N otimes k}
\end{equation}
is non-degenerate and $g_{ij}$ is invertible.
The inverse matrix is used to define elements dual to $b_i$ which we denote $b_i^*$
\begin{equation}
	\boxed{b_i^{*} = \sum_{j=1}^{B(2k)} (g^{-1})_{ij}b_j.} \label{eq: algebra dual}
\end{equation}
Dual elements satisfy
\begin{equation}
	\langle b_i^*, b_j \rangle = \delta_{ij}. \label{eq: inner prod b_i dual b_j}
\end{equation}
The dual elements are essential for proving orthogonality of matrix elements.
The proof also uses the following property of the bilinear form
\begin{equation}
	\langle b_i, b_j b_k \rangle = \langle b_i b_k^T, b_j \rangle = \langle b_j^T b_i, b_k \rangle. \label{eq: frob associ}
\end{equation}
The first step uses $(b_j b_k)^T = b_k^T b_j^T$ and the second step uses cyclicity of the trace.
%This is a slightly modified version of the associativity of a Frobenius pairing (see 2.3.10 in \cite{Kock2004}).

\subsection{Orthogonality of matrix elements}
The matrix elements $D_{\alpha \beta}^{\Lambda_1}(b_i)$ of irreducible representations of $P_k(N)$ are orthogonal. This is a generalization of the corresponding orthogonality theorem for group algebras (see section 3.15 in \cite{Hamermesh1962}). As we will now prove, the definition of dual elements given in the previous subsection is such that 
\begin{equation}
	\sum_{i=1}^{B(2k)} D_{\alpha \beta}^{\Lambda_1}(b_i)D_{\rho \sigma}^{\Lambda_1'}((b_i^*)^T) \propto \delta_{\alpha \sigma}\delta_{\beta \rho}\delta^{\Lambda_1 \Lambda_1'}. \label{eq: almost grand orth theorem}
\end{equation}

As we now prove, we can always choose irreducible representations satisfying
\begin{equation}
	D_{\alpha \beta}^{\Lambda_1}(d^T) = D_{\beta \alpha}^{\Lambda_1}(d), \qq{for $d \in P_k(N)$,}
\end{equation}
where $d^T$ is as in \eqref{eq: transpose of diagram}.
Starting from the Clebsch-Gordan coefficients $C_{a,i_1 \dots i_k}^{\Lambda_1, \alpha}$ for the decomposition of $V_N^{\otimes k}$ and using Schur-Weyl duality, we identify the multiplicity index $\alpha$ with an orthogonal basis for $V_{\Lambda_1}^{P_k(N)}$. Specifically, define
\begin{equation}
	D_{\alpha \beta}^{\Lambda_1}(d) = \sum_{a} C_{a,i_{1'} \dots i_{k'}}^{\Lambda_1, \alpha}  C_{a,i_1 \dots i_k}^{\Lambda_1, \beta} (d)^{i_{1'} \dots i_{k'}}_{i_1 \dots i_k}.
\end{equation}
Here we are using the fact that Clebsch-Gordan coefficients for $S_N$ can be chosen real \cite[Section 7.14]{Hamermesh1962}.
It follows that
\begin{equation}
	D_{\alpha \beta}^{\Lambda_1}(d^T) = \sum_{a} C_{a,i_{1'} \dots i_{k'}}^{\Lambda_1, \alpha}  C_{a,i_1 \dots i_k}^{\Lambda_1, \beta} (d^T)^{i_{1'} \dots i_{k'}}_{i_1 \dots i_k} = \sum_{a} C_{a,i_{1'} \dots i_{k'}}^{\Lambda_1, \alpha}  C_{a,i_1 \dots i_k}^{\Lambda_1, \beta} (d)_{i_{1'} \dots i_{k'}}^{i_1 \dots i_k} = D_{\beta \alpha}^{\Lambda_1}(d). \label{eq: transpose diagram is transpose matrix element}
\end{equation}

Because the above bilinear form \eqref{eq: bilinear form V_N otimes k} includes a transpose, the symmetrisation theorem (proposition 2.6 in \cite{AR90DissertCh1}) is modified accordingly. Let $C$ be a $\DimPk{\Lambda_1}$ by $\DimPk{\Lambda_1'}$ matrix, and $D^{\Lambda_1}(d), D^{\Lambda_1'}(d)$ be two irreducible matrix representations of $P_k(N)$ with dimension $\DimPk{\Lambda_1}, \DimPk{\Lambda_1'}$ respectively. We have the following version of the symmetrisation theorem. The matrix
\begin{equation}
	[C] = \sum_{i=1}^{B(2k)} D^{\Lambda_1}(b_i)CD^{\Lambda_1'}((b_i^*)^T)
\end{equation}
satisfies
\begin{equation}
	D^{\Lambda_1}(d)[C] = [C]D^{\Lambda_1'}(d), \label{eq: sym theorem}
\end{equation}
for all $d \in P_k(N)$.
The proof is essentially identical to the original case,
\begin{equation}
	\begin{aligned}
		D^{\Lambda_1}(d)[C] &= \sum_{i} D^{\Lambda_1}(db_i)CD^{\Lambda_1'}((b_i^*)^T) = \sum_{i} D^{\Lambda_1}(\sum_j \langle b_j^*, db_i\rangle b_j)CD^{\Lambda_1'}((b_i^*)^T) \\
		&=\sum_j  D^{\Lambda_1}(b_j)CD^{\Lambda_1'}(\sum_{i} (b_i^*)^T \langle b_j^*, db_i\rangle ) \\
		&=\sum_j  D^{\Lambda_1}(b_j)CD^{\Lambda_1'}(\sum_{i} (b_i^*)^T \langle d^Tb_j^*, b_i\rangle ) \\
		&=\sum_j  D^{\Lambda_1}(b_j)CD^{\Lambda_1'}( (d^T b_j^*)^T) \\
		&=[C]D^{\Lambda_1'}(d),
	\end{aligned}
\end{equation}
where in the third line we used the modified Frobenius associativity in equation \eqref{eq: frob associ}.

By Schur's lemma, $[C]$ is proportional to the identity matrix if and only if $\Lambda_1 = \Lambda_1'$ and zero otherwise. For some constant $c^{\Lambda_1}$,
\begin{equation}
	[C]_{\alpha \sigma} = \delta^{\Lambda_1 \Lambda_1'} c^{\Lambda_1} \delta_{\alpha \sigma}.
\end{equation}
The LHS of equation \eqref{eq: almost grand orth theorem} is equal to
\begin{equation}
	\sum_{i} \big( D^{\Lambda_1}(b_i)E_{\beta \rho}D^{\Lambda_1'}( (b_i^*)^T) ) \big)_{\alpha \sigma} = [E_{\beta \rho}]_{\alpha \sigma},
\end{equation}
where $E_{\beta \rho}$ is the elementary matrix with 0 everywhere except in row $\beta$, column $\rho$ which has a 1. It follows from the symmetrisation theorem \eqref{eq: sym theorem} that,
\begin{equation}
	\sum_{i} D_{\alpha \beta}^{\Lambda_1}(b_i)D_{\rho \sigma}^{\Lambda_1'}((b_i^*)^T) = [E_{\beta \rho}]_{\alpha \sigma} = C^{\Lambda_1}_{\beta \rho} \delta_{\alpha \sigma}\delta^{\Lambda_1 \Lambda_1'},
\end{equation}
where $C^{\Lambda_1}_{\beta \rho}$ is a constant that a priori depends on the choice of elementary matrix.
We now show that it only depends on $\Lambda_1$.

Using the property
\begin{equation}
	D_{\alpha \beta}^{\Lambda_1}(d^T) = D_{\beta \alpha}^{\Lambda_1}(d), \label{eq: step 1}
\end{equation}
we derive
\begin{equation}
	\begin{aligned}
		C^{\Lambda_1}_{\beta \rho} \delta_{\alpha \sigma}\delta^{\Lambda_1 \Lambda_1'} = [E_{\beta \rho}]_{\alpha \sigma} &= \sum_{i} D_{\alpha \beta}^{\Lambda_1}(b_i)D_{\rho \sigma}^{\Lambda_1'}((b_i^*)^T)  \\
		&= \sum_{i} D_{\beta \alpha}^{\Lambda_1}(b_i^T)D_{\sigma \rho}^{\Lambda_1'}(b_i^*) = \sum_{i} D_{\beta \alpha}^{\Lambda_1}(b_i)D_{\sigma \rho}^{\Lambda_1'}((b_i^*)^T) \\
		&= [E_{\alpha \sigma}]_{\beta \rho} = C^{\Lambda_1}_{\alpha \sigma} \delta_{\beta \rho} \delta^{\Lambda_1 \Lambda_1'}.
	\end{aligned}
\end{equation}
Going from the first line to the second uses \eqref{eq: step 1}.
The RHS of the second line follows by summing over $b_i^T$ instead of $b_i$ (transposition maps $\mathcal{B}$ to $\mathcal{B}$ bijectively, this is particularly clear in the diagram basis).
Comparing the LHS of the first line to the RHS of the last line gives
\begin{equation}
	C^{\Lambda_1}_{\beta \rho} = C^{\Lambda_1}\delta_{\beta \rho},
\end{equation}
which proves equation \eqref{eq: almost grand orth theorem}.

The normalization constant $C^{\Lambda_1}$ is important for constructing matrix units. We will prove that
\begin{equation}
	C^{\Lambda_1} = \frac{1}{\Dim V_{\Lambda_1}^{S_N}}. \label{eq: normalization constant}
\end{equation}
The normalization constant is determined by contracting all the indices in equation \eqref{eq: almost grand orth theorem}. Set $\Lambda_1 = \Lambda_1'$, then
\begin{equation}
	(\Dim V_{\Lambda_1}^{P_k(N)})^2C^{\Lambda_1} = \sum_{\alpha}D_{\alpha \alpha}^{\Lambda_1}(\sum_i b_i (b_i^*)^T), \label{eq: normalization constant from trace}
\end{equation}
and all that remains is to compute the element $\sum_i b_i (b_i^*)^T$.

As we will now see,
\begin{equation}
	\sum_{i=1}^{B(2k)} b_i (b_i^*)^T = \Omega^{-1} = \sum_{\Lambda_1} \frac{\DimPk{\Lambda_1}}{\DimSN{\Lambda_1}} p_{\Lambda_1}, \label{eq: sum dd-dual is Q}
\end{equation}
where $\Omega^{-1}$ is the inverse of the element defined in equation \eqref{eq: omega}.
Using the relationship \eqref{eq: traces relation omega factor} between the two traces we have that
\begin{equation}
	\begin{aligned}
		\tr(d\sum_i b_i (b_i^*)^T) &= \Tr_{V_N^{\otimes k}}(\Omega^{-1} d\sum_i b_i (b_i^*)^T) \\
		&=\sum_i \langle \Omega^{-1} db_i,b_i^* \rangle \\
		&=\sum_i\mathrm{Coeff}(b_i, \Omega^{-1}db_i) \\
		&=\tr(\Omega^{-1}d),
	\end{aligned}
\end{equation}
holds for all $d \in P_k(N)$, from which it follows that
\begin{equation}
	\tr(d(\sum_i b_i (b_i^*)^T-\Omega^{-1})) = 0,
\end{equation}
holds for all $d \in P_k(N)$.
Since the trace in the regular representation is non-degenerate we must have
\begin{equation}
	\sum_{i=1}^{B(2k)} b_i (b_i^*)^T - \Omega^{-1} = 0.
\end{equation} 
Inserting this expression into equation \eqref{eq: normalization constant from trace} gives
\begin{equation}
	\begin{aligned}
		(\Dim V_{\Lambda_1}^{P_k(N)})^2C^{\Lambda_1} &= \sum_{\alpha}D_{\alpha \alpha}^{\Lambda_1}(\Omega^{-1}) =\sum_{\alpha}\sum_{\Lambda_1' \vdash N} \frac{\Dim V_{\Lambda_1'}^{P_k(N)}}{\Dim V_{\Lambda_1'}^{S_N}}D_{\alpha \alpha}^{\Lambda_1}(p_{\Lambda_1'}) \\
		&=\sum_{\Lambda_1' \vdash N} \frac{\Dim V_{\Lambda_1'}^{P_k(N)}}{\Dim V_{\Lambda_1'}^{S_N}}\delta^{\Lambda_1 \Lambda_1'} \Dim V_{\Lambda_1}^{P_k(N)} \\
		&=\frac{\Dim V_{\Lambda_1'}^{P_k(N)}}{\Dim V_{\Lambda_1'}^{S_N}}\Dim V_{\Lambda_1'}^{P_k(N)} 
	\end{aligned}
\end{equation}
which gives equation \eqref{eq: normalization constant}.
To summarize, we have proven that
\begin{equation}\boxed{
		\sum_{i=1}^{B(2k)} D_{\alpha \beta}^{\Lambda_1}(b_i)D_{\rho \sigma}^{\Lambda_1'}((b_i^*)^T) = \frac{1}{\Dim V_{\Lambda_1}^{S_N}}\delta_{\beta \rho} \delta_{\alpha \sigma}\delta^{\Lambda_1 \Lambda_1'}. }\label{eq: orthogonality Pk matrix elements}
\end{equation}

\subsection{Matrix units for $P_k(N)$} \label{apx: subsec matrix unit PkN}
In this subsection we want to use orthogonality of matrix elements to show that
\begin{equation}
	\boxed{Q_{\alpha \beta}^{\Lambda_1} = \sum_{i} \Dim(V_{\Lambda_1}^{S_N}) D_{\beta \alpha}^{\Lambda_1}((b_i^*)^T)b_i,} \label{eq: fourier inversion}
\end{equation}
multiply like a generalized matrix algebra.
That is, 
\begin{equation}
	Q_{\alpha \beta}^{\Lambda_1}Q_{\rho \sigma}^{\Lambda_1'} = \delta^{\Lambda_1 \Lambda_1'}\delta_{\beta \rho}Q_{\alpha \sigma}^{{\Lambda_1}}.
\end{equation}
This is straight-forward given the results in the previous subsections.
The proof goes as follows.
By definition we have
\begin{equation}
	\begin{aligned}
		Q_{\alpha \beta}^{\Lambda_1}Q_{\rho \sigma}^{\Lambda_1'} 
		&= \sum_{i,j} \Dim(V_{\Lambda_1}^{S_N})\Dim(V_{\Lambda_1'}^{S_N})D_{\beta \alpha}^{\Lambda_1}((b_i^*)^T)D_{\sigma \rho}^{\Lambda_1'}((b_j^*)^T)b_ib_j \\
		&=\sum_{i,j,k} \Dim(V_{\Lambda_1}^{S_N})\Dim(V_{\Lambda_1'}^{S_N})D_{\beta \alpha}^{\Lambda_1}((b_i^*)^T)D_{\sigma \rho}^{\Lambda_1'}((b_j^*)^T) \langle b_ib_j, b_k^* \rangle b_k,
	\end{aligned}
\end{equation}
where the second equality expands the product $b_i b_j$ in the basis $b_k$ using equation \eqref{eq: inner prod b_i dual b_j}.
Using the modified Frobenius associativity \eqref{eq: frob associ} we have
\begin{equation}
	\begin{aligned}
		&\sum_{i,j,k} \Dim(V_{\Lambda_1}^{S_N})\Dim(V_{\Lambda_1'}^{S_N})D_{\beta \alpha}^{\Lambda_1}((b_i^*)^T)D_{\sigma \rho}^{\Lambda_1'}((b_j^*)^T) \langle b_ib_j, b_k^* \rangle b_k \\
		&=\sum_{i,j,k} \Dim(V_{\Lambda_1}^{S_N})\Dim(V_{\Lambda_1'}^{S_N})D_{\beta \alpha}^{\Lambda_1}((b_i^*)^T)D_{\sigma \rho}^{\Lambda_1'}((b_j^*)^T)\langle b_j, b_i^Tb_k^* \rangle b_k\\
		&=\sum_{i,k} \Dim(V_{\Lambda_1}^{S_N})\Dim(V_{\Lambda_1'}^{S_N})D_{\beta \alpha}^{\Lambda_1}((b_i^*)^T)D_{\sigma \rho}^{\Lambda_1'}(\sum_j(b_j^*)^T\langle b_j, b_i^Tb_k^* \rangle) b_k,
	\end{aligned}
\end{equation}
where in the last line we have pulled the coefficient $\langle b_j, b_i^Tb_k^* \rangle$ inside the matrix representation (using linearity).
This prepares us for the next step, where we use the fact that
\begin{equation}
	\sum_j(b_j^*)^T\langle b_j, b_i^Tb_k^* \rangle = (b_i^Tb_k^*)^T,
\end{equation}
which follows from \eqref{eq: inner prod b_i dual b_j},
\begin{equation}
	\begin{aligned}
		&\sum_{i,k} \Dim(V_{\Lambda_1}^{S_N})\Dim(V_{\Lambda_1'}^{S_N})D_{\beta \alpha}^{\Lambda_1}((b_i^*)^T)D_{\sigma \rho}^{\Lambda_1'}(\sum_j(b_j^*)^T\langle b_j, b_i^Tb_k^* \rangle) b_k \\
		&=\sum_{i,k} \Dim(V_{\Lambda_1}^{S_N})\Dim(V_{\Lambda_1'}^{S_N})D_{\beta \alpha}^{\Lambda_1}((b_i^*)^T)D_{\sigma \rho}^{\Lambda_1'}((b_i^T b_k^*)^T) b_k \\
		&=\sum_{i,k}\sum_\gamma \Dim(V_{\Lambda_1}^{S_N})\Dim(V_{\Lambda_1'}^{S_N})D_{\beta \alpha}^{\Lambda_1}((b_i^*)^T)D_{\sigma \gamma}^{\Lambda_1'}((b_k^*)^T)D_{\gamma \rho}^{\Lambda_1'}(b_i) b_k.
	\end{aligned}
\end{equation}
In the last line we can use orthogonality of matrix elements \eqref{eq: orthogonality Pk matrix elements} to find
\begin{equation}
	\begin{aligned}
		&\sum_{i,k}\sum_\gamma \Dim(V_{\Lambda_1}^{S_N})\Dim(V_{\Lambda_1'}^{S_N})D_{\beta \alpha}^{\Lambda_1}((b_i^*)^T)D_{\sigma \gamma}^{\Lambda_1'}((b_k^*)^T)D_{\gamma \rho}^{\Lambda_1'}(b_i) b_k \\
		&=\sum_{k}\sum_\gamma \Dim(V_{\Lambda_1}^{S_N})\delta^{\Lambda_1 \Lambda_1'}\delta_{\rho \beta}\delta_{\gamma \alpha} D_{\sigma \gamma}^{\Lambda_1'}((b_k^*)^T)b_k \\
		&=\delta^{\Lambda_1 \Lambda_1'}\delta_{\beta \rho}Q_{\alpha \sigma}^{{\Lambda_1}},
	\end{aligned}
\end{equation}
which concludes the proof.

Equipped with a matrix unit basis of $P_k(N)$ we use this to show
\begin{align} \label{eq: d on Q left}
	\boxed{ d Q^{\Lambda_1}_{\alpha \beta} = D^{\Lambda_1}_{\alpha \sigma} (d^T) Q^{\Lambda_1}_{\sigma \beta}
	}
\end{align}
Expanding $Q^{\Lambda_1}_{\alpha \beta}$ on the RHS of this expression as per \eqref{eq: fourier inversion} we find
\begin{align} \nonumber
	d Q^{\Lambda_1}_{\alpha \beta} &= \sum_{i} \Dim(V_{\Lambda_1}^{S_N}) D_{\beta \alpha}^{\Lambda_1} \big( (b_i^*)^T \big) d b_i \\ \nonumber
	&= \sum_{i,k} \Dim(V_{\Lambda_1}^{S_N}) D_{\beta \alpha}^{\Lambda_1} \big( (b_i^*)^T \big) \langle d b_i, b_k \rangle b_k^* \\ \nonumber
	&= \sum_{i,k} \Dim(V_{\Lambda_1}^{S_N}) D_{\beta \alpha}^{\Lambda_1} \big( \langle d b_i, b_k \rangle (b_i^*)^T \big) b_k^* \\ \nonumber
	&= \sum_{i,k} \Dim(V_{\Lambda_1}^{S_N}) D_{\beta \alpha}^{\Lambda_1} \big( \langle b_k^T d , b_i^T \rangle (b_i^*)^T \big) b_k^* \\ \nonumber
	&= \sum_{k} \Dim(V_{\Lambda_1}^{S_N}) D_{\beta \alpha}^{\Lambda_1}(b_k^T d ) b_k^* \\ \nonumber
	&= \sum_{k} \Dim(V_{\Lambda_1}^{S_N}) D_{\beta \sigma}^{\Lambda_1}(b_k^T) D_{\sigma \alpha}^{\Lambda_1}(d) b_k^* \\
	&= D_{\alpha \sigma}^{\Lambda_1}(d^T) Q^{\Lambda_1}_{\sigma \beta}.
\end{align}
In the third line we have used \eqref{eq: frob associ}, and in the fourth line we have used the following property of the bilinear form
\begin{align}
\langle d b_i, b_k \rangle = \Tr_{V_N^{\otimes k}}(d b_i b_k^T) =  \Tr_{V_N^{\otimes k}}( b_i^T (b_k^T d)^T) = \langle b_k^T d , b_i^T \rangle.
\end{align}
For the sake of brevity we omit the analogous proof of the action of $d$ on the RHS,
\begin{align} \label{eq: d on Q right}
	\boxed{ Q^{\Lambda_1}_{\alpha \beta} d = Q^{\Lambda_1}_{\alpha \sigma} D^{\Lambda_1}_{\sigma \beta} (d^T).
	}
\end{align}

\subsection{Matrix units for $SP_k(N)$ and normalization constants}
The matrix units for $SP_k(N)$ are constructed from $Q_{\alpha \beta}^{\Lambda_1}$ using Branching coefficients. 
\begin{equation}
	\boxed{Q^{\Lambda_1}_{\Lambda_2, \mu \nu} = \sum_{\alpha, \beta, p} Q^{\Lambda_1}_{\alpha \beta}B^{P_k(N) \rightarrow \mathbb{C}[S_k]}_{\Lambda_1, \alpha \rightarrow \Lambda_2, p; \mu}B^{P_k(N) \rightarrow \mathbb{C}[S_k]}_{\Lambda_1, \beta \rightarrow \Lambda_2, p; \nu}}. \label{eq: spkn units}
\end{equation}
Branching coefficients are understood as follows.
The partition algebra $P_k(N)$ has a subalgebra (isomorphic to) $\mathbb{C}[S_k]$ (for example, see equation \eqref{eq: CS3 subalgebra diagrams}). For any given irreducible representation $V_{\Lambda_1}^{P_k(N)}$ there exists a basis where the action of $\mathbb{C}[S_k] \subset P_k(N)$ is manifest and irreducible. That is, we consider the decomposition
\begin{equation}
	V_{\Lambda_1}^{P_k(N)} \cong \bigoplus_{\Lambda_2 \vdash k} V_{\Lambda_2}^{\mathbb{C}[S_k]} \otimes V_{\Lambda_1 \Lambda_2}^{P_k(N) \rightarrow \mathbb{C}[S_k]}.
\end{equation}
On the LHS we have a basis
\begin{equation}
	E^{\Lambda_1}_{\alpha}, \quad \alpha \in \{1, \dots \Dim(V_{\Lambda_1}^{P_k(N)})\},
\end{equation}
where the representation of $d \in P_k(N)$ is irreducible,
\begin{equation}
	d(E^{\Lambda_1}_{\alpha}) = \sum_\beta D_{\beta \alpha}^{\Lambda_1}(d)E^{\Lambda_1}_{\beta}. \label{eq: PkN irrep2}
\end{equation}
The RHS has a basis
\begin{equation}
	\begin{aligned}
		E^{\Lambda_1, \mu}_{\Lambda_2, p}, \quad &p \in \{1, \dots, \Dim(V_{\Lambda_1}^{\mathbb{C}[S_k]}) \}, \\
		&\mu \in \{1,\dots, \Dim(V_{\Lambda_1 \Lambda_2}^{P_k(N) \rightarrow \mathbb{C}[S_k]})\}, 
	\end{aligned}
\end{equation}
where $\mu$ is a multiplicity label for $V_{\Lambda_2}^{\mathbb{C}[S_k]}$ in the decomposition.
We demand that the representation of $\tau \in \mathbb{C}[S_k]$ is irreducible in this basis,
\begin{equation}
	\tau(E^{\Lambda_1, \mu}_{\Lambda_2, p}) = \sum_q D_{qp}^{\Lambda_2}(\tau)E^{\Lambda_1, \mu}_{\Lambda_2, q},\label{eq: Sk branching irrep2}
\end{equation}
where $D_{qp}^{\Lambda_2}(\tau)$ is an irreducible representation of $\tau \in \mathbb{C}[S_k]$. The change of basis coefficients are called Branching coefficients
\begin{equation}
	E^{\Lambda_1, \mu}_{\Lambda_2, p} = \sum_{\alpha}B^{P_k(N)\rightarrow \mathbb{C}[S_k]}_{\Lambda_1,\alpha \rightarrow \Lambda_2, p; \mu} E^{\Lambda_1}_{\alpha}
\end{equation}

The matrix unit property
\begin{equation}
	Q^{\Lambda_1}_{\Lambda_2, \mu \nu}Q^{\Lambda_1'}_{\Lambda_2', \mu' \nu'} = \delta^{\Lambda_1 \Lambda_1'}\delta^{\Lambda_2 \Lambda_2'}\delta_{\nu \mu'} Q^{\Lambda_1}_{\Lambda_2,\mu \nu'},
\end{equation}
of the $SP_k(N)$ basis follows from that of the $P_k(N)$ units together with orthogonality of $E^{\Lambda_1, \mu}_{\Lambda_2, p}$,
\begin{equation}
	\begin{aligned}
		Q^{\Lambda_1}_{\Lambda_2, \mu \nu}Q^{\Lambda_1'}_{\Lambda_2', \mu' \nu'} &=  \sum_{\substack{\alpha, \beta, p \\ \alpha', \beta', p'}}B^{P_k(N) \rightarrow \mathbb{C}[S_k]}_{\Lambda_1, \alpha \rightarrow \Lambda_2, p; \mu}B^{P_k(N) \rightarrow \mathbb{C}[S_k]}_{\Lambda_1, \beta \rightarrow \Lambda_2, p; \nu} B^{P_k(N) \rightarrow \mathbb{C}[S_k]}_{\Lambda_1', \alpha' \rightarrow \Lambda_2', p'; \mu'}B^{P_k(N) \rightarrow \mathbb{C}[S_k]}_{\Lambda_1', \beta' \rightarrow \Lambda_2', p'; \nu'} Q^{\Lambda_1}_{\alpha \beta} Q^{\Lambda_1'}_{\alpha' \beta'} \\
		&=\sum_{\substack{\alpha, \beta, p \\ \alpha', \beta', p'}}B^{P_k(N) \rightarrow \mathbb{C}[S_k]}_{\Lambda_1, \alpha \rightarrow \Lambda_2, p; \mu}B^{P_k(N) \rightarrow \mathbb{C}[S_k]}_{\Lambda_1, \beta \rightarrow \Lambda_2, p; \nu} B^{P_k(N) \rightarrow \mathbb{C}[S_k]}_{\Lambda_1', \alpha' \rightarrow \Lambda_2', p'; \mu'}B^{P_k(N) \rightarrow \mathbb{C}[S_k]}_{\Lambda_1', \beta' \rightarrow \Lambda_2', p'; \nu'} \delta^{\Lambda_1 \Lambda_1'} \delta_{\beta \alpha'} Q_{\alpha \beta'}^{\Lambda_1} \\
		&=\sum_{\substack{\alpha, p \\ \beta', p'}}B^{P_k(N) \rightarrow \mathbb{C}[S_k]}_{\Lambda_1, \alpha \rightarrow \Lambda_2, p; \mu}B^{P_k(N) \rightarrow \mathbb{C}[S_k]}_{\Lambda_1', \beta' \rightarrow \Lambda_2', p'; \nu'}\delta^{\Lambda_1 \Lambda_1'}\delta^{\Lambda_2 \Lambda_2'}\delta_{pp'}\delta_{\nu \mu'}Q_{\alpha \beta'}^{\Lambda_1}  \\
		&=\sum_{\substack{\alpha,\beta',p}}B^{P_k(N) \rightarrow \mathbb{C}[S_k]}_{\Lambda_1, \alpha \rightarrow \Lambda_2, p; \mu}B^{P_k(N) \rightarrow \mathbb{C}[S_k]}_{\Lambda_1', \beta' \rightarrow \Lambda_2', p; \nu'}\delta^{\Lambda_1 \Lambda_1'}\delta^{\Lambda_2 \Lambda_2'}\delta_{\nu \mu'}Q_{\alpha \beta'}^{\Lambda_1} \\
		&=\delta^{\Lambda_1 \Lambda_1'}\delta^{\Lambda_2 \Lambda_2'}\delta_{\nu \mu'} Q^{\Lambda_1}_{\Lambda_2,\mu \nu'}.
	\end{aligned} \label{eq: SPkN matrix unit product}
\end{equation}
Going from the first line to the second we used the matrix unit property of $Q^{\Lambda_1}_{\alpha \beta}$. Going from the second line to the third line uses orthogonality
\begin{equation}
	\sum_{\alpha} B^{P_k(N) \rightarrow \mathbb{C}[S_k]}_{\Lambda_1, \alpha \rightarrow \Lambda_2, p; \mu} B^{P_k(N) \rightarrow \mathbb{C}[S_k]}_{\Lambda_1, \alpha \rightarrow \Lambda_2', q; \nu} = \delta^{\Lambda_2 \Lambda_2'} \delta_{pq} \delta_{\mu \nu}.
\end{equation}

Further, we prove equation \eqref{eq: Normalization Constant Two Point Function} for the normalization of the two point function.
The orthogonality of matrix elements implies
\begin{equation}
	\chi^{{\Lambda_1'}}(Q_{\alpha \beta}^{\Lambda_1}) =\sum_i D_{\beta \alpha}^{{\Lambda_1}}((b_i^*)^T)\sum_{\gamma} D_{\gamma \gamma}^{\Lambda_1'}(b_i) =\delta^{\Lambda_1 \Lambda_1'} \delta_{\alpha \beta}.
\end{equation}
We use this fact together with Schur-Weyl duality to compute $\Tr_{V_N^{\otimes k}}(Q^{\Lambda_1}_{\alpha \beta})$
\begin{equation}
	\Tr_{V_N^{\otimes k}}(Q^{\Lambda_1}_{\alpha \beta}) = \sum_{\Lambda_1' \vdash N} \Dim V_{\Lambda_1'}^{S_N}\chi^{\Lambda_1'}(Q^{\Lambda_1}_{\alpha \beta}) = \sum_{\Lambda_1' \vdash N} \Dim V_{\Lambda_1'}^{S_N}\delta_{\alpha \beta} \delta^{\Lambda_1 \Lambda_1'} = \Dim V_{\Lambda_1}^{S_N}\delta_{\alpha \beta}. \label{eq: trace of matrix units}
\end{equation}
Consequently,
\begin{equation}\boxed{
		\begin{aligned}
			\Tr_{V_N^{\otimes k}}(Q^{ \Lambda_1}_{\Lambda_2, \mu \nu}) 
			&= \sum_{\alpha, \beta, p} B^{P_k(N) \rightarrow S_k}_{\Lambda_1, \alpha \rightarrow \Lambda_2, p; \mu}B^{P_k(N) \rightarrow S_k}_{\Lambda_1, \beta \rightarrow \Lambda_2, p; \nu} \Tr_{V_N^{\otimes k}}(Q^{\Lambda_1}_{\alpha \beta}) \\
			&=\sum_{\alpha, \beta, p} B^{P_k(N) \rightarrow S_k}_{\Lambda_1, \alpha \rightarrow \Lambda_2, p; \mu}B^{P_k(N) \rightarrow S_k}_{\Lambda_1, \beta \rightarrow \Lambda_2, p; \nu} \delta_{\alpha \beta}\Dim V_{\Lambda_1}^{S_N} \\
			&=\sum_{p} \delta_{pp}\delta_{\mu \nu}\Dim V_{\Lambda_1}^{S_N} = \delta_{\mu \nu}\Dim V_{\Lambda_1}^{S_N}\Dim V_{\Lambda_2}^{S_k},
	\end{aligned}} \label{eq: normalization constant as dimensions}
\end{equation}
where the last two equalities hold if and only if the branching coefficients are non-zero.

Finally, we check that this construction gives $S_k$ invariant elements:
\begin{equation}\label{Qskinv} 
\boxed{ 	\tau Q^{\Lambda_1}_{\Lambda_2, \mu {\nu}}  \tau^{-1}  = Q^{\Lambda_1}_{\Lambda_2, \mu {\nu}}  \qq{for $\tau \in S_k$.}}
\end{equation}
From the definition \eqref{eq: spkn units} and \eqref{eq: d on Q left} we have
\begin{equation}
	\begin{aligned}
		\tau Q^{\Lambda_1}_{\Lambda_2, \mu \nu} &= \sum_{\alpha, \beta, \gamma, p} D_{\gamma \alpha}^{\Lambda_1}(\tau)Q^{\Lambda_1}_{\gamma \beta}B^{P_k(N) \rightarrow \mathbb{C}[S_k]}_{\Lambda_1, \alpha \rightarrow \Lambda_2, p; \mu}B^{P_k(N) \rightarrow \mathbb{C}[S_k]}_{\Lambda_1, \beta \rightarrow \Lambda_2, p; \nu}.
	\end{aligned} \label{eq: tau on Q step 1}
\end{equation}
We re-write $D_{\gamma \alpha}^{\Lambda_1}(\tau)$ by inserting the completeness relation
\begin{equation}
	\sum_{\Lambda_2', p', \mu'} B^{P_k(N) \rightarrow \mathbb{C}[S_k]}_{\Lambda_1, \gamma \rightarrow \Lambda_2', p'; \mu'} B^{P_k(N) \rightarrow \mathbb{C}[S_k]}_{\Lambda_1, \gamma' \rightarrow \Lambda_2', p'; \mu'} = \delta_{\gamma \gamma'}, \label{eq: branching completeness}
\end{equation}
on both sides. This gives
\begin{equation}
	D_{\gamma \alpha}^{\Lambda_1}(\tau) = \sum_{\Lambda_2',p',p'',\mu'} B^{P_k(N) \rightarrow \mathbb{C}[S_k]}_{\Lambda_1, \gamma \rightarrow \Lambda_2', p'; \mu'} D_{p' p''}^{\Lambda_2'}(\tau)  B^{P_k(N) \rightarrow \mathbb{C}[S_k]}_{\Lambda_1, \alpha \rightarrow \Lambda_2', p''; \mu'}.
\end{equation}
Inserting this into \eqref{eq: tau on Q step 1} gives
\begin{equation}
	\begin{aligned}
		&\sum_{\alpha, \beta, \gamma, p} D_{\gamma \alpha}^{\Lambda_1}(\tau)Q^{\Lambda_1}_{\gamma \beta}B^{P_k(N) \rightarrow \mathbb{C}[S_k]}_{\Lambda_1, \alpha \rightarrow \Lambda_2, p; \mu}B^{P_k(N) \rightarrow \mathbb{C}[S_k]}_{\Lambda_1, \beta \rightarrow \Lambda_2, p; \nu} \\
= 		&\sum_{\alpha, \beta, \gamma, p}  \sum_{\Lambda_2',p',p'',\mu'} B^{P_k(N) \rightarrow \mathbb{C}[S_k]}_{\Lambda_1, \gamma \rightarrow \Lambda_2', p'; \mu'} D_{p' p''}^{\Lambda_2'}(\tau)  B^{P_k(N) \rightarrow \mathbb{C}[S_k]}_{\Lambda_1, \alpha \rightarrow \Lambda_2', p''; \mu'} Q^{\Lambda_1}_{\gamma \beta}B^{P_k(N) \rightarrow \mathbb{C}[S_k]}_{\Lambda_1, \alpha \rightarrow \Lambda_2, p; \mu}B^{P_k(N) \rightarrow \mathbb{C}[S_k]}_{\Lambda_1, \beta \rightarrow \Lambda_2, p; \nu}  \\
	= 	&\sum_{\beta, \gamma, p}  \sum_{\Lambda_2',p',p'',\mu'} B^{P_k(N) \rightarrow \mathbb{C}[S_k]}_{\Lambda_1, \gamma \rightarrow \Lambda_2', p'; \mu'} D_{p' p''}^{\Lambda_2'}(\tau)  \delta_{\Lambda_2 \Lambda_2'}\delta_{p'' p} \delta_{\mu' \mu} Q^{\Lambda_1}_{\gamma \beta}B^{P_k(N) \rightarrow \mathbb{C}[S_k]}_{\Lambda_1, \beta \rightarrow \Lambda_2, p; \nu},
	\end{aligned}	
\end{equation}
where we used completeness in the last line. Eliminating the Kronecker deltas by carrying out the sums gives
\begin{equation}
	\tau Q^{\Lambda_1}_{\Lambda_2, \mu \nu} = \sum_{\beta, \gamma, p,p'} B^{P_k(N) \rightarrow \mathbb{C}[S_k]}_{\Lambda_1, \gamma \rightarrow \Lambda_2, p'; \mu} D_{p' p}^{\Lambda_2}(\tau) Q^{\Lambda_1}_{\gamma \beta}B^{P_k(N) \rightarrow \mathbb{C}[S_k]}_{\Lambda_1, \beta \rightarrow \Lambda_2, p; \nu}.
\end{equation}
To finish the proof we follow the same steps for the right action: using \eqref{eq: d on Q right} gives
\begin{equation}
	Q^{\Lambda_1}_{\Lambda_2, \mu \nu}  \tau = \sum_{\alpha, \beta, \gamma, p} Q^{\Lambda_1}_{\alpha \gamma}D_{\beta \gamma}^{\Lambda_1}(\tau)B^{P_k(N) \rightarrow \mathbb{C}[S_k]}_{\Lambda_1, \alpha \rightarrow \Lambda_2, p; \mu}B^{P_k(N) \rightarrow \mathbb{C}[S_k]}_{\Lambda_1, \beta \rightarrow \Lambda_2, p; \nu}.
\end{equation}
Inserting \eqref{eq: branching completeness} and carrying out the sums yields
\begin{equation}
	Q^{\Lambda_1}_{\Lambda_2, \mu \nu}  \tau  = \sum_{\alpha, \gamma,p,p''}B^{P_k(N) \rightarrow \mathbb{C}[S_k]}_{\Lambda_1, \alpha \rightarrow \Lambda_2, p; \mu} D_{p p''}^{\Lambda_2}(\tau)Q_{\alpha \gamma}^{\Lambda_1}B^{P_k(N) \rightarrow \mathbb{C}[S_k]}_{\Lambda_1, \gamma \rightarrow \Lambda_2, p''; \nu} = \tau Q^{\Lambda_1}_{\Lambda_2, \mu \nu},
\end{equation}
which immediately leads to \eqref{Qskinv}. 

\section{Orbit basis} \label{sec: orbit basis}

In section \ref{sec: Perm subspace} we described two  bases for the partition algebra $P_k(N)$: a diagram basis and a representation basis. Here we describe another basis, in terms of combinatorially explicit linear combination of the diagrams from section 
\ref{sec: diagram basis}. This basis is called the orbit basis \cite{Jones1994}. 
 In this appendix we describe this basis and  show that it is orthogonal for any $N$ and $k$. This makes it a suitable basis to describe permutation invariant matrix quantum mechanics in the $N < 2k$ regime, a preliminary discussion of which concludes this subsection. A possible future direction is to use the orbit basis to describe how the representation basis is modified, in this regime. 

As in the diagram basis, this basis is indexed by the set partitions $\Pi_{2k}$ of $\{1, \dots, k, 1',\dots,k'\}$. These are partially ordered under the relation
\begin{equation}
	\pi \preceq \pi' \qquad \text{if every block of } \pi \text{ is contained within a block of } \pi' , 
\end{equation}
in this case we say that $\pi$ is a refinement of $\pi'$ or equivalently that $\pi'$ is a coarsening of $\pi$. Since we are already familiar with the diagram basis of $P_k(N)$ we express the orbit basis in terms of the diagram basis using the above partial ordering 
\begin{equation} \label{eq: diagram to orbit basis}
	d_{\pi} = \sum_{\pi \preceq \pi'} x_{\pi'}
\end{equation}
with  $\{x_{\pi} | \pi \in \Pi_{2k} \}$. The diagram basis element $d_{\pi}$ is a sum of all orbit basis elements labelled by set partitions equal to or coarser than $\pi$, for example
\begin{equation}
	\PAdiagramcurve{2}{1/-1} = \PAdiagramOrbit{2}{1/-1} +  \PAdiagramOrbit{2}{1/-1, 2/-2} + \PAdiagramOrbit[2/1]{2}{1/-1}  + \PAdiagramOrbit[-1/-2]{2}{1/-1} + \PAdiagramOrbit[2/1]{2}{1/-1, 2/-2}.
\end{equation}
We will continue to distinguish the diagram and orbit bases by drawing diagram basis elements with black vertices and labelling them with the letter $d$, and drawing orbit basis elements with white vertices and labelling them with the letter $x$.
The transition matrix determined by \eqref{eq: diagram to orbit basis} is $\zeta_{2k}$ and is called the zeta matrix of the partially ordered set $\Pi_{2k}$. It is upper triangular, with ones on the diagonal and hence invertible.

The inverse of $\zeta_{2k}$ is given in \cite{HalversonBenk2017}. It is the matrix $\mu_{2k}$ 
\begin{equation} \label{eq: orbit basis expansion}
	x_{\pi} = \sum_{\pi \preceq \pi'} \mu_{2k}(\pi, \pi') d_{\pi'}.
\end{equation}
If $\pi \preceq \pi'$ and $\pi'$ consists of $l$ blocks such that the $i$th block of $\pi'$ is the union of $b_i$ blocks of $\pi$ then
\begin{equation}
	\mu_{2k}(\pi, \pi') = \prod_{i=1}^l (-1)^{b_i -1} (b_i - 1)!
\end{equation}
For example, this gives the following expansion of the orbit basis element labelled by $\pi = \{1|2|3|4\}$
\begin{align} \nonumber \label{eq: o1 in diagram basis}
	\PAdiagramOrbit[]{2}{} = &\PAdiagramcurve{2}{} - \PAdiagramcurve{2}{1/-1} - \PAdiagramcurve{2}{2/-2} - \PAdiagramcurve[2/1]{2}{} - \PAdiagramcurve[-1/-2]{2}{} - \PAdiagramcurve{2}{1/-2} - \PAdiagramcurve{2}{-1/2} + \PAdiagramcurve{2}{1/-1, 2/-2} + \PAdiagramcurve{2}{1/-2, -1/2} + \PAdiagramcurve[2/1, -1/-2]{2}{} \\
	&+ 2 \PAdiagramcurve[2/1]{2}{1/-1} + 2 \PAdiagramcurve[-1/-2]{2}{1/-1} + 2 \PAdiagramcurve[2/1]{2}{2/-2} + 2 \PAdiagramcurve[-1/-2]{2}{2/-2} - 6 \PAdiagramcurve[2/1]{2}{1/-1, 2/-2}.
\end{align}

The orbit basis is orthogonal with respect to the inner product \eqref{eq: inner product is trace}. We will prove,
\begin{equation} \boxed{
	\bra{x_\pi}\ket{x_{\pi'}} =  \begin{cases}
		\abs{G_{\pi}}N_{(\abs{\pi})} \qq{if $[x_{\pi'}] = [x_\pi]$,}\\
		0 \qq{otherwise.}
	\end{cases}
	}
\end{equation}
where $\pi, \pi'$ are set partitions of $\{1,\dots,k,1',\dots,k'\}$, $N_{(l)} = N(N-1) \dots (N-l+1)$ is the falling factorial, $\abs{\pi}$ is the number of blocks in $\pi$, and $ \abs{G_{\pi} }$ is the order of the subgroup of $S_k$ that leaves $x_\pi$ invariant. As was the case in the diagram basis, we note that
\begin{align}
 \ket{[x_{\pi'}]} = \ket{x_{\pi'}}
\end{align}
and use the RHS ket labels for the sake of notational efficiency. 

First consider the simpler proposition
\begin{equation} \boxed{
	\Tr_{V_N^{\otimes k}}(x_{\pi} x_{\pi'}^T) = N_{(\abs{\pi})}\delta_{\pi \pi'}.
	}
\end{equation}
The proof of this follows from the definition (see section 5.2 in \cite{HalversonBenk2017}) of $x_{\pi}$ acting on $V_N^{\otimes k}$
\begin{equation}
	(x_{\pi})^{i_{1'} \dots i_{k'}}_{i_{1} \dots i_{k}} = \begin{cases}
		1 \qq{if $i_a = i_b$ if and only if a and b are in the same block of $\pi$,} \\
		0 \qq{otherwise.}
	\end{cases} \label{eq: orbit basis action}
\end{equation}
The trace is equal to
\begin{equation}
	\Tr_{V_N^{\otimes k}}(x_{\pi} x_{\pi'}^T) = \sum_{\substack{i_1 \dots i_k \\ i_{1'} \dots i_{k'}}} (x_{\pi})^{i_{1'} \dots i_{k'}}_{i_{1} \dots i_{k}} (x_{\pi'})^{i_{1'} \dots i_{k'}}_{i_{1} \dots i_{k}}.
\end{equation}
Equation \eqref{eq: orbit basis action} implies
\begin{equation}
	(x_{\pi})^{i_{1'} \dots i_{k'}}_{i_{1} \dots i_{k}} (x_{\pi'})^{i_{1'} \dots i_{k'}}_{i_{1} \dots i_{k}} = \begin{cases}
		1 \qq{\parbox[t]{8cm}{if $i_a = i_b$ if and only if $a$ and $b$ are in the same block of $\pi$ and the same block of $\pi'$,}} \\
		0 \qq{otherwise.}
	\end{cases}\label{eq: orbit basis matrix elements product}
\end{equation}

If $\pi \neq \pi'$ two situations exist. Consider the set of all pairs $(a,b)$ for $a,b=1,\dots,k,1',\dots,k'$ such that $a$ and $b$ are in the same block of $\pi$. Since $\pi \neq \pi'$ at least one of these pairs are such that $a$ and $b$ are in different blocks of $\pi'$. The second case is the reverse. Consider the set of all $(a,b)$ such that $a$ and $b$ are in the same block of $\pi'$. Then $\pi' \neq \pi$ implies that there exists at least one pair such that $a$ and $b$ are not in the same block of $\pi$.
In that case, there are no choices of $i_a, i_b$ which satisfy the first criteria in \eqref{eq: orbit basis matrix elements product}.
For example, take $a,b$ to be in the same block of $\pi$ but different blocks of $\pi'$. The matrix elements $(x_{\pi})^{i_{1'} \dots i_{k'}}_{i_{1} \dots i_{k}}$ vanish if $i_a \neq i_b$ while the matrix elements $(x_{\pi'})^{i_{1'} \dots i_{k'}}_{i_{1} \dots i_{k}}$ vanish unless $i_a \neq i_b$. Therefore, the product identically vanishes,
\begin{equation}
	(x_{\pi})^{i_{1'} \dots i_{k'}}_{i_{1} \dots i_{k}} (x_{\pi'})^{i_{1'} \dots i_{k'}}_{i_{1} \dots i_{k}} = \delta_{\pi \pi'} (x_{\pi})^{i_{1'} \dots i_{k'}}_{i_{1} \dots i_{k}}
\end{equation}
and
\begin{equation}
	\Tr_{V_N^{\otimes k}}(x_{\pi} x_{\pi'}^T)  = \sum_{\substack{i_1 \dots i_k \\ i_{1'} \dots i_{k'}}} (x_{\pi})^{i_{1'} \dots i_{k'}}_{i_{1} \dots i_{k}}  \delta_{\pi \pi'} = \delta_{\pi \pi'} N(N-1) \dots (N-|\pi|+1).
\end{equation}
The last equality is a consequence of \eqref{eq: orbit basis action}. For example, consider the set partition $12|1'2'$. The trace of $x_{12|1'2'}$ is
\begin{equation}
	\Tr_{V_N^{\otimes 2}}(x_{12|1'2'}) = \sum_{i_1 = i_2 \neq i_3, i_3 = i_4} = N(N-1),
\end{equation}
since we have $N$ choices of indices for $i_1$ and $(N-1)$ choices for $i_3$ (for every choice of $i_1$). The general case is analogous,
\begin{equation}
	\Tr_{V_N^{\otimes k}}(x_{\pi}) = N_{(\abs{\pi})}.
\end{equation}
We have $N$ choices for the indices of the first block of $\pi$, $N-1$ choices for the indices of the second block and so on.

The inner product of two orbit basis elements of $SP_k(N)$ is given by \eqref{eq: inner product is trace}
\begin{equation}
	\bra{x_\pi}\ket{x_{\pi'}} = \sum_{\gamma \in S_k} \Tr_{V_N^{\otimes k}}(\gamma x_{\pi} \gamma^{-1} x_{\pi'}^T).
\end{equation}
We re-write
\begin{equation}
	\sum_{\gamma \in S_k} \gamma x_{\pi} \gamma^{-1}  = \abs{G_{\pi}}\sum_{\lambda \in [\pi]} x_{\lambda},
\end{equation}
where the sum on the RHS is over the distinct elements in the $S_k$ orbit of $x_{\pi}$. Substituting this into the trace gives
\begin{equation}
	\bra{x_\pi}\ket{x_{\pi'}} = \abs{G_{\pi}}\sum_{\lambda \in [\pi]} \Tr_{V_N^{\otimes k}}(x_{\lambda}x_{\pi'}) = \abs{G_{\pi}}\sum_{\lambda \in [\pi]} N_{(\abs{\pi})} \delta_{\lambda \pi'} = \begin{cases}
		\abs{G_{\pi}}N_{(\abs{\pi})} \qq{if $[x_{\pi'}] = [x_\pi]$,} \\
		0 \qq{otherwise,}
	\end{cases}
\end{equation}
where $[x_{\pi}]$ denotes $S_k$ symmetrisation as in equation \eqref{eq: SP_kN basis}.

For the majority of this paper we assume $N \geq 2k$ in order to take advantage of the many simplifications that occur in this limit. However, utilising results from the partition algebra literature we are able to say something about what happens below this limit, in which we expect to encounter finite $N$ effects. 

In the limit $N \geq 2k$ the map from the partition algebra to $\text{End}_{S_N}(V_N^{\otimes k})$ is bijective. When $N < 2k$ this map acquires a non-trivial kernel (but remains surjective). Accordingly, we expect a reduction in the size of the state space $\Hilbertspace$. This reduction is most easily expressed in the orbit basis of $P_k(N)$. Theorem 5.17 (a) in \cite{HalversonBenk2017} states that if $N \in \mathbb{Z}_{\geq 1}$ and $\{ x_{\pi} | \pi \in \Pi_{2k} \}$ is the orbit basis for $P_k(N)$ then for $k \in \mathbb{Z}_{\geq 1}$, the representation $\Phi_{k, N} : P_k(N) \rightarrow \text{End}(V_N^{\otimes k})$ has the following image and kernel
\begin{align} \label{eq: thm 5.17} \nonumber
	\text{im}( \Phi_{k, N} ) &= \text{End}_{S_N}(V_N^{\otimes k}) = \text{span}_{\mathbb{C}} \{\Phi_{k,N} (x_{\pi}) | \pi \in \Pi_{2k} \text{ has $\leq N$ blocks} \}, \\
	\text{ker}( \Phi_{k, N} ) &= \text{span}_{\mathbb{C}} \{x_{\pi} | \pi \in \Pi_{2k} \text{ has $> N$ blocks} \}.
\end{align}

Due to the bosonic symmetry of our theory we are actually interested in the map from the symmetrised partition algebra $SP_k(N)$, defined in \eqref{eq: SP_kN basis}, to $\text{End} (V_N^{\otimes k})$. To this end we note that the definition of the kernel of $\Phi_{k, N}$ given in \eqref{eq: thm 5.17} is $S_k$ invariant. If one element of an $S_k$ orbit is in the kernel then \eqref{eq: thm 5.17} tells us that the entire orbit belongs to the kernel - the action of $S_k$ does not change the number of blocks in a partition $\pi$. The image and kernel of this map are the following
\begin{align} \nonumber
	\text{im}(\Phi_{k, N}) &=  \text{span}_{\mathbb{C}} \Big\{ [b] = \frac{1}{k!} \sum_{\gamma \in S_k} \gamma b \gamma^{-1} \medspace | \medspace b = \Phi_{k,N}(x_{\pi}), \forall \pi \in \Pi_{2k} \text{ with $\leq N$ blocks} \Big\}, \\
	\text{ker}(\Phi_{k, N}) &= \text{span}_{\mathbb{C}} \Big\{ [x_{\pi}] | \pi \in \Pi_{2k}, \pi \text{ has $> N $ blocks} \Big\}.
\end{align}
Therefore a state basis is given by $\ket{[x_{\pi}]}$ for $\pi$ having $N$ or fewer blocks, this basis is orthogonal for all $N$, including for $N < 2k$.

The original statement \eqref{eq: thm 5.17} applies to multi-matrix theories in which observables are constructed from distinct matrices - in this case there is no bosonic $S_k$ symmetry to account for. If a state in this theory is null then all states generated by the action of $S_k$ on this state will also be null. The equivalent of \eqref{eq: diagram state def} for the multi-matrix case is
\begin{align}
	| d \rangle = \sum_{\substack{i_1, \dots, i_k \\ i_{1'}, \dots, i_{k'}}} (d)^{i_{1'} \dots i_{k'}}_{i_1 \dots i_k} (a^{\dagger}_1)^{i_1}_{i_{1'}} \dots (a^{\dagger}_k)^{i_k}_{i_{k'}}\ket{0} = \Tr_{V_N^{\otimes k}}[d(a^{\dagger}_1 \otimes \dots \otimes a^{\dagger}_k)]\ket{0} \label{eq: diagram state def k flavs}
\end{align}
in which we have $k$ distinct oscillators and each element $d$ in the full partition algebra $P_k(N)$ corresponds to a unique state, instead of $S_k$ equivalence classes $[d] \in SP_k(N)$. We illustrate with the following examples that under the map \eqref{eq: diagram state def k flavs} elements $d \in P_k(N)$ that are in the kernel of $\Phi_{k,N}$ label zero vectors in the Hilbert space $\mathcal{H}$. For $k=2$ and $N=1$ we see 
\begin{align} \nonumber
	\ket{\PAdiagramOrbit[]{2}{1/-1,2/-2}} &= \ket{\PAdiagramcurve[]{2}{1/-1,2/-2}} - \ket{\PAdiagramcurve[2/1]{2}{1/-1,2/-2}} \\ \nonumber
	&= \Bigg[ \sum_{i,j} (a^{\dagger}_1)^i_i (a^{\dagger}_2)^j_j - \sum_{i} (a^{\dagger}_1)^i_i (a^{\dagger}_2)^i_i \Bigg] \ket{0} \\ \nonumber
	&= \Bigg[ (a^{\dagger}_1)^1_1 (a^{\dagger}_2)^1_1 -  (a^{\dagger}_1)^1_1 (a^{\dagger}_2)^1_1 \Bigg] \ket{0} \\
	&= 0
\end{align}
in the first line we have used \eqref{eq: orbit basis expansion} to express the orbit basis element in terms of the diagram basis. Similarly, taking $k=2$ and $N=2$ we have
\begin{align} \label{eq: k, N = 2 example} \nonumber
	\ket{\PAdiagramOrbit[]{2}{2/-2}} &= \ket{\PAdiagramcurve[]{2}{2/-2}} - \ket{\PAdiagramcurve{2}{1/-1, 2/-2}} - \ket{\PAdiagramcurve[2/1]{2}{2/-2}} - \ket{\PAdiagramcurve[-1/-2]{2}{2/-2}} + 2 \ket{\PAdiagramcurve[2/1]{2}{1/-1, 2/-2}} \\ \nonumber
	&= \Bigg[ \sum_{i,j,k} (a_1^{\dagger})^i_j (a_2^{\dagger})^k_k - \sum_{i,j} (a_1^{\dagger})^i_i (a_2^{\dagger})^j_j - \sum_{i,j} (a_1^{\dagger})^i_j (a_2^{\dagger})^j_j  - \sum_{i,j} (a_1^{\dagger})^i_j (a_2^{\dagger})^i_i + 2 \sum_{i} (a_1^{\dagger})^i_i (a_2^{\dagger})^i_i \Bigg] \ket{0} \\
	&= 0.
\end{align}
We can split the first term by imposing different restrictions on the ranges of the sum 
\begin{align} \label{eq: range refinement}
	\sum_{i,j,k} = \sum_{i=j=k} + \sum_{\substack{i = j \\ j \neq k}} + \sum_{\substack{i = k \\ k \neq j}} + \sum_{\substack{j = k \\ k \neq i}} + \sum_{i \neq j \neq k}
\end{align}
Similarly, we can split the second, third and fourth terms
\begin{align}
	\sum_{i,j} = \sum_{i = j} + \sum_{i \neq j}.
\end{align}
The terms in \eqref{eq: k, N = 2 example} cancel due to the equivalence of coarsening diagrams and restricting summation ranges - adding edges to a diagram $d \in P_k(N)$ is equivalent to evaluating the original diagram $d$ over a restricted summation range. Another way of saying this is that \eqref{eq: diagram to orbit basis} and \eqref{eq: range refinement} encode identical expansions, in fact the five terms in each expansion give equivalent contributions. Orbit basis elements label states in which the oscillator indices are summed over the restricted range $i_1 \neq i_2 \neq \dots \neq i_m$ where $m$ is the number of blocks in the orbit basis element. From this perspective it is easy to see that these states must be zero when $N < m$ as there are not enough distinct values in $[1,N]$ to satisfy the inequality defining the summation range. Contrastingly, the diagram basis produces states corresponding to sums with unrestricted indices. Although at finite $N$ there is a stark difference between states in the orbit and diagram bases, at large $N$ the two descriptions are equivalent.

Elements of $SP_k(N)$ are $S_k$ orbits on $P_k(N)$ and so states in $\mathcal{H}^{(k)}_{\text{inv}}$ are linear combinations of states in $\mathcal{H}$. If a state in $\mathcal{H}$ is labelled by a partition algebra element in the kernel of $\Phi_{k, N}$, the state in $\mathcal{H}^{(k)}_{\text{inv}}$ generated by the action of $S_k$ on this zero $\mathcal{H}$ state will also be zero. It is clear that if an element $d \in P_k(N)$ produces a zero vector under \eqref{eq: diagram state def k flavs} then the equivalence class $[d] \in SP_k(N)$ containing that element $d \in P_k(N)$ also produces a zero vector under the map to $\mathcal{H}^{(k)}_{\text{inv}}$
\begin{equation}
	| d \rangle = \sum_{\substack{i_1, \dots, i_k \\ i_{1'}, \dots, i_{k'}}} ([d])^{i_{1'} \dots i_{k'}}_{i_1 \dots i_k} (a^{\dagger})^{i_1}_{i_{1'}} \dots (a^{\dagger})^{i_k}_{i_{k'}}\ket{0} = \Tr_{V_N^{\otimes k}}([d](a^\dagger)^{\otimes k})\ket{0}. \label{eq: diagram state def 2}
\end{equation}

We can also check that for suitably low values of $N$ the norm of the orbit basis states vanishes. For $x_{\pi_1} = \PAdiagramOrbit[]{2}{} $ we expect 
\begin{align} \label{eq: N=3 k=2 null state}
	\langle x_{\pi_1} | x_{\pi_1} \rangle \big\rvert_{N < 4} = g_{x_{\pi_1} x_{\pi_1}}\big\rvert_{N < 4} =  0 
\end{align}
Indeed, substituting \eqref{eq: o1 in diagram basis} into this expression gives
\begin{align} \nonumber \label{eq: norm o_1}
	\langle x_{\pi_1} | x_{\pi_1} \rangle &= \langle {d}_{\pi_1} | {d}_{\pi_1} \rangle - \langle {d}_{\pi_1} | {d}_{\pi_2} \rangle - \langle {d}_{\pi_2} | {d}_{\pi_1} \rangle + \langle {d}_{\pi_2} | {d}_{\pi_2} \rangle  + \dots - 12 \hspace{2pt} \langle {d}_{\pi_{14}} | {d}_{\pi_{15}} \rangle + 36 \hspace{2pt} \langle {d}_{\pi_{15}} | {d}_{\pi_{15}} \rangle \\ \nonumber
	&= N (N-1) (N-2) (N-3),
\end{align}
which is zero for $N < 4$.

Similarly, we consider $x_{\pi_2} = \PAdiagramOrbit[2/1]{2}{}$, which we expect to vanish for $N < 3$. This has a diagram basis expansion
\begin{align}
	\PAdiagramOrbit[2/1]{2}{} = \PAdiagramcurve[2/1]{2}{} - \PAdiagramcurve[2/1]{2}{2/-2} - \PAdiagramcurve[2/1]{2}{1/-1} - \PAdiagramcurve[2/1, -1/-2]{2}{} + 2 \PAdiagramcurve[2/1]{2}{1/-1,2/-2}.
\end{align}
The norm of this state is
\begin{align} \nonumber
	\langle x_{\pi_2} | x_{\pi_2} \rangle &= \Big\langle \PAdiagramcurve[2/1]{2}{} \Big| \PAdiagramcurve[2/1]{2}{} \Big\rangle - \Big\langle \PAdiagramcurve[2/1]{2}{} \Big| \PAdiagramcurve[2/1]{2}{2/-2} \Big\rangle - \dots + 4 \Big\langle \PAdiagramcurve[2/1]{2}{1/-1,2/-2} \Big| \PAdiagramcurve[2/1]{2}{1/-1,2/-2} \Big\rangle \\
	&= N(N-1)(N-2),
\end{align}
which does vanish for $N < 3$. For a general orbit basis state $x_{\pi}$ we expect the norm to be some polynomial in $N$ which vanishes for any $N < |\pi|$.

\section{Computing low degree matrix units}\label{apx: matrix units}
In this appendix we find the full set of matrix units for $k=2$ and the subset of multiplicity free matrix units for $k=3$. These results can be reproduced using the accompanying Sage code.
\subsection{Degree two}
As discussed in \ref{sec: degree two basis}, we use the following elements of $SP_2(N)$ to distinguish the full set of labels on matrix units $Q^{{\Lambda_1}}_{\Lambda_2, \mu \nu}$. The irreducible representation $\Lambda_1 \vdash N$ is distinguished by
\begin{equation}
	\SWdual{T_2}^{(2)} = \begin{aligned}
		\frac{(N-2)(N-3)-4}{2}&\PAdiagram[]{2}{1/-1,2/-2} + \PAdiagram[]{2}{1/-1} + \PAdiagram[]{2}{2/-2} + \PAdiagram[2/1, -1/-2]{2}{} + \PAdiagram[]{2}{1/-2,2/-1} + N\PAdiagram[2/1]{2}{1/-1,2/-2} \\
		& - \PAdiagram[2/1]{2}{2/-2} - \PAdiagram[2/1]{2}{2/-2} -\PAdiagram[-1/-2]{2}{1/-1} - \PAdiagram[2/1]{2}{1/-1},
	\end{aligned}
\end{equation}
while $\Lambda_2 \vdash k$ is distinguished by
\begin{equation}
	t_2^{(2)} = \PAdiagram[]{2}{1/-2,2/-1},
\end{equation}
and multiplicity labels $\mu, \nu$ are distinguished by acting with
\begin{equation}
	\SWdual{T}_{2,1}^{(2)} = \PAdiagram[]{2}{-1/1} + \PAdiagram[]{2}{-2/2},
\end{equation}
on the left and right. It will be useful to know that $\SWdual{T}_{2,1}^{(2)}$ is related to
\begin{equation}
	\SWdual{T}_2^{(1)} = \frac{N(N-3)}{2}\PAdiagram[]{1}{-1/1}+\PAdiagram[]{1}{},
\end{equation}
since
\begin{equation}
	\SWdual{T}_2^{(1)} \otimes 1 + 1 \otimes \SWdual{T}_2^{(1)} = \PAdiagram[]{2}{-1/1} + \PAdiagram[]{2}{-2/2} + N(N-3)\PAdiagram[]{2}{-2/2,-1/1}.
\end{equation}

As we will now explain, the eigenvalues of $\SWdual{T}_{2,1}^{(2)}$ uniquely determine the labels $\mu, \nu$ by left and right action respectively.
For fixed $\Lambda_1, \Lambda_2$ the multiplicity labels correspond to basis elements for $V_{\Lambda_1, \Lambda_2}^{P_2(N) \rightarrow \mathbb{C}[S_2]}$, appearing in the decomposition
\begin{equation}
\begin{aligned}
		V_N \otimes V_N \cong &\qty(V_{[N]}^{S_N} \otimes V^{S_2}_{[2]} \otimes V_{[N], [2]}^{P_2(N) \rightarrow \mathbb{C}[S_2]}) \oplus \qty(V_{[N-1,1]}^{S_N} \otimes V^{S_2}_{[2]} \otimes V_{[N-1,1],[2]}^{P_2(N) \rightarrow \mathbb{C}[S_2]}) \oplus \\
		&\qty(V_{[N-1,1]}^{S_N} \otimes V^{S_2}_{[1,1]} \otimes V_{[N-1,1], [1,1]}^{P_2(N) \rightarrow \mathbb{C}[S_2]} )\oplus \qty(V_{[N-2,2]}^{S_N} \otimes V^{S_2}_{[2]} \otimes V_{[N-2,2], [2]}^{P_2(N) \rightarrow \mathbb{C}[S_2]} )\oplus \\
		&\qty(V_{[N-2,1,1]}^{S_N} \otimes V^{S_2}_{[1,1]} \otimes V_{[N-2,1,1], [1,1]}^{P_2(N) \rightarrow \mathbb{C}[S_2]}).
\end{aligned} \label{eq: vnvn apx}
\end{equation}
On the right hand side, $\SWdual{T}_{2,1}^{(2)}$ acts on the vector spaces $V_{\Lambda_1 \Lambda_2}^{P_2(N) \rightarrow \mathbb{C}[S_2]}$ with dimensions
\begin{align} \nonumber
	&\Dim V_{[N],[2]}^{P_2(N) \rightarrow \mathbb{C}[S_2]} = 2, \hspace{4pt} \Dim V_{[N-1,1], [2]}^{P_2(N) \rightarrow \mathbb{C}[S_2]} = 2, \hspace{4pt} \Dim V_{[N-1,1], [1,1]}^{P_2(N) \rightarrow \mathbb{C}[S_2]} = 1\\
	&\Dim V_{[N-2,2], [2]}^{P_2(N) \rightarrow \mathbb{C}[S_2]} = 1, \hspace{4pt} \Dim V_{[N-2,1,1], [1,1]}^{P_2(N) \rightarrow \mathbb{C}[S_2]} = 1.
\end{align}
We will find that $\SWdual{T}_{2,1}^{(2)}$ has precisely as many distinct eigenvalues (in each subspace) as the corresponding dimension.
%We will now show how $\SWdual{T}_{2,1}^{(2)}$ distinguishes the states in these vector spaces

To confirm that this is the case, note that $\SWdual{T}_{2,1}^{(2)}$ acts on $V_N^{\otimes 2}$ as
\begin{equation}
	\SWdual{T}_{2,1}^{(2)}(e_{i_1} \otimes e_{i_2}) = \SWdual{T}_2^{(1)} e_{i_1} \otimes e_{i_2} + e_{i_1} \otimes \SWdual{T}_2^{(1)} e_{i_2}-N(N-3)e_{i_1} \otimes e_{i_2}.
\end{equation}
It follows that the eigenvalues are directly related to the eigenvalues of $\SWdual{T}_2^{(1)}$ defined in \eqref{eq: T2 P1N}. These are known by the decompsition
\begin{equation}
	V_N \cong V_{[N]}^{S_N}  \oplus V_{[N-1,1]}^{S_N},
\end{equation}
where $\SWdual{T}_2^{(1)}$ acts on each summand by a normalized character.
Using this on the left hand side of \eqref{eq: vnvn apx} gives
\begin{equation}
	V_N \otimes V_N \cong \qty(V_{[N]}^{S_N} \otimes V_{[N]}^{S_N}) \oplus \qty(V_{[N]}^{S_N} \otimes V_{[N-1,1]}^{S_N}) \oplus \qty(V_{[N-1,1]}^{S_N} \otimes V_{[N]}^{S_N}) \oplus \qty(V_{[N-1,1]}^{S_N} \otimes V_{[N-1,1]}^{S_N}). \label{eq: decomp VN2}
\end{equation}
Consequently, the three distinct eigenvalues of $\SWdual{T}_{2,1}^{(2)}$ are (one for each summand, but the vectors in the second and third space have the same eigenvalue)
\begin{align}
	2\frac{\chi^{[N]}(T_2)}{\Dim V_{[N]}^{S_N}}-N(N-3) &= N(N-1)-N(N-3) = 2N, \\
	2\frac{\chi^{[N-1,1]}(T_2)}{\Dim V_{[N-1,1]}^{S_N}}--N(N-3) &= N(N-3)-N(N-3) = 0, \\
	\frac{\chi^{[N]}(T_2)}{\Dim V_{[N]}^{S_N}} + \frac{\chi^{[N-1,1]}(T_2)}{\Dim V_{[N-1,1]}^{S_N}}-N(N-3) &=
	\frac{1}{2}N(N-1) + \frac{1}{2}N(N-3) -N(N-3)
	= N.\label{eq: eigenvals of T21}
\end{align}

By decomposing \eqref{eq: decomp VN2} into $S_N \times S_k$ representations we will see that the multiplicities in \eqref{eq: vnvn apx} are uniquely associated with one of the above eigenvalues. We start by considering the multiplicities of $V_{[N]}^{S_N} \otimes V_{[2]}^{S_2}$. The representation $V_{[N]}^{S_N}$ occurs in the decomposition \eqref{eq: decomp VN2} as subspaces
\begin{equation}
	V_{[N]}^{S_N} \cong V_{[N]}^{S_N} \otimes V_{[N]}^{S_N} \qq{and} V_{[N]}^{S_N} \subset V_{[N-1,1]}^{S_N} \otimes V_{[N-1,1]}^{S_N}.
\end{equation}
The first subspace has eigenvalue $2N$, while the second subspace has eigenvalue $0$ with respect to $\SWdual{T}_{2,1}^{(2)}$. Therefore, the two multiplicities are distinguished. Next we consider multiple occurances of $V_{[N-1,1]}^{S_N}$. The two spaces
\begin{equation}
	\qty(V_{[N]}^{S_N} \otimes V_{[N-1,1]}^{S_N}) \oplus \qty(V_{[N-1,1]}^{S_N} \otimes V_{[N]}^{S_N})
\end{equation}
combine into representations of $S_N \times S_2$ as
\begin{equation}
	\qty(V_{[N-1,1]}^{S_N} \otimes V^{S_2}_{[2]}) \oplus \qty(V_{[N-1,1]}^{S_N} \otimes V^{S_2}_{[1,1]}).
\end{equation}
Both of these spaces have eigenvalue $N$ with respect to $\SWdual{T}_{2,1}^{(2)}$, but they are distinguished by their $S_2$ representation (or equivalently eigenvalue of $t_2^{(2)}$).
The symmetric part of $V_{[N-1,1]}^{S_N} \otimes V_{[N-1,1]}^{S_N}$ has a subspace
\begin{equation}
	V_{[N-1,1]}^{S_N} \otimes S_{[2]}^{S_2} \subset V_{[N-1,1]}^{S_N} \otimes V_{[N-1,1]}^{S_N},
\end{equation}
with eigenvalue $0$. We have found that the two subspaces $V_{[N-1,1]}^{S_N} \otimes V^{S_2}_{[2]}$ are distinguished by the eigenvalues $N$ and $0$ with respect to $\SWdual{T}_{2,1}^{(2)}$. The last two terms in \eqref{eq: vnvn apx} are multiplicity free and uniquely determined by their eigenvalue with respect to $T_2^{(2)}$,

In the Sage code, we simultaneously diagonalized all the operators by considering a linear combination
\begin{equation}
	T = a\SWdual{T_{2}}^{(2)} + bt_2^{(2)} + c \SWdual{T}_{2,1}^{(2), L} +  f\SWdual{T}_{2,1}^{(2), R},
\end{equation}
with $a,b,c,f \in \mathbb{R}$ such that there is no eigenvalue degeneracy in $T$. The superscript $L$ means left action and $R$ means right action.
An eigenbasis for $T$ will be a simultaneous eigenbasis for $\{\SWdual{T_{2}}^{(2)}, t_2^{(2)}, \SWdual{T}_{2,1}^{(2),L}, \SWdual{T}_{2,1}^{(2), R}\}$, which corresponds to a basis of matrix units.
In the implementation, these operators act on $P_2(N)$, as opposed to $SP_2(N)$.
The projection to $SP_2(N)$ was implemented by adding a fifth operator $P^{SP_2(N)}$ to $T$.
The action of $P^{SP_2(N)}$ on $d\in P_{2}(N)$ is $d \mapsto [d]$.
It commutes with all of the previous operators. This was useful in practice, since elements in $SP_2(N)$ will have eigenvalue 1 with respect to $P^{SP_2(N)}$ (the orthogonal complement has eigenvalue 0).

The matrix units for $k=2$ are given below. The multiplicity labels have been chosen to correspond to eigenvalues of $\SWdual{T}_{2,1}^{(2),L}$ and $\SWdual{T}_{2,1}^{(2),R}$ as follows
\begin{equation}
	\begin{aligned}
		&1 \leftrightarrow 2N, \\
		&2 \leftrightarrow 0, \\
		&3 \leftrightarrow N.
	\end{aligned}
\end{equation}
The elements below have not gone through the final step of being normalized.
\begin{align}
	&(Q^{[N]}_{[2]})_{11} = \PAdiagram[]{2}{}, \\
	&(Q^{[N]}_{[2]})_{21} = -\frac{1}{N} \PAdiagram[]{2}{} + \PAdiagram[2/1]{2}{}, \\
	&(Q^{[N-1,1]}_{[2]})_{33}=-\frac{4}{N} \PAdiagram[]{2}{} + \PAdiagram[]{2}{1/-1} +  \PAdiagram[]{2}{2/-1} +  \PAdiagram[]{2}{1/-2} +  \PAdiagram[]{2}{2/-2}, \\
	&(Q^{[N-1,1]}_{[1,1]})_{33} = -\PAdiagram[]{2}{1/-1} +  \PAdiagram[]{2}{2/-1} +  \PAdiagram[]{2}{1/-2} - \PAdiagram[]{2}{2/-2}, \\
	&(Q^{[N-1,1]}_{[2]})_{23} = \frac{4}{N^{2}} \PAdiagram[]{2}{} - \frac{2}{N} \PAdiagram[2/1]{2}{} - \frac{1}{N} \PAdiagram[]{2}{1/-1} + \PAdiagram[2/1]{2}{1/-1} - \frac{1}{N} \PAdiagram[]{2}{2/-1} - \frac{1}{N} \PAdiagram[]{2}{1/-2} + \PAdiagram[2/1]{2}{2/-2} - \frac{1}{N} \PAdiagram[]{2}{2/-2} , \\
	&(Q^{[N]}_{[2]})_{12} = -\frac{1}{N} \PAdiagram[]{2}{} + \PAdiagram[]{2}{-1/-2} , \\
	&(Q^{[N-1,1]}_{[2]})_{32} = \frac{4}{N^{2}} \PAdiagram[]{2}{} - \frac{2}{N} \PAdiagram[-1/-2]{2}{} - \frac{1}{N} \PAdiagram[]{2}{1/-1} + \PAdiagram[-1/-2]{2}{1/-1} - \frac{1}{N} \PAdiagram[]{2}{2/-1} - \frac{1}{N} \PAdiagram[]{2}{1/-2} + \PAdiagram[-1/-2]{2}{2/-2} - \frac{1}{N} \PAdiagram[]{2}{2/-2} , \\
	&(Q^{[N]}_{[2]})_{22} = \frac{1}{N^{2}} \PAdiagram[]{2}{} - \frac{1}{N} \PAdiagram[2/1]{2}{} - \frac{1}{N} \PAdiagram[-1/-2]{2}{} + \PAdiagram[2/1, -1/-2]{2}{} , \\
	&(Q^{[N-1,1]}_{[2]})_{22} = \begin{aligned}[t]
		&-\frac{4}{N^{3}} \PAdiagram[]{2}{} + \frac{2}{N^{2}} \PAdiagram[2/1]{2}{} + \frac{1}{N^{2}} \PAdiagram[]{2}{1/-1} - \frac{1}{N} \PAdiagram[2/1]{2}{1/-1} + \frac{1}{N^{2}}\PAdiagram[]{2}{2/-1} + \frac{2}{N^{2}} \PAdiagram[-1/-2]{2}{} \\
		&- \frac{1}{N} \PAdiagram[2/1, -1/-2]{2}{} - \frac{1}{N} \PAdiagram[-1/-2]{2}{1/-1} + \PAdiagram[2/1, -1/-2]{2}{1/-1} - \frac{1}{N} \PAdiagram[-1/-2]{2}{2/-2} + \frac{1}{N^{2}} \PAdiagram[]{2}{1/-2} - \frac{1}{N} \PAdiagram[2/1]{2}{2/-2} + \frac{1}{N^{2}} \PAdiagram[]{2}{2/-2},
	\end{aligned} \\
	&(Q^{[N-2,2]}_{[2]})_{22} = \begin{aligned}[t]
		&-\left(\frac{1}{N^{2} - N}\right) \PAdiagram[]{2}{} + \left(\frac{1}{N^{2} - N}\right) \PAdiagram[2/1]{2}{} + \frac{1}{2N}\PAdiagram[]{2}{1/-1} - \frac{1}{N} \PAdiagram[2/1]{2}{1/-1} \\
		&+ \frac{1}{2N} \PAdiagram[]{2}{2/-1} + \left(\frac{1}{N^{2} - N}\right) \PAdiagram[-1/-2]{2}{} - \left(\frac{1}{N^{2} - N}\right) \PAdiagram[2/1, -1/-2]{2}{} - \frac{1}{N} \PAdiagram[]{2}{1/-1, 1/-2} \\
		&+ \PAdiagram[2/1, -1/-2]{2}{1/-1} - \frac{1}{N} \PAdiagram[-1/-2]{2}{2/-2} + \frac{1}{2N} \PAdiagram[]{2}{1/-2} + \left(\frac{-\frac{1}{2} N + 1}{N}\right) \PAdiagram[]{2}{1/-2, 2/-1} \\
		&- \frac{1}{N} \PAdiagram[2/1]{2}{2/-2} + \frac{1}{2N} \PAdiagram[]{2}{2/-2} + \left(\frac{-\frac{1}{2} N + 1}{N}\right)\PAdiagram[]{2}{1/-1, 2/-2},
	\end{aligned} \\
	&(Q^{[N-2,1,1]}_{[1,1]})_{22} = \frac{1}{N} \PAdiagram[]{2}{1/-1} - \frac{1}{N} \PAdiagram[]{2}{2/-1} - \frac{1}{N} \PAdiagram[]{2}{1/-2} + \PAdiagram[]{2}{1/-2, 2/-1} + \frac{1}{N} \PAdiagram[]{2}{2/-2} - \PAdiagram[]{2}{1/-1, 2/-2}. \label{eq: k=2 matrix unit example}
\end{align}
For example,
\begin{equation}
	(Q^{[N]}_{[2]})_{11}(Q^{[N]}_{[2]})_{11} = N^2 (Q^{[N]}_{[2]})_{11},
\end{equation}
and the properly normalized matrix unit is given by
\begin{equation}
	\frac{(Q^{[N]}_{[2]})_{11}}{N^2}.
\end{equation}

\subsection{Degree three}
For degree $k=3$, we find the multiplicity free matrix units $Q^{\Lambda_1}_{\Lambda_2}$, where $\Lambda_1 = [N-3,3], [N-3,2,1], [N-3,1,1,1]$. It is sufficient to use
\begin{align} \nonumber
	\frac{1}{3!}\SWdual{T}_2^{(3)} = &\SPAdiagram[2/1, 3/2, -1/-2, -2/-3]{3}{} + \SPAdiagram[2/1, -1/-2, -2/-3]{3}{1/-1} - N \SPAdiagram[2/1, 3/2, -1/-2, -2/-3]{3}{1/-1} - \SPAdiagram[2/1, -1/-2, -2/-3]{3}{3/-3} - \SPAdiagram[2/1, -2/-3]{3}{3/-1, 2/-3} + \SPAdiagram[2/1, 3/2, -2/-3]{3}{3/-3} \\ \nonumber
	- &\SPAdiagram[-1/-3]{3}{1/-1,2/-2} + \SPAdiagram[2/1, -1/-3]{3}{2/-2, 3/-3} + \SPAdiagram[]{3}{1/-3, 2/-2, 3/-1} + \SPAdiagram[2/1, -1/-2]{3}{2/-3, 3/-2} - \SPAdiagram[2/1, 3/2, -1/-2]{3}{3/-3} + \SPAdiagram[2/1, -1/-2]{3}{3/-3} \\
	+ &(N-1) \SPAdiagram[2/1, -1/-2]{3}{1/-1, 3/-3} - \SPAdiagram[2/1]{3}{2/-2, 3/-3} + \SPAdiagram[]{3}{2/-2, 3/-3} + \frac{(N-1)(N-6)}{2} \SPAdiagram[]{3}{1/-1, 2/-2, 3/-3},
\end{align}
which distinguishes $\Lambda_1$. The square brackets denote $S_3$ symmetrization as in \eqref{eq: SP_kN basis}.

The multiplicity free matrix units for $k=3$ are given below, where $(N)_m$ is the falling factorial $N(N-1) \dots (N-m+1)$,
%\begingroup
%\allowdisplaybreaks
\begin{align} \nonumber
		&Q_{[3]}^{[n-3,3]} = \\ \nonumber
		& -\frac{1}{(N)_{5}} \SPAdiagram[]{3}{} + \frac{1}{(N)_{5}} \SPAdiagram[2/1]{3}{} - \frac{2}{(N)_{5}} \SPAdiagram[2/1, 3/2]{3}{} + \frac{1}{(N)_{5}} \SPAdiagram[2/1, -1/-2]{3}{} \\ \nonumber
		&+ \frac{(N+1)(N-2)}{3(N)_{5}} \SPAdiagram[2/1, -1/-2]{3}{1/-1} + \frac{2(N-2)}{3(N)_{5}} \SPAdiagram[2/1, -1/-2]{3}{3/-2} - \frac{2(N-2)}{3(N)_{5}} \SPAdiagram[2/1]{3}{2/-2}  \\ \nonumber
		& - \frac{(N-2)}{3(N)_{5}} \SPAdiagram[2/1]{3}{3/-2} + \frac{1}{(N)_{5}} \SPAdiagram[-2/-3]{3}{} - \frac{1}{(N)_{5}} \SPAdiagram[2/1, -2/-3]{3}{} + \frac{2}{(N)_{5}} \SPAdiagram[3/2, 2/1, -2/-3]{3}{} \\ \nonumber
		&- \frac{2}{(N)_{5}} \SPAdiagram[-1/-2, -2/-3]{3}{} + \frac{2}{(N)_{5}} \SPAdiagram[-1/-2, -2/-3, 2/1]{3}{} - \frac{4}{(N)_{5}} \SPAdiagram[3/2, 2/1, -1/-2, -2/-3]{3}{} +\frac{2(N-2)}{(N)_{5}} \SPAdiagram[-1/-2, -2/-3]{3}{1/-1} \\ \nonumber
		&- \frac{2(N-1)(N-2)}{(N)_{5}} \SPAdiagram[-1/-2, -2/-3, 2/1]{3}{1/-1}  + \frac{2N(N-1)(N-2)}{(N)_{5}} \SPAdiagram[3/2, 2/1, -1/-2, -2/-3]{3}{1/-1}  - \frac{2(N-2)}{(N)_{5}}\SPAdiagram[2/1, -1/-2, -2/-3]{3}{3/-3}  \\ \nonumber
		&- \frac{2(N-2)}{3(N)_{5}} \SPAdiagram[-2/-3]{3}{1/-2}  + \frac{(N-2)(N-3)}{3(N)_{5}} \SPAdiagram[-2/-3]{3}{1/-2, 2/-1} + \frac{(N+1)(N-2)}{3(N)_{5}} \SPAdiagram[2/1, -2/-3]{3}{1/-2}  \\ \nonumber
		&- \frac{(N-2)^2(N-3)}{3(N)_{5}} \SPAdiagram[2/1, -2/-3]{3}{1/-2, 3/-1} - \frac{2(N-1)(N-2)}{(N)_{5}} \SPAdiagram[3/2, 2/1, -2/-3]{3}{3/-3} + \frac{2(N-2)}{3(N)_{5}}\SPAdiagram[2/1, -2/-3]{3}{3/-3} \\ \nonumber
		&+ \frac{2(N-2)}{3(N)_{5}} \SPAdiagram[-1/-3, 2/1]{3}{2/-2} + \frac{(N-2)}{3(N)_{5}} \SPAdiagram[2/1, -1/-3]{3}{3/-2} - \frac{2(N-2)}{3(N)_{5}} \SPAdiagram[-1/-3]{3}{1/-1} \\ \nonumber
		&+ \frac{(N-2)(N-3)}{3(N)_{5}} \SPAdiagram[-1/-3]{3}{1/-1, 2/-2} - \frac{2(N-2)(N-3)}{3(N)_{5}} \SPAdiagram[2/1, -1/-3]{3}{2/-2, 3/-3} + \frac{(N-2)}{3(N)_{5}} \SPAdiagram[]{3}{1/-3} \\ \nonumber
		&- \frac{(N-2)(N-3)}{6(N)_{5}} \SPAdiagram[]{3}{1/-3, 3/-1} - \frac{(N-2)}{3(N)_{5}} \SPAdiagram[-1/-2]{3}{1/-3} + \frac{(N-2)(N-3)}{3(N)_{5}} \SPAdiagram[-1/-2]{3}{1/-3, 2/-2}  \\ \nonumber
		&- \frac{(N-2)(N-3)}{6(N)_{5}} \SPAdiagram[]{3}{1/-3, 2/-2} + \frac{(N-2)(N-3)(N-4)}{6(N)_{5}} \SPAdiagram[]{3}{1/-3, 2/-2, 3/-1}  -\frac{(N-2)(N-3)}{6(N)_{5}} \SPAdiagram[]{3}{1/-3, 3/-2}  \\ \nonumber
		&+ \frac{(N-2)(N-3)(N-4)}{6(N)_{5}} \SPAdiagram[]{3}{1/-3, 2/-1, 3/-2} - \frac{2(N-2)}{3(N)_{5}} \SPAdiagram[2/1]{3}{2/-3} + \frac{2(N-2)}{3(N)_{5}} \SPAdiagram[2/1, -1/-2]{3}{2/-3}  \\ \nonumber
		&- \frac{2(N-2)(N-3)}{3(N)_{5}} \SPAdiagram[2/1, -1/-2]{3}{2/-3, 3/-2}  + \frac{(N-2)(N-3)}{3(N)_{5}} \SPAdiagram[2/1]{3}{2/-3, 3/-2} + \frac{2(N-2)}{(N)_{5}} \SPAdiagram[3/2, 2/1]{3}{3/-3}  \\ \nonumber
		&-\frac{2(N-2)}{(N)_{5}} \SPAdiagram[3/2, 2/1, -1/-2]{3}{3/-3} + \frac{(N-2)}{3(N)_{5}} \SPAdiagram[]{3}{3/-3} - \frac{(N-2)}{3(N)_{5}} \SPAdiagram[2/1]{3}{3/-3} \\ \nonumber
		&+ \frac{(N-2)}{3(N)_{5}} \SPAdiagram[-1/-2]{3}{3/-3} + \frac{(N-2)}{3(N)_{5}} \SPAdiagram[2/1, -1/-2]{3}{3/-3} - \frac{(N-2)^2(N-3)}{3(N)_{5}}\SPAdiagram[2/1, -1/-2]{3}{1/-1, 3/-3} \\
		&+ \frac{(N-2)(N-3)}{3(N)_{5}} \SPAdiagram[2/1]{3}{2/-2, 3/-3} - \frac{(N-2)(N-3)}{6(N)_{5}} \SPAdiagram[]{3}{2/-2, 3/-3} 
		+ \frac{(N-2)(N-3)(N-4)}{6(N)_{5}} \SPAdiagram[]{3}{1/-1, 2/-2, 3/-3},
	\end{align}
%\endgroup

\begin{equation}
	\begin{aligned}
		Q_{[2,1]}^{[n-3,2,1]} &= \frac{2(N-2)}{N_{(5)}} \SPAdiagram[2/1]{3}{1/-1, 2/-2} -\frac{2}{N_{(5)}} \SPAdiagram[2/1, -1/-2]{3}{3/-2} - \frac{1}{N_{(5)}} \SPAdiagram[2/1]{3}{2/-2} \\
		&-\frac{(N-3)}{N_{(5)}} \SPAdiagram[2/1]{3}{2/-2, 3/-1} + \frac{1}{N_{(5)}} \SPAdiagram[2/1]{3}{3/-2} + \frac{2}{N_{(5)}} \SPAdiagram[-2/-3]{3}{1/-2} \\
		&- \frac{N}{N_{(5)}} \SPAdiagram[-2/-3]{3}{2/-1, 1/-2} - \frac{(N-2)}{N_{(5)}} \SPAdiagram[2/1, -2/-3]{3}{1/-2} + \frac{N(N-2)}{N_{(5)}} \SPAdiagram[2/1, -2/-3]{3}{3/-1, 1/-2} \\
		&+ \frac{1}{N_{(5)}} \SPAdiagram[2/1, -2/-3]{3}{3/-3} + \frac{1}{N_{(5)}} \SPAdiagram[2/1, -1/-3]{3}{2/-2} -\frac{1}{N_{(5)}} \SPAdiagram[2/1, -1/-3]{3}{3/-2} \\
		&-\frac{1}{N_{(5)}} \SPAdiagram[-1/-3]{3}{1/-1} + \frac{2N-3}{N_{(5)}} \SPAdiagram[-1/-3]{3}{1/-1, 2/-2} - \frac{N}{N_{(5)}} \SPAdiagram[2/1, -1/-3]{3}{2/-2, 3/-3} \\
		&-\frac{1}{N_{(5)}} \SPAdiagram[]{3}{1/-3} + \frac{2}{N_{(5)}} \SPAdiagram[]{3}{1/-3, 3/-1} + \frac{1}{N_{(5)}} \SPAdiagram[-1/-2]{3}{1/-3} \\
		&-\frac{(N-3)}{N_{(5)}} \SPAdiagram[-1/-2]{3}{2/-2, 1/-3} -\frac{1}{N_{(5)}} \SPAdiagram[]{3}{1/-3, 2/-2} + \frac{(N-2)}{N_{(5)}} \SPAdiagram[]{3}{1/-3, 3/-2} \\
		&-\frac{(N-1)(N-3)}{N_{(5)}} \SPAdiagram[]{3}{1/-3, 2/-1, 3/-2} + \frac{2}{N_{(5)}} \SPAdiagram[2/1]{3}{2/-3} - \frac{2}{N_{(5)}} \SPAdiagram[2/1, -1/-2]{3}{2/-3} \\
		&+ \frac{2N}{N_{(5)}} \SPAdiagram[2/1, -1/-2]{3}{2/-3, 3/-2} - \frac{N}{N_{(5)}} \SPAdiagram[3/2]{3}{2/-3, 3/-2} + \frac{2}{N_{(5)}} \SPAdiagram[]{3}{3/-3} \\
		&-\frac{2}{N_{(5)}} \SPAdiagram[2/1]{3}{3/-3} - \frac{2}{N_{(5)}} \SPAdiagram[-1/-2]{3}{3/-3} + \frac{2}{N_{(5)}} \SPAdiagram[2/1, -1/-2]{3}{3/-3} \\
		&-\frac{2N(N-2)}{N_{(5)}} \SPAdiagram[2/1, -1/-2]{3}{1/-1, 3/-3} + \frac{2N-3}{N_{(5)}} \SPAdiagram[2/1]{3}{2/-2, 3/-3} - \frac{2(N-2)}{N_{(5)}} \SPAdiagram[]{3}{2/-2, 3/-3} \\
		&+ \frac{2(N-1)(N-3)}{N_{(5)}} \SPAdiagram[]{3}{1/-1, 2/-2, 3/-3},
	\end{aligned}
\end{equation}

\begin{equation}
	\begin{aligned}
		Q_{[1,1,1]}^{[n-3,1,1,1]} &= -\frac{1}{N_{(5)}} \SPAdiagram[]{3}{1/-3, 3/-1} -\frac{1}{N_{(5)}} \SPAdiagram[]{3}{1/-3, 2/-2} + \frac{N}{N_{(5)}} \SPAdiagram[]{3}{1/-3, 2/-2, 3/-1} +\frac{1}{N_{(5)}} \SPAdiagram[]{3}{1/-3, 3/-2} \\
		&-\frac{N}{N_{(5)}} \SPAdiagram[]{3}{1/-3, 2/-1, 3/-2} +\frac{1}{N_{(5)}}\SPAdiagram[]{3}{2/-2, 3/-3} -\frac{N}{N_{(5)}}\SPAdiagram[]{3}{1/-1, 2/-2, 3/-3}.
	\end{aligned}
\end{equation}

\section{The metric and its inverse} \label{apx: Inverse Metric}
We would like to be more explicit about the form of the metric on $P_k(N)$ as well as its inverse, defined by our inner product on observables
\begin{equation}
	\expval{\normord{\mathcal{O}_{d_1}}\normord{\mathcal{O}_{d_2}}} = \sum_{\gamma \in S_k} \Tr_{V_N^{\otimes k}}(d_1 \gamma d_2^T \gamma^{-1}). \label{eq: Two Point Function of Partition Algebra Observables}
\end{equation} 
We note that similar results hold for the metric on $SP_k(N)$. First of all we write an explicit form for the metric. It was shown in \cite{PIMO_Factor} that in the large $N$ limit the inner product on normalised PIMOs
\begin{equation}
	\mathcal{\hat{O}}_{d} = \frac{\mathcal{O}_{d}}{\sqrt{\expval{\normord{\mathcal{O}_{d}}\normord{\mathcal{O}_{d}}}_{\con}}},
\end{equation}
factorises, and so the metric is given by a delta function at leading order
\begin{equation}
	\hat{g}_{d_1 d_2} = 
	\expval{\normord{\mathcal{\hat{O}}_{d_1}}\normord{\mathcal{\hat{O}}_{d_2}}}_{\con} = \begin{cases}
		1 + O(1/\sqrt{N}) \qq{if $[d_1] = [d_2]$,} \\
		0 + O(1/\sqrt{N}) \qq{if $[d_1] \neq [d_2]$.}
	\end{cases}
\end{equation}
Furthermore, it was shown that the $\frac{1}{\sqrt{N}}$ corrections are given by the inclusion of a second term
\begin{equation}
	\hat{g} = \mathds{1} + \sum_{d_1 \neq d_2} N^{ c(d_1 \merge d_2) - \frac{1}{2} \qty(c(d_1) + c(d_2))} E_{d_1 d_2},
\end{equation}
with $E_{d_1 d_2}$ the matrix consisting of a 1 in the $(d_1, d_2)$ position and zeros elsewhere. Setting
\begin{equation}
	X = \sum_{d_1 \neq d_2} N^{ c(d_1 \merge d_2) - \frac{1}{2} \qty(c(d_1) + c(d_2))} E_{d_1 d_2}
\end{equation} 
we have
\begin{equation} \label{eq: inverse metric}
	\hat{g}^{-1} = ( \mathds{1} +X)^{-1} = \mathds{1} - X + X^2 - X^3 + \dots.
\end{equation}

We now calculate the inverse metric for $P_1(N)$ explicitly. $P_1(N)$ is spanned by just two diagrams
\begin{align}
	P_1(N) = \text{Span} \Big\{ \PAdiagram{1}{} \hspace{4pt}, \hspace{4pt} \PAdiagram{1}{1/-1}  \Big\}.
\end{align}
Using our expression for off diagonal elements of the metric
\begin{equation}
	X_{P_1(N)} = \sum_{d_1 \neq d_2} N^{c(d_1 \merge d_2) - \frac{1}{2} \qty( c(d_1) + c(d_2))} E_{d_1 d_2}
\end{equation} 
we find the one independent element
\begin{equation}
	N^{ c\qty(\PAdiagram{1}{} \merge \PAdiagram{1}{1/-1}) - \frac{1}{2} \qty(c\qty(\PAdiagram{1}{}) + c\qty(\PAdiagram{1}{1/-1}))} = N^{\qty(1 - 1 - \frac{1}{2} )} = N^{-\frac{1}{2}}
\end{equation}
and therefore
\begin{align}
	\hat{g}_{P_1(N)} = \mathds{1} + X_{P_1(N)} = 
	\begin{bmatrix}
		1 & 0 \\ 
		0 & 1 
	\end{bmatrix}
	+
	\begin{bmatrix}
		0 & N^{-\frac{1}{2}} \\ 
		N^{-\frac{1}{2}} & 0 
	\end{bmatrix}.
\end{align}
Substituting this into our equation for the inverse metric \eqref{eq: inverse metric} we find
The inverse is given by
\begin{align} \nonumber
	\hat{g}^{-1}_{P_1(N)} &= 
	\begin{bmatrix} 1 & 0 \\ 0 & 1 \end{bmatrix} - N^{-\frac{1}{2}} \begin{bmatrix} 0 & 1 \\  1 & 0 \end{bmatrix} 
	+ N^{-1} \begin{bmatrix} 1 & 0 \\ 0 & 1 \end{bmatrix} - N^{-\frac{3}{2}} \begin{bmatrix} 0 & 1 \\  1 & 0 \end{bmatrix} + \dots \\ \nonumber
	&= \sum_{i = 0}^{\infty} \qty( N^{-i} \begin{bmatrix} 1 & 0 \\ 0 & 1 \end{bmatrix} - N^{-(i+\frac{1}{2})} \begin{bmatrix} 0 & 1 \\  1 & 0 \end{bmatrix}) \\
	&= \frac{N}{N-1}
	\begin{bmatrix}
		1&     - N^{-\frac{1}{2} } \\ 
		- N^{-\frac{1}{2} }   & 1
	\end{bmatrix}
\end{align}
as
\begin{align} \nonumber
	&\sum_{i=0}^{\infty} N^{-i} = \frac{1}{1-\frac{1}{N}} = \frac{N}{N-1}  , \\
	&\sum_{i=0}^{\infty}N^{-(i+\frac{1}{2})} = N^{-\frac{1}{2}} \sum_{i=0}^{\infty}N^{-i} = N^{-\frac{1}{2}} \frac{1}{1-\frac{1}{N}} = \frac{N^{\frac{1}{2}}}{N-1}.
\end{align}

As a further example we calculate the inverse metric for $P_2(N)$, which is spanned by 15 diagrams. We are interested in the off-diagonal elements of our metric and so we have $\frac{15(15-1)}{2} = 105$ independent elements. Using Sage we find the metric is given by
\begin{align} \nonumber
	\hat{g}_{P_2(N)} &=  \mathds{1} + X_{P_2(N)} = \\
	&\begin{tiny}
		\begin{bmatrix}
			1&N^{-\frac{1}{2}}&N^{-\frac{1}{2}}&N^{-\frac{1}{2}}&N^{-\frac{1}{2}}&N^{-\frac{1}{2}}&N^{-\frac{1}{2}}&N^{-\frac{1}{2}}&N^{-1}&N^{-1}&N^{-1}&N^{-1}&N^{-1}&N^{-1}&N^{-\frac{3}{2}} \\
			N^{-\frac{1}{2}}&1&N^{-1}&N^{-1}&N^{-1}&N^{-1}&N^{-1}&N^{-1}&N^{-\frac{1}{2}}&N^{-\frac{1}{2}}&N^{-\frac{1}{2}}&N^{-\frac{3}{2}}&N^{-\frac{3}{2}}&N^{-\frac{3}{2}}&N^{-1} \\
			N^{-\frac{1}{2}}&N^{-1}&1&N^{-1}&N^{-1}&N^{-1}&N^{-1}&N^{-1}&N^{-\frac{1}{2}}&N^{-\frac{3}{2}}&N^{-\frac{3}{2}}&N^{-\frac{1}{2}}&N^{-\frac{1}{2}}&N^{-\frac{3}{2}}&N^{-1} \\
			N^{-\frac{1}{2}}&N^{-1}&N^{-1}&1&N^{-1}&N^{-1}&N^{-1}&N^{-1}&N^{-\frac{1}{2}}&N^{-\frac{3}{2}}&N^{-\frac{3}{2}}&N^{-\frac{3}{2}}&N^{-\frac{3}{2}}&N^{-\frac{1}{2}}&N^{-1} \\
			N^{-\frac{1}{2}}&N^{-1}&N^{-1}&N^{-1}&1&N^{-1}&N^{-1}&N^{-1}&N^{-\frac{3}{2}}&N^{-\frac{1}{2}}&N^{-\frac{3}{2}}&N^{-\frac{1}{2}}&N^{-\frac{3}{2}}&N^{-\frac{1}{2}}&N^{-1} \\
			N^{-\frac{1}{2}}&N^{-1}&N^{-1}&N^{-1}&N^{-1}&1&N^{-1}&N^{-1}&N^{-\frac{3}{2}}&N^{-\frac{1}{2}}&N^{-\frac{3}{2}}&N^{-\frac{3}{2}}&N^{-\frac{1}{2}}&N^{-\frac{3}{2}}&N^{-1} \\
			N^{-\frac{1}{2}}&N^{-1}&N^{-1}&N^{-1}&N^{-1}&N^{-1}&1&N^{-1}&N^{-\frac{3}{2}}&N^{-\frac{3}{2}}&N^{-\frac{1}{2}}&N^{-\frac{1}{2}}&N^{-\frac{3}{2}}&N^{-\frac{3}{2}}&N^{-1} \\
			N^{-\frac{1}{2}}&N^{-1}&N^{-1}&N^{-1}&N^{-1}&N^{-1}&N^{-1}&1&N^{-\frac{3}{2}}&N^{-\frac{3}{2}}&N^{-\frac{1}{2}}&N^{-\frac{3}{2}}&N^{-\frac{1}{2}}&N^{-\frac{1}{2}}&N^{-1} \\
			N^{-1}&N^{-\frac{1}{2}}&N^{-\frac{1}{2}}&N^{-\frac{1}{2}}&N^{-\frac{3}{2}}&N^{-\frac{3}{2}}&N^{-\frac{3}{2}}&N^{-\frac{3}{2}}&1&N^{-1}&N^{-1}&N^{-1}&N^{-1}&N^{-1}&N^{-\frac{1}{2}} \\
			N^{-1}&N^{-\frac{1}{2}}&N^{-\frac{3}{2}}&N^{-\frac{3}{2}}&N^{-\frac{1}{2}}&N^{-\frac{1}{2}}&N^{-\frac{3}{2}}&N^{-\frac{3}{2}}&N^{-1}&1&N^{-1}&N^{-1}&N^{-1}&N^{-1}&N^{-\frac{1}{2}} \\
			N^{-1}&N^{-\frac{1}{2}}&N^{-\frac{3}{2}}&N^{-\frac{3}{2}}&N^{-\frac{3}{2}}&N^{-\frac{3}{2}}&N^{-\frac{1}{2}}&N^{-\frac{1}{2}}&N^{-1}&N^{-1}&1&N^{-1}&N^{-1}&N^{-1}&N^{-\frac{1}{2}} \\
			N^{-1}&N^{-\frac{3}{2}}&N^{-\frac{1}{2}}&N^{-\frac{3}{2}}&N^{-\frac{1}{2}}&N^{-\frac{3}{2}}&N^{-\frac{1}{2}}&N^{-\frac{3}{2}}&N^{-1}&N^{-1}&N^{-1}&1&N^{-1}&N^{-1}&N^{-\frac{1}{2}} \\
			N^{-1}&N^{-\frac{3}{2}}&N^{-\frac{1}{2}}&N^{-\frac{3}{2}}&N^{-\frac{3}{2}}&N^{-\frac{1}{2}}&N^{-\frac{3}{2}}&N^{-\frac{1}{2}}&N^{-1}&N^{-1}&N^{-1}&N^{-1}&1&N^{-1}&N^{-\frac{1}{2}} \\
			N^{-1}&N^{-\frac{3}{2}}&N^{-\frac{3}{2}}&N^{-\frac{1}{2}}&N^{-\frac{1}{2}}&N^{-\frac{3}{2}}&N^{-\frac{3}{2}}&N^{-\frac{1}{2}}&N^{-1}&N^{-1}&N^{-1}&N^{-1}&N^{-1}&1&N^{-\frac{1}{2}} \\
			N^{-\frac{3}{2}}&N^{-1}&N^{-1}&N^{-1}&N^{-1}&N^{-1}&N^{-1}&N^{-1}&N^{-\frac{1}{2}}&N^{-\frac{1}{2}}&N^{-\frac{1}{2}}&N^{-\frac{1}{2}}&N^{-\frac{1}{2}}&N^{-\frac{1}{2}}&1
		\end{bmatrix}
	\end{tiny}
\end{align}
Inverting this metric directly gives
\begin{align} \nonumber
	\hat{g}^{-1} &= \frac{N}{(N-1)(N-2)(N-3)} && \\
	&\begin{tiny}
		\scalemath{0.6}{
			\begin{bmatrix}
				N^2+N &((-N)-1) N^{\frac{1}{2}} &((-N)-1) N^{\frac{1}{2}} &(1-N) N^{\frac{1}{2}} &((-N)-1) N^{\frac{1}{2}} &(1-N) N^{\frac{1}{2}} &(1-N) N^{\frac{1}{2}} &((-N)-1) N^{\frac{1}{2}} &2 N &2 N &2 N &2 N &2 N &2 N &-6 N^{\frac{1}{2}} \\
				((-N)-1) N^{\frac{1}{2}} &(N-1)^2 &N+1 &N-1 &N+1 &N-1 &N-1 &N+1 &(1-N) N^{\frac{1}{2}} &(1-N) N^{\frac{1}{2}} &(1-N) N^{\frac{1}{2}} &-2 N^{\frac{1}{2}} &-2 N^{\frac{1}{2}} &-2 N^{\frac{1}{2}} &2 N \\
				((-N)-1) N^{\frac{1}{2}} &N+1 &(N-1)^2 &N-1 &N+1 &N-1 &N-1 &N+1 &(1-N) N^{\frac{1}{2}} &-2 N^{\frac{1}{2}} &-2 N^{\frac{1}{2}} &(1-N) N^{\frac{1}{2}} &(1-N) N^{\frac{1}{2}} &-2 N^{\frac{1}{2}} &2 N \\
				(1-N) N^{\frac{1}{2}} &N-1 &N-1 &N^2-3 N+1 &N-1 &1 &1 &N-1 &(2-N) N^{\frac{1}{2}} &-N^{\frac{1}{2}} &-N^{\frac{1}{2}} &-N^{\frac{1}{2}} &-N^{\frac{1}{2}} &(2-N) N^{\frac{1}{2}} &N \\
				((-N)-1) N^{\frac{1}{2}} &N+1 &N+1 &N-1 &(N-1)^2 &N-1 &N-1 &N+1 &-2 N^{\frac{1}{2}} &(1-N) N^{\frac{1}{2}} &-2 N^{\frac{1}{2}} &(1-N) N^{\frac{1}{2}} &-2 N^{\frac{1}{2}} &(1-N) N^{\frac{1}{2}} &2 N \\ 
				(1-N) N^{\frac{1}{2}} &N-1 &N-1 &1 &N-1 &N^2-3 N+1 &1 &N-1 &-N^{\frac{1}{2}} &(2-N) N^{\frac{1}{2}} &-N^{\frac{1}{2}} &-N^{\frac{1}{2}} &(2-N) N^{\frac{1}{2}} &-N^{\frac{1}{2}} &N \\ 
				(1-N) N^{\frac{1}{2}} &N-1 &N-1 &1 &N-1 &1 &N^2-3 N+1 &N-1 &-N^{\frac{1}{2}} &-N^{\frac{1}{2}} &(2-N) N^{\frac{1}{2}} &(2-N) N^{\frac{1}{2}} &-N^{\frac{1}{2}} &-N^{\frac{1}{2}} &N \\
				((-N)-1) N^{\frac{1}{2}} &N+1 &N+1 &N-1 &N+1 &N-1 &N-1 &(N-1)^2 &-2 N^{\frac{1}{2}} &-2 N^{\frac{1}{2}} &(1-N) N^{\frac{1}{2}} &-2 N^{\frac{1}{2}} &(1-N) N^{\frac{1}{2}} &(1-N) N^{\frac{1}{2}} &2 N \\
				2 N &(1-N) N^{\frac{1}{2}} &(1-N) N^{\frac{1}{2}} &(2-N) N^{\frac{1}{2}} &-2 N^{\frac{1}{2}} &-N^{\frac{1}{2}} &-N^{\frac{1}{2}} &-2 N^{\frac{1}{2}} &N^2-2 N &N &N &N &N &N &- N^{\frac{3}{2}} \\
				2 N &(1-N) N^{\frac{1}{2}} &-2 N^{\frac{1}{2}} &-N^{\frac{1}{2}} &(1-N) N^{\frac{1}{2}} &(2-N) N^{\frac{1}{2}} &-N^{\frac{1}{2}} &-2 N^{\frac{1}{2}} &N &N^2-2 N &N &N &N &N &- N^{\frac{3}{2}} \\
				2 N &(1-N) N^{\frac{1}{2}} &-2 N^{\frac{1}{2}} &-N^{\frac{1}{2}} &-2 N^{\frac{1}{2}} &-N^{\frac{1}{2}} &(2-N) N^{\frac{1}{2}} &(1-N) N^{\frac{1}{2}} &N &N &N^2-2 N &N &N &N &- N^{\frac{3}{2}} \\
				2 N &-2 N^{\frac{1}{2}} &(1-N) N^{\frac{1}{2}} &-N^{\frac{1}{2}} &(1-N) N^{\frac{1}{2}} &-N^{\frac{1}{2}} &(2-N) N^{\frac{1}{2}} &-2 N^{\frac{1}{2}} &N &N &N &N^2-2 N &N &N &- N^{\frac{3}{2}} \\
				2 N &-2 N^{\frac{1}{2}} &(1-N) N^{\frac{1}{2}} &-N^{\frac{1}{2}} &-2 N^{\frac{1}{2}} &(2-N) N^{\frac{1}{2}} &-N^{\frac{1}{2}} &(1-N) N^{\frac{1}{2}} &N &N &N &N &N^2-2 N &N &- N^{\frac{3}{2}} \\
				2 N &-2 N^{\frac{1}{2}} &-2 N^{\frac{1}{2}} &(2-N) N^{\frac{1}{2}} &(1-N) N^{\frac{1}{2}} &-N^{\frac{1}{2}} &-N^{\frac{1}{2}} &(1-N) N^{\frac{1}{2}} &N &N &N &N &N &N^2-2 N &- N^{\frac{3}{2}} \\
				-6 N^{\frac{1}{2}} &2 N &2 N &N &2 N &N &N &2 N &- N^{\frac{3}{2}} &- N^{\frac{3}{2}} &- N^{\frac{3}{2}} &- N^{\frac{3}{2}} &- N^{\frac{3}{2}} &- N^{\frac{3}{2}}&N^2
			\end{bmatrix}
		}
	\end{tiny} &&
\end{align}

\subsection{First and second order corrections}
The leading corrections to this are of order $N^{-\frac{1}{2}}$ and occur precisely when $d_2$ is obtained from $d_1$ by removing a single edge (or vice-versa). We recognise this as a method of constructing the Hasse diagram for $P_k(N)$. Accordingly the leading order corrections to the metric are given precisely by the elements $d_1, d_2 \in P_k(N)$ that share a connection in the relevant Hasse diagram.
For example, the following is the Hasse diagram for $k=2$
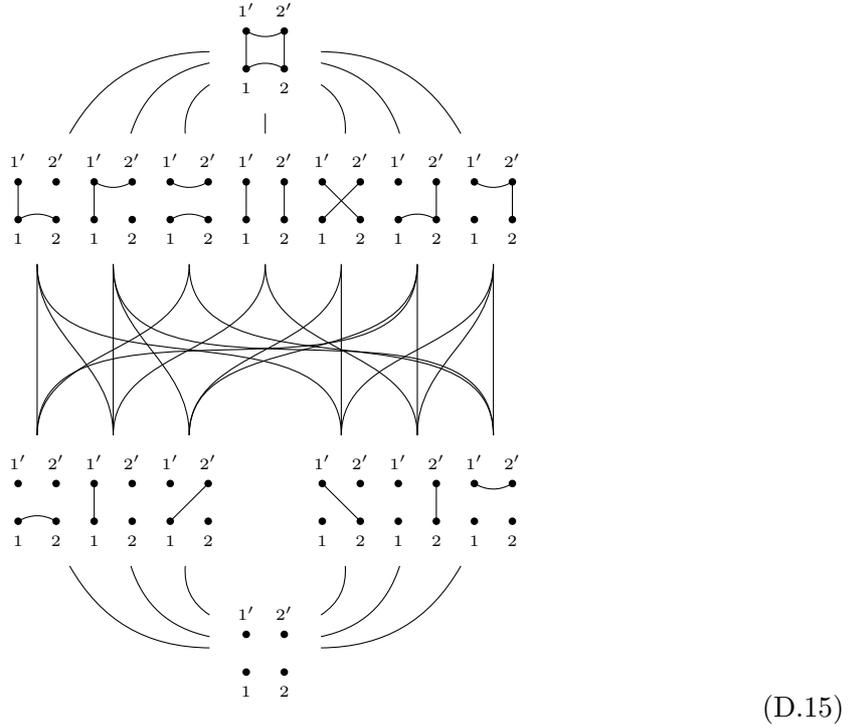
\begin{equation}
	\begin{tikzpicture}[scale=2]
		\node (min) at (0,0) {$\PAdiagramLabeled[]{2}{}$};
		\node (l1a) at (-1.5,1) {$\PAdiagramLabeled[-1/-2]{2}{}$};
		\node (l1b) at (-1,1) {$\PAdiagramLabeled[]{2}{-1/1}$};
		\node (l1c) at (-.5,1) {$\PAdiagramLabeled[]{2}{-1/2}$};
		\node (l1d) at (.5,1) {$\PAdiagramLabeled[]{2}{-2/1}$};	
		\node (l1e) at (1,1) {$\PAdiagramLabeled[]{2}{-2/2}$};
		\node (l1f) at (1.5,1) {$\PAdiagramLabeled[2/1]{2}{}$};
		\node (l2a) at (-1.5,3) {$\PAdiagramLabeled[-1/-2]{2}{-1/1}$};
		\node (l2b) at (-1,3) {$\PAdiagramLabeled[2/1]{2}{-1/1}$};
		\node (l2c) at (-.5,3) {$\PAdiagramLabeled[-1/-2, 2/1]{2}{}$};
		\node (l2d) at (0,3) {$\PAdiagramLabeled[]{2}{-1/1,-2/2}$};
		\node (l2e) at (.5,3) {$\PAdiagramLabeled[]{2}{-1/2,-2/1}$};
		\node (l2f) at (1,3) {$\PAdiagramLabeled[-1/-2]{2}{2/-2}$};
		\node (l2g) at (1.5,3) {$\PAdiagramLabeled[2/1]{2}{-2/2}$};
		\node (max) at (0,4) {$\PAdiagramLabeled[2/1,-1/-2]{2}{-1/1,-2/2}$};
		\draw (min) to[bend left] (l1a); \draw (min) to[bend left] (l1b); \draw (min) to[bend left] (l1c); \draw (min) to[bend right] (l1d); \draw (min) to[bend right] (l1e); \draw (min) to[bend right] (l1f);
		\draw (max) to[bend right] (l2a); \draw (max) to[bend right] (l2b); \draw (max) to[bend right] (l2c); \draw (max) -- (l2d); \draw (max) to[bend left] (l2e); \draw (max) to[bend left] (l2f); \draw (max) to[bend left] (l2g);
		\draw (l1a) to[out=90, in=-90]  (l2a); \draw (l1a) to[out=90, in=-90]  (l2c); \draw (l1a) to[out=90, in=-90] (l2f);
		\draw (l1b) to[out=90, in=-90]  (l2a); \draw (l1b) to[out=90, in=-90]  (l2b); \draw (l1b) to[out=90, in=-90] (l2d);
		\draw (l1c) to[out=90, in=-90]  (l2b); \draw (l1c) to[out=90, in=-90]  (l2e); \draw (l1c) to[out=90, in=-90] (l2f);
		\draw (l1d) to[out=90, in=-90]  (l2a); \draw (l1d) to[out=90, in=-90]  (l2e); \draw (l1d) to[out=90, in=-90] (l2g);
		\draw (l1e) to[out=90, in=-90]  (l2f); \draw (l1e) to[out=90, in=-90]  (l2g); \draw (l1e) to[out=90, in=-90] (l2d);
		\draw (l1f) to[out=90, in=-90]  (l2c); \draw (l1f) to[out=90, in=-90]  (l2b); \draw (l1f) to[out=90, in=-90] (l2g);
	\end{tikzpicture}
\end{equation}
Indeed, every connection in this diagram corresponds to an $N^{\frac{1}{2}}$ element in $\hat{g}_{P_2(N)}$ and all $N^{\frac{1}{2}}$ elements of $\hat{g}_{P_2(N)}$ are given by a connection in the diagram. We call each row in the Hasse diagram a level $L_i$ and index them by $i$ - the number of connected components in the partition diagrams on that level. For example $\PAdiagram[2/1, -1/-2]{2}{-1/1,-2/2} \in L_1$ , $\PAdiagram[-1/-2]{2}{-1/1} \in L_2$, $\PAdiagram[-1/-2]{2}{} \in L_3$ and $\PAdiagram{2}{} \in L_4$.

Ordering our basis according to the levels in the Hasse diagram we see that the metric has block diagonal contributions from within any given level of the Hasse diagram. As the leading order corrections are generated by $d_1$ and $d_2$ in different levels these occur outside of the diagonal blocks. Everything we have said here about the metric applies equally well to the inverse metric, as to first order this is given by
\begin{align}
	\hat{g}^{-1} \thicksim \mathds{1} - X
\end{align}

The $N^{-1}$ corrections to the metric are again easily described with reference to the Hasse diagram. There are two ways in which we can get $N^{-1}$ contributions:
\begin{enumerate}
	\item{For any $d_1, d_2 \in L_i$ if $d_1 \merge d_2 \in L_{i-1}$ then $\hat{g}_{d_1 , d_2} = N^{-1}$.}
	\item{For and $d_1 \in L_i$, $d_2 \in L_{i-2}$ if $d_1 < d_2$, that is if $d_1$ is contained within $d_2$, then $\hat{g}_{d_1 , d_2} = N^{-1}$. If $d_1$ and $d_2$ are incomparable then their inner product will be a larger negative power of $N$ as this incomparability will only reduce the number of connected components of the merge of $d_1$ and $d_2$. }
\end{enumerate}
More generally for $d_1 \in L_i$, $d_2 \in L_{i-2n}$ for $n \in \mathbb{Z}^+$ and $d_1 < d_2$ we have
\begin{equation}
	\hat{g}_{d_1 , d_2} = N^{ c(d_1 \merge d_2) - \frac{1}{2} \qty(c(d_1) + c(d_2))} = N^{ (i-2n) - \frac{1}{2} \qty(i + i-2n)} = N^{-n}.
\end{equation}

\printbibliography
\end{document}